\documentclass[12pt,a4paper]{article}
\usepackage{amssymb}

\usepackage{graphicx}
\usepackage{amsmath}

\begin{document}

\title{The asymptotic quasi-stationary states of the two-dimensional magnetically
confined plasma and of the planetary atmosphere}
\author{F. Spineanu and M. Vlad \\
%EndAName
Association EURATOM-MEdC Romania,\\
NILPRP, MG-36, Magurele, Bucharest, Romania}
\maketitle

\begin{abstract}
We derive the differential equation governing the asymptotic
quasi-stationary states of the two dimensional plasma immersed in a strong
confining magnetic field and of the planetary atmosphere. These two systems
are related by the property that there is an intrinsic constant length: the
Larmor radius and respectively the Rossby radius and a condensate of the
vorticity field in the unperturbed state related to the cyclotronic gyration
and respectively to the Coriolis frequency. Although the closest physical
model is the Charney-Hasegawa-Mima (CHM) equation, our model is more general
and is related to the system consisting of a discrete set of point-like
vortices interacting in plane by a short range potential. A
field-theoretical formalism is developed for describing the continuous
version of this system. The action functional can be written in the
Bogomolnyi form (emphasizing the role of Self-Duality of the asymptotic
states) but the minimum energy is no more topological and the asymptotic
structures appear to be non-stationary, which is a major difference with
respect to traditional topological vortex solutions. Versions of this field
theory are discussed and \ we find arguments in favor of a particular form
of the equation. We comment upon the significant difference between the CHM
fluid/plasma and the Euler fluid and respectively the Abelian-Higgs vortex
models.
\end{abstract}

\tableofcontents

\section{Introduction}

The instabilities of a plasma embedded in a confining magnetic field (\emph{%
e.g.} experimental fusion devices, like tokamak) evolve in a geometry that
is strongly anisotropic. The motion of the electrons along the magnetic
field lines is in many cases sufficiently fast to produce a density
perturbation that has the Boltzmann distribution in the electric potential.
In these cases a two dimensional approximation (the dynamical equations are
written in the plane that is transversal to the magnetic field) may be
satisfactory. The Boltzmann electron distribution suppresses the convective
nonlinearity, leaving the ion polarization drift as the essential
nonlinearity. This is of high differential degree (therefore it is enhanced
at small spatial scales) and basically describes the advection of the
fluctuating vorticity by the velocity fluctuations. The differential
equation for the electrostatic potential has been derived by Hasegawa and
Mima \cite{HM}. A similar situation appears in the physics of the atmosphere
and the differential equation for the streamfunction of the velocity field
has been derived by Charney \cite{Charney}.

Many plasma instabilities (in particular in tokamak) depend on this
nonlinear term and will likely show some common aspects. In numerical
simulations of the Charney-Hasegawa-Mima (CHM) equation it has been proved
that the plasma is evolving at large times (in the absence of external
driving forces and starting from an irregular flow pattern) to states that
are characterised by a regular form of the potential. The evolution toward a
very regular pattern of vortical flow is also characteristic to the
incompressible ideal fluid, described by the Euler equation \cite
{Montgomery1}, \cite{Montgomery2}, \cite{KraichnanMontgomery}, \cite{Montg2}%
, \cite{Montg3}, \cite{Joyce}, \cite{Smith}. In this case the
time-asymptotic states consists of few vortices, of regular shape, with very
slow motion. The streamfunction obeys, in these states, the \emph{sinh}%
-Poisson equation. In the case of CHM equation, the cuasi-stationary states
also consist of structures but it has not been possible to derive an
equation for the streamfunction \cite{Seyler}. There are seveal studies of
the CHM (or very similar) equations showing that at large time the flow is
strongly organized and dominated by structures \cite{HTK}, \cite{KMcWT}, 
\cite{KTMcWP}, \cite{HH} (and references therein). At the oposite limit the
turbulent regime can be treated with renormalization group methods \cite
{DiamondKim}.

The formulation of the problem exclusively in terms of
experimentally-accessible quantities, velocity and vorticity seems to not
allow too much freedom in elaborating a theoretical model from which the
stationary states to be determined. In all our considerations we will be
guided by the analogous experience in the case of the Euler fluid. In that
case the existence of a parallel formulation (although the mathematical
equivalence is still not fully proven) has been decissive. That model
consists of the discrete set of point-like vortices evolving in plane due to
a potential given as the natural logarithm of the relative distance.
Comparing with the differential equation of the Euler fluid this alternative
formulation provides something fundamentally new: it splits the dynamics
into two objects of distinct nature: point-like vortices and potential, or,
in other words, matter and field carrying the interaction. Going along this
model we are led to consider the standard treatment in these cases. We must
assume that each of these objects can evolve freely (\emph{i.e.} independent
of the other) and in addition there is an interaction between them. This
formulation is standard in electrodynamics (classical or quantum). Going to
continuum, the discrete set of point-like vortices becomes a field of
``matter'' (which must be assumed in general complex), the potential will be
a free field (called ``gauge'', similar with the free electromagnetic field)
and, in addition, there is the interaction between the matter and the gauge
fields (similar to the classical $j_{\mu }A_{\mu }$ charge-field
interaction). Basic properties of the original, \emph{i.e.} physical, model
impose constraints to this field-theoretical formulation. The presence of
vorticity requires a particular form of the potential between discrete
vortices: it is the \emph{curl} of a sum of natural logarithms. This can
only be derived from a gauge-field Lagrangean density of Chern-Simons type,
instead of Maxwell type. Therefore the field-theoretical model will contain
the Chern-Simons Lagrangean. The matter field has a nonlinear
self-interaction that reflects the stationary structure of the free matter
field. The coupling of the two fields (the interaction) is minimal, via the
covariant derivatives. In a previous work we have formulated this model and
have obtained, in this way, a purely analytic derivation of the \emph{sinh}%
-Poisson equation for the Euler fluid \cite{FlorinMadi1}.

This paper will develop a model for the Charney-Hasegawa-Mima equation along
the same lines. However several differences will impose new features of the
model.

\bigskip

In a previous work \cite{Toki2003} we have investigated a model based on the
similarity with the superfluid field theory, the Abelian-Higgs model. This
is able to describe \emph{positive} fluid vortices. We have also proposed,
without details, a more extended model, which seemed able to describe the
physical vortices of plasma and atmosphere within the regime of the CHM
equation.

The full development of this model is the main objective of the present work.

\section{The physical problem and the Charney-Hasegawa-Mima equation}

The analytical model we develop in the present work is intended to describe
a system characterized by the following elements:

\begin{enumerate}
\item  the existence of an intrinsic length; this is the Larmor radius $\rho
_{s}$ for the two-dimensional plasma immersed in a strong, confining,
transversal magnetic field; and the Rossby radius for the two-dimensional
quasi-geostrophical approximation of the planetary atmosphere;

\item  the existence of a condensate of vorticity as a background in the
system, in the absence of any perturbation. This background consists of the
cyclotronic gyration of ions, for the plasma; and of the Coriolis effect
resulting from the planetary rotation, for the atmosphere.
\end{enumerate}

These basic elements are very general and other systems belong to the class
that is defined by them. In particular the non-neutral plasma produced in
laboratory experiments and the vortices produced on a fluid in a rotating
tank (see  Schecter, Nezlin, Hopfinger and Van Heijst, etc.)

\bigskip

To make the discussion more specific we will refer in the following to the
physical model developed for the two-dimensional plasma and atmosphere,
which leads to the equation of Charney-Hasegawa-Mima (CHM): 
\begin{equation}
\left( 1-\mathbf{\nabla }_{\perp }^{2}\right) \frac{\partial \phi }{\partial
t}-\kappa \frac{\partial \phi }{\partial y}-\left[ \left( -\mathbf{\nabla }%
_{\perp }\phi \times \widehat{\mathbf{n}}\right) \cdot \mathbf{\nabla }%
_{\perp }\right] \mathbf{\nabla }_{\perp }^{2}\phi =0  \label{cce}
\end{equation}
where $\kappa \widehat{\mathbf{e}}_{y}=-\widehat{\mathbf{n}}\times \mathbf{%
\nabla }_{\perp }\ln n_{0}$. For simplicity we will refer to the problem of
plasma physics, the adaptation to the problem in the physics of atmosphere
being easily done (see \cite{HH}). The quantities appearing in the Eq.(\ref
{cce}) are the \emph{physical} ones after having been normalized : 
\begin{eqnarray}
\phi &=&\frac{\left| e\right| \phi ^{phys}}{T_{e}}  \label{scal} \\
\left( x,y\right) &=&\left( x^{phys}/\rho _{s},y^{phys}/\rho _{s}\right) 
\notag \\
t &=&t^{phys}\Omega _{ci}  \notag
\end{eqnarray}
where $\Omega _{ci}=\left| e\right| B/m_{i}$, $\rho _{s}=c_{s}/\Omega _{ci}$%
, $c_{s}^{2}=T_{e}/m_{i}$. The derivation of the equation, in the drift
instability in tokamaks is done in the Appendix A.

For comparison, the Euler equation is 
\begin{eqnarray}
\frac{d\mathbf{\omega }}{dt} &=&0\;\text{or}\;  \label{eulereq} \\
\frac{\partial }{\partial t}\left( \mathbf{\nabla }_{\perp }^{2}\phi \right)
+\left[ \left( -\mathbf{\nabla }_{\perp }\phi \times \widehat{\mathbf{n}}%
\right) \cdot \mathbf{\nabla }_{\perp }\right] \mathbf{\nabla }_{\perp
}^{2}\phi &=&0  \notag
\end{eqnarray}
where $\phi $ is the \emph{streamfunction}.

The similarity between the two equation, Eq.(\ref{cce}) and Eq.(\ref{eulereq}%
) is apparent. However, with regard to the form and properties of the
stationary states of the two equations we should not attempt of simply
taking any previous conclusion derived from the Euler fluid context into the
CHM context. This is because the naive stationarity $\left( \partial
/\partial t=0\right) $ imposed in the two equations leads to an equation
with a vast degree of generality and does not provide, by itself, a clear
identification of the final vortex shapes. We may represent the family of
all possible solutions of the naive stationary limit as a subset in a
function space. In the time evolution the two equations produces two series
of configurations representing functions belonging to two distinct paths,
ending in this set at distinct points (\emph{i.e.} configurations).

Compared with the equation for the ideal fluid (Euler equation) the CHM eq.
is not scale invariant \cite{horton1}. To see this we make the rescaling of
the space variables 
\begin{equation*}
\left( x,y\right) \rightarrow \left( x^{\prime },y^{\prime }\right) =\left(
\lambda x,\lambda y\right)
\end{equation*}
Expressing the Euler equation in the new variables, we have 
\begin{equation*}
\frac{\partial }{\partial t}\left( \mathbf{\nabla }_{\perp }^{\prime 2}\phi
\right) +\lambda ^{2}\left[ \left( -\mathbf{\nabla }_{\perp }^{\prime }\phi
\times \widehat{\mathbf{n}}\right) \cdot \mathbf{\nabla }_{\perp }^{\prime }%
\right] \mathbf{\nabla }_{\perp }^{\prime 2}\phi =0
\end{equation*}
The factor $\lambda ^{2}$ can be absorbed in a rescaling of the time
variable and the equation preserves its form. By contrast the equation CHM
becomes (for simplicity we take $\kappa =0$) 
\begin{equation*}
\left( 1-\lambda ^{2}\mathbf{\nabla }_{\perp }^{\prime 2}\right) \frac{%
\partial \phi }{\partial t}-\lambda ^{4}\left[ \left( -\mathbf{\nabla }%
_{\perp }^{\prime }\phi \times \widehat{\mathbf{n}}\right) \cdot \mathbf{%
\nabla }_{\perp }^{\prime }\right] \mathbf{\nabla }_{\perp }^{\prime 2}\phi
=0
\end{equation*}
While the factor $\lambda ^{4}$ can be absorbed by time rescaling, the
factor $\lambda ^{2}$ in the first paranthesis cannot be absorbed. The form
of the equation is invariant only for $\lambda =1$ which means the space is
measured in units of $\rho _{s}$. The equation CHM exhibits an intrinsic
spatial scale, which is $\rho _{s}$.

\section{An equivalent discrete model}

In the case of the Euler fluid, there is an equivalent model whose dynamical
evolution is considered to be identical with that of the physical system
(for a list of references, see \cite{FlorinMadi1}). It consists of a
collection of discrete point-like vortices in two-dimensions evolving from
mutual interaction defined in terms of a potential. The potential is the
natural logarithm of the relative distance of two vortices. This model has
been proposed and used long ago (Kirchoff, Onsager, etc.) but the rigurous
proof of the equivalence between it and the physical Euler description is a
difficult mathematical problem \cite{math1}.

For the CHM equation there is a similar model: a collection of point-like
vortices interacting by a potential that has a short range. This model has
been proposed in meteorology by Morikawa \cite{Morikawa} and Stewart \cite
{Stewart} (see Horton and Hasegawa \cite{HH}). For a set of $N$ vortices
with strength $\omega _{j}$, $j=1,N$ with instantaneous positions $\mathbf{r}%
_{j}$ the streamfunction $\psi \left( x,y\right) $ has the following
expression 
\begin{equation}
\psi \left( \mathbf{r}\right) =\sum_{j}\psi _{j}\left( \mathbf{r}\right)
=\sum_{j}\omega _{j}K_{0}\left( m\left| \mathbf{r-r}_{j}\right| \right)
\label{phik0}
\end{equation}
where $m$ is a constant. The differential equation from which the
contributions to the streamfunction $\psi $ in Eq.(\ref{phik0}) are derived,
is 
\begin{equation}
\left( \Delta -m^{2}\right) \psi _{j}\left( \mathbf{r}\right) =-2\pi \omega
_{j}\delta \left( \mathbf{r-r}_{j}\right)  \label{kgreen}
\end{equation}
in two dimensions. For a single vortex of strength $\omega $ placed in the
origin, the azimuthal component of the velocity can be derived from the
streamfunction $\psi \left( \mathbf{r}\right) $%
\begin{equation}
v_{\theta }=\frac{\partial \psi }{\partial r}=-\omega mK_{1}\left( mr\right)
\label{vtheta}
\end{equation}
At small distances 
\begin{equation}
v_{\theta }\sim \frac{1}{r}\;\text{for}\;r\rightarrow 0  \label{vth1}
\end{equation}
which is the same as for the Euler case, where $\psi \left( \mathbf{r}%
\right) $ is given by the natural logarithm. The streamfunction decays fast
at large $r$ since the modified Bessel function of the second kind $K_{0}$
decays exponentially at large argument 
\begin{equation}
\psi \sim \frac{1}{\sqrt{mr}}\exp \left( -mr\right) \;\text{for}%
\;r\rightarrow \infty  \label{vth2}
\end{equation}
This means that the vortices are shielded. The elementary vortex of the
Charney-Hasegawa-Mima equation is localised by $m^{-1}$ and one can
associate a finite spatial extension to it, $\rho _{s}$ in plasma physics, $%
\rho _{g}$ in the physics of atmosphere \cite{HH}.

The equations of motion for the vortex $\omega _{k}$ at $\left(
x_{k},y_{k}\right) $ under the effect of the others are \cite{Morikawa} 
\begin{eqnarray}
-2\pi \omega _{k}\frac{dx_{k}}{dt} &=&\frac{\partial W}{\partial y_{k}}
\label{eqmot} \\
-2\pi \omega _{k}\frac{dy_{k}}{dt} &=&-\frac{\partial W}{\partial x_{k}} 
\notag
\end{eqnarray}
where 
\begin{equation}
W=\pi \underset{i\neq j}{\sum_{i=1}^{N}\sum_{j=1}^{N}}\omega _{i}\omega
_{j}K_{0}\left( m\left| \mathbf{r}_{i}-\mathbf{r}_{j}\right| \right)
\label{w}
\end{equation}
This is the Kirchhoff function for the system of interacting point-like
vortices in plane. It is the Hamiltonian for the system of $N$ vortices. If
we introduce the versor of the normal to the plane, $\widehat{\mathbf{n}}$,
the equations can be expressed 
\begin{equation}
\frac{d\mathbf{r}}{dt}=-\mathbf{\nabla }\psi \times \widehat{\mathbf{n}}
\label{10}
\end{equation}

We will develop a formalism for this discrete system which in the continuum
becomes a field theory.

\section{The main components and the main steps of construction of the
continuum model}

We will develop a continuum model whose equations of motion could reproduce,
in the discrete approximation, the Eqs.(\ref{phik0}), (\ref{eqmot}) and the
energy (\ref{w}). The model must be defined in terms of a Lagrangean density
for two interacting fields:

\begin{itemize}
\item  the field associated with the density of point-like vortices $\phi
\left( x,y\right) $; we will call it the \emph{matter} (or \emph{scalar}, or 
\emph{Higgs}) field; and

\item  the field associated with the potential carrying the interaction
between the vortices; we will call it the \emph{gauge} field.
\end{itemize}

\subsection{The gauge field}

We note that the interaction potential $K_{0}\left( \left| \mathbf{r-r}%
_{j}\right| \right) $ (assume that $m$ is normalised as $m=1$) appearing in
the discrete model proposed in meteorology is similar to the potential
appearing in the Euler problem, $\ln \left( \left| \mathbf{r-r}_{j}\right|
\right) $ in the following sense: they both have topological properties, in
sense to be explained. For the Euler equation, the potential can be
represented using the angle made by the line connecting the reference ($%
\mathbf{r}_{j}$) and the current ($\mathbf{r}$) points with a fixed line,
and in order to remove the multivaluedness one has to make a cut in the
plane from the center \ (where is singular) to infinity \cite{JackiwPi}. 
\begin{eqnarray}
\text{Euler fluid} &:&\text{{}}\;\frac{dr_{i}^{\alpha }}{dt}=\left( -\mathbf{%
\nabla }\phi \times \widehat{\mathbf{n}}\right) ^{\alpha }  \label{12} \\
&=&\varepsilon ^{\alpha \beta }\sum_{j\neq i}^{N}\omega _{j}\frac{r^{\beta
}-r_{j}^{\beta }}{\left| \mathbf{r}-\mathbf{r}_{j}\right| ^{2}}  \notag
\end{eqnarray}
(Here $i$, $j$ label the point-like vortices and $\alpha $ , $\beta $ label
the coordinates of the position vectors, $r_{i}^{\alpha }$, $\alpha =1,2$).
Since 
\begin{equation}
\varepsilon ^{\alpha \beta }\frac{r^{\beta }}{r^{2}}=\varepsilon ^{\alpha
\beta }\partial _{\beta }\ln r  \label{13}
\end{equation}
we see that the potential in Eq.(\ref{12}) is expressed through the Green
function of the $2D$ Laplacian, defined by the equation 
\begin{equation}
\mathbf{\nabla }^{2}\ln r=2\pi \delta ^{2}\left( r\right)  \label{14}
\end{equation}
The potential is obtained by applying the rotational operator $\varepsilon
^{\alpha \beta }\partial _{\beta }$ on this Green function.

The CHM case is similar, with the difference that the $K_{0}\left( mr\right) 
$ is the Green function of the Helmholtz operator, as results from Eq.(\ref
{kgreen}). The series representation of the function $K_{0}$ is 
\begin{equation*}
K_{0}\left( z\right) =-I_{0}\left( z\right) \ln \frac{z}{2}%
+\sum_{k=0}^{\infty }\frac{z^{2k}}{2^{2k}\left( k!\right) ^{2}}\psi \left(
k+1\right)
\end{equation*}
we see that close to the origin the two potentials are similar 
\begin{equation}
K_{0}\left( r\rightarrow 0\right) =-\ln \frac{r}{2}+...  \label{1408}
\end{equation}

We note that the potential in the Euler fluid case may be presented as a
singular \emph{pure gauge} \cite{JackiwPi}: 
\begin{eqnarray}
\frac{1}{2\pi }\varepsilon ^{\alpha \beta }\frac{r^{\beta }}{r^{2}} &=&-%
\frac{1}{2\pi }\frac{\partial }{\partial r^{\alpha }}\arctan \frac{y}{x}
\label{angle} \\
&=&-\frac{1}{2\pi }\frac{\partial }{\partial r^{\alpha }}\theta  \notag
\end{eqnarray}
The individual contributions to the potential are the derivatives of the
angle $\theta $ made by the particle position vector with an arbitrary fixed
direction in plane. In the case of the CHM equation we have from Eqs.(\ref
{phik0}), (\ref{eqmot}), (\ref{kgreen}) 
\begin{eqnarray}
\text{CHM plasma} &:&\text{{}}\;\frac{dr_{i}^{\alpha }}{dt}=\left( -\mathbf{%
\nabla }\phi \times \widehat{\mathbf{n}}\right) ^{\alpha }  \label{16} \\
&=&\varepsilon ^{\alpha \beta }\sum_{j\neq i}^{N}\omega _{j}\frac{r^{\beta
}-r_{j}^{\beta }}{\left| \mathbf{r}-\mathbf{r}_{j}\right| ^{2}}\left[
m\left| \mathbf{r}-\mathbf{r}_{j}\right| K_{1}\left( m\left| \mathbf{r-r}%
_{j}\right| \right) \right]  \notag
\end{eqnarray}
Using the small argument expansion (see formula 8.446 in \cite{GR} or,
alternatively, the Eq.(\ref{1408})) 
\begin{equation*}
m\left| \mathbf{r}-\mathbf{r}_{j}\right| K_{1}\left( m\left| \mathbf{r-r}%
_{j}\right| \right) \rightarrow 1\;\text{for }\left| \mathbf{r}-\mathbf{r}%
_{j}\right| \rightarrow 0
\end{equation*}
we note that Eq.(\ref{16}) can be written like Eq.(\ref{angle}). The
function in the right paranthesis only changes the spatial decay.

The angle $\theta \left( x,y\right) $ is a (multivalued) scalar function and
we have in both cases a typical situation of the type 
\begin{equation}
\left( -\mathbf{\nabla }\phi \times \widehat{\mathbf{n}}\right) ^{\alpha
}\sim \partial _{\alpha }\theta  \label{17}
\end{equation}
so that the potential can be considered at large distances a pure gauge 
\begin{equation}
\left( -\mathbf{\nabla }\phi \times \widehat{\mathbf{n}}\right) ^{\alpha
}\sim g^{-1}dg  \label{18}
\end{equation}
with $g\in U\left( 1\right) $ \emph{i.e.} $g=\exp \left( i\theta \right) $.
In other words, for every point $\left( x,y\right) $ on a large circle on
the plane, we have a value of the angle $\theta $. These potentials have
therefore a \emph{topological} nature, since they map the circle at infinity
in $2D$, ($r\rightarrow \infty $) onto the set of values of the angle $%
\theta $, which is also a circle. This is a typical homotopic classification
of states and the potentials $\phi $ are classified into distinct sets
characterised by an integer, representing how many times the circle in the
plane is covered by the circle representing the values of $\theta $.

The main difference between the Euler and the CHM cases is the short range
of the potential in the latter case. If we use formally the concept of
photon (a ``particle'' that mediates the gauge interaction as in
electrodynamics), one can say that in the Euler case we have the usual (two
dimensional) \emph{massless photon}, whereas in the CHM case we have a \emph{%
massive photon}. The fact that the photon is massive is another way to
express the fast spatial decay of the potential function, \emph{i.e.} the
short range and we will often use this formulation. From physical reasons we
know that this short spacial range must be of the order of $\rho _{s}$, the
intrinsic length in the CHM equation. The need that the equations of motion
lead to a short-range potential (finite-mass photon) and that the potential
has a topological nature represent constraints for the part of the
Lagrangean density coming from the gauge field.

As we have shown in the case of Euler fluid \cite{FlorinMadi1}, the \emph{%
topological nature} of the potential is provided by the Chern-Simons (CS)
Lagrangean 
\begin{equation}
\mathcal{L}_{CS}=\frac{\kappa }{2}\varepsilon ^{\alpha \beta \gamma
}A_{\alpha }\partial _{\beta }A_{\gamma }  \label{19}
\end{equation}
where $\varepsilon ^{\alpha \beta \gamma }$ is the totally antisymmetric
tensor in $2+1$ dimensions ($\alpha $, $\beta $ and $\gamma $ can take three
values: $0,1,2$, corresponding to the time and the two coordinates $\left(
x,y\right) $) and $\kappa $ is a constant. This Lagrangean is essentially
the density of ``magnetic'' helicity. It is known that this Lagrangean does
not lead by itself to dynamical equations for the potential $A_{\mu }$ \
since it is first order in the time derivatives; it only represents a
constraint on the dynamics, analogous to the Lorentz force in an external
magnetic field given by the combination of $\kappa $ with the other
constants of the model. We can heuristically say that the Chern-Simons
Lagrangean induce vortical effects on the dynamics. This will become more
clear later.

The gauge field dynamics can be introduced either by coupling the
Chern-Simons potential with the matter field, or by including the Maxwell
Lagrangean density 
\begin{equation}
\mathcal{L}_{M}=-\frac{1}{4}F_{\mu \nu }F^{\mu \nu }  \label{20}
\end{equation}
or both. We note that any of these combinations provide a finite mass for
the photon but the way they do that is different.

For the combination Maxwell and Chern-Simons, the vortical effect of the
Chern-Simons part induces decay of the field on structures of small spatial
scale, with an extension governed by the ``external magnetic field'', $%
\kappa $. This corresponds to the gyration motion.

For the combination Maxwell and matter field the generation of the photon
mass is due to a classical Higgs mechanism. The matter field has a nonlinear
self-interaction which vanishes at certain non-zero values of the field.
Therefore the extremum of the action implies the minimum of this
self-interaction potential and the matter field will take one of these
values (called: the vacuum value) at infinity. This is the symmetry breaking
leading to the Higgs mechanism. The motion of the photon in a polarisable
medium consisting of this background matter density induces a finite mass
effect for the photon. Then the value of the mass for the photon (the short
range of the spatial decay) is determined by this vacuum value of the matter
field and by the coefficient of the Maxwell contribution in the Lagrangean
(the electric charge).

The combination Maxwell, Chern-Simons and matter has therefore two possible
ways to obtain a finite mass for the photon: the gyration (due to
Chern-Simons) and the Higgs mechanism, due to the finite background of
matter field corresponding to one of its ``vacuum'' values. There is a
mixing of these two ways and there are two possible masses, or short ranges
for the gauge potential. At the limit where the Maxwell term is suppressed
from the Lagrangean, the short range of Chern-Simons with matter is
recovered.

We argue that including the Maxwell term is not necessary in our model
describing the flow governed by the CHM equation. This would only provide
for the gauge field an independent dynamical evolution since, even when the
matter field is absent, the Maxwell Lagrangean leads to plane waves, \emph{%
i.e.} a propagating field, without any meaning or justification in our case.

\subsection{The matter field}

The matter field $\phi $ is associated to (without being identical with) the
density of point-like vortices in the discrete model. The matter field must
be complex since the vorticity carried by any point-like vortex appears as a
sort of electrical charge (only complex fields can represent charged
particles). The kinematical part of the matter field in the Lagrangean
consists as usual in the squared momentum but with the covariant
derivatives, to reflect the so-called \emph{minimal coupling} with the gauge
field 
\begin{equation}
\mathcal{L}_{kin}=-\frac{1}{2}\left( D^{\mu }\phi \right) ^{\dagger }\left(
D_{\mu }\phi \right)  \label{21}
\end{equation}
where 
\begin{equation}
D_{\mu }=\partial _{\mu }+A_{\mu }  \label{22}
\end{equation}

In the Hamiltonian formulation of the discrete vortices model for the Euler
equation it has been derived an equation connecting the gauge field with the
``density'' of the point-like vortices. Going to the continuum (\emph{i.e.}
field-theoretical) version it appeared that the only way to keep this
constraint was to assume a self-interaction of the matter field. The same
reasons act in our present case, but now the problem is more complicated.
The self-interaction potential $V\left( \phi \right) $ must have a minimum
at a nonzero value of the matter field such as to ensure the background that
will induce (together with the CS term) the short-range of the gauge field.
Comparing with the Euler case (we neglect the various constant factors) 
\begin{equation}
V_{Euler}\left( \phi \right) \sim \left| \phi \right| ^{4}  \label{23}
\end{equation}
the simplest form would be 
\begin{equation}
V_{CHM}\left( \phi \right) \sim \left( \left| \phi \right| ^{2}-v^{2}\right)
^{2}  \label{24}
\end{equation}
where $v$ is the vacuum value of the matter field. However, it will be shown
below that this form cannot provide, for the Lagrangean density consisting
of Chern-Simons and matter part, the most symmetrical extremum of the action
functional for the system. This particular state is called self-duality and
we adopt the point of view that this is a fundamental requirement on the
model. In particular in our previous paper \cite{FlorinMadi1} it was shown
that the stationary states of the ideal fluid obeying the \emph{sinh}%
-Poisson equation correspond to the self-duality.

In a different context \cite{coreenii}, \cite{JackiwWeinberg} the form of
the self-interaction potential able to support self-duality has been found
as 
\begin{equation}
V_{CHM}\left( \phi \right) \sim \left| \phi \right| ^{2}\left( \left| \phi
\right| ^{2}-v^{2}\right) ^{2}  \label{25}
\end{equation}
and we will work with this form.

\subsection{The necessity to consider ``pairing'' of the fields}

The main physical content of the states governed by the CHM equation is the
vorticity, which is organizing in large scale vortical structures. The
vorticity and the velocity fields generate the kinematic helicity, which is
a topological invariant for a dissipationless fluid/plasma. It has been
shown \cite{Jackiw2} that the helicity is determined from the boundary
condition (this becomes evident in the Clebsch representation of the
velocity): the values of helicity on the boundary of the volume are
sufficient to determine the value at any internal point. We understand that
besides the fields defined in the internal points, we need to consider
fields that carry the information from the boundary toward the interior, on
equal foot with fields that carry informations from the points of the
internal volume to the boundary. This \emph{pairing} of functions suggests
that all quantities involved in our model will be matrices. The model
becomes non-Abelian and the quantities are elements of the algebra of the
group $SU\left( 2\right) $.

\bigskip

A more formal explanation of the necessity to adopt a non-Abelian algebraic
structure of the theory results from the consideration of the spinorial
nature of the elementary point-like vortices and from the Parity, Charge
Conjugation and Time inversion invariances of the theory.

\section{The field theoretical formalism}

The continuum limit of the system of discrete point-like vortices is a field
theory. From the discussion of the previous Section, we have the field
theoretical model: covariant, $SU\left( 2\right) $, Chern-Simons for the
gauge field and $6^{th}$ order self-interaction for the matter field.

\begin{itemize}
\item  gauge field, with ``potential'' $A^{\mu }$, ($\mu =0,1,2$ for $\left(
t,x,y\right) $) described by the Chern-Simons Lagrangean;

\item  matter (``Higgs'' or ``scalar'') field $\phi $ described by the
covariant kinematic Lagrangean (\emph{i.e.} covariant derivatives,
implementing the minimal coupling of the gauge and matter fields)

\item  matter-field self-interaction given by a potential $V\left( \phi
,\phi ^{\dagger }\right) $ with $6^{th}$ power of $\phi $;

\item  the matter and gauge fields belong to the \emph{adjoint}
representation of the algebra $SU\left( 2\right) $
\end{itemize}

The Lagrangean density for such a model has been used in $\left( 2+1\right) $
field theories and reads 
\begin{eqnarray}
\mathcal{L} &=&-\kappa \varepsilon ^{\mu \nu \rho }\mathrm{tr}\left(
\partial _{\mu }A_{\nu }A_{\rho }+\frac{2}{3}A_{\mu }A_{\nu }A_{\rho }\right)
\label{26} \\
&&-\mathrm{tr}\left[ \left( D^{\mu }\phi \right) ^{\dagger }\left( D_{\mu
}\phi \right) \right]  \notag \\
&&-V\left( \phi ,\phi ^{\dagger }\right)  \notag
\end{eqnarray}
What follows is already exposed in field-theoretical literature, in
particular in \textbf{Dunne} \cite{Dunne1}, \cite{Dunne5}, \cite{Dunne4}.
For the Abelian version, see \cite{JackiwLeeW}.

The transformations of the space and time variables must be connected
through the condition of the general covariance of the theory. The metric of
the space-time is 
\begin{equation}
g_{\mu \nu }=g^{\mu \nu }=\left( 
\begin{array}{ccc}
-1 & 0 & 0 \\ 
0 & 1 & 0 \\ 
0 & 0 & 1
\end{array}
\right)  \label{27}
\end{equation}
This means that we have to take account of the covariant and contravariant
coordinates of vectors, tensors and operators. We can use both notations $%
\left( A_{1},A_{2}\right) $ and $\left( A_{x},A_{y}\right) $ since no
confusion is possible. We have 
\begin{eqnarray}
x^{\mu } &\equiv &\left( t,x,y\right)  \label{274} \\
x_{\mu } &=&g_{\mu \nu }x^{\nu }=\left( -t,x,y\right)  \notag
\end{eqnarray}
\begin{eqnarray}
A^{\mu } &\equiv &\left( A^{0},A^{1},A^{2}\right)  \label{275} \\
A_{\mu } &\equiv &\left( A_{0},A_{1},A_{2}\right) =g_{\mu \nu }A^{\nu
}=\left( -A^{0},A^{1},A^{2}\right)  \notag
\end{eqnarray}
the derivation operator is 
\begin{equation}
\partial _{\mu }\equiv \frac{\partial }{\partial x^{\mu }}=\left( \frac{%
\partial }{\partial t},\frac{\partial }{\partial x},\frac{\partial }{%
\partial y}\right)  \label{2706}
\end{equation}
and 
\begin{eqnarray}
\partial ^{\mu } &=&g^{\mu \nu }\partial _{\nu }  \label{2708} \\
&=&\left( -\frac{\partial }{\partial t},\frac{\partial }{\partial x},\frac{%
\partial }{\partial y}\right)  \notag
\end{eqnarray}
The covariant derivatives are 
\begin{equation}
D_{\mu }=\partial _{\mu }+\left[ A_{\mu },\right]  \label{28}
\end{equation}
(note that we need not introduce an electric charge, $e$). We write the
detailed expression 
\begin{eqnarray}
D^{\mu }\phi &=&g^{\mu \nu }D_{\nu }\phi  \label{2826} \\
&=&g^{\mu \nu }\partial _{\nu }\phi +g^{\mu \nu }\left[ A_{\nu },\phi \right]
\notag \\
&=&\partial ^{\mu }\phi +A^{\mu }\phi -\phi A^{\mu }  \notag
\end{eqnarray}
For comparison we write them in detail (see also Eq.(\ref{275}) ) 
\begin{equation*}
D_{\mu }\phi =\left\{ \frac{\partial \phi }{\partial t}+A_{0}\phi -\phi
A_{0},\frac{\partial \phi }{\partial x}+A_{1}\phi -\phi A_{1},\frac{\partial
\phi }{\partial y}+A_{2}\phi -\phi A_{2}\right\}
\end{equation*}
\begin{equation}
D^{\mu }\phi =\left\{ -\frac{\partial \phi }{\partial t}+A^{0}\phi -\phi
A^{0},\frac{\partial \phi }{\partial x}+A^{1}\phi -\phi A^{1},\frac{\partial
\phi }{\partial y}+A^{2}\phi -\phi A^{2}\right\}  \label{2827}
\end{equation}
The \emph{Hermitean conjugate} of a matrix is the transpose matrix with
complex conjugated entries. For Eq.(\ref{2826}) the Hermitian conjugate is 
\begin{eqnarray}
\left( D^{\mu }\phi \right) ^{\dagger } &=&\left( \partial ^{\mu }\phi
\right) ^{\dagger }+\left[ A^{\mu },\phi \right] ^{\dagger }  \label{2828} \\
&=&\partial _{\mu }\phi ^{\dagger }+\left[ \phi ^{\dagger },A^{\mu \dagger }%
\right]  \notag \\
&=&\partial _{\mu }\phi ^{\dagger }+\phi ^{\dagger }A^{\mu \dagger }-A^{\mu
\dagger }\phi ^{\dagger }  \notag
\end{eqnarray}
or, in detail 
\begin{eqnarray}
\left( D^{\mu }\phi \right) ^{\dagger } &=&\left\{ -\frac{\partial \phi
^{\dagger }}{\partial t}+\phi ^{\dagger }A^{0\dagger }-A^{0\dagger }\phi
^{\dagger },\right.  \label{2831} \\
&&+\frac{\partial \phi ^{\dagger }}{\partial x}+\phi ^{\dagger }A^{1\dagger
}-A^{1\dagger }\phi ^{\dagger },  \notag \\
&&\left. +\frac{\partial \phi ^{\dagger }}{\partial y}+\phi ^{\dagger
}A^{2\dagger }-A^{2\dagger }\phi ^{\dagger }\right\}  \notag
\end{eqnarray}
keeping the following rules ($\ast $ is complex conjugate and $T$ is
transpose) 
\begin{eqnarray}
A^{\mu \dagger } &=&\left( A_{\mu }\right) ^{\ast T}  \label{2835} \\
A_{\mu }^{\dagger } &=&\left( A^{\mu }\right) ^{\ast T}  \notag
\end{eqnarray}
This means 
\begin{eqnarray}
A^{0\dagger } &=&A^{0\ast T}=\left( -A_{0}\right) ^{\ast T}  \label{2837} \\
A^{k\dagger } &=&A^{k\ast T}=\left( A_{k}\right) ^{\ast T}\;,\;k=1,2  \notag
\end{eqnarray}

It has been found that the only possiblity this model has to reach self-dual
states is to choose a matter field nonlinear self-interaction given by a
sixth order potential \cite{coreenii} 
\begin{equation}
V\left( \phi ,\phi ^{\dagger }\right) =\frac{1}{4\kappa ^{2}}\mathrm{tr}%
\left[ \left( \left[ \left[ \phi ,\phi ^{\dagger }\right] ,\phi \right]
-v^{2}\phi \right) ^{\dagger }\left( \left[ \left[ \phi ,\phi ^{\dagger }%
\right] ,\phi \right] -v^{2}\phi \right) \right] .  \label{vpot}
\end{equation}

The trace is taken in a finite dimensional representation of the compact
simple Lie algebra $\mathcal{G}$ to which the gauge field $A_{\mu }$ and the
charged matter field $\phi $ and $\phi ^{\dagger }$ belong.

The Euler Lagrange equations are 
\begin{equation}
D_{\mu }D^{\mu }\phi =\frac{\partial V}{\partial \phi ^{\dagger }}
\label{eleq1}
\end{equation}
\begin{equation}
-\kappa \varepsilon ^{\nu \mu \rho }F_{\mu \rho }=iJ^{\nu }  \label{eleq2}
\end{equation}
where the field strength is 
\begin{equation}
F_{\mu \nu }=\partial _{\mu }A_{\nu }-\partial _{\nu }A_{\mu }+\left[ A_{\mu
},A_{\nu }\right]  \label{field}
\end{equation}
(We may note that the second equation of motion shows a typical
Chern-Simons' action property, \emph{i.e.} the proportionality between the
magnetic field and the current density, which in physical systems would be
called \textbf{force-free}).

The current is 
\begin{equation}
J^{\mu }=-i\left( \left[ \phi ^{\dagger },D^{\mu }\phi \right] -\left[
\left( D^{\mu }\phi \right) ^{\dagger },\phi \right] \right)  \label{jmiu}
\end{equation}
with the conservation (covariant) 
\begin{equation}
D^{\mu }J_{\mu }=0  \label{29}
\end{equation}

The Gauss law constraint is the $0$-component of the second equation of
motion 
\begin{eqnarray}
-\kappa \left( \varepsilon ^{012}F_{12}+\varepsilon ^{021}F_{21}\right)
&=&iJ^{0}  \label{2964} \\
-2\kappa F_{12} &=&iJ^{0}  \notag \\
&=&-iJ_{0}  \notag
\end{eqnarray}
since 
\begin{equation}
J^{0}=g^{0\mu }J_{\mu }=g^{00}J_{0}=-J_{0}  \label{2965}
\end{equation}
Using the Eq.(\ref{jmiu}) we have the Gauss law 
\begin{equation}
2\kappa F_{12}=\left[ \phi ^{\dagger },D_{0}\phi \right] -\left[ \left(
D_{0}\phi \right) ^{\dagger },\phi \right]  \label{gacon}
\end{equation}
in the nonabelian form. The equation written above is the $0$-th component ($%
\varepsilon ^{012}=1$, $\varepsilon ^{021}=-1$) of the equation of motions
connecting the field tensor with the current of matter, Eq.(\ref{eleq2}). We
identify in the right hand side the magnetic field, since 
\begin{eqnarray}
F_{\mu \nu } &\equiv &\partial _{\mu }A_{\nu }-\partial _{\nu }A_{\mu }+ 
\left[ A_{\mu },A_{\nu }\right]  \label{295} \\
&=&\left( 
\begin{array}{ccc}
0 & -E_{y} & -E_{x} \\ 
E_{y} & 0 & -B \\ 
E_{x} & B & 0
\end{array}
\right)  \notag
\end{eqnarray}
($\mu ,\nu =0,1,2$) \emph{i.e.} 
\begin{equation}
F_{12}=-B  \label{297}
\end{equation}

The energy density of the system is 
\begin{eqnarray}
\mathcal{E} &=&\mathrm{tr}\left( \left( D_{0}\phi \right) ^{\dagger }\left(
D_{0}\phi \right) \right) +  \label{hamil} \\
&&+\mathrm{tr}\left( \left( D_{k}\phi \right) ^{\dagger }\left( D_{k}\phi
\right) \right)  \notag \\
&&+V\left( \phi ,\phi ^{\dagger }\right)  \notag
\end{eqnarray}

The energy density can be rewritten in the Bogomolnyi form, \emph{i.e.} as a
sum of squares plus a quantity that integrated over the plane becomes a
lower bound for the energy. The space part of the Lagrangean containing
covariant derivatives can be written \cite{Lee}, \cite{Dunne5} 
\begin{equation}
\mathrm{tr}\left( \left( D_{k}\phi \right) ^{\dagger }\left( D_{k}\phi
\right) \right) =\mathrm{tr}\left( \left( D_{-}\phi \right) ^{\dagger
}\left( D_{-}\phi \right) \right) -i\mathrm{tr}\left( \phi ^{\dagger }\left[
F_{12},\phi \right] \right)  \label{eq30}
\end{equation}
with the notation 
\begin{equation}
D_{\pm }=D_{1}\pm iD_{2}  \label{31}
\end{equation}
In the following we will verify the equality 
\begin{equation}
-i\mathrm{tr}\left( \phi ^{\dagger }\left[ F_{12},\phi \right] \right) =%
\frac{i}{2\kappa }\mathrm{tr}\left( \left[ \left[ \phi ,\phi ^{\dagger }%
\right] ,\phi \right] ^{\dagger }D_{0}\phi -\left[ \left[ \phi ,\phi
^{\dagger }\right] ,\phi \right] \left( D_{0}\phi \right) ^{\dagger }\right)
\label{3151}
\end{equation}
Using Eq.(\ref{gacon}) we replace $F_{12}$ and obtain 
\begin{eqnarray}
&&-i\mathrm{tr}\left( \phi ^{\dagger }\left[ F_{12},\phi \right] \right)
\label{3152} \\
&=&-i\mathrm{tr}\left( \phi ^{\dagger }\left[ \frac{1}{2\kappa }\left( \left[
\phi ^{\dagger },D_{0}\phi \right] -\left[ \left( D_{0}\phi \right)
^{\dagger },\phi \right] \right) ,\phi \right] \right)  \notag \\
&=&-\frac{i}{2\kappa }\mathrm{tr}\left\{ \phi ^{\dagger }\left\{ \left( %
\left[ \phi ^{\dagger },D_{0}\phi \right] -\left[ \left( D_{0}\phi \right)
^{\dagger },\phi \right] \right) \phi -\phi \left( \left[ \phi ^{\dagger
},D_{0}\phi \right] -\left[ \left( D_{0}\phi \right) ^{\dagger },\phi \right]
\right) \right\} \right\}  \notag \\
&=&-\frac{i}{2\kappa }\mathrm{tr}\left\{ \phi ^{\dagger }\left\{ \left[ \phi
^{\dagger },D_{0}\phi \right] \phi -\left[ \left( D_{0}\phi \right)
^{\dagger },\phi \right] \phi \right. \right.  \notag \\
&&\left. \left. -\phi \left[ \phi ^{\dagger },D_{0}\phi \right] +\phi \left[
\left( D_{0}\phi \right) ^{\dagger },\phi \right] \right\} \right\}  \notag
\\
&=&-\frac{i}{2\kappa }\mathrm{tr}\left\{ \phi ^{\dagger }\left[ \phi
^{\dagger },D_{0}\phi \right] \phi \right.  \notag \\
&&-\phi ^{\dagger }\left[ \left( D_{0}\phi \right) ^{\dagger },\phi \right]
\phi  \notag \\
&&-\phi ^{\dagger }\phi \left[ \phi ^{\dagger },D_{0}\phi \right]  \notag \\
&&\left. +\phi ^{\dagger }\phi \left[ \left( D_{0}\phi \right) ^{\dagger
},\phi \right] \right\}  \notag
\end{eqnarray}
We expand the commutators in order to collect together the factors of $%
D_{0}\phi $ and respectively $\left( D_{0}\phi \right) ^{\dagger }$. 
\begin{eqnarray}
&&-i\mathrm{tr}\left( \phi ^{\dagger }\left[ F_{12},\phi \right] \right)
\label{3153} \\
&=&-\frac{i}{2\kappa }\mathrm{tr}\left\{ \phi ^{\dagger }\left( \phi
^{\dagger }\left( D_{0}\phi \right) -\left( D_{0}\phi \right) \phi ^{\dagger
}\right) \phi \right.  \notag \\
&&-\phi ^{\dagger }\left( \left( D_{0}\phi \right) ^{\dagger }\phi -\phi
\left( D_{0}\phi \right) ^{\dagger }\right) \phi  \notag \\
&&-\phi ^{\dagger }\phi \left( \phi ^{\dagger }\left( D_{0}\phi \right)
-\left( D_{0}\phi \right) \phi ^{\dagger }\right)  \notag \\
&&\left. +\phi ^{\dagger }\phi \left( \left( D_{0}\phi \right) ^{\dagger
}\phi -\phi \left( D_{0}\phi \right) ^{\dagger }\right) \right\}  \notag
\end{eqnarray}
\begin{eqnarray}
&&-i\mathrm{tr}\left( \phi ^{\dagger }\left[ F_{12},\phi \right] \right)
\label{3154} \\
&=&-\frac{i}{2\kappa }\mathrm{tr}\left\{ \phi ^{\dagger }\phi ^{\dagger
}\left( D_{0}\phi \right) \phi \right.  \notag \\
&&-\phi ^{\dagger }\left( D_{0}\phi \right) \phi ^{\dagger }\phi  \notag \\
&&-\phi ^{\dagger }\left( D_{0}\phi \right) ^{\dagger }\phi \phi  \notag \\
&&+\phi ^{\dagger }\phi \left( D_{0}\phi \right) ^{\dagger }\phi  \notag \\
&&-\phi ^{\dagger }\phi \phi ^{\dagger }\left( D_{0}\phi \right)  \notag \\
&&+\phi ^{\dagger }\phi \left( D_{0}\phi \right) \phi ^{\dagger }  \notag \\
&&+\phi ^{\dagger }\phi \left( D_{0}\phi \right) ^{\dagger }\phi  \notag \\
&&\left. -\phi ^{\dagger }\phi \phi \left( D_{0}\phi \right) ^{\dagger
}\right\}  \notag
\end{eqnarray}
We take separately the terms containing $D_{0}\phi $ and use the cyclic
symmetry of the Trace operator 
\begin{eqnarray}
&&\phi ^{\dagger }\phi ^{\dagger }\left( D_{0}\phi \right) \phi  \label{3155}
\\
&&-\phi ^{\dagger }\left( D_{0}\phi \right) \phi ^{\dagger }\phi  \notag \\
&&-\phi ^{\dagger }\phi \phi ^{\dagger }\left( D_{0}\phi \right)  \notag \\
&&+\phi ^{\dagger }\phi \left( D_{0}\phi \right) \phi ^{\dagger }  \notag \\
&\rightarrow &\phi \phi ^{\dagger }\phi ^{\dagger }\left( D_{0}\phi \right) 
\notag \\
&&-\phi ^{\dagger }\phi \phi ^{\dagger }\left( D_{0}\phi \right)  \notag \\
&&-\phi ^{\dagger }\phi \phi ^{\dagger }\left( D_{0}\phi \right)  \notag \\
&&+\phi ^{\dagger }\phi ^{\dagger }\phi \left( D_{0}\phi \right)  \notag \\
&=&\left\{ \left( \phi \phi ^{\dagger }-\phi ^{\dagger }\phi \right) \phi
^{\dagger }-\phi ^{\dagger }\left( \phi \phi ^{\dagger }-\phi ^{\dagger
}\phi \right) \right\} \left( D_{0}\phi \right)  \notag \\
&=&\left[ \left[ \phi ,\phi ^{\dagger }\right] ,\phi ^{\dagger }\right]
\left( D_{0}\phi \right)  \notag
\end{eqnarray}
and analogously for the factors of $\left( D_{0}\phi \right) ^{\dagger }$. 
\begin{eqnarray}
&&-\phi ^{\dagger }\left( D_{0}\phi \right) ^{\dagger }\phi \phi
\label{3156} \\
&&+\phi ^{\dagger }\phi \left( D_{0}\phi \right) ^{\dagger }\phi  \notag \\
&&+\phi ^{\dagger }\phi \left( D_{0}\phi \right) ^{\dagger }\phi  \notag \\
&&-\phi ^{\dagger }\phi \phi \left( D_{0}\phi \right) ^{\dagger }  \notag \\
&\rightarrow &-\phi \phi \phi ^{\dagger }\left( D_{0}\phi \right) ^{\dagger }
\notag \\
&&+\phi \phi ^{\dagger }\phi \left( D_{0}\phi \right) ^{\dagger }  \notag \\
&&+\phi \phi ^{\dagger }\phi \left( D_{0}\phi \right) ^{\dagger }  \notag \\
&&-\phi ^{\dagger }\phi \phi \left( D_{0}\phi \right) ^{\dagger }  \notag \\
&=&\left\{ \left( \phi \phi ^{\dagger }-\phi ^{\dagger }\phi \right) \phi
-\phi \left( \phi \phi ^{\dagger }-\phi ^{\dagger }\phi \right) \right\}
\left( D_{0}\phi \right) ^{\dagger }  \notag \\
&&\left\{ \left[ \phi ,\phi ^{\dagger }\right] \phi -\phi \left[ \phi ,\phi
^{\dagger }\right] \right\} \left( D_{0}\phi \right) ^{\dagger }  \notag \\
&=&\left[ \left[ \phi ,\phi ^{\dagger }\right] ,\phi \right] \left(
D_{0}\phi \right) ^{\dagger }  \notag
\end{eqnarray}
It results 
\begin{eqnarray}
&&-i\mathrm{tr}\left( \phi ^{\dagger }\left[ F_{12},\phi \right] \right)
\label{3157} \\
&=&-\frac{i}{2\kappa }\mathrm{tr}\left\{ \left[ \left[ \phi ,\phi ^{\dagger }%
\right] ,\phi ^{\dagger }\right] \left( D_{0}\phi \right) +\left[ \left[
\phi ,\phi ^{\dagger }\right] ,\phi \right] \left( D_{0}\phi \right)
^{\dagger }\right\}  \notag \\
&=&\frac{i}{2\kappa }\mathrm{tr}\left\{ -\left[ \left[ \phi ,\phi ^{\dagger }%
\right] ,\phi ^{\dagger }\right] \left( D_{0}\phi \right) -\left[ \left[
\phi ,\phi ^{\dagger }\right] ,\phi \right] \left( D_{0}\phi \right)
^{\dagger }\right\}  \notag
\end{eqnarray}
and we will prove that 
\begin{equation}
\mathrm{tr}\left\{ -\left[ \left[ \phi ,\phi ^{\dagger }\right] ,\phi
^{\dagger }\right] \right\} =\mathrm{tr}\left\{ \left[ \left[ \phi ,\phi
^{\dagger }\right] ,\phi \right] ^{\dagger }\right\}  \label{3158}
\end{equation}
The right hand side is 
\begin{eqnarray}
&&\left[ \left[ \phi ,\phi ^{\dagger }\right] ,\phi \right] ^{\dagger }
\label{3159} \\
&=&\left\{ \left( \phi \phi ^{\dagger }-\phi ^{\dagger }\phi \right) \phi
-\phi \left( \phi \phi ^{\dagger }-\phi ^{\dagger }\phi \right) \right\}
^{\dagger }  \notag \\
&=&\left\{ \phi \phi ^{\dagger }\phi -\phi ^{\dagger }\phi \phi -\phi \phi
\phi ^{\dagger }+\phi \phi ^{\dagger }\phi \right\} ^{\dagger }  \notag \\
&=&\phi ^{\dagger }\phi \phi ^{\dagger }-\phi ^{\dagger }\phi ^{\dagger
}\phi -\phi \phi ^{\dagger }\phi ^{\dagger }+\phi ^{\dagger }\phi \phi
^{\dagger }  \notag
\end{eqnarray}
and the left hand side 
\begin{eqnarray}
&&-\left[ \left[ \phi ,\phi ^{\dagger }\right] ,\phi ^{\dagger }\right]
\label{3160} \\
&=&-\left\{ \left( \phi \phi ^{\dagger }-\phi ^{\dagger }\phi \right) \phi
^{\dagger }-\phi ^{\dagger }\left( \phi \phi ^{\dagger }-\phi ^{\dagger
}\phi \right) \right\}  \notag \\
&=&-\left\{ \phi \phi ^{\dagger }\phi ^{\dagger }-\phi ^{\dagger }\phi \phi
^{\dagger }-\phi ^{\dagger }\phi \phi ^{\dagger }+\phi ^{\dagger }\phi
^{\dagger }\phi \right\}  \notag \\
&=&-\phi \phi ^{\dagger }\phi ^{\dagger }+\phi ^{\dagger }\phi \phi
^{\dagger }+\phi ^{\dagger }\phi \phi ^{\dagger }-\phi ^{\dagger }\phi
^{\dagger }\phi  \notag
\end{eqnarray}
and one can see the identity of the two expressions. Then 
\begin{eqnarray}
&&-i\mathrm{tr}\left( \phi ^{\dagger }\left[ F_{12},\phi \right] \right)
\label{3161} \\
&=&\frac{i}{2\kappa }\mathrm{tr}\left\{ \left[ \left[ \phi ,\phi ^{\dagger }%
\right] ,\phi \right] ^{\dagger }\left( D_{0}\phi \right) -\left[ \left[
\phi ,\phi ^{\dagger }\right] ,\phi \right] \left( D_{0}\phi \right)
^{\dagger }\right\}  \notag
\end{eqnarray}
This is the expression that is used in Eq.(\ref{eq30}). We have 
\begin{eqnarray}
\mathrm{tr}\left( \left( D_{k}\phi \right) ^{\dagger }\left( D_{k}\phi
\right) \right) &=&\mathrm{tr}\left( \left( D_{-}\phi \right) ^{\dagger
}\left( D_{-}\phi \right) \right) -i\mathrm{tr}\left( \phi ^{\dagger }\left[
F_{12},\phi \right] \right)  \label{30d} \\
&=&\mathrm{tr}\left( \left( D_{-}\phi \right) ^{\dagger }\left( D_{-}\phi
\right) \right)  \notag \\
&&+\frac{i}{2\kappa }\mathrm{tr}\left\{ \left[ \left[ \phi ,\phi ^{\dagger }%
\right] ,\phi \right] ^{\dagger }\left( D_{0}\phi \right) -\left[ \left[
\phi ,\phi ^{\dagger }\right] ,\phi \right] \left( D_{0}\phi \right)
^{\dagger }\right\}  \notag
\end{eqnarray}

\section{The term containing the time-derivatives}

We remind that the objective of this calculation is to find the
configurations for which the energy Eq.(\ref{hamil}) is minimum. According
to the usual approach to this problem, we will try to reexpress Eq.(\ref
{hamil}) as a sum of squared terms plus a ``residual'' term. The squared
contributions, being always positive, are minimum when the corresponding
expressions are zero, while the additional term in general has a topological
meaning. In the particular case of the CHM fluids, the residual energy
cannot be associated to a topological quantity, for reasons that will be
discussed later. Since we are no more guided by the physical significance of
the additional energy as resulting from a topological property of the system
we must accept that there is no unique way of separating in Eq.(\ref{hamil})
the square terms and the additional term. We present in the following two
such formulations and discuss them comparatively.

\subsection{First mode of separating the squared terms in the energy
expression}

Now we have to write in detail the term from the Lagrangian containing the
zero-th covariant derivatives. This is done by including the constant $v$
representing the asymptotic value of the charged field. 
\begin{eqnarray}
&&\mathrm{tr}\left( \left( D_{0}\phi \right) ^{\dagger }\left( D_{0}\phi
\right) \right)  \label{32} \\
&=&\mathrm{tr}\left( \left( D_{0}\phi -\frac{i}{2\kappa }\left( \left[ \left[
\phi ,\phi ^{\dagger }\right] ,\phi \right] -v^{2}\phi \right) \right)
^{\dagger }\right.  \notag \\
&&\hspace*{1cm}\times \left( D_{0}\phi -\frac{i}{2\kappa }\left( \left[ %
\left[ \phi ,\phi ^{\dagger }\right] ,\phi \right] -v^{2}\phi \right) \right)
\notag \\
&&-\frac{i}{2\kappa }\mathrm{tr}\left( \left( \left[ \left[ \phi ,\phi
^{\dagger }\right] ,\phi \right] -v^{2}\phi \right) ^{\dagger }D_{0}\phi
-\left( \left[ \left[ \phi ,\phi ^{\dagger }\right] ,\phi \right] -v^{2}\phi
\right) \left( D_{0}\phi \right) ^{\dagger }\right)  \notag \\
&&-\frac{1}{4\kappa ^{2}}\mathrm{tr}\left( \left( \left[ \left[ \phi ,\phi
^{\dagger }\right] ,\phi \right] -v^{2}\phi \right) ^{\dagger }\left( \left[ %
\left[ \phi ,\phi ^{\dagger }\right] ,\phi \right] -v^{2}\phi \right) \right)
\notag
\end{eqnarray}
This expression is obtained after several cicliv translations of the
factors, which is allowed under the Trace operator. The last term, when we
return to Eq.(\ref{hamil}), cancels the potential $V\left( \phi ,\phi
^{\dagger }\right) $ given in (\ref{vpot}). We now understand that only this
choice allows to write the energy in the Bogomolnyi form.

The total energy in the system, Eq.(\ref{hamil}), written in Bogomolnyi
form, results from the expressions (\ref{30d}) and (\ref{32}) 
\begin{eqnarray}
\mathcal{E} &=&\mathrm{tr}\left( \left( D_{0}\phi -\frac{i}{2\kappa }\left( %
\left[ \left[ \phi ,\phi ^{\dagger }\right] ,\phi \right] -v^{2}\phi \right)
\right) ^{\dagger }\right.  \label{inth} \\
&&\hspace*{1cm}\left. \times \left( D_{0}\phi -\frac{i}{2\kappa }\left( %
\left[ \left[ \phi ,\phi ^{\dagger }\right] ,\phi \right] -v^{2}\phi \right)
\right) \right)  \notag \\
&&+\mathrm{tr}\left( \left( D_{-}\phi \right) ^{\dagger }\left( D_{-}\phi
\right) \right)  \notag \\
&&+\frac{iv^{2}}{2\kappa }\mathrm{tr}\left( \phi ^{\dagger }\left( D_{0}\phi
\right) -\left( D_{0}\phi \right) ^{\dagger }\phi \right)  \notag
\end{eqnarray}
The energy contains a sum of positive quantities (squares) and is minimised
by those states where these terms are vanishing. The last term shows that
there is a lower bound the energy 
\begin{equation}
\mathcal{E}\geqslant \frac{iv^{2}}{2\kappa }\mathrm{tr}\left( \phi ^{\dagger
}\left( D_{0}\phi \right) -\left( D_{0}\phi \right) ^{\dagger }\phi \right)
\label{egreater}
\end{equation}

\subsubsection{The first form of the self-duality equations}

The vanishing of the squared terms in the energy leads to the \emph{%
self-duality equations} 
\begin{eqnarray}
D_{-}\phi &=&0  \label{sd1} \\
D_{0}\phi &=&\frac{i}{2\kappa }\left( \left[ \left[ \phi ,\phi ^{\dagger }%
\right] ,\phi \right] -v^{2}\phi \right)  \notag
\end{eqnarray}
Combining these two equations such as to put in evidence the gauge field $%
F_{+-}$, whose expression is in Eq.(\ref{3114}), (the calculation is
presented in detail in \emph{Appendix C}) 
\begin{eqnarray}
D_{-}\phi &=&0  \label{sd} \\
F_{+-} &=&\frac{1}{\kappa ^{2}}\left[ v^{2}\phi -\left[ \left[ \phi ,\phi
^{\dagger }\right] ,\phi \right] ,\phi ^{\dagger }\right]  \notag
\end{eqnarray}
Using Eqs.(\ref{sd1}) we can derive a new expression for the energy in the%
\textbf{\ }self-dual state 
\begin{equation}
\mathcal{E}_{SD}=\frac{v^{2}}{2\kappa ^{2}}\mathrm{tr}\left( \phi ^{\dagger
}\left( v^{2}\phi -\left[ \left[ \phi ,\phi ^{\dagger }\right] ,\phi \right]
\right) \right)  \label{ebound}
\end{equation}
which is the saturated lower bound shown above.

\subsection{Second mode of separating squared terms in the expression of the
energy}

We look for an expression of a square term that differs from the previous
one by a change of sign within the first two terms of Eq.(\ref{32})

\begin{eqnarray}
&&\hspace*{-1cm}\mathrm{tr}\left\{ \left( D_{0}\phi +\frac{i}{2\kappa }%
\left( \left[ \left[ \phi ,\phi ^{\dagger }\right] ,\phi \right] -v^{2}\phi
\right) \right) ^{\dagger }\left( D_{0}\phi +\frac{i}{2\kappa }\left( \left[ %
\left[ \phi ,\phi ^{\dagger }\right] ,\phi \right] -v^{2}\phi \right)
\right) \right\}  \label{350} \\
&=&\mathrm{tr}\left\{ \left( \left( D_{0}\phi \right) ^{\dagger }-\frac{i}{%
2\kappa }\left( \left[ \left[ \phi ,\phi ^{\dagger }\right] ,\phi \right]
-v^{2}\phi \right) ^{\dagger }\right) \left( D_{0}\phi +\frac{i}{2\kappa }%
\left( \left[ \left[ \phi ,\phi ^{\dagger }\right] ,\phi \right] -v^{2}\phi
\right) \right) \right\}  \notag \\
&=&\mathrm{tr}\left\{ \left( D_{0}\phi \right) ^{\dagger }\left( D_{0}\phi
\right) \right.  \notag \\
&&-\frac{i}{2\kappa }\left( \left[ \left[ \phi ,\phi ^{\dagger }\right]
,\phi \right] -v^{2}\phi \right) ^{\dagger }\left( D_{0}\phi \right)  \notag
\\
&&+\left( D_{0}\phi \right) ^{\dagger }\frac{i}{2\kappa }\left( \left[ \left[
\phi ,\phi ^{\dagger }\right] ,\phi \right] -v^{2}\phi \right)  \notag \\
&&\left. +\frac{1}{4\kappa ^{2}}\left( \left[ \left[ \phi ,\phi ^{\dagger }%
\right] ,\phi \right] -v^{2}\phi \right) ^{\dagger }\left( \left[ \left[
\phi ,\phi ^{\dagger }\right] ,\phi \right] -v^{2}\phi \right) \right\} 
\notag
\end{eqnarray}
The two median lines are expanded and the full expression is rewritten 
\begin{eqnarray}
&&\hspace*{-1cm}\mathrm{tr}\left\{ \left( D_{0}\phi +\frac{i}{2\kappa }%
\left( \left[ \left[ \phi ,\phi ^{\dagger }\right] ,\phi \right] -v^{2}\phi
\right) \right) ^{\dagger }\left( D_{0}\phi +\frac{i}{2\kappa }\left( \left[ %
\left[ \phi ,\phi ^{\dagger }\right] ,\phi \right] -v^{2}\phi \right)
\right) \right\}  \notag \\
&=&\mathrm{tr}\left\{ \left( D_{0}\phi \right) ^{\dagger }\left( D_{0}\phi
\right) \right.  \notag \\
&&-\frac{i}{2\kappa }\left( \left[ \left[ \phi ,\phi ^{\dagger }\right]
,\phi \right] \right) ^{\dagger }\left( D_{0}\phi \right) +\frac{i}{2\kappa }%
v^{2}\phi ^{\dagger }\left( D_{0}\phi \right)  \notag \\
&&+\frac{i}{2\kappa }\left( D_{0}\phi \right) ^{\dagger }\left( \left[ \left[
\phi ,\phi ^{\dagger }\right] ,\phi \right] \right) -\frac{i}{2\kappa }%
\left( D_{0}\phi \right) ^{\dagger }v^{2}\phi  \notag \\
&&\left. +\frac{1}{4\kappa ^{2}}\left( \left[ \left[ \phi ,\phi ^{\dagger }%
\right] ,\phi \right] -v^{2}\phi \right) ^{\dagger }\left( \left[ \left[
\phi ,\phi ^{\dagger }\right] ,\phi \right] -v^{2}\phi \right) \right\}
\label{351}
\end{eqnarray}
and we can now get an expression for the first contribution to the energy 
\begin{eqnarray}
&&\mathrm{tr}\left\{ \left( D_{0}\phi \right) ^{\dagger }\left( D_{0}\phi
\right) \right\}  \label{352} \\
&=&\mathrm{tr}\left\{ \left( D_{0}\phi +\frac{i}{2\kappa }\left( \left[ %
\left[ \phi ,\phi ^{\dagger }\right] ,\phi \right] -v^{2}\phi \right)
\right) ^{\dagger }\left( D_{0}\phi +\frac{i}{2\kappa }\left( \left[ \left[
\phi ,\phi ^{\dagger }\right] ,\phi \right] -v^{2}\phi \right) \right)
\right\}  \notag \\
&&-\mathrm{tr}\left\{ -\frac{i}{2\kappa }\left( \left[ \left[ \phi ,\phi
^{\dagger }\right] ,\phi \right] \right) ^{\dagger }\left( D_{0}\phi \right)
+\frac{i}{2\kappa }v^{2}\phi ^{\dagger }\left( D_{0}\phi \right) \right. 
\notag \\
&&\left. +\frac{i}{2\kappa }\left( D_{0}\phi \right) ^{\dagger }\left( \left[
\left[ \phi ,\phi ^{\dagger }\right] ,\phi \right] \right) -\frac{i}{2\kappa 
}\left( D_{0}\phi \right) ^{\dagger }v^{2}\phi \right\}  \notag \\
&&-\mathrm{tr}\left\{ \frac{1}{4\kappa ^{2}}\left( \left[ \left[ \phi ,\phi
^{\dagger }\right] ,\phi \right] -v^{2}\phi \right) ^{\dagger }\left( \left[ %
\left[ \phi ,\phi ^{\dagger }\right] ,\phi \right] -v^{2}\phi \right)
\right\}  \notag
\end{eqnarray}

\bigskip

With the two expressions detailed above, we can rewrite the energy Eq.(\ref
{hamil}) 
\begin{eqnarray}
\mathcal{E} &=&\mathrm{tr}\left( \left( D_{0}\phi \right) ^{\dagger }\left(
D_{0}\phi \right) \right) +\mathrm{tr}\left( \left( D_{k}\phi \right)
^{\dagger }\left( D_{k}\phi \right) \right) +V\left( \phi ,\phi ^{\dagger
}\right)  \label{353} \\
&=&\mathrm{tr}\left\{ \left( D_{0}\phi +\frac{i}{2\kappa }\left( \left[ %
\left[ \phi ,\phi ^{\dagger }\right] ,\phi \right] -v^{2}\phi \right)
\right) ^{\dagger }\left( D_{0}\phi +\frac{i}{2\kappa }\left( \left[ \left[
\phi ,\phi ^{\dagger }\right] ,\phi \right] -v^{2}\phi \right) \right)
\right\}  \notag \\
&&-\mathrm{tr}\left\{ -\frac{i}{2\kappa }\left( \left[ \left[ \phi ,\phi
^{\dagger }\right] ,\phi \right] \right) ^{\dagger }\left( D_{0}\phi \right)
+\frac{i}{2\kappa }v^{2}\phi ^{\dagger }\left( D_{0}\phi \right) \right. 
\notag \\
&&\left. +\frac{i}{2\kappa }\left( D_{0}\phi \right) ^{\dagger }\left( \left[
\left[ \phi ,\phi ^{\dagger }\right] ,\phi \right] \right) -\frac{i}{2\kappa 
}\left( D_{0}\phi \right) ^{\dagger }v^{2}\phi \right\}  \notag \\
&&-\mathrm{tr}\left\{ \frac{1}{4\kappa ^{2}}\left( \left[ \left[ \phi ,\phi
^{\dagger }\right] ,\phi \right] -v^{2}\phi \right) ^{\dagger }\left( \left[ %
\left[ \phi ,\phi ^{\dagger }\right] ,\phi \right] -v^{2}\phi \right)
\right\}  \notag \\
&&+\mathrm{tr}\left( \left( D_{-}\phi \right) ^{\dagger }\left( D_{-}\phi
\right) \right) +\frac{i}{2\kappa }\mathrm{tr}\left\{ \left[ \left[ \phi
,\phi ^{\dagger }\right] ,\phi \right] ^{\dagger }\left( D_{0}\phi \right) -%
\left[ \left[ \phi ,\phi ^{\dagger }\right] ,\phi \right] \left( D_{0}\phi
\right) ^{\dagger }\right\}  \notag \\
&&+\frac{1}{4\kappa ^{2}}\mathrm{tr}\left\{ \left( \left[ \left[ \phi ,\phi
^{\dagger }\right] ,\phi \right] -v^{2}\phi \right) ^{\dagger }\left( \left[ %
\left[ \phi ,\phi ^{\dagger }\right] ,\phi \right] -v^{2}\phi \right)
\right\}  \notag
\end{eqnarray}
We note the cancellation of the term which is the potential $V\left( \phi
^{\dagger },\phi \right) $ and we obtain 
\begin{eqnarray}
\hspace*{-1cm}\mathcal{E} &=&\mathrm{tr}\left\{ \left( D_{0}\phi +\frac{i}{%
2\kappa }\left( \left[ \left[ \phi ,\phi ^{\dagger }\right] ,\phi \right]
-v^{2}\phi \right) \right) ^{\dagger }\left( D_{0}\phi +\frac{i}{2\kappa }%
\left( \left[ \left[ \phi ,\phi ^{\dagger }\right] ,\phi \right] -v^{2}\phi
\right) \right) \right\}  \notag \\
&&+\mathrm{tr}\left( \left( D_{-}\phi \right) ^{\dagger }\left( D_{-}\phi
\right) \right)  \notag \\
&&+\mathcal{E}^{a}  \label{354}
\end{eqnarray}
where the additional term in the energy expression is 
\begin{eqnarray}
\mathcal{E}^{a} &=&-\mathrm{tr}\left\{ -\frac{i}{2\kappa }\left( \left[ %
\left[ \phi ,\phi ^{\dagger }\right] ,\phi \right] \right) ^{\dagger }\left(
D_{0}\phi \right) +\frac{i}{2\kappa }v^{2}\phi ^{\dagger }\left( D_{0}\phi
\right) \right.  \label{355} \\
&&\left. +\frac{i}{2\kappa }\left( D_{0}\phi \right) ^{\dagger }\left[ \left[
\phi ,\phi ^{\dagger }\right] ,\phi \right] -\frac{i}{2\kappa }\left(
D_{0}\phi \right) ^{\dagger }v^{2}\phi \right\}  \notag \\
&&+\frac{i}{2\kappa }\mathrm{tr}\left\{ \left[ \left[ \phi ,\phi ^{\dagger }%
\right] ,\phi \right] ^{\dagger }\left( D_{0}\phi \right) -\left[ \left[
\phi ,\phi ^{\dagger }\right] ,\phi \right] \left( D_{0}\phi \right)
^{\dagger }\right\}  \notag \\
&=&\frac{i}{\kappa }\mathrm{tr}\left\{ \left[ \left[ \phi ,\phi ^{\dagger }%
\right] ,\phi \right] ^{\dagger }\left( D_{0}\phi \right) \right\}  \notag \\
&&-\frac{i}{\kappa }\mathrm{tr}\left\{ \left( D_{0}\phi \right) ^{\dagger }%
\left[ \left[ \phi ,\phi ^{\dagger }\right] ,\phi \right] \right\}
\;\;\left( \text{using cyclic permutation in Trace}\right)  \notag \\
&&-\frac{iv^{2}}{2\kappa }\mathrm{tr}\left\{ \phi ^{\dagger }\left(
D_{0}\phi \right) -\left( D_{0}\phi \right) ^{\dagger }\phi \right\}  \notag
\end{eqnarray}
We write 
\begin{equation}
\mathcal{E}^{a}=\mathcal{E}^{a\left( 1\right) }+\mathcal{E}^{a\left(
2\right) }+\mathcal{E}^{a\left( 3\right) }  \label{356}
\end{equation}
for the last three lines of the equation above and these will be calculated
below. At this point it is more useful to focus on the set of equations at
self-duality that are derived from this choice adopted in Eq.(\ref{350}).

\subsubsection{Second form of the Self-Duality equations}

The NEW equations as they result from the alternative Bogomolny form of the
action are 
\begin{eqnarray}
D_{-}\phi &=&0  \label{357} \\
D_{0}\phi &=&-\frac{i}{2\kappa }\left( \left[ \left[ \phi ,\phi ^{\dagger }%
\right] ,\phi \right] -v^{2}\phi \right)  \notag
\end{eqnarray}
We introduce the field tensor $F_{+-}$ \ and it is shown in Appendix C that 
\begin{eqnarray}
F_{+-} &=&-\frac{J^{0}}{\kappa }=\frac{J_{0}}{\kappa }  \label{358} \\
&=&\frac{1}{\kappa }\left\{ -i\left( \left[ \phi ^{\dagger },D_{0}\phi %
\right] -\left[ \left( D_{0}\phi \right) ^{\dagger },\phi \right] \right)
\right\}  \notag
\end{eqnarray}
Now, we use the NEW equations 
\begin{eqnarray}
D_{0}\phi &=&-\frac{i}{2\kappa }\left( \left[ \left[ \phi ,\phi ^{\dagger }%
\right] ,\phi \right] -v^{2}\phi \right)  \label{359} \\
\left( D_{0}\phi \right) ^{\dagger } &=&\frac{i}{2\kappa }\left( \left( %
\left[ \left[ \phi ,\phi ^{\dagger }\right] ,\phi \right] \right) ^{\dagger
}-v^{2}\phi ^{\dagger }\right)  \notag
\end{eqnarray}
Inserting the two operators from the equations above, and after finding that
the two commutators in the Eq.(\ref{3106}) are equal and opposite, we get an
expression for $F_{+-}$ as 
\begin{eqnarray}
F_{+-} &=&-\frac{i}{\kappa }\left\{ \left( \left[ \phi ^{\dagger },D_{0}\phi %
\right] -\left[ \left( D_{0}\phi \right) ^{\dagger },\phi \right] \right)
\right\}  \label{360} \\
&=&-\frac{2i}{\kappa }\left[ \phi ^{\dagger },D_{0}\phi \right]  \notag
\end{eqnarray}
where we replace the NEW expression of $D_{0}\phi $ , \emph{i.e.} Eq.(\ref
{359}), obtaining 
\begin{eqnarray}
F_{+-} &=&-\frac{2i}{\kappa }\left[ \phi ^{\dagger },-\frac{i}{2\kappa }%
\left( \left[ \left[ \phi ,\phi ^{\dagger }\right] ,\phi \right] -v^{2}\phi
\right) \right]  \label{361} \\
&=&-\frac{1}{\kappa ^{2}}\left[ \phi ^{\dagger },\left[ \left[ \phi ,\phi
^{\dagger }\right] ,\phi \right] -v^{2}\phi \right]  \notag \\
&=&-\frac{1}{\kappa ^{2}}\left[ v^{2}\phi -\left[ \left[ \phi ,\phi
^{\dagger }\right] ,\phi \right] ,\phi ^{\dagger }\right]  \notag
\end{eqnarray}
Then the NEW equations at self-duality are 
\begin{eqnarray}
D_{-}\phi &=&0  \label{362} \\
F_{+-} &=&-\frac{1}{\kappa ^{2}}\left[ v^{2}\phi -\left[ \left[ \phi ,\phi
^{\dagger }\right] ,\phi \right] ,\phi ^{\dagger }\right]  \notag
\end{eqnarray}

We note that this form of the self-duality equations differs from the
previous one by the opposite sign of the right-hand side term of the second
equation.

Using the definition of Eq.(\ref{31}) the left hand side of the first
equation of motion (\ref{eleq1}) can be written 
\begin{equation}
D_{\mu }D^{\mu }\phi =-D_{0}D_{0}\phi +D_{+}D_{-}\phi +i\left[ F_{12},\phi %
\right]  \label{3011}
\end{equation}

\section{The group theoretical ansatz}

\subsection{Elements of the $SU\left( 2\right) $ algebra structure}

In order to solve the self-duality equations it is considered, as in the
case of the Euler equation, the Lie algebra of the group $SU\left( 2\right) $%
. Then the Chevalley basis is \cite{Slansky} 
\begin{eqnarray}
\left[ E_{+},E_{-}\right] &=&H  \label{34} \\
\left[ H,E_{\pm }\right] &=&\pm 2E_{\pm }  \notag \\
\mathrm{tr}\left( E_{+}E_{-}\right) &=&1  \notag \\
\mathrm{tr}\left( H^{2}\right) &=&2  \notag
\end{eqnarray}
where 
\begin{equation*}
H\;\text{is the Cartan subalgebra generator}
\end{equation*}
Since the \emph{rank} of $SU\left( 2\right) $ is $r=1$ the generator $H$ is
unique. 
\begin{equation*}
E_{\pm }\;\text{are step (ladder) operators}
\end{equation*}
The $2\times 2$ representation is 
\begin{equation}
E_{+}=\left( 
\begin{array}{cc}
0 & 1 \\ 
0 & 0
\end{array}
\right)  \label{45}
\end{equation}
\begin{equation}
E_{-}=\left( 
\begin{array}{cc}
0 & 0 \\ 
1 & 0
\end{array}
\right)  \label{36}
\end{equation}
\begin{equation}
H=\left( 
\begin{array}{cc}
1 & 0 \\ 
0 & -1
\end{array}
\right)  \label{37}
\end{equation}
The Hemitian conjugates of the generators are the transposed complex
conjugated matrices 
\begin{eqnarray}
E_{+}^{\dagger } &=&E_{-}  \label{38} \\
E_{-}^{\dagger } &=&E_{+}  \notag \\
H^{\dagger } &=&H  \notag
\end{eqnarray}
These adjoint generators will be used to express the adjoint fields in the
calculations based on a particular \emph{ansatz}.

\subsection{The fields within the algebraic \emph{ansatz}}

According to Dunne \cite{Dunne2}, \cite{Dunne3}, the following \emph{ansatz}
can be adopted 
\begin{eqnarray}
\phi &=&\sum_{a=1}^{r}\phi _{a}E_{a}+\phi _{-M}E_{-M}  \label{39} \\
&=&\phi _{1}E_{+}+\phi _{2}E_{-}  \notag
\end{eqnarray}
since the rank of $SU\left( 2\right) $ is $r=1$. We take the Hermitian
conjugate, which is 
\begin{eqnarray}
\phi ^{\dagger } &=&\phi _{1}^{\ast }E_{+}^{\dagger }+\phi _{2}^{\ast
}E_{-}^{\dagger }  \label{40} \\
&=&\phi _{1}^{\ast }E_{-}+\phi _{2}^{\ast }E_{+}  \notag
\end{eqnarray}
In this ansatz the matter Higgs field is represented by a linear combination
of the ladder generators plus the generator associated with minus the
maximal root.

The gauge potential is taken as 
\begin{eqnarray}
A_{+} &=&aH  \label{41} \\
A_{-} &=&-a^{\ast }H  \notag
\end{eqnarray}

The notations with $+$ and $-$ correspond to the combinations of the $x$ and 
$y$ components, with the coefficient $i$ for the $y$ component.

\subsubsection{The explicit form of the equations with the \emph{ansatz}}

The gauge field tensor 
\begin{eqnarray}
F_{+-} &=&\partial _{+}A_{-}-\partial _{-}A_{+}+\left[ A_{+},A_{-}\right]
\label{42} \\
&=&\partial _{+}\left( -a^{\ast }H\right) -\partial _{-}\left( aH\right) + 
\left[ aH,-a^{\ast }H\right]  \notag \\
&=&\left( -\partial _{+}a^{\ast }-\partial _{-}a\right) H  \notag
\end{eqnarray}
where 
\begin{eqnarray}
\partial _{+} &=&\partial _{x}+i\partial _{y}=2\frac{\partial }{\partial
z^{\ast }}  \label{43} \\
\partial _{-} &=&\partial _{x}-i\partial _{y}=2\frac{\partial }{\partial z} 
\notag
\end{eqnarray}

We will have to calculate, with this ansatz, the terms of the equations. The
right hand side of the second equation 
\begin{equation}
\left[ v^{2}\phi -\left[ \left[ \phi ,\phi ^{\dagger }\right] ,\phi \right]
,\phi ^{\dagger }\right]  \label{44}
\end{equation}
will be calculated using the commutator 
\begin{eqnarray}
\left[ \phi ,\phi ^{\dagger }\right] &=&\left[ \phi _{1}E_{+}+\phi
_{2}E_{-},\phi _{1}^{\ast }E_{-}+\phi _{2}^{\ast }E_{+}\right]  \label{46} \\
&=&\phi _{1}\phi _{1}^{\ast }\left[ E_{+},E_{-}\right] +\phi _{2}\phi
_{1}^{\ast }\left[ E_{-},E_{-}\right]  \notag \\
&&+\phi _{1}\phi _{2}^{\ast }\left[ E_{+},E_{+}\right] +\phi _{2}\phi
_{2}^{\ast }\left[ E_{-},E_{+}\right]  \notag
\end{eqnarray}
\begin{eqnarray}
\left[ \phi ,\phi ^{\dagger }\right] &=&\phi _{1}^{\ast }\phi _{1}\left[
E_{+},E_{-}\right]  \label{47} \\
&&+\phi _{2}^{\ast }\phi _{2}\left[ E_{-},E_{+}\right]  \notag \\
&=&\phi _{1}^{\ast }\phi _{1}H-\phi _{2}^{\ast }\phi _{2}H  \notag
\end{eqnarray}
\begin{eqnarray}
\left[ \phi ,\phi ^{\dagger }\right] &=&\left( \phi _{1}^{\ast }\phi
_{1}-\phi _{2}^{\ast }\phi _{2}\right) H  \label{48} \\
&=&\left( \rho _{1}-\rho _{2}\right) H  \notag
\end{eqnarray}
where we have introduced the notations 
\begin{eqnarray}
\rho _{1} &\equiv &\left| \phi _{1}\right| ^{2}  \label{49} \\
\rho _{2} &\equiv &\left| \phi _{2}\right| ^{2}  \notag
\end{eqnarray}
The next step is to calculate 
\begin{equation}
\left[ \left[ \phi ,\phi ^{\dagger }\right] ,\phi \right] =\left[ \left(
\rho _{1}-\rho _{2}\right) H,\phi _{1}E_{+}+\phi _{2}E_{-}\right]  \label{50}
\end{equation}
This is 
\begin{eqnarray}
\left[ \left[ \phi ,\phi ^{\dagger }\right] ,\phi \right] &=&\left( \rho
_{1}-\rho _{2}\right) \phi _{1}\left[ H,E_{+}\right] +\left( \rho _{1}-\rho
_{2}\right) \phi _{2}\left[ H,E_{-}\right]  \label{51} \\
&=&2\left( \rho _{1}-\rho _{2}\right) \left( \phi _{1}E_{+}-\phi
_{2}E_{-}\right)  \notag
\end{eqnarray}
The next level in the commutator is 
\begin{eqnarray}
v^{2}\phi -\left[ \left[ \phi ,\phi ^{\dagger }\right] ,\phi \right]
&=&v^{2}\phi _{1}E_{+}+v^{2}\phi _{2}E_{-}  \label{52} \\
&&-2\left( \rho _{1}-\rho _{2}\right) \phi _{1}E_{+}+2\left( \rho _{1}-\rho
_{2}\right) \phi _{2}E_{-}  \notag \\
&\equiv &PE_{+}+QE_{-}  \notag
\end{eqnarray}
where 
\begin{eqnarray}
P &\equiv &v^{2}\phi _{1}-2\left( \rho _{1}-\rho _{2}\right) \phi _{1}
\label{53} \\
Q &\equiv &v^{2}\phi _{2}+2\left( \rho _{1}-\rho _{2}\right) \phi _{2} 
\notag
\end{eqnarray}

Returning to the Eq.(\ref{sd}), the full right hand side term is 
\begin{eqnarray}
\left[ v^{2}\phi -\left[ \left[ \phi ,\phi ^{\dagger }\right] ,\phi \right]
,\phi ^{\dagger }\right] &=&\left[ PE_{+}+QE_{-},\phi _{1}^{\ast }E_{-}+\phi
_{2}^{\ast }E_{+}\right]  \label{54} \\
&=&P\phi _{1}^{\ast }\left[ E_{+},E_{-}\right] +Q\phi _{2}^{\ast }\left[
E_{-},E_{+}\right]  \notag \\
&=&\left( P\phi _{1}^{\ast }-Q\phi _{2}^{\ast }\right) \left[ E_{+},E_{-}%
\right]  \notag \\
&=&\left( P\phi _{1}^{\ast }-Q\phi _{2}^{\ast }\right) H  \notag
\end{eqnarray}
or 
\begin{eqnarray}
&&\left[ v^{2}\phi -\left[ \left[ \phi ,\phi ^{\dagger }\right] ,\phi \right]
,\phi ^{\dagger }\right]  \label{55} \\
&=&\left\{ \left( v^{2}\phi _{1}-2\left( \rho _{1}-\rho _{2}\right) \phi
_{1}\right) \phi _{1}^{\ast }-\left( v^{2}\phi _{2}+2\left( \rho _{1}-\rho
_{2}\right) \phi _{2}\right) \phi _{2}^{\ast }\right\} H  \notag \\
&=&\left\{ \left( v^{2}-2\left( \rho _{1}-\rho _{2}\right) \right) \rho
_{1}-\left( v^{2}+2\left( \rho _{1}-\rho _{2}\right) \right) \rho
_{2}\right\} H  \notag \\
&=&\left( v^{2}-2\left( \rho _{1}+\rho _{2}\right) \right) \left( \rho
_{1}-\rho _{2}\right) H  \notag
\end{eqnarray}

\subsection{Using the algebraic ansatz in the first version of the SD
equations}

The second self-duality equation Eq.(\ref{sd}) becomes, using Eqs.(\ref{42})
and (\ref{55}) 
\begin{equation}
-\frac{\partial a^{\ast }}{\partial x_{+}}-\frac{\partial a}{\partial x_{-}}=%
\frac{1}{k^{2}}\left( \rho _{1}-\rho _{2}\right) \left[ v^{2}-2\left( \rho
_{1}+\rho _{2}\right) \right]  \label{56}
\end{equation}

Now we turn to the first self-duality equation 
\begin{equation}
D_{-}\phi =0  \label{57}
\end{equation}
and its adjoint form. It has been defined 
\begin{equation}
D_{-}\equiv D_{1}-iD_{2}  \label{58}
\end{equation}
then 
\begin{equation}
D_{-}\phi =\frac{\partial \phi }{\partial x}+\left[ A_{x},\phi \right] -i%
\frac{\partial \phi }{\partial y}-i\left[ A_{y},\phi \right]  \label{59}
\end{equation}
To proceed further we express the components of the potential 
\begin{eqnarray}
A_{+} &=&A_{x}+iA_{y}=aH  \label{60} \\
A_{-} &=&A_{x}-iA_{y}=-a^{\ast }H  \notag
\end{eqnarray}
Then 
\begin{eqnarray}
A_{x} &=&\frac{1}{2}\left( a-a^{\ast }\right) H  \label{61} \\
A_{y} &=&\frac{1}{2i}\left( a+a^{\ast }\right) H  \notag
\end{eqnarray}
Then 
\begin{eqnarray}
D_{-}\phi &=&\left( \frac{\partial \phi _{1}}{\partial x}-i\frac{\partial
\phi _{1}}{\partial y}\right) E_{+}+\left( \frac{\partial \phi _{2}}{%
\partial x}-i\frac{\partial \phi _{2}}{\partial y}\right) E_{-}  \label{62}
\\
&&+\frac{1}{2}\left( a-a^{\ast }\right) \phi _{1}\left[ H,E_{+}\right] 
\notag \\
&&+\frac{1}{2}\left( a-a^{\ast }\right) \phi _{2}\left[ H,E_{-}\right] 
\notag \\
&&-i\frac{1}{2i}\left( a+a^{\ast }\right) \phi _{1}\left[ H,E_{+}\right] 
\notag \\
&&-i\frac{1}{2i}\left( a+a^{\ast }\right) \phi _{2}\left[ H,E_{-}\right] 
\notag
\end{eqnarray}
\begin{eqnarray}
D_{-}\phi &=&\left( \frac{\partial \phi _{1}}{\partial x}-i\frac{\partial
\phi _{1}}{\partial y}+2\frac{1}{2}\left( a-a^{\ast }\right) \phi _{1}-2%
\frac{1}{2}\left( a+a^{\ast }\right) \phi _{1}\right) E_{+}  \label{63} \\
&&+\left( \frac{\partial \phi _{2}}{\partial x}-i\frac{\partial \phi _{2}}{%
\partial y}-2\frac{1}{2}\left( a-a^{\ast }\right) \phi _{2}+2\frac{1}{2}%
\left( a+a^{\ast }\right) \phi _{2}\right) E_{-}  \notag \\
&=&0  \notag
\end{eqnarray}
From the explicit form of the ladder generators we obtain the equations
derived from the first self-duality equation 
\begin{equation}
\frac{\partial \phi _{1}}{\partial x}-i\frac{\partial \phi _{1}}{\partial y}%
-2\phi _{1}a^{\ast }=0  \label{64}
\end{equation}
\begin{equation}
\frac{\partial \phi _{2}}{\partial x}-i\frac{\partial \phi _{2}}{\partial y}%
+2\phi _{2}a^{\ast }=0  \label{645}
\end{equation}

\bigskip

\subsubsection{The explicit form of the \emph{adjoint} equations with the
algebraic \emph{ansatz}}

Now we consider the \emph{adjoint} equation (also derived from the extremum
of the corresponding part of the action expressed in the Bogomolnyi form) 
\begin{equation}
\left( D_{-}\phi \right) ^{\dagger }=0  \label{65}
\end{equation}
We have 
\begin{equation}
D_{-}^{\dagger }=\frac{\partial }{\partial x}+\left[ ,A_{x}^{\dagger }\right]
+i\frac{\partial }{\partial y}+i\left[ ,A_{y}^{\dagger }\right]  \label{66}
\end{equation}
where the adjoint is taken for any matrix as the transpose complex
conjugated. The change of the order in the commutators is due to the
property that for any two matrices $R_{1}$ and $R_{2}$ the Hermitian
conjugate of their commutator is 
\begin{eqnarray}
\left[ R_{1},R_{2}\right] ^{\dagger } &=&\left( R_{1}R_{2}-R_{2}R_{1}\right)
^{\dagger }  \label{67} \\
&=&\left( R_{2}^{T}R_{1}^{T}-R_{1}^{T}R_{2}^{T}\right) ^{\ast }  \notag \\
&=&R_{2}^{\dagger }R_{1}^{\dagger }-R_{1}^{\dagger }R_{2}^{\dagger }  \notag
\\
&=&\left[ R_{2}^{\dagger },R_{1}^{\dagger }\right]  \notag
\end{eqnarray}
($^{\ast }$ is complex conjugate and $^{T}$ is the transpose operators) and
we take into account that in the expression of $\phi ^{\dagger }$ we have
already used the Hermitian conjugated matrices of $E_{\pm }$.

The Hermitian conjugates of the gauge field matrices are 
\begin{eqnarray}
A_{x}^{\dagger } &=&\frac{1}{2}\left( a^{\ast }-a\right) H^{\dagger }=\frac{1%
}{2}\left( a^{\ast }-a\right) H  \label{68} \\
A_{y}^{\dagger } &=&-\frac{1}{2i}\left( a^{\ast }+a\right) H^{\dagger }=-%
\frac{1}{2i}\left( a^{\ast }+a\right) H  \notag
\end{eqnarray}
Then 
\begin{equation}
D_{-}^{\dagger }\equiv \frac{\partial }{\partial x}+i\frac{\partial }{%
\partial y}+\frac{1}{2}\left( a^{\ast }-a\right) \left[ ,H\right] -\frac{1}{2%
}\left( a^{\ast }+a\right) \left[ ,H\right]  \label{69}
\end{equation}
We recall that 
\begin{equation}
\phi ^{\dagger }=\phi _{1}^{\ast }E_{-}+\phi _{2}^{\ast }E_{+}  \label{70}
\end{equation}
The we have 
\begin{eqnarray}
\left( D_{-}\phi \right) ^{\dagger } &=&\left\{ \frac{\partial }{\partial x}%
+i\frac{\partial }{\partial y}+\frac{1}{2}\left( a^{\ast }-a\right) \left[ ,H%
\right] -\frac{1}{2}\left( a^{\ast }+a\right) \left[ ,H\right] \right\}
\label{71} \\
&&\times \left( \phi _{1}^{\ast }E_{-}+\phi _{2}^{\ast }E_{+}\right)  \notag
\\
&=&\left( \frac{\partial \phi _{1}^{\ast }}{\partial x}+i\frac{\partial \phi
_{1}^{\ast }}{\partial y}\right) E_{-}+\left( \frac{\partial \phi _{2}^{\ast
}}{\partial x}+i\frac{\partial \phi _{2}^{\ast }}{\partial y}\right) E_{+} 
\notag \\
&&+\frac{1}{2}\left( a^{\ast }-a\right) \phi _{1}^{\ast }\left[ E_{-},H%
\right]  \notag \\
&&+\frac{1}{2}\left( a^{\ast }-a\right) \phi _{2}^{\ast }\left[ E_{+},H%
\right]  \notag \\
&&-\frac{1}{2}\left( a^{\ast }+a\right) \phi _{1}^{\ast }\left[ E_{-},H%
\right]  \notag \\
&&-\frac{1}{2}\left( a^{\ast }+a\right) \phi _{2}^{\ast }\left[ E_{+},H%
\right]  \notag
\end{eqnarray}
or 
\begin{eqnarray}
\left( D_{-}\phi \right) ^{\dagger } &=&2\frac{\partial \phi _{1}^{\ast }}{%
\partial z^{\ast }}E_{-}+2\frac{\partial \phi _{2}^{\ast }}{\partial z^{\ast
}}E_{+}  \label{72} \\
&&+\frac{1}{2}\left( a^{\ast }-a\right) \phi _{1}^{\ast }\left( 2E_{-}\right)
\notag \\
&&+\frac{1}{2}\left( a^{\ast }-a\right) \phi _{2}^{\ast }\left(
-2E_{+}\right)  \notag \\
&&-\frac{1}{2}\left( a^{\ast }+a\right) \phi _{1}^{\ast }\left( 2E_{-}\right)
\notag \\
&&-\frac{1}{2}\left( a^{\ast }+a\right) \phi _{2}^{\ast }\left(
-2E_{+}\right)  \notag
\end{eqnarray}
The equation becomes 
\begin{eqnarray}
\left( D_{-}\phi \right) ^{\dagger } &=&\left( 2\frac{\partial \phi
_{1}^{\ast }}{\partial z^{\ast }}+\left( a^{\ast }-a\right) \phi _{1}^{\ast
}-\left( a^{\ast }+a\right) \phi _{1}^{\ast }\right) E_{-}  \label{73} \\
&&+\left( 2\frac{\partial \phi _{2}^{\ast }}{\partial z^{\ast }}-\left(
a^{\ast }-a\right) \phi _{2}^{\ast }+\left( a^{\ast }+a\right) \phi
_{2}^{\ast }\right) E_{+}  \notag \\
&=&0  \notag
\end{eqnarray}
Here we have made use of the identifications 
\begin{equation}
\frac{\partial }{\partial x}+i\frac{\partial }{\partial y}\equiv 2\frac{%
\partial }{\partial z^{\ast }}  \label{74}
\end{equation}
and 
\begin{equation}
\frac{\partial }{\partial x}-i\frac{\partial }{\partial y}\equiv 2\frac{%
\partial }{\partial z}  \label{75}
\end{equation}
The resulting equations are 
\begin{equation}
2\frac{\partial \phi _{1}^{\ast }}{\partial z^{\ast }}-2a\phi _{1}^{\ast }=0
\label{76}
\end{equation}
and 
\begin{equation}
2\frac{\partial \phi _{2}}{\partial z^{\ast }}+2a\phi _{2}^{\ast }=0
\label{77}
\end{equation}
which represent the adjoints of the first set, Eqs(\ref{64}), as expected.

\subsubsection{Using the two sets of equations}

Now we consider the first equations (\emph{i.e.} those refering to $\phi
_{1} $) in the two sets, Eqs.(\ref{64}) and (\ref{76}) 
\begin{eqnarray}
2\frac{\partial \phi _{1}}{\partial z}-2a^{\ast }\phi _{1} &=&0  \label{78}
\\
2\frac{\partial \phi _{1}^{\ast }}{\partial z^{\ast }}-2a\phi _{1}^{\ast }
&=&0  \notag
\end{eqnarray}
From here we obtain the expressions of $a$ and $a^{\ast }$%
\begin{equation}
a=\frac{\partial }{\partial z^{\ast }}\ln \left( \phi _{1}^{\ast }\right)
\label{79}
\end{equation}
\begin{equation}
a^{\ast }=\frac{\partial }{\partial z}\ln \left( \phi _{1}\right)  \label{80}
\end{equation}
The left hand side of the second self-duality equation (\ref{56}) is 
\begin{eqnarray}
-2\frac{\partial a^{\ast }}{\partial z^{\ast }}-2\frac{\partial a}{\partial z%
} &=&-2\frac{\partial }{\partial z^{\ast }}\frac{\partial }{\partial z}\ln
\left( \phi _{1}\right) -2\frac{\partial }{\partial z}\frac{\partial }{%
\partial z^{\ast }}\ln \left( \phi _{1}^{\ast }\right)  \label{81} \\
&=&-2\frac{\partial ^{2}}{\partial z\partial z^{\ast }}\left[ \ln \left(
\phi _{1}\right) +\ln \left( \phi _{1}^{\ast }\right) \right]  \notag \\
&=&-2\frac{\partial ^{2}}{\partial z\partial z^{\ast }}\ln \left( \left|
\phi _{1}\right| ^{2}\right)  \notag
\end{eqnarray}
In the differential operator we recognize the Laplacean, 
\begin{equation*}
\Delta =4\frac{\partial ^{2}}{\partial z\partial z^{\ast }}
\end{equation*}
Equating the expressions that we have obtained for the left hand side and
respectively for right hand side of the second self-duality equation (\ref
{56}) we obtain 
\begin{equation}
-\frac{1}{2}\Delta \ln \rho _{1}=-\frac{1}{\kappa ^{2}}\left( \rho _{1}-\rho
_{2}\right) \left[ 2\left( \rho _{1}+\rho _{2}\right) -v^{2}\right]
\label{82}
\end{equation}

The second equations (those refering to $\phi _{2}$) in the two sets Eqs.(%
\ref{645}) and (\ref{77}) give the result 
\begin{equation}
a^{\ast }=-\frac{\partial }{\partial z}\ln \phi _{2}  \label{83}
\end{equation}
and 
\begin{equation}
a=-\frac{\partial }{\partial z^{\ast }}\ln \phi _{2}^{\ast }  \label{84}
\end{equation}
from where we obtain the form of the right hand side in the second
self-duality equation, (\ref{56}) 
\begin{eqnarray}
-2\frac{\partial a^{\ast }}{\partial z^{\ast }}-2\frac{\partial a}{\partial z%
} &=&2\frac{\partial }{\partial z^{\ast }}\frac{\partial }{\partial z}\ln
\left( \phi _{2}\right) +2\frac{\partial }{\partial z}\frac{\partial }{%
\partial z^{\ast }}\ln \left( \phi _{2}^{\ast }\right)  \label{85} \\
&=&2\frac{\partial ^{2}}{\partial z\partial z^{\ast }}\left[ \ln \left( \phi
_{2}\right) +\ln \left( \phi _{2}^{\ast }\right) \right]  \notag \\
&=&2\frac{\partial ^{2}}{\partial z\partial z^{\ast }}\ln \left( \left| \phi
_{2}\right| ^{2}\right)  \notag
\end{eqnarray}
The final form is 
\begin{equation}
\frac{1}{2}\Delta \ln \rho _{2}=-\frac{1}{\kappa ^{2}}\left( \rho _{1}-\rho
_{2}\right) \left[ 2\left( \rho _{1}+\rho _{2}\right) -v^{2}\right]
\label{86}
\end{equation}
The right hand side in Eqs.(\ref{82}) and (\ref{86}) is the same and if we
substract the equations we obtain 
\begin{eqnarray}
\Delta \ln \rho _{1}+\Delta \ln \rho _{2} &=&0  \label{87} \\
\Delta \ln \left( \rho _{1}\rho _{2}\right) &=&0  \notag
\end{eqnarray}
The function $\ln \left( \rho _{1}\rho _{2}\right) $ is an arbitrary
harmonic function and this aspect will be discussed later. For the moment we
simply take a constant, convenient for normalization, 
\begin{equation}
\rho _{1}\rho _{2}=v^{4}/16  \label{88}
\end{equation}

With this relation we return to the equation for $\rho _{1}$, (\ref{82}) 
\begin{equation}
-\frac{1}{2}\Delta \ln \rho _{1}=-\frac{1}{\kappa ^{2}}\left( \rho _{1}-%
\frac{v^{4}/16}{\rho _{1}}\right) \left[ 2\left( \rho _{1}+\frac{v^{4}/16}{%
\rho _{1}}\right) -v^{2}\right]  \label{89}
\end{equation}
We add the zero-valued Laplacean of a constant to the left side and
factorise in the right hand side 
\begin{equation}
\frac{1}{2}\Delta \ln \rho _{1}-\frac{1}{2}\Delta \ln \left( v^{2}/4\right)
=4\frac{\left( v^{2}/4\right) ^{2}}{\kappa ^{2}}\left( \frac{\rho _{1}}{%
v^{2}/4}-\frac{v^{2}/4}{\rho _{1}}\right) \left[ \frac{1}{2}\left( \frac{%
\rho _{1}}{v^{2}/4}+\frac{v^{2}/4}{\rho _{1}}\right) -1\right]  \label{893}
\end{equation}
Now we introduce a single variable 
\begin{equation}
\rho \equiv \frac{\rho _{1}}{v^{2}/4}=\frac{v^{2}/4}{\rho _{2}}  \label{894}
\end{equation}
and obtain 
\begin{equation}
\frac{1}{2}\Delta \ln \rho =\frac{1}{4}\left( \frac{v^{2}}{\kappa }\right)
^{2}\left( \rho -\frac{1}{\rho }\right) \left[ \frac{1}{2}\left( \rho +\frac{%
1}{\rho }\right) -1\right]  \label{895}
\end{equation}
We make the substitution 
\begin{equation}
\psi \equiv \ln \rho  \label{90}
\end{equation}
and we obtain 
\begin{eqnarray}
\frac{1}{2}\Delta \psi &=&\frac{1}{4}\left( \frac{v^{2}}{\kappa }\right) ^{2}%
\left[ \exp \left( \psi \right) -\exp \left( -\psi \right) \right]
\label{91} \\
&&\times \left\{ \frac{1}{2}\left[ \exp \left( \psi \right) +\exp \left(
-\psi \right) \right] -1\right\}  \notag \\
&=&\frac{1}{2}\left( \frac{v^{2}}{\kappa }\right) ^{2}\sinh \psi \left(
\cosh \psi -1\right)  \notag
\end{eqnarray}
\begin{equation}
\left( \frac{\kappa }{v^{2}}\right) ^{2}\Delta \psi -\sinh \psi \left( \cosh
\psi -1\right) =0  \label{92}
\end{equation}
Exactly the same equation would have been obtained starting from the one for 
$\rho _{2}$, (\ref{86}) after a change of the unknown function, $\psi
\rightarrow -\psi $.

After normalizing the coordinates by the length $\kappa /v^{2}$, we obtain 
\begin{equation}
\Delta \psi -\sinh \psi \left( \cosh \psi -1\right) =0  \label{94}
\end{equation}

This is the equation governing the stationary states of the CHM equation,
resulting from the first form of the SD equations.

\subsubsection{Calculation of the additional energy for the first version of
the SD equations}

We start from the energy as integral of the density of the Hamiltonian Eq.(%
\ref{hamil}). Since all other terms in the expression of the energy are
positive (they vanish after adopting the self-duality and the particular $%
6^{th}$ order potential), the energy is bounded from below 
\begin{equation}
\mathcal{E}\geqslant \frac{iv^{2}}{2\kappa }\mathrm{tr}\left( \phi ^{\dagger
}\left( D_{0}\phi \right) -\left( D_{0}\phi \right) ^{\dagger }\phi \right)
\label{131}
\end{equation}
where the second of the equations at self-duality Eq.(\ref{sd1}) is 
\begin{equation}
D_{0}\phi =\frac{i}{2\kappa }\left( \left[ \left[ \phi ,\phi ^{\dagger }%
\right] ,\phi \right] -v^{2}\phi \right)  \label{132}
\end{equation}
and 
\begin{equation*}
\left( D_{0}\phi \right) ^{\dagger }=-\frac{i}{2\kappa }\left( \left[ \left[
\phi ,\phi ^{\dagger }\right] ,\phi \right] -v^{2}\phi \right) ^{\dagger }
\end{equation*}
Then we have 
\begin{equation}
\mathcal{E}\geqslant -\frac{v^{2}}{4\kappa ^{2}}\mathrm{tr}\left\{ \phi
^{\dagger }\left( \left[ \left[ \phi ,\phi ^{\dagger }\right] ,\phi \right]
-v^{2}\phi \right) +\left( \left[ \left[ \phi ,\phi ^{\dagger }\right] ,\phi %
\right] -v^{2}\phi \right) ^{\dagger }\phi \right\}  \label{esum}
\end{equation}
and we will prove that the second term in the curly brackets is equal with
the first. 
\begin{eqnarray}
\left( \left[ \left[ \phi ,\phi ^{\dagger }\right] ,\phi \right] -v^{2}\phi
\right) ^{\dagger }\phi &=&\left[ \left[ \phi ,\phi ^{\dagger }\right] ,\phi %
\right] ^{\dagger }\phi -v^{2}\phi ^{\dagger }\phi  \label{133} \\
&=&\left[ \phi ^{\dagger },\left[ \phi ,\phi ^{\dagger }\right] ^{\dagger }%
\right] \phi -v^{2}\phi ^{\dagger }\phi  \notag \\
&=&\left[ \phi ^{\dagger },\left[ \phi ,\phi ^{\dagger }\right] \right] \phi
-v^{2}\phi ^{\dagger }\phi  \notag \\
&=&\left( \phi ^{\dagger }\left[ \phi ,\phi ^{\dagger }\right] -\left[ \phi
,\phi ^{\dagger }\right] \phi ^{\dagger }\right) \phi -v^{2}\phi ^{\dagger
}\phi  \notag \\
&=&\phi ^{\dagger }\left[ \phi ,\phi ^{\dagger }\right] \phi -\underline{%
\left[ \phi ,\phi ^{\dagger }\right] \phi ^{\dagger }\phi }-v^{2}\phi
^{\dagger }\phi  \notag
\end{eqnarray}
We can apply in the second term (underlined), the cyclic symmetry of the $%
\mathrm{tr}$ operator, moving successively the factors $\phi $ and $\phi
^{\dagger }$ in the first position and have 
\begin{eqnarray}
&&\mathrm{tr}\left\{ \left( \left[ \left[ \phi ,\phi ^{\dagger }\right]
,\phi \right] -v^{2}\phi \right) ^{\dagger }\phi \right\}  \label{134} \\
&=&\mathrm{tr}\left\{ \phi ^{\dagger }\left[ \phi ,\phi ^{\dagger }\right]
\phi -\phi ^{\dagger }\phi \left[ \phi ,\phi ^{\dagger }\right] -v^{2}\phi
^{\dagger }\phi \right\}  \notag \\
&=&\mathrm{tr}\left\{ \phi ^{\dagger }\left[ \left[ \phi ,\phi ^{\dagger }%
\right] ,\phi \right] -v^{2}\phi ^{\dagger }\phi \right\}  \notag \\
&=&\mathrm{tr}\left\{ \phi ^{\dagger }\left( \left[ \left[ \phi ,\phi
^{\dagger }\right] ,\phi \right] -v^{2}\phi \right) \right\}  \notag
\end{eqnarray}
and the equality with the first term in Eq.(\ref{esum}) is proved. It
results 
\begin{equation}
\mathcal{E}\geqslant -\frac{v^{2}}{2\kappa ^{2}}\mathrm{tr}\left\{ \phi
^{\dagger }\left( \left[ \left[ \phi ,\phi ^{\dagger }\right] ,\phi \right]
-v^{2}\phi \right) \right\}  \label{135}
\end{equation}
but at self-duality (since we have already used the equations derived from
self-duality) the limit is saturated 
\begin{equation}
\mathcal{E}_{SD}=\frac{v^{2}}{2\kappa ^{2}}\mathrm{tr}\left\{ \phi ^{\dagger
}\left( v^{2}\phi -\left[ \left[ \phi ,\phi ^{\dagger }\right] ,\phi \right]
\right) \right\}  \label{136}
\end{equation}

We can obtain the explicit formula using the algebraic representation of the
fields 
\begin{eqnarray}
\phi &=&\phi _{1}E_{+}+\phi _{2}E_{-}  \label{137} \\
\phi ^{\dagger } &=&\phi _{1}^{\ast }E_{-}+\phi _{2}^{\ast }E_{+}  \notag
\end{eqnarray}
and recall the previous result 
\begin{equation}
v^{2}\phi -\left[ \left[ \phi ,\phi ^{\dagger }\right] ,\phi \right]
=PE_{+}+QE_{-}  \label{138}
\end{equation}
where 
\begin{eqnarray}
P &\equiv &v^{2}\phi _{1}-2\left( \rho _{1}-\rho _{2}\right) \phi _{1}
\label{139} \\
Q &\equiv &v^{2}\phi _{2}+2\left( \rho _{1}-\rho _{2}\right) \phi _{2} 
\notag
\end{eqnarray}

A detailed calculation, starting from Eq.(\ref{136}): 
\begin{eqnarray}
\mathcal{E}_{SD} &=&\frac{v^{2}}{2\kappa ^{2}}\mathrm{tr}\left( \phi
^{\dagger }\left( v^{2}\phi -\left[ \left[ \phi ,\phi ^{\dagger }\right]
,\phi \right] \right) \right)  \label{143} \\
&=&\frac{v^{2}}{2\kappa ^{2}}\mathrm{tr}\left\{ \left( \phi _{1}^{\ast
}E_{-}+\phi _{2}^{\ast }E_{+}\right) \left( PE_{+}+QE_{-}\right) \right\} 
\notag \\
&=&\frac{v^{2}}{2\kappa ^{2}}\mathrm{tr}\left\{ \phi _{1}^{\ast
}PE_{-}E_{+}+\phi _{1}^{\ast }QE_{-}E_{-}+\phi _{2}^{\ast }PE_{+}E_{+}+\phi
_{2}^{\ast }QE_{+}E_{-}\right\}  \notag
\end{eqnarray}
We calculate separately 
\begin{eqnarray}
\mathrm{tr}\left( \phi _{1}^{\ast }PE_{-}E_{+}\right) &=&\mathrm{tr}\left[
\phi _{1}^{\ast }\left( v^{2}\phi _{1}-2\left( \rho _{1}-\rho _{2}\right)
\phi _{1}\right) E_{-}E_{+}\right]  \label{144} \\
&=&\left[ v^{2}\rho _{1}-2\left( \rho _{1}-\rho _{2}\right) \rho _{1}\right] 
\mathrm{tr}\left( 
\begin{array}{cc}
0 & 0 \\ 
0 & 1
\end{array}
\right)  \notag \\
&=&v^{2}\rho _{1}-2\left( \rho _{1}-\rho _{2}\right) \rho _{1}  \notag
\end{eqnarray}
\begin{eqnarray}
\mathrm{tr}\left( \phi _{1}^{\ast }QE_{-}E_{-}\right) &=&\mathrm{tr}\left[
\phi _{1}^{\ast }\left( v^{2}\phi _{2}+2\left( \rho _{1}-\rho _{2}\right)
\phi _{2}\right) E_{-}E_{-}\right]  \label{145} \\
&=&\phi _{1}^{\ast }\left( v^{2}\phi _{2}+2\left( \rho _{1}-\rho _{2}\right)
\phi _{2}\right) \mathrm{tr}\left( 
\begin{array}{cc}
0 & 0 \\ 
0 & 0
\end{array}
\right)  \notag \\
&=&0  \notag
\end{eqnarray}
\begin{eqnarray}
\mathrm{tr}\left( \phi _{2}^{\ast }PE_{+}E_{+}\right) &=&\mathrm{tr}\left[
\phi _{2}^{\ast }\left( v^{2}\phi _{1}-2\left( \rho _{1}-\rho _{2}\right)
\phi _{1}\right) E_{+}E_{+}\right]  \label{146} \\
&=&\phi _{2}^{\ast }\left( v^{2}\phi _{1}-2\left( \rho _{1}-\rho _{2}\right)
\phi _{1}\right) \mathrm{tr}\left( 
\begin{array}{cc}
0 & 0 \\ 
0 & 0
\end{array}
\right)  \notag \\
&=&0  \notag
\end{eqnarray}
\begin{eqnarray}
\mathrm{tr}\left( \phi _{2}^{\ast }QE_{+}E_{-}\right) &=&\mathrm{tr}\left[
\phi _{2}^{\ast }\left( v^{2}\phi _{2}+2\left( \rho _{1}-\rho _{2}\right)
\phi _{2}\right) E_{+}E_{-}\right]  \label{147} \\
&=&\left[ v^{2}\rho _{2}+2\left( \rho _{1}-\rho _{2}\right) \rho _{2}\right] 
\mathrm{tr}\left( 
\begin{array}{cc}
1 & 0 \\ 
0 & 0
\end{array}
\right)  \notag \\
&=&v^{2}\rho _{2}+2\left( \rho _{1}-\rho _{2}\right) \rho _{2}  \notag
\end{eqnarray}
Summing up the contributions 
\begin{eqnarray}
\mathcal{E}_{SD} &=&\frac{v^{2}}{2\kappa ^{2}}\left[ v^{2}\rho _{1}-2\left(
\rho _{1}-\rho _{2}\right) \rho _{1}+v^{2}\rho _{2}+2\left( \rho _{1}-\rho
_{2}\right) \rho _{2}\right]  \label{148} \\
&=&\frac{v^{2}}{2\kappa ^{2}}\left[ v^{2}\left( \rho _{1}+\rho _{2}\right)
-2\left( \rho _{1}-\rho _{2}\right) ^{2}\right]  \notag
\end{eqnarray}
Expressed in the normalised variable, the energy is 
\begin{eqnarray}
\mathcal{E}_{SD} &=&\frac{v^{2}}{2\kappa ^{2}}\left[ v^{2}\left( \rho
_{1}+\rho _{2}\right) -2\left( \rho _{1}-\rho _{2}\right) ^{2}\right]
\label{1588} \\
&=&\frac{v^{2}}{2\kappa ^{2}}\left[ v^{2}\left( \rho \frac{v^{2}}{4}+\frac{%
v^{2}}{4}\frac{1}{\rho }\right) -2\left( \rho \frac{v^{2}}{4}-\frac{v^{2}}{4}%
\frac{1}{\rho }\right) ^{2}\right]  \notag \\
&=&\frac{v^{2}}{2\kappa ^{2}}4\left( \frac{v^{2}}{4}\right) ^{2}\left[ \rho +%
\frac{1}{\rho }-\frac{1}{2}\left( \rho -\frac{1}{\rho }\right) ^{2}\right] 
\notag \\
&=&-\frac{v^{2}}{8}\frac{1}{\rho _{s}^{2}}\left[ \frac{1}{2}\left( \rho -%
\frac{1}{\rho }\right) ^{2}-\left( \rho +\frac{1}{\rho }\right) \right] 
\notag
\end{eqnarray}
Introducing the streamfunction $\rho \equiv \exp \left( \psi \right) $ we
get 
\begin{eqnarray}
\mathcal{E}_{SD} &=&-\frac{v^{2}}{8}\frac{1}{\rho _{s}^{2}}\left[ \frac{1}{2}%
\left( \rho -\frac{1}{\rho }\right) ^{2}-\left( \rho +\frac{1}{\rho }\right) 
\right]  \label{15884} \\
&=&-\frac{v^{2}}{8}\frac{1}{\rho _{s}^{2}}\left[ 2\left( \sinh \psi \right)
^{2}-2\cosh \psi \right]  \notag \\
&=&-\frac{v^{2}}{4}\frac{1}{\rho _{s}^{2}}\left[ \left( \cosh \psi \right)
^{2}-\cosh \psi -1\right]  \notag
\end{eqnarray}
or 
\begin{equation}
\mathcal{E}_{SD}=v^{2}\frac{1}{\rho _{s}^{2}}\frac{1}{4}\left[ -\left( \cosh
\psi \right) ^{2}+\cosh \psi +1\right]  \label{15885}
\end{equation}
We note that this expression must be integrated over the plane (the factor $%
1/\rho _{s}^{2}$ will ensure the correct dimension) and the dimension of the 
\emph{energy} is actually given by $v^{2}\equiv \Omega _{ci}$.

\subsection{Using the algebraic ansatz in the second version of the SD
equations}

We now turn to the second version of the SD equations (\ref{362}) and
introduce the algebraic ansatz. Then the second equation of the second
version of the Self-Duality becomes 
\begin{eqnarray}
F_{+-} &=&-\frac{1}{\kappa ^{2}}\left[ v^{2}\phi -\left[ \left[ \phi ,\phi
^{\dagger }\right] ,\phi \right] ,\phi ^{\dagger }\right]  \label{451} \\
&=&-\frac{1}{\kappa ^{2}}\left[ v^{2}-2\left( \rho _{1}+\rho _{2}\right) %
\right] \left( \rho _{1}-\rho _{2}\right) H  \notag
\end{eqnarray}
Using Eq.(\ref{42}) 
\begin{equation}
-\frac{\partial a^{\ast }}{\partial x_{+}}-\frac{\partial a}{\partial x_{-}}%
=-\frac{1}{\kappa ^{2}}\left( \rho _{1}-\rho _{2}\right) \left[
v^{2}-2\left( \rho _{1}+\rho _{2}\right) \right]  \label{452}
\end{equation}

From the first equation of self duality, which is common to the two choices 
\begin{equation}
a=\frac{\partial }{\partial z^{\ast }}\ln \left( \phi _{1}^{\ast }\right)
\label{453}
\end{equation}
\begin{equation}
a^{\ast }=\frac{\partial }{\partial z}\ln \left( \phi _{1}\right)
\label{454}
\end{equation}
The left hand side of the second self-duality equation (\ref{56}) is 
\begin{eqnarray}
-2\frac{\partial a^{\ast }}{\partial z^{\ast }}-2\frac{\partial a}{\partial z%
} &=&-2\frac{\partial }{\partial z^{\ast }}\frac{\partial }{\partial z}\ln
\left( \phi _{1}\right) -2\frac{\partial }{\partial z}\frac{\partial }{%
\partial z^{\ast }}\ln \left( \phi _{1}^{\ast }\right)  \label{455} \\
&=&-2\frac{\partial ^{2}}{\partial z\partial z^{\ast }}\left[ \ln \left(
\phi _{1}\right) +\ln \left( \phi _{1}^{\ast }\right) \right]  \notag \\
&=&-2\frac{\partial ^{2}}{\partial z\partial z^{\ast }}\ln \left( \left|
\phi _{1}\right| ^{2}\right)  \notag
\end{eqnarray}
Since the Laplace operator is defined as 
\begin{equation}
\Delta =4\frac{\partial ^{2}}{\partial z\partial z^{\ast }}  \label{456}
\end{equation}
we get 
\begin{equation}
-\frac{1}{2}\Delta \ln \left( \left| \phi _{1}\right| ^{2}\right) =-\frac{1}{%
\kappa ^{2}}\left( \rho _{1}-\rho _{2}\right) \left[ v^{2}-2\left( \rho
_{1}+\rho _{2}\right) \right]  \label{457}
\end{equation}
or 
\begin{equation}
\frac{1}{2}\Delta \ln \rho _{1}=\frac{1}{\kappa ^{2}}\left( \rho _{1}-\rho
_{2}\right) \left[ v^{2}-2\left( \rho _{1}+\rho _{2}\right) \right]
\label{458}
\end{equation}
and we can now replace 
\begin{equation}
\rho \equiv \frac{\rho _{1}}{v^{2}/4}=\frac{v^{2}/4}{\rho _{2}}  \label{459}
\end{equation}
\begin{eqnarray}
\frac{1}{2}\Delta \ln \rho &=&\frac{1}{\kappa ^{2}}\left( \frac{v^{2}}{4}%
\right) ^{2}\left( \rho -\frac{1}{\rho }\right) \left[ 4-2\left( \rho +\frac{%
1}{\rho }\right) \right]  \label{460} \\
\frac{1}{2}\Delta \ln \rho &=&\frac{v^{4}}{2\kappa ^{2}}\left( \frac{1}{2}%
\right) \left( \rho -\frac{1}{\rho }\right) \left[ 1-\frac{1}{2}\left( \rho +%
\frac{1}{\rho }\right) \right]  \notag
\end{eqnarray}
and introduce the streamfunction $\psi $%
\begin{equation}
\rho =\exp \left( \psi \right)  \label{461}
\end{equation}
\begin{equation}
\frac{1}{2}\Delta \psi =\frac{1}{2}\left( \frac{v^{2}}{\kappa }\right)
^{2}\sinh \psi \left( 1-\cosh \psi \right)  \label{462}
\end{equation}
or 
\begin{equation}
\Delta \psi +\left( \frac{v^{2}}{\kappa }\right) ^{2}\sinh \psi \left( \cosh
\psi -1\right) =0  \label{463}
\end{equation}
The unit of space is 
\begin{equation}
\frac{1}{\rho _{s}}=\frac{v^{2}}{\kappa }  \label{464}
\end{equation}
and the equation results 
\begin{equation}
\Delta \psi +\sinh \psi \left( \cosh \psi -1\right) =0  \label{465}
\end{equation}

All the other calculations, in particular those implying the function $\phi
_{2}$ and the complex conjugated, $\phi _{1}^{\ast }$ and $\phi _{2}^{\ast }$
are similar to the calculations made for the first version of the SD
equations.

\subsubsection{Calculation of the additional energy for the second version
of the self-duality}

The additional term in the Bogomolnyi form of the energy, in the second
version, Eq.(\ref{356}) consists of three contributions. The first
contribution is 
\begin{equation}
\mathcal{E}^{a\left( 1\right) }\equiv \frac{i}{\kappa }\mathrm{tr}\left\{ %
\left[ \left[ \phi ,\phi ^{\dagger }\right] ,\phi \right] ^{\dagger }\left(
D_{0}\phi \right) \right\}  \label{466}
\end{equation}
and we use the previously derived expression 
\begin{equation}
\left[ \left[ \phi ,\phi ^{\dagger }\right] ,\phi \right] =2\left( \rho
_{1}-\rho _{2}\right) \left( \phi _{1}E_{+}-\phi _{2}E_{-}\right)
\label{467}
\end{equation}
and 
\begin{equation}
\left[ \left[ \phi ,\phi ^{\dagger }\right] ,\phi \right] ^{\dagger
}=2\left( \rho _{1}-\rho _{2}\right) \left( \phi _{1}^{\ast }E_{-}-\phi
_{2}^{\ast }E_{+}\right)  \label{468}
\end{equation}
Also we use the following relation 
\begin{equation}
v^{2}\phi -\left[ \left[ \phi ,\phi ^{\dagger }\right] ,\phi \right]
=PE_{+}+QE_{-}  \label{469}
\end{equation}
where 
\begin{eqnarray}
P &\equiv &v^{2}\phi _{1}-2\left( \rho _{1}-\rho _{2}\right) \phi _{1}
\label{470} \\
Q &\equiv &v^{2}\phi _{2}+2\left( \rho _{1}-\rho _{2}\right) \phi _{2} 
\notag
\end{eqnarray}
Using the second (new) equation of self-duality we have 
\begin{eqnarray}
\mathcal{E}^{a\left( 1\right) } &\equiv &\frac{i}{\kappa }\mathrm{tr}\left\{ %
\left[ \left[ \phi ,\phi ^{\dagger }\right] ,\phi \right] ^{\dagger }\left(
D_{0}\phi \right) \right\}  \label{471} \\
&=&\frac{i}{\kappa }\mathrm{tr}\left\{ 2\left( \rho _{1}-\rho _{2}\right)
\left( \phi _{1}^{\ast }E_{-}-\phi _{2}^{\ast }E_{+}\right) \left( -\frac{i}{%
2\kappa }\left( \left[ \left[ \phi ,\phi ^{\dagger }\right] ,\phi \right]
-v^{2}\phi \right) \right) \right\}  \notag \\
&=&\frac{1}{2\kappa ^{2}}2\left( \rho _{1}-\rho _{2}\right) \mathrm{tr}%
\left\{ \left( \phi _{1}^{\ast }E_{-}-\phi _{2}^{\ast }E_{+}\right) \left(
-PE_{+}-QE_{-}\right) \right\}  \notag
\end{eqnarray}
The trace is 
\begin{eqnarray}
&&\mathrm{tr}\left\{ \left( \phi _{1}^{\ast }E_{-}-\phi _{2}^{\ast
}E_{+}\right) \left( -PE_{+}-QE_{-}\right) \right\}  \label{472} \\
&=&\phi _{1}^{\ast }\left( -P\right) \mathrm{tr}\left( E_{-}E_{+}\right) \;\
\left( \text{trace is }1\right)  \notag \\
&&+\left( -\phi _{2}^{\ast }\right) \left( -P\right) \mathrm{tr}\left(
E_{+}E_{+}\right) \;\;\left( \text{trace is }0\right)  \notag \\
&&+\phi _{1}^{\ast }\left( -Q\right) \mathrm{tr}\left( E_{-}E_{-}\right)
\;\;\left( \text{trace is }0\right)  \notag \\
&&+\left( -\phi _{2}^{\ast }\right) \left( -Q\right) \mathrm{tr}\left(
E_{+}E_{-}\right) \;\;\left( \text{trace is }1\right)  \notag \\
&=&\phi _{1}^{\ast }\left( -P\right) +\left( -\phi _{2}^{\ast }\right)
\left( -Q\right)  \notag \\
&=&-\phi _{1}^{\ast }\left[ v^{2}\phi _{1}-2\left( \rho _{1}-\rho
_{2}\right) \phi _{1}\right] +\phi _{2}^{\ast }\left[ v^{2}\phi _{2}+2\left(
\rho _{1}-\rho _{2}\right) \phi _{2}\right]  \notag \\
&=&-\rho _{1}\left[ v^{2}-2\left( \rho _{1}-\rho _{2}\right) \right] +\rho
_{2}\left[ v^{2}+2\left( \rho _{1}-\rho _{2}\right) \right]  \notag \\
&=&-v^{2}\left( \rho _{1}-\rho _{2}\right) +2\left( \rho _{1}-\rho
_{2}\right) \left( \rho _{1}+\rho _{2}\right)  \notag \\
&=&-\left( \rho _{1}-\rho _{2}\right) \left[ v^{2}-2\left( \rho _{1}+\rho
_{2}\right) \right]  \notag
\end{eqnarray}
and the first contribution to the residual energy becomes 
\begin{eqnarray}
\mathcal{E}^{a\left( 1\right) } &=&\frac{1}{\kappa ^{2}}\left( \rho
_{1}-\rho _{2}\right) \left\{ -\left( \rho _{1}-\rho _{2}\right) \left[
v^{2}-2\left( \rho _{1}+\rho _{2}\right) \right] \right\}  \label{473} \\
&=&-\frac{1}{\kappa ^{2}}\left( \rho _{1}-\rho _{2}\right) ^{2}\left[
v^{2}-2\left( \rho _{1}+\rho _{2}\right) \right]  \notag
\end{eqnarray}

\bigskip

Now we calculate the second contribution to the residual energy 
\begin{eqnarray}
\mathcal{E}^{a\left( 2\right) } &=&-\frac{i}{\kappa }\mathrm{tr}\left\{
\left( D_{0}\phi \right) ^{\dagger }\left[ \left[ \phi ,\phi ^{\dagger }%
\right] ,\phi \right] \right\}  \label{474} \\
&=&-\frac{i}{\kappa }\mathrm{tr}\left\{ \left( D_{0}\phi \right) ^{\dagger
}2\left( \rho _{1}-\rho _{2}\right) \left( \phi _{1}E_{+}-\phi
_{2}E_{-}\right) \right\}  \notag
\end{eqnarray}
and we replace, according to the new second SD equation 
\begin{eqnarray}
\left( D_{0}\phi \right) ^{\dagger } &=&\left( -\frac{i}{2\kappa }\right)
^{\ast }\left( \left[ \left[ \phi ,\phi ^{\dagger }\right] ,\phi \right]
-v^{2}\phi \right) ^{\dagger }  \label{475} \\
&=&\frac{i}{2\kappa }\left( -PE_{+}-QE_{-}\right) ^{\dagger }  \notag \\
&=&-\frac{i}{2\kappa }\left( P^{\ast }E_{-}+Q^{\ast }E_{+}\right)  \notag
\end{eqnarray}
and replacing in the previous equation 
\begin{eqnarray}
\mathcal{E}^{a\left( 2\right) } &=&-\frac{i}{\kappa }\mathrm{tr}\left\{
\left( D_{0}\phi \right) ^{\dagger }2\left( \rho _{1}-\rho _{2}\right)
\left( \phi _{1}E_{+}-\phi _{2}E_{-}\right) \right\}  \label{476} \\
&=&-\frac{i}{\kappa }\mathrm{tr}\left\{ \left( -\frac{i}{2\kappa }\left(
P^{\ast }E_{-}+Q^{\ast }E_{+}\right) \right) 2\left( \rho _{1}-\rho
_{2}\right) \left( \phi _{1}E_{+}-\phi _{2}E_{-}\right) \right\}  \notag \\
&=&-\frac{1}{2\kappa ^{2}}2\left( \rho _{1}-\rho _{2}\right) \mathrm{tr}%
\left\{ \left( P^{\ast }E_{-}+Q^{\ast }E_{+}\right) \left( \phi
_{1}E_{+}-\phi _{2}E_{-}\right) \right\}  \notag
\end{eqnarray}
The trace is calculated separately 
\begin{eqnarray}
&&\mathrm{tr}\left\{ \left( P^{\ast }E_{-}+Q^{\ast }E_{+}\right) \left( \phi
_{1}E_{+}-\phi _{2}E_{-}\right) \right\}  \label{477} \\
&=&P^{\ast }\phi _{1}\mathrm{tr}\left\{ E_{-}E_{+}\right\} \;\;\left( \text{%
trace is }1\right)  \notag \\
&&+Q^{\ast }\phi _{1}\mathrm{tr}\left\{ E_{+}E_{+}\right\} \;\;\left( \text{%
trace is }0\right)  \notag \\
&&+P^{\ast }\left( -\phi _{2}\right) \mathrm{tr}\left\{ E_{-}E_{-}\right\}
\;\;\left( \text{trace is }0\right)  \notag \\
&&+Q^{\ast }\left( -\phi _{2}\right) \mathrm{tr}\left\{ E_{+}E_{-}\right\}
\;\;\left( \text{trace is }1\right)  \notag \\
&=&P^{\ast }\phi _{1}+Q^{\ast }\left( -\phi _{2}\right)  \notag \\
&=&\left[ v^{2}\phi _{1}-2\left( \rho _{1}-\rho _{2}\right) \phi _{1}\right]
^{\ast }\phi _{1}-\left[ v^{2}\phi _{2}+2\left( \rho _{1}-\rho _{2}\right)
\phi _{2}\right] ^{\ast }\phi _{2}  \notag \\
&=&\rho _{1}\left[ v^{2}-2\left( \rho _{1}-\rho _{2}\right) \right] -\rho
_{2}\left[ v^{2}+2\left( \rho _{1}-\rho _{2}\right) \right]  \notag \\
&=&v^{2}\left( \rho _{1}-\rho _{2}\right) -2\left( \rho _{1}-\rho
_{2}\right) \left( \rho _{1}+\rho _{2}\right)  \notag \\
&=&\left( \rho _{1}-\rho _{2}\right) \left[ v^{2}-2\left( \rho _{1}+\rho
_{2}\right) \right]  \notag
\end{eqnarray}
and the second contribution is 
\begin{eqnarray}
\mathcal{E}^{a\left( 2\right) } &=&-\frac{1}{\kappa ^{2}}\left( \rho
_{1}-\rho _{2}\right) \left( \rho _{1}-\rho _{2}\right) \left[ v^{2}-2\left(
\rho _{1}+\rho _{2}\right) \right]  \label{478} \\
&=&-\frac{1}{\kappa ^{2}}\left( \rho _{1}-\rho _{2}\right) ^{2}\left[
v^{2}-2\left( \rho _{1}+\rho _{2}\right) \right]  \notag
\end{eqnarray}
We note that the two contributions $\mathcal{E}^{a\left( 1\right) }$ and $%
\mathcal{E}^{a\left( 2\right) }$ are equal.

\bigskip

The third contribution is 
\begin{equation}
\mathcal{E}^{a\left( 3\right) }=-\frac{iv^{2}}{2\kappa }\mathrm{tr}\left\{
\phi ^{\dagger }\left( D_{0}\phi \right) -\left( D_{0}\phi \right) ^{\dagger
}\phi \right\}  \label{479}
\end{equation}
We can use the new second SD equation 
\begin{equation}
D_{0}\phi =-\frac{i}{2\kappa }\left( \left[ \left[ \phi ,\phi ^{\dagger }%
\right] ,\phi \right] -v^{2}\phi \right)  \label{480}
\end{equation}
together with 
\begin{equation}
v^{2}\phi -\left[ \left[ \phi ,\phi ^{\dagger }\right] ,\phi \right]
=PE_{+}+QE_{-}  \label{481}
\end{equation}
We find, as in previous cases, 
\begin{eqnarray}
D_{0}\phi &=&-\frac{i}{2\kappa }\left( -PE_{+}-QE_{-}\right)  \label{482} \\
&=&\frac{i}{2\kappa }\left( PE_{+}+QE_{-}\right)  \notag
\end{eqnarray}
and 
\begin{eqnarray}
\left( D_{0}\phi \right) ^{\dagger } &=&\frac{-i}{2\kappa }\left(
PE_{+}+QE_{-}\right) ^{\dagger }  \label{483} \\
&=&\frac{-i}{2\kappa }\left( P^{\ast }E_{-}+Q^{\ast }E_{+}\right)  \notag
\end{eqnarray}
Then 
\begin{eqnarray}
\mathcal{E}^{a\left( 3\right) } &=&-\frac{iv^{2}}{2\kappa }\mathrm{tr}%
\left\{ \phi ^{\dagger }\left( D_{0}\phi \right) -\left( D_{0}\phi \right)
^{\dagger }\phi \right\}  \label{484} \\
&=&-\frac{iv^{2}}{2\kappa }\mathrm{tr}\left\{ \phi ^{\dagger }\frac{i}{%
2\kappa }\left( PE_{+}+QE_{-}\right) -\frac{-i}{2\kappa }\left( P^{\ast
}E_{-}+Q^{\ast }E_{+}\right) \phi \right\}  \notag \\
&=&-\frac{iv^{2}}{2\kappa }\left( \frac{i}{2\kappa }\right) \mathrm{tr}%
\left\{ \left( \phi _{1}^{\ast }E_{-}+\phi _{2}^{\ast }E_{+}\right) \left(
PE_{+}+QE_{-}\right) +\left( P^{\ast }E_{-}+Q^{\ast }E_{+}\right) \left(
\phi _{1}E_{+}+\phi _{2}E_{-}\right) \right\}  \notag \\
&=&\frac{v^{2}}{4\kappa ^{2}}\left\{ \phi _{1}^{\ast }P\mathrm{tr}\left(
E_{-}E_{+}\right) \right. \;\;\left( \text{trace is }1\right)  \notag \\
&&+\phi _{1}^{\ast }Q\mathrm{tr}\left( E_{-}E_{-}\right) \;\;\left( \text{%
trace is }0\right)  \notag \\
&&+\phi _{2}^{\ast }P\mathrm{tr}\left( E_{+}E_{+}\right) \;\;\left( \text{%
trace is }0\right)  \notag \\
&&+\phi _{2}^{\ast }Q\mathrm{tr}\left( E_{+}E_{-}\right) \;\;\left( \text{%
trace is }1\right)  \notag \\
&&+P^{\ast }\phi _{1}\mathrm{tr}\left( E_{-}E_{+}\right) \;\;\left( \text{%
trace is }1\right)  \notag \\
&&+Q^{\ast }\phi _{1}\mathrm{tr}\left( E_{+}E_{+}\right) \;\;\left( \text{%
trace is }0\right)  \notag \\
&&+P^{\ast }\phi _{2}\mathrm{tr}\left( E_{-}E_{-}\right) \;\;\left( \text{%
trace is }0\right)  \notag \\
&&+Q^{\ast }\phi _{2}\mathrm{tr}\left( E_{+}E_{-}\right) \;\;\left( \text{%
trace is }1\right)  \notag
\end{eqnarray}
We obtain 
\begin{eqnarray}
\mathcal{E}^{a\left( 3\right) } &=&\frac{v^{2}}{4\kappa ^{2}}\left\{ \phi
_{1}^{\ast }P+P^{\ast }\phi _{1}+\phi _{2}^{\ast }Q+Q^{\ast }\phi
_{2}\right\}  \label{485} \\
&=&\frac{v^{2}}{2\kappa ^{2}}\left\{ \phi _{1}^{\ast }\left[ v^{2}\phi
_{1}-2\left( \rho _{1}-\rho _{2}\right) \phi _{1}\right] +\phi _{2}^{\ast }%
\left[ v^{2}\phi _{2}+2\left( \rho _{1}-\rho _{2}\right) \phi _{2}\right]
\right\}  \notag \\
&=&\frac{v^{2}}{2\kappa ^{2}}\left\{ \rho _{1}\left[ v^{2}-2\left( \rho
_{1}-\rho _{2}\right) \right] +\rho _{2}\left[ v^{2}+2\left( \rho _{1}-\rho
_{2}\right) \right] \right\}  \notag \\
&=&\frac{v^{2}}{2\kappa ^{2}}\left[ v^{2}\left( \rho _{1}+\rho _{2}\right)
-2\left( \rho _{1}-\rho _{2}\right) ^{2}\right]  \notag
\end{eqnarray}
\begin{equation}
\mathcal{E}^{a\left( 3\right) }=\frac{v^{2}}{2\kappa ^{2}}\left[ v^{2}\left(
\rho _{1}+\rho _{2}\right) -2\left( \rho _{1}-\rho _{2}\right) ^{2}\right]
\label{486}
\end{equation}

\bigskip

Now we collect all results 
\begin{eqnarray}
\mathcal{E}^{a} &=&\mathcal{E}^{a\left( 1\right) }+\mathcal{E}^{a\left(
2\right) }+\mathcal{E}^{a\left( 3\right) }  \label{487} \\
&=&-\frac{1}{\kappa ^{2}}\left( \rho _{1}-\rho _{2}\right) ^{2}\left[
v^{2}-2\left( \rho _{1}+\rho _{2}\right) \right]  \notag \\
&&-\frac{1}{\kappa ^{2}}\left( \rho _{1}-\rho _{2}\right) ^{2}\left[
v^{2}-2\left( \rho _{1}+\rho _{2}\right) \right]  \notag \\
&&+\frac{v^{2}}{2\kappa ^{2}}\left[ v^{2}\left( \rho _{1}+\rho _{2}\right)
-2\left( \rho _{1}-\rho _{2}\right) ^{2}\right]  \notag
\end{eqnarray}
there are three powers of $v$ and we separate the coefficients 
\begin{eqnarray}
v^{0} &:&  \label{488} \\
&&\frac{1}{\kappa ^{2}}\left( \rho _{1}-\rho _{2}\right) ^{2}2\left( \rho
_{1}+\rho _{2}\right) +\frac{1}{\kappa ^{2}}\left( \rho _{1}-\rho
_{2}\right) ^{2}2\left( \rho _{1}+\rho _{2}\right)  \notag \\
&=&\frac{4}{\kappa ^{2}}\left( \rho _{1}-\rho _{2}\right) ^{2}\left( \rho
_{1}+\rho _{2}\right)  \notag
\end{eqnarray}
\begin{eqnarray}
v^{2} &:&  \label{489} \\
&&-\frac{1}{\kappa ^{2}}\left( \rho _{1}-\rho _{2}\right) ^{2}-\frac{1}{%
\kappa ^{2}}\left( \rho _{1}-\rho _{2}\right) ^{2}-\frac{1}{2\kappa ^{2}}%
2\left( \rho _{1}-\rho _{2}\right) ^{2}  \notag \\
&=&-\frac{3}{\kappa ^{2}}\left( \rho _{1}-\rho _{2}\right) ^{2}  \notag
\end{eqnarray}
\begin{eqnarray}
v^{4} &:&  \label{490} \\
&&\frac{1}{2\kappa ^{2}}\left( \rho _{1}+\rho _{2}\right)  \notag
\end{eqnarray}
and the total expression is 
\begin{equation}
\mathcal{E}^{a}=\frac{4}{\kappa ^{2}}\left( \rho _{1}-\rho _{2}\right)
^{2}\left( \rho _{1}+\rho _{2}\right) -\frac{3v^{2}}{\kappa ^{2}}\left( \rho
_{1}-\rho _{2}\right) ^{2}+\frac{v^{4}}{2\kappa ^{2}}\left( \rho _{1}+\rho
_{2}\right)  \label{491}
\end{equation}
or 
\begin{equation}
\mathcal{E}^{a}=\frac{1}{\kappa ^{2}}\left( \rho _{1}-\rho _{2}\right) ^{2}%
\left[ 4\left( \rho _{1}+\rho _{2}\right) -3v^{2}\right] +\frac{v^{4}}{%
2\kappa ^{2}}\left( \rho _{1}+\rho _{2}\right)  \label{492}
\end{equation}
Introducing the normalization 
\begin{equation}
\rho \equiv \frac{\rho _{1}}{v^{2}/4}=\frac{v^{2}/4}{\rho _{2}}  \label{493}
\end{equation}
\begin{eqnarray}
\mathcal{E}^{a} &=&\frac{1}{\kappa ^{2}}\left( \frac{v^{2}}{4}\right)
^{2}\left( \rho -\frac{1}{\rho }\right) ^{2}\left( \frac{v^{2}}{4}\right) %
\left[ 4\left( \rho +\frac{1}{\rho }\right) -12\right]  \label{494} \\
&&+\frac{v^{4}}{2\kappa ^{2}}\left( \frac{v^{2}}{4}\right) \left( \rho +%
\frac{1}{\rho }\right)  \notag
\end{eqnarray}
\begin{eqnarray}
\mathcal{E}^{a} &=&\frac{v^{6}}{16\kappa ^{2}}\left( \rho -\frac{1}{\rho }%
\right) ^{2}\left[ \left( \rho +\frac{1}{\rho }\right) -3\right]  \label{495}
\\
&&+\frac{v^{6}}{8\kappa ^{2}}\left( \rho +\frac{1}{\rho }\right)  \notag
\end{eqnarray}
\begin{eqnarray}
\mathcal{E}^{a} &=&\frac{v^{6}}{4\kappa ^{2}}\left( \sinh \psi \right) ^{2}%
\left[ 2\cosh \psi -3\right] +\frac{v^{6}}{4\kappa ^{2}}\cosh \psi
\label{496} \\
&=&\frac{v^{6}}{4\kappa ^{2}}\left[ 2\left( \sinh \psi \right) ^{2}\cosh
\psi -3\left( \sinh \psi \right) ^{2}+\cosh \psi \right]  \notag \\
&=&\frac{v^{6}}{4\kappa ^{2}}\left\{ 2\cosh \psi \left[ \left( \sinh \psi
\right) ^{2}+1\right] -\cosh \psi -3\left[ \left( \sinh \psi \right) ^{2}+1%
\right] +3\right\}  \notag \\
&=&v^{2}\left( \frac{v^{2}}{\kappa }\right) ^{2}\frac{1}{4}\left[ 2\left(
\cosh \psi \right) ^{3}-3\left( \cosh \psi \right) ^{2}-\cosh \psi +3\right]
\notag
\end{eqnarray}
As in the case of the Eq.(\ref{15885}) we will note that the expression must
be integrated over the plane, which removes from the coefficient the
dimensional factor 
\begin{equation*}
\left( \frac{v^{2}}{\kappa }\right) ^{2}=\frac{1}{\rho _{s}^{2}}
\end{equation*}
and the dimension of the \emph{energy }is given by $v^{2}=\Omega _{ci}$.

\section{Discussion on the versions of the SD equations}

The possibility of formulating the expression of the energy as a sum of
squared terms plus a (additional) term with topological significance (known
as Bogomolnyi formulation) is fundamental for the self-duality. In our case,
the CHM fluid/plasma cannot associate a topological significance to the
additional energy, a characteristic signalized by \textbf{Lee 1991} and by 
\textbf{Dunne}. This induces a certain imprecision in the choice of the way
of separating the squared terms, with consequences on the form of the
equations, etc. This aspect will only be discussed briefly here, with the
only intention to compare few possible choices.

\bigskip

We have shown how to derive two versions of writting the total energy of the
system as a sum of squared terms plus an additional (\emph{residual}) term,
while this one has no topological significance. After adopting the algebraic
ansatz we arrive at two different equations for the scalar function $\psi $
which we associate with the physical streamfunction of the CHM fluid.

The choice of the version that has the correct physical significance should
be done on the basis of the supersymmetric invariance of the extended field
theoretical model. However, even from this advanced point of view, we can
expect at most an indication which will not be applicable directly to our
problem. The field theroetical model for the CHM equation shows significant
differences compared with topological theories in $2D$. The SD equations, in
both versions, lead to time dependent solutions, therefore the stationarity
typical for the solutions obtained from the Bogomonlyi form in other
theories is here lost. The topological aspect is also lost, the additional
energy, in both versions, is not proportional with the total winding number
induced by the vortices present in the plane. We must note however that the
point of view that results from the SUSY extension of the theory (a natural
extension) favorizes the Eq.(\ref{94}) or possibly, as mentioned below, the
Abelian version, Eq.(\ref{497}). This even if the additional energy does not
have a topological meaning.

\bigskip

One possible help comes from looking at the theory of the CHM fluid as being
a development of the theory for the Euler fluid. Since the asymptotic states
of the latter are goverend by the \emph{sinh}-Poisson equation, we expect
that the nonlinear term of the equation for CHM fluid to be of a similar
nature. Or, we see that the first set of SD equations leads to a sign which
is opposite to the one appearing in the \emph{sinh}-Poisson equation. The
physical form and properties of the solutions would be, in that case,
completely different. Or, one expects that an ideal fluid without an
intrinsic length will transform smootly into a fluid which has an intrinsic
length, as the CHM fluid. Although this is not an argument, we take this as
a sort of indication that the second choice is more appropriate and we adopt
Eq.(\ref{465}) as the equation governing the asymptotic states of the CHM
fluid.

%%%%%%%%%%%%%%%%%%%%%%%%%%%%%%%%%%%%%%%%%%%%%%%%%%%%%%%%%%%%%%%%%%%%% 
\begin{figure}[tbph]
\centerline{\includegraphics[height=5cm]{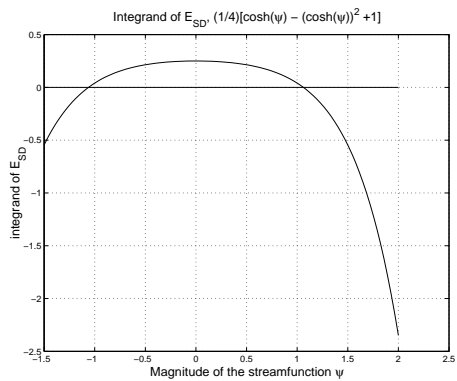}}
\caption{The integrand of the energy ${\mathcal{E}_{SD}}$ which is the
additional term in the \textit{first} Bogomolnyi form.}
\label{E_SD_1}
\end{figure}
%%%%%%%%%%%%%%%%%%%%%%%%%%%%%%%%%%%%%%%%%%%%%%%%%%%%%%%%%%%%%%%%%%%%%

%%%%%%%%%%%%%%%%%%%%%%%%%%%%%%%%%%%%%%%%%%%%%%%%%%%%%%%%%%%%%%%%%%%%% 
\begin{figure}[tbph]
\centerline{\includegraphics[height=5cm]{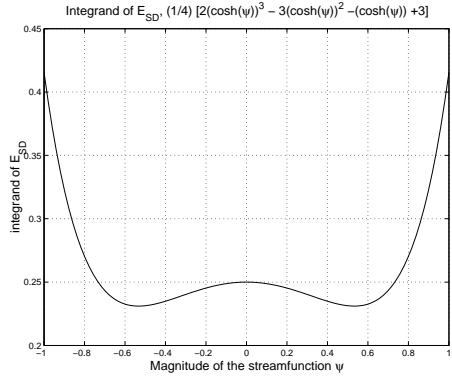}}
\caption{The integrand of the energy ${\mathcal{E}_{SD}}$ which is the
additional term in the \textit{second} Bogomolnyi form.}
\label{E_SD_2}
\end{figure}
%%%%%%%%%%%%%%%%%%%%%%%%%%%%%%%%%%%%%%%%%%%%%%%%%%%%%%%%%%%%%%%%%%%%%
\bigskip

A different approach can be developed on the basis of the analysis made by 
\textbf{Lee (1991)} of the first set of equations at SD. After adopting the
algebraic ansatz, Lee finds that the boundary conditions on the two
functions $\phi _{1}$ and $\phi _{2}$ lead to restrictive choices :
basically it results that the second ladder generator should not be present
in the algebraic form of $\phi $. Then the equation resulting from the first
set of SD equations is identical with the equation which is derived from the
Abelian version of the theory 
\begin{equation}
\Delta \psi =\exp \left( 2\psi \right) -\exp \left( \psi \right)  \label{497}
\end{equation}
Contrary to the Equation (\ref{94}), Eq.(\ref{497}) has physically
interesting solutions, consisting of rings of vorticity. The way this theory
can arise from the \emph{Abelian dominance} in the full field-theoretical
(FT) model of the CHM fluid and the physical consequences of this theory
will be analysed elsewhere.

\bigskip

\begin{figure}[tbph]
\centerline{\includegraphics[height=5cm]{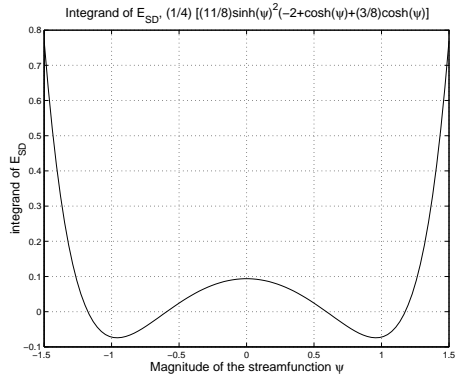}}
\caption{The integrand of the energy ${\mathcal{E}_{SD}}$ which is the
additional term in the \textit{third} (for $\protect\lambda =1/2$)
Bogomolnyi form.}
\label{E_SD_3}
\end{figure}

Another possible criterion for the choice of one or another Bogomolnyi form
is the behavior of the additional energy. In some cases, the solution of the
equations at self-duality (\emph{i.e.} by taking to zero the squared terms)
show a higher energy due to the residual term, when compared with other
choices. In particular we have found that the following form has apparently
lower energy for many of the solutions of its associated self-duality
equation, in practical applications like the tropical cyclone, etc.: 
\begin{equation}
\Delta \psi +\frac{1}{2}\sinh \psi \left( \cosh \psi -1\right) =0
\label{4973}
\end{equation}
or a version in which the harmonic function is a constant different of $1$, 
\emph{i.e.} taking 
\begin{equation}
\rho \equiv \frac{\rho _{1}}{v^{2}/\left( 4p\right) }  \label{4974}
\end{equation}
instead of Eq.(\ref{459}) we have 
\begin{equation}
\Delta \psi +\frac{1}{2p^{2}}\sinh \psi \left( \cosh \psi -p\right) =0
\label{4975}
\end{equation}
This form arises by taking 
\begin{equation*}
D_{0}\phi +\lambda \frac{i}{2\kappa }\left( \left[ \left[ \phi ,\phi
^{\dagger }\right] ,\phi \right] -v^{2}\phi \right)
\end{equation*}
in the Eq.(\ref{353}), with 
\begin{equation*}
\lambda =\frac{1}{2}
\end{equation*}
and has an associated \emph{residual}, non-topological energy 
\begin{eqnarray*}
\mathcal{E}_{3}^{a} &=&v^{2}\left( \frac{v^{2}}{\kappa }\right) ^{2} \\
&&\times \frac{1}{4}\left[ \frac{11}{8}\left( \sinh \psi \right) ^{2}\left(
-2+\cosh \psi \right) +\frac{3}{8}\cosh \psi \right]
\end{eqnarray*}
which is shown in the Fig.(\ref{E_SD_3}).

%%%%%%%%%%%%%%%%%%%%%%%%%%%%%%%%%%%%%%%%%%%%%%%%%%%%%%%%%%%%%%%%%%%%% 
%%%%%%%%%%%%%%%%%%%%%%%%%%%%%%%%%%%%%%%%%%%%%%%%%%%%%%%%%%%%%%%%%%%%%
\bigskip

\section{Various forms of the equation}

The crucial step in the derivation of the form Eq.(\ref{94}) or Eq.(\ref{465}%
) is the choice of a solution for the equation (\ref{87}) 
\begin{equation}
\Delta \ln \left( \rho _{1}\rho _{2}\right) =0  \label{711}
\end{equation}
with the consequence that the product $\rho _{1}\rho _{2}$ is the
exponential of a harmonic function. This leads to the conclusion that the
asymptotic states of the CHM fluid can be described by a \emph{class of
differential equations}, parametrized by the harmonic functions. Without
developing this aspect here we mention that the fact that instead one
equation we find a class of equations has a simple physical significance.
The difference resides in the background motion and the equations from this
class describes vortices on a background of fluid/plasma motion which has
zero \emph{physical} vorticity. Then the ``streamfunctions'' $\psi _{1}$ and 
$\psi _{2}$ that can be introduced separately for $\rho _{1}$ respectively
for $\rho _{2}$ will differ by a function whose Laplacean is zero, \emph{i.e.%
} a potential flow. We note that by \emph{physical} vorticity we understand
the vorticity perturbation above the condensate background, \emph{i.e.}
above $\Omega _{ci}$.

In the following we mention few examples illustrating this freedom of choice
of the zero-vorticity background flow.

\bigskip

The separation of variables in polar coordinates 
\begin{eqnarray}
\Delta \Phi &=&0  \label{712} \\
\Phi \left( r,\theta \right) &\equiv &R\left( r\right) \Theta \left( \theta
\right)  \notag
\end{eqnarray}
with 
\begin{eqnarray}
\frac{d^{2}R}{dr^{2}}+\frac{1}{r}\frac{dR}{dr}-\frac{n^{2}}{r^{2}}R &=&0
\label{713} \\
\frac{d^{2}\Theta }{d\theta ^{2}}+n^{2}\Theta &=&0  \notag
\end{eqnarray}

\subsection{Solution 1}

The first choice was 
\begin{equation}
\ln \left( \rho _{1}\rho _{2}\right) =0  \label{714}
\end{equation}
with the consequence 
\begin{equation}
\rho _{1}\rho _{2}=1  \label{715}
\end{equation}

\subsection{Solution 2}

A different (almost arbitrary) choice 
\begin{equation}
\Delta h=0  \label{716}
\end{equation}
\begin{equation}
h=\exp \left( x\right) \sin \left( y\right)  \label{717}
\end{equation}
Then 
\begin{equation}
\ln \left( \rho _{1}\rho _{2}\right) =\exp \left( x\right) \sin \left(
y\right)  \label{718}
\end{equation}
\begin{equation}
\rho _{1}\rho _{2}=\exp \left[ \exp \left( x\right) \sin \left( y\right) %
\right]  \label{719}
\end{equation}

\subsection{Solution 3 (general solution in cylindrical coordinates)}

A different choice for the cylindrical harmonic function $\Phi =\ln \left(
\rho _{1}\rho _{2}\right) $%
\begin{eqnarray}
\Phi \left( r,\theta \right) &=&\left[ Ar^{n}+B\frac{1}{r^{n}}\right]
\label{720} \\
&&\times \left[ C\cos \left( n\theta \right) +D\sin \left( n\theta \right) %
\right]  \notag
\end{eqnarray}
and for $n=0$, 
\begin{eqnarray}
\Phi \left( r,\theta \right) &=&\left( A\theta +B\right)  \label{721} \\
&&\times \left[ C\ln r+D\right]  \notag
\end{eqnarray}

\subsubsection{Example polar 1}

For example, 
\begin{equation}
\Phi \left( r,\theta \right) =ar\cos \theta +b  \label{722}
\end{equation}
Consider the choice with a particular value for $b$ 
\begin{equation}
\ln \left( \rho _{1}\rho _{2}\right) =ar\cos \theta +b  \label{723}
\end{equation}
\begin{eqnarray}
\rho _{1}\rho _{2} &=&\exp \left( ar\cos \theta +b\right)  \label{724} \\
&=&\frac{v^{4}}{16p^{2}}\exp \left( ar\cos \theta \right)  \notag
\end{eqnarray}
for $p$ a positive constant. Define $\rho $ as before 
\begin{eqnarray}
\rho &\equiv &\frac{\rho _{1}}{\frac{v^{2}}{4p}\exp \left( \frac{1}{2}ar\cos
\theta \right) }  \label{725} \\
&=&\frac{1}{\rho _{2}}\frac{v^{2}}{4p}\exp \left( \frac{1}{2}ar\cos \theta
\right)  \notag
\end{eqnarray}
\begin{eqnarray}
&&-\frac{1}{2}\Delta \ln \left[ \rho \frac{v^{2}}{4p}\exp \left( \frac{1}{2}%
ar\cos \theta \right) \right]  \label{726} \\
&=&-\frac{1}{\kappa ^{2}}\left( \rho _{1}-\rho _{2}\right) \left[ 2\left(
\rho _{1}+\rho _{2}\right) -v^{2}\right]  \notag
\end{eqnarray}
where 
\begin{eqnarray}
RHS &\equiv &-\frac{1}{\kappa ^{2}}\left( \rho _{1}-\rho _{2}\right) \left[
2\left( \rho _{1}+\rho _{2}\right) -v^{2}\right]  \label{727} \\
&=&-\frac{1}{\kappa ^{2}}\left[ \rho \frac{v^{2}}{4p}\exp \left( \frac{1}{2}%
ar\cos \theta \right) -\frac{1}{\rho }\frac{v^{2}}{4p}\exp \left( \frac{1}{2}%
ar\cos \theta \right) \right]  \notag \\
&&\times \left\{ 2\left[ \rho \frac{v^{2}}{4p}\exp \left( \frac{1}{2}ar\cos
\theta \right) -\frac{1}{\rho }\frac{v^{2}}{4p}\exp \left( \frac{1}{2}ar\cos
\theta \right) \right] -v^{2}\right\}  \notag
\end{eqnarray}
\begin{eqnarray}
RHS &=&-\frac{1}{\kappa ^{2}}\left[ \frac{v^{2}}{4p}\exp \left( \frac{1}{2}%
ar\cos \theta \right) \right] \left( \rho -\frac{1}{\rho }\right)
\label{728} \\
&&\times \frac{v^{2}}{4p}\exp \left( \frac{1}{2}ar\cos \theta \right)
\left\{ 2\left( \rho +\frac{1}{\rho }\right) -4p\exp \left( -\frac{1}{2}%
ar\cos \theta \right) \right\}  \notag
\end{eqnarray}
\begin{eqnarray}
RHS &=&-\frac{1}{4p^{2}}\left( \frac{v^{2}}{\kappa }\right) ^{2}\exp \left(
ar\cos \theta \right)  \label{729} \\
&&\times \left( \rho -\frac{1}{\rho }\right)  \notag \\
&&\times \left[ \frac{1}{2}\left( \rho +\frac{1}{\rho }\right) -p\exp \left(
-\frac{1}{2}ar\cos \theta \right) \right]  \notag
\end{eqnarray}
and the Right Hand Side 
\begin{eqnarray}
LHS &=&-\frac{1}{2}\Delta \ln \left[ \rho \frac{v^{2}}{4p}\exp \left( \frac{1%
}{2}ar\cos \theta \right) \right]  \label{730} \\
&=&-\frac{1}{2}\Delta \ln \rho +\Delta \ln \left( \frac{v^{2}}{4p}\right) 
\notag \\
&&-\frac{1}{2}\Delta \left( \frac{1}{2}ar\cos \theta \right)  \notag \\
&=&-\frac{1}{2}\Delta \rho  \notag
\end{eqnarray}
The equation is 
\begin{eqnarray}
&&-\frac{1}{2}\Delta \rho  \label{731} \\
&=&-\frac{1}{4p^{2}}\left( \frac{v^{2}}{\kappa }\right) ^{2}\exp \left(
ar\cos \theta \right) \left( \rho -\frac{1}{\rho }\right) \left[ \frac{1}{2}%
\left( \rho +\frac{1}{\rho }\right) -p\exp \left( -\frac{1}{2}ar\cos \theta
\right) \right]  \notag
\end{eqnarray}

However, since we will note 
\begin{equation}
\ln \rho \equiv \psi  \label{732}
\end{equation}
it would have been easier to do that from the beginning.

The equation 
\begin{eqnarray}
\exp \left( \psi \right) &\equiv &\frac{\rho _{1}}{\frac{v^{2}}{4p}\exp
\left( \frac{1}{2}ar\cos \theta \right) }  \label{733} \\
&=&\frac{1}{\rho _{2}}\frac{v^{2}}{4p}\exp \left( \frac{1}{2}ar\cos \theta
\right)  \notag
\end{eqnarray}
generates explicit forms for $\rho _{1,2}$ that will be inserted in the
equation 
\begin{eqnarray}
\rho _{1} &=&\frac{v^{2}}{4p}\exp \left( \psi +\frac{1}{2}ar\cos \theta
\right)  \label{734} \\
\rho _{2} &=&\frac{v^{2}}{4p}\exp \left( -\psi +\frac{1}{2}ar\cos \theta
\right)  \notag
\end{eqnarray}
with the same result.

The equation now reads 
\begin{equation}
-\Delta \psi +\frac{1}{p^{2}}\exp \left( ar\cos \theta \right) \sinh \psi %
\left[ \cosh \psi -p\exp \left( -\frac{1}{2}ar\cos \theta \right) \right] =0
\label{735}
\end{equation}
where the space unit is now $\kappa /v^{2}\equiv \rho _{s}$.

\subsubsection{Example polar 2}

The second simple polar choice of $\Delta \Phi =0$%
\begin{equation}
\Phi \left( r,\theta \right) =a\ln r+b  \label{736}
\end{equation}
\begin{eqnarray}
\ln \left( \rho _{1}\rho _{2}\right) &=&a\ln r+b  \label{737} \\
\rho _{1}\rho _{2} &=&\exp \left( a\ln r+b\right)  \notag \\
&=&\exp \left( b\right) r^{a}  \notag
\end{eqnarray}
The normalization suggets 
\begin{equation}
\rho _{1}\rho _{2}=\frac{v^{4}}{16p^{2}}r^{a}  \label{738}
\end{equation}
\begin{eqnarray}
\rho &\equiv &\frac{\rho _{1}}{\frac{v^{2}}{4p}r^{a/2}}  \label{739} \\
&=&\frac{1}{\rho _{2}}\frac{v^{2}}{4p}r^{a/2}  \notag
\end{eqnarray}
\begin{eqnarray}
\rho _{1} &=&\rho \frac{v^{2}}{4p}r^{a/2}  \label{740} \\
\rho _{2} &=&\frac{1}{\rho }\frac{v^{2}}{4p}r^{a/2}  \notag
\end{eqnarray}
The equation is 
\begin{eqnarray}
&&-\frac{1}{2}\Delta \ln \rho -\frac{1}{2}\Delta \ln \left( r^{a/2}\right)
\label{741} \\
&=&-\frac{1}{\kappa ^{2}}\left[ \rho \frac{v^{2}}{4p}r^{a/2}-\frac{1}{\rho }%
\frac{v^{2}}{4p}r^{a/2}\right]  \notag \\
&&\times \left\{ 2\left[ \rho \frac{v^{2}}{4p}r^{a/2}-\frac{1}{\rho }\frac{%
v^{2}}{4p}r^{a/2}\right] -v^{2}\right\}  \notag
\end{eqnarray}
\begin{eqnarray}
&&-\frac{1}{2}\Delta \ln \rho  \label{742} \\
&=&-\frac{1}{4p^{2}}\left( \frac{v^{2}}{\kappa }\right) ^{2}r^{a}\left( \rho
-\frac{1}{\rho }\right)  \notag \\
&&\times \left[ \frac{1}{2}\left( \rho +\frac{1}{\rho }\right) -pr^{-a/2}%
\right]  \notag
\end{eqnarray}
After introducing the substitution $\ln \rho \equiv \psi $ and measuring the
space in units of $\kappa /v^{2}=\rho _{s}$ we have 
\begin{equation}
-\Delta \psi +\frac{1}{p^{2}}r^{a}\sinh \psi \left( \cosh \psi -\frac{p}{%
r^{a}}\right) =0  \label{743}
\end{equation}
or, better 
\begin{equation}
-\Delta \psi +\frac{1}{p^{2}}\sinh \psi \left( r^{a}\cosh \psi -p\right) =0
\label{744}
\end{equation}

\section{Discussion on the physical meaning of the model}

\subsection{The short range of the potential}

It is considered that the scalar field is very close to the vacuum value 
\begin{equation}
\phi \sim v  \label{95}
\end{equation}
We calculate the current in the region of vanishing space-variation. 
\begin{eqnarray}
J^{\mu } &=&-i\left( \left[ \phi ^{\dagger },D^{\mu }\phi \right] -\left[
\left( D^{\mu }\phi \right) ^{\dagger },\phi \right] \right)  \label{96} \\
&=&-i\left\{ \phi ^{\dagger }\left( \partial ^{\mu }\phi +\left[ A^{\mu
},\phi \right] \right) -\left( \partial ^{\mu }\phi +\left[ A^{\mu },\phi %
\right] \right) \phi ^{\dagger }\right.  \notag \\
&&\left. -\left( \partial _{\mu }\phi ^{\dagger }+\left[ \phi ^{\dagger
},A^{\mu \dagger }\right] \right) \phi +\phi \left( \partial _{\mu }\phi
^{\dagger }+\left[ \phi ^{\dagger },A^{\mu \dagger }\right] \right) \right\}
\notag \\
&=&-i\left\{ \phi ^{\dagger }\left( \partial ^{\mu }\phi \right) -\left(
\partial ^{\mu }\phi \right) \phi ^{\dagger }-\left( \partial _{\mu }\phi
^{\dagger }\right) \phi +\phi \left( \partial _{\mu }\phi ^{\dagger }\right)
\right.  \notag \\
&&\left. +\phi ^{\dagger }\left[ A^{\mu },\phi \right] -\left[ A^{\mu },\phi %
\right] \phi ^{\dagger }-\left[ \phi ^{\dagger },A^{\mu \dagger }\right]
\phi +\phi \left[ \phi ^{\dagger },A^{\mu \dagger }\right] \right\}  \notag
\end{eqnarray}
Since we consider that the field $\phi $ is almost constant (and equal to $v$%
) we can negelct all terms on the first line and obtain 
\begin{equation}
J^{\mu }\simeq -i\left( \left[ \phi ^{\dagger },\left[ A^{\mu },\phi \right] %
\right] +\left[ \phi ,\left[ \phi ^{\dagger },A^{\mu \dagger }\right] \right]
\right)  \label{97}
\end{equation}
Let us consider the explicit expressions for the fields 
\begin{eqnarray}
A_{x} &=&A^{x}=\frac{1}{2}\left( a-a^{\ast }\right) H\;,\;A^{x\dagger }=%
\frac{1}{2}\left( a^{\ast }-a\right) H  \label{98} \\
A_{y} &=&A^{y}=\frac{1}{2i}\left( a+a^{\ast }\right) H\;,\;A^{y\dagger }=-%
\frac{1}{2i}\left( a^{\ast }+a\right) H  \notag
\end{eqnarray}
and 
\begin{eqnarray}
\phi &=&\phi _{1}E_{+}+\phi _{2}E_{-}  \label{99} \\
\phi ^{\dagger } &=&\phi _{1}^{\ast }E_{-}+\phi _{2}^{\ast }E_{+}  \notag
\end{eqnarray}
Then, for $x$, the first part of the formula for the current $j^{x}$ is 
\begin{eqnarray}
\left[ \phi ^{\dagger },\left[ A^{x},\phi \right] \right] &=&\left[ \phi
^{\dagger },\frac{1}{2}\left( a-a^{\ast }\right) \left( \phi _{1}\left[
H,E_{+}\right] +\phi _{2}\left[ H,E_{-}\right] \right) \right]  \label{100}
\\
&=&\frac{1}{2}\left( a-a^{\ast }\right) \left[ \phi ^{\dagger },\left( \phi
_{1}2E_{+}-\phi _{2}2E_{-}\right) \right]  \notag \\
&=&\left( a-a^{\ast }\right) \left[ \phi _{1}^{\ast }E_{-}+\phi _{2}^{\ast
}E_{+},\phi _{1}E_{+}-\phi _{2}E_{-}\right]  \notag \\
&=&\left( a-a^{\ast }\right) \left\{ \phi _{1}^{\ast }\phi _{1}\left[
E_{-},E_{+}\right] -\phi _{2}^{\ast }\phi _{2}\left[ E_{+},E_{-}\right]
\right\}  \notag \\
&=&\left( a-a^{\ast }\right) \left( -\left| \phi _{1}\right| ^{2}H-\left|
\phi _{2}\right| ^{2}H\right)  \notag \\
&=&-\left( a-a^{\ast }\right) \left( \rho _{1}+\rho _{2}\right) H  \notag
\end{eqnarray}
and the second part 
\begin{eqnarray}
\left[ \phi ,\left[ \phi ^{\dagger },A^{x\dagger }\right] \right] &=&\left[
\phi ,\left[ \phi _{1}^{\ast }E_{-}+\phi _{2}^{\ast }E_{+},\frac{1}{2}\left(
a^{\ast }-a\right) H\right] \right]  \label{101} \\
&=&\frac{1}{2}\left( a^{\ast }-a\right) \left[ \phi ,\phi _{1}^{\ast }\left[
E_{-},H\right] +\phi _{2}^{\ast }\left[ E_{+},H\right] \right]  \notag \\
&=&\frac{1}{2}\left( a^{\ast }-a\right) \left[ \phi _{1}E_{+}+\phi
_{2}E_{-},\phi _{1}^{\ast }2E_{-}-\phi _{2}^{\ast }2E_{+}\right]  \notag \\
&=&\left( a^{\ast }-a\right) \left\{ \phi _{1}\phi _{1}^{\ast }\left[
E_{+},E_{-}\right] -\phi _{2}\phi _{2}^{\ast }\left[ E_{-},E_{+}\right]
\right\}  \notag \\
&=&\left( a^{\ast }-a\right) \left( \left| \phi _{1}\right| ^{2}H+\left|
\phi _{2}\right| ^{2}H\right)  \notag \\
&=&\left( a^{\ast }-a\right) \left( \rho _{1}+\rho _{2}\right) H  \notag
\end{eqnarray}
The $x$ component of the current is 
\begin{eqnarray}
J_{x} &\simeq &-i\left\{ -\left( a-a^{\ast }\right) \left( \rho _{1}+\rho
_{2}\right) H+\left( a^{\ast }-a\right) \left( \rho _{1}+\rho _{2}\right)
H\right\}  \label{102} \\
&=&2i(a-a^{\ast })\left( \rho _{1}+\rho _{2}\right) H  \notag
\end{eqnarray}
For the far regions we take the value 
\begin{equation}
\rho _{1}+\rho _{2}\sim v^{2}/2  \label{112}
\end{equation}
We return to the potential notation, $(a-a^{\ast })H=2A_{x}$, 
\begin{equation}
J_{x}=2iv^{2}A_{x}  \label{103}
\end{equation}
Analogous for the $y$ component of the current, 
\begin{eqnarray}
\left[ \phi ^{\dagger },\left[ A^{y},\phi \right] \right] &=&\left[ \phi
^{\dagger },\frac{1}{2i}\left( a+a^{\ast }\right) \left( \phi _{1}\left[
H,E_{+}\right] +\phi _{2}\left[ H,E_{-}\right] \right) \right]  \label{1035}
\\
&=&\frac{1}{2i}\left( a+a^{\ast }\right) \left[ \phi ^{\dagger },\left( \phi
_{1}2E_{+}-\phi _{2}2E_{-}\right) \right]  \notag \\
&=&-i\left( a+a^{\ast }\right) \left[ \phi _{1}^{\ast }E_{-}+\phi _{2}^{\ast
}E_{+},\phi _{1}E_{+}-\phi _{2}E_{-}\right]  \notag \\
&=&-i\left( a+a^{\ast }\right) \left\{ \phi _{1}^{\ast }\phi _{1}\left[
E_{-},E_{+}\right] -\phi _{2}^{\ast }\phi _{2}\left[ E_{+},E_{-}\right]
\right\}  \notag \\
&=&-i\left( a+a^{\ast }\right) \left( -\left| \phi _{1}\right| ^{2}H-\left|
\phi _{2}\right| ^{2}H\right)  \notag \\
&=&i\left( a+a^{\ast }\right) \left( \rho _{1}+\rho _{2}\right) H  \notag
\end{eqnarray}
and the second part 
\begin{eqnarray}
\left[ \phi ,\left[ \phi ^{\dagger },A^{y\dagger }\right] \right] &=&\left[
\phi ,\left[ \phi _{1}^{\ast }E_{-}+\phi _{2}^{\ast }E_{+},-\frac{1}{2i}%
\left( a^{\ast }+a\right) H\right] \right]  \label{1037} \\
&=&-\frac{1}{2i}\left( a^{\ast }+a\right) \left[ \phi ,\phi _{1}^{\ast }%
\left[ E_{-},H\right] +\phi _{2}^{\ast }\left[ E_{+},H\right] \right]  \notag
\\
&=&-\frac{1}{2i}\left( a^{\ast }+a\right) \left[ \phi _{1}E_{+}+\phi
_{2}E_{-},\phi _{1}^{\ast }2E_{-}-\phi _{2}^{\ast }2E_{+}\right]  \notag \\
&=&i\left( a^{\ast }+a\right) \left\{ \phi _{1}\phi _{1}^{\ast }\left[
E_{+},E_{-}\right] -\phi _{2}\phi _{2}^{\ast }\left[ E_{-},E_{+}\right]
\right\}  \notag \\
&=&i\left( a^{\ast }+a\right) \left( \left| \phi _{1}\right| ^{2}H+\left|
\phi _{2}\right| ^{2}H\right)  \notag \\
&=&i\left( a^{\ast }+a\right) \left( \rho _{1}+\rho _{2}\right) H  \notag
\end{eqnarray}
The $x$ component of the current is 
\begin{eqnarray}
J_{y} &\simeq &-i\left\{ i\left( a+a^{\ast }\right) \left( \rho _{1}+\rho
_{2}\right) H+i\left( a^{\ast }+a\right) \left( \rho _{1}+\rho _{2}\right)
H\right\}  \label{1039} \\
&=&2(a+a^{\ast })\left( \rho _{1}+\rho _{2}\right) H  \notag
\end{eqnarray}
As before, we replace here $\rho _{1}+\rho _{2}\sim v^{2}$ and $(a+a^{\ast
})H=2iA_{y}$, 
\begin{equation}
J_{y}=2iv^{2}A_{y}  \label{104}
\end{equation}

We take the temporal component of the potential in the form 
\begin{eqnarray}
A^{0} &=&bH  \label{105} \\
A^{0\dagger } &=&\left( A^{0}\right) ^{\ast T}=b^{\ast }H  \notag
\end{eqnarray}
Then the temporal component of the current density is (cf. Eq.(\ref{97})) 
\begin{equation}
J^{0}\simeq -i\left\{ \left[ \phi ^{\dagger },\left[ A^{0},\phi \right] %
\right] +\left[ \phi ,\left[ \phi ^{\dagger },A^{0\dagger }\right] \right]
\right\}  \label{106}
\end{equation}
and we calculate again the terms, with the particular choice Eq.(\ref{105}) 
\begin{eqnarray}
\left[ \phi ^{\dagger },\left[ A^{0},\phi \right] \right] &=&\left[ \phi
^{\dagger },b\left( \phi _{1}\left[ H,E_{+}\right] +\phi _{2}\left[ H,E_{-}%
\right] \right) \right]  \label{107} \\
&=&b\left[ \phi ^{\dagger },\left( \phi _{1}2E_{+}-\phi _{2}2E_{-}\right) %
\right]  \notag \\
&=&2b\left[ \phi _{1}^{\ast }E_{-}+\phi _{2}^{\ast }E_{+},\phi
_{1}E_{+}-\phi _{2}E_{-}\right]  \notag \\
&=&2b\left\{ \phi _{1}^{\ast }\phi _{1}\left[ E_{-},E_{+}\right] -\phi
_{2}^{\ast }\phi _{2}\left[ E_{+},E_{-}\right] \right\}  \notag \\
&=&2b\left( -\left| \phi _{1}\right| ^{2}H-\left| \phi _{2}\right|
^{2}H\right)  \notag \\
&=&-2b\left( \rho _{1}+\rho _{2}\right) H  \notag
\end{eqnarray}
and the second part 
\begin{eqnarray}
\left[ \phi ,\left[ \phi ^{\dagger },A^{0\dagger }\right] \right] &=&\left[
\phi ,\left[ \phi _{1}^{\ast }E_{-}+\phi _{2}^{\ast }E_{+},b^{\ast }H\right] %
\right]  \label{108} \\
&=&b^{\ast }\left[ \phi ,\phi _{1}^{\ast }\left[ E_{-},H\right] +\phi
_{2}^{\ast }\left[ E_{+},H\right] \right]  \notag \\
&=&b^{\ast }\left[ \phi _{1}E_{+}+\phi _{2}E_{-},\phi _{1}^{\ast
}2E_{-}-\phi _{2}^{\ast }2E_{+}\right]  \notag \\
&=&2b^{\ast }\left\{ \phi _{1}\phi _{1}^{\ast }\left[ E_{+},E_{-}\right]
-\phi _{2}\phi _{2}^{\ast }\left[ E_{-},E_{+}\right] \right\}  \notag \\
&=&2b^{\ast }\left( \left| \phi _{1}\right| ^{2}H+\left| \phi _{2}\right|
^{2}H\right)  \notag \\
&=&2b^{\ast }\left( \rho _{1}+\rho _{2}\right) H  \notag
\end{eqnarray}
The result for the current in the approximation of the constant background
is 
\begin{eqnarray}
J^{0} &\simeq &-i\left\{ \left[ \phi ^{\dagger },\left[ A^{0},\phi \right] %
\right] +\left[ \phi ,\left[ \phi ^{\dagger },A^{0\dagger }\right] \right]
\right\}  \label{109} \\
&=&-i\left\{ -2b\left( \rho _{1}+\rho _{2}\right) H+2b^{\ast }\left( \rho
_{1}+\rho _{2}\right) H\right\}  \notag \\
&=&2i\left( b-b^{\ast }\right) \left( \rho _{1}+\rho _{2}\right) H  \notag
\end{eqnarray}
We then obtain 
\begin{equation}
J^{0}=2iv^{2}A^{0}  \label{110}
\end{equation}
Then 
\begin{equation}
J^{\mu }\equiv \left( J^{0},J^{x},J^{y}\right) =2iv^{2}\left[
A^{0},A^{x},A^{y}\right]  \label{113}
\end{equation}
or 
\begin{equation}
J^{\mu }\simeq 2iv^{2}A^{\mu }  \label{114}
\end{equation}

With this value of the current density we return to equation connecting the
gauge field with the matter current. The equation is 
\begin{equation}
-\kappa \varepsilon ^{\mu \nu \rho }F_{\nu \rho }=iJ^{\mu }  \label{115}
\end{equation}
To transfer the antisymmetric tensor $\varepsilon ^{\mu \nu \rho }$ in the
other side, we multiply by $\varepsilon _{\mu \sigma \tau }$ and sum over
repeated indices 
\begin{eqnarray}
\varepsilon _{\mu \sigma \tau }\varepsilon ^{\mu \nu \rho }F_{\nu \rho } &=&-%
\frac{i}{\kappa }\varepsilon _{\mu \sigma \tau }J^{\mu }  \label{116} \\
\left( \delta _{\sigma \nu }\delta _{\tau \rho }-\delta _{\sigma \rho
}\delta _{\tau \nu }\right) F_{\nu \rho } &=&-\frac{i}{\kappa }\varepsilon
_{\mu \sigma \tau }J^{\mu }  \notag \\
F_{\sigma \tau }-F_{\tau \sigma } &=&-\frac{i}{\kappa }\varepsilon _{\mu
\sigma \tau }J^{\mu }  \notag \\
F_{\sigma \tau } &=&-\frac{i}{2\kappa }\varepsilon _{\mu \sigma \tau }J^{\mu
}  \notag
\end{eqnarray}
A direct relation with the previous result is obtained taking the explicit
form of $J^{\mu }$ from Eq.(\ref{114}) 
\begin{eqnarray}
F_{\sigma \tau } &=&-\frac{i}{2\kappa }\varepsilon _{\mu \sigma \tau }J^{\mu
}  \label{117} \\
&=&\frac{v^{2}}{\kappa }\varepsilon _{\mu \sigma \tau }A^{\mu }  \notag
\end{eqnarray}
Introducing the expression for the field 
\begin{equation}
F_{\sigma \tau }=\partial _{\sigma }A_{\tau }-\partial _{\tau }A_{\sigma }=-%
\frac{i}{2\kappa }\varepsilon _{\mu \sigma \tau }J^{\mu }  \label{118}
\end{equation}
we apply the derivative operator $\partial _{\tau }$ and sum over the index $%
\tau $ 
\begin{eqnarray}
&&\partial _{\tau }\partial _{\sigma }A_{\tau }-\partial _{\tau }\partial
_{\tau }A_{\sigma }  \label{119} \\
&=&-\frac{i}{2\kappa }\varepsilon _{\mu \sigma \tau }\partial _{\tau }J^{\mu
}  \notag \\
&=&-\frac{i}{2\kappa }\varepsilon _{\mu \sigma \tau }\partial _{\tau }\left(
2iv^{2}A^{\mu }\right) \;\text{(from Eq.(\ref{114}))}  \notag \\
&=&\frac{v^{2}}{\kappa }\varepsilon _{\mu \sigma \tau }\partial _{\tau
}A^{\mu }  \notag
\end{eqnarray}
The term on the right hand side is 
\begin{equation}
\varepsilon _{\sigma \tau \mu }\partial _{\tau }A^{\mu }=\frac{1}{2}%
\varepsilon _{\sigma \tau \mu }F_{\tau }^{\mu }  \label{120}
\end{equation}
and here we replace, from Eq.(\ref{117}) 
\begin{equation}
F_{\tau }^{\mu }=g^{\mu \alpha }F_{\tau \alpha }=g^{\mu \alpha }\frac{v^{2}}{%
\kappa }\varepsilon _{\tau \alpha \eta }A^{\eta }  \label{121}
\end{equation}
Further the product of the two antisymmetric tensors $\varepsilon $ is
expanded 
\begin{eqnarray}
\partial _{\tau }\partial _{\sigma }A_{\tau }-\partial _{\tau }\partial
_{\tau }A_{\sigma } &=&\frac{v^{2}}{\kappa }\varepsilon _{\mu \sigma \tau
}\partial _{\tau }A^{\mu }\;\text{(from Eq.(\ref{119}))}  \label{122} \\
&=&\frac{v^{2}}{\kappa }\frac{1}{2}\varepsilon _{\sigma \tau \mu }F_{\tau
}^{\mu }\;\;\text{(from Eq.(\ref{120}))}  \notag \\
&=&\frac{v^{2}}{\kappa }\frac{1}{2}\varepsilon _{\sigma \tau \mu }g^{\mu
\alpha }\frac{v^{2}}{\kappa }\varepsilon _{\tau \alpha \eta }A^{\eta }\;%
\text{(from Eq.(\ref{121}))}  \notag \\
&=&\frac{1}{2}\left( \frac{v^{2}}{\kappa }\right) ^{2}g^{\mu \alpha
}\varepsilon _{\sigma \tau \mu }\varepsilon _{\tau \alpha \eta }A^{\eta } 
\notag
\end{eqnarray}
The explicit expression for the sum 
\begin{eqnarray}
&&g^{\mu \alpha }\varepsilon _{\sigma \tau \mu }\varepsilon _{\tau \alpha
\eta }  \label{1228} \\
&=&g^{00}\varepsilon _{\sigma \tau 0}\varepsilon _{\tau 0\eta
}+g^{11}\varepsilon _{\sigma \tau 1}\varepsilon _{\tau 1\eta
}+g^{22}\varepsilon _{\sigma \tau 2}\varepsilon _{\tau 2\eta }  \notag \\
&=&g^{00}\left( \varepsilon _{\sigma 10}\varepsilon _{10\eta }+\varepsilon
_{\sigma 20}\varepsilon _{20\eta }\right)  \notag \\
&&+g^{11}\left( \varepsilon _{\sigma 01}\varepsilon _{01\eta }+\varepsilon
_{\sigma 21}\varepsilon _{21\eta }\right)  \notag \\
&&+g^{22}\left( \varepsilon _{\sigma 02}\varepsilon _{02\eta }+\varepsilon
_{\sigma 12}\varepsilon _{12\eta }\right)  \notag \\
&=&g^{00}\left( \delta _{\sigma 2}\delta _{\eta 2}+\delta _{\sigma 1}\delta
_{\eta 1}\right)  \notag \\
&&+g^{11}\left( \delta _{\sigma 2}\delta _{\eta 2}+\delta _{\sigma 0}\delta
_{\eta 0}\right)  \notag \\
&&+g^{22}\left( \delta _{\sigma 1}\delta _{\eta 1}+\delta _{\sigma 0}\delta
_{\eta 0}\right)  \notag \\
&=&-\delta _{\sigma 2}\delta _{\eta 2}-\delta _{\sigma 1}\delta _{\eta 1} 
\notag \\
&&+\delta _{\sigma 2}\delta _{\eta 2}+\delta _{\sigma 0}\delta _{\eta
0}+\delta _{\sigma 1}\delta _{\eta 1}+\delta _{\sigma 0}\delta _{\eta 0} 
\notag \\
&=&\delta _{\sigma 0}\delta _{\eta 0}+\delta _{\sigma 0}\delta _{\eta 0} 
\notag \\
&=&2\delta _{\sigma 0}\delta _{\eta 0}  \notag
\end{eqnarray}
This is replaced in the Eq.(\ref{122}) 
\begin{eqnarray*}
\partial _{\tau }\partial _{\sigma }A_{\tau }-\partial _{\tau }\partial
_{\tau }A_{\sigma } &=&\frac{1}{2}\left( \frac{v^{2}}{\kappa }\right)
^{2}g^{\mu \alpha }\varepsilon _{\sigma \tau \mu }\varepsilon _{\tau \alpha
\eta }A^{\eta } \\
&=&\frac{1}{2}\left( \frac{v^{2}}{\kappa }\right) ^{2}2\delta _{\sigma
0}\delta _{\eta 0}A^{\eta } \\
&=&\left( \frac{v^{2}}{\kappa }\right) ^{2}\delta _{\sigma 0}A^{0} \\
&=&-\left( \frac{v^{2}}{\kappa }\right) ^{2}\delta _{\sigma 0}A_{0}
\end{eqnarray*}
Taking the gauge condition 
\begin{equation}
\partial _{\tau }A_{\tau }=0  \label{123}
\end{equation}
it results the equation 
\begin{equation}
\partial _{\tau }\partial _{\tau }A_{0}-\left( \frac{v^{2}}{\kappa }\right)
^{2}A_{0}=0  \label{massa}
\end{equation}
The solution of this equation is, in cylindrical geometry, 
\begin{equation*}
A_{0}\left( r\right) =K_{0}\left( mr\right)
\end{equation*}

From here we conclude that the mass of the photon is 
\begin{equation}
m=\frac{v^{2}}{\kappa }  \label{124}
\end{equation}
and this mass is generated via the Higgs mechanism adapted to the
Chern-Simons action. The photon gets a mass because it moves in a background
where the scalar field is equal with the vacuum value, nonzero value.

From physical considerations, we know that 
\begin{equation}
m=\frac{v^{2}}{\kappa }=\frac{1}{\rho _{s}}  \label{1245}
\end{equation}

\subsection{A bound on the energy}

We can express the total energy of the system as the space integral of the 
\emph{time-time} component of the energy-momentum tensor 
\begin{equation}
\mathcal{E}_{tot}=\int d^{2}rT^{00}  \label{152}
\end{equation}

A useful formula (Gradshtein 6.561 formula16, \cite{GR}) is 
\begin{equation}
\int_{0}^{\infty }x^{\mu }K_{\nu }\left( ax\right) dx=2^{\mu -1}a^{-\mu
-1}\Gamma \left( \frac{1+\mu +\nu }{2}\right) \Gamma \left( \frac{1+\mu -\nu 
}{2}\right)  \label{125}
\end{equation}
for 
\begin{eqnarray}
\mathrm{Re}\left( \mu +1\pm \nu \right) &>&0  \label{126} \\
\mathrm{Re}a &>&0  \notag
\end{eqnarray}
Then, for $\mu =1$ and $a=1$, $\nu =0$, 
\begin{eqnarray}
\int_{0}^{\infty }xK_{0}\left( x\right) dx &=&\left[ \Gamma \left( 1\right) %
\right] ^{2}  \label{127} \\
&=&1  \notag
\end{eqnarray}
(cf. Gradshtein 8.338). This must be used with Eq.(\ref{w}) to calculate the
total energy of a system of vortices in plane. 
\begin{eqnarray}
W^{cont} &=&2\pi \int d^{2}r\omega ^{2}K\left( m\left| \mathbf{r}_{1}-%
\mathbf{r}_{2}\right| \right)  \label{128} \\
&=&\omega ^{2}4\pi ^{2}\frac{1}{m^{2}}\int_{0}^{\infty }\left( mr\right)
d\left( mr\right) K_{0}\left( mr\right)  \notag \\
&=&4\pi ^{2}\frac{\omega ^{2}}{m^{2}}=4\pi ^{2}\omega ^{2}\rho _{s}^{2} 
\notag
\end{eqnarray}
Then we have that the $2D$ integral over the plane of the continuum version
of the energy of a system with discrete vortices is constant multiplying the
square of the elementary quantity of vorticity, which was before associated
to each elementary vortex. This corresponds actually to the value of the
energy in the field theoretical model, precisely at the self-dual limit, Eq.(%
\ref{egreater}).

\subsection{Calculation of the flux of the ``magnetic field'' through the
plane}

Start with the second differential equation of self-duality Eq.(\ref{sd}) 
\begin{equation}
F_{+-}=\frac{1}{\kappa ^{2}}\left[ v^{2}\phi -\left[ \left[ \phi ,\phi
^{\dagger }\right] ,\phi \right] ,\phi ^{\dagger }\right]  \label{149}
\end{equation}
This has been calculated previously, with the result 
\begin{equation}
\left[ v^{2}\phi -\left[ \left[ \phi ,\phi ^{\dagger }\right] ,\phi \right]
,\phi ^{\dagger }\right] =\left( v^{2}-2\left( \rho _{1}+\rho _{2}\right)
\right) \left( \rho _{1}-\rho _{2}\right) H  \label{150}
\end{equation}
Then 
\begin{equation}
F_{+-}=\frac{1}{\kappa ^{2}}\left( v^{2}-2\left( \rho _{1}+\rho _{2}\right)
\right) \left( \rho _{1}-\rho _{2}\right) H  \label{151}
\end{equation}

We return to Eq.(\ref{gacon}) (the Gauss law constraint) 
\begin{equation}
F_{12}=\frac{1}{2\kappa }\left( \left[ \phi ^{\dagger },D_{0}\phi \right] -%
\left[ \left( D_{0}\phi \right) ^{\dagger },\phi \right] \right)
\label{f12e}
\end{equation}
We can express in detail this constraint, using the Eqs.(\ref{140}) and (\ref
{1488}) 
\begin{eqnarray}
F_{12} &=&\frac{1}{2\kappa }\left\{ \left[ \phi _{1}^{\ast }E_{-}+\phi
_{2}^{\ast }E_{+},-\frac{i}{2\kappa }\left( PE_{+}+QE_{-}\right) \right]
\right.  \label{153} \\
&&\hspace*{1cm}\left. -\left[ \left( -\frac{i}{2\kappa }\left(
PE_{+}+QE_{-}\right) \right) ^{\dagger },\phi _{1}E_{+}+\phi _{2}E_{-}\right]
\right\}  \notag
\end{eqnarray}
The first term is 
\begin{eqnarray}
&&\left[ \phi _{1}^{\ast }E_{-}+\phi _{2}^{\ast }E_{+},-\frac{i}{2\kappa }%
\left( PE_{+}+QE_{-}\right) \right]  \label{154} \\
&=&\left( -\frac{i}{2\kappa }\right) \left( \phi _{1}^{\ast }P\left[
E_{-},E_{+}\right] +\phi _{2}^{\ast }Q\left[ E_{+},E_{-}\right] \right) 
\notag \\
&=&\frac{i}{2\kappa }\left( \phi _{1}^{\ast }P-\phi _{2}^{\ast }Q\right) H 
\notag \\
&=&\frac{i}{2\kappa }\left( v^{2}-2\left( \rho _{1}+\rho _{2}\right) \right)
\left( \rho _{1}-\rho _{2}\right) H  \notag
\end{eqnarray}
The second term is 
\begin{eqnarray}
&&\left[ \left( -\frac{i}{2\kappa }\left( PE_{+}+QE_{-}\right) \right)
^{\dagger },\phi _{1}E_{+}+\phi _{2}E_{-}\right]  \label{155} \\
&=&\frac{i}{2\kappa }\left[ P^{\ast }E_{-}+Q^{\ast }E_{+},\phi
_{1}E_{+}+\phi _{2}E_{-}\right]  \notag \\
&=&\frac{i}{2\kappa }\left( P^{\ast }\phi _{1}\left[ E_{-},E_{+}\right]
+Q^{\ast }\phi _{2}\left[ E_{+},E_{-}\right] \right)  \notag \\
&=&-\frac{i}{2\kappa }\left( P^{\ast }\phi _{1}-Q^{\ast }\phi _{2}\right) H 
\notag \\
&=&-\frac{i}{2\kappa }\left( v^{2}-2\left( \rho _{1}+\rho _{2}\right)
\right) \left( \rho _{1}-\rho _{2}\right) H  \notag
\end{eqnarray}
Then 
\begin{eqnarray}
F_{12} &=&\frac{1}{2\kappa }\left\{ \frac{i}{2\kappa }\left( v^{2}-2\left(
\rho _{1}+\rho _{2}\right) \right) \left( \rho _{1}-\rho _{2}\right) H\right.
\label{156} \\
&&\hspace*{1cm}\left. +\frac{i}{2\kappa }\left( v^{2}-2\left( \rho _{1}+\rho
_{2}\right) \right) \left( \rho _{1}-\rho _{2}\right) H\right\}  \notag
\end{eqnarray}

The result gives us the magnetic field 
\begin{eqnarray}
F_{12} &=&-B  \label{157} \\
&=&\frac{i}{2\kappa ^{2}}\left( v^{2}-2\left( \rho _{1}+\rho _{2}\right)
\right) \left( \rho _{1}-\rho _{2}\right) H  \notag
\end{eqnarray}
Comparing with 
\begin{eqnarray}
F_{+-} &=&\frac{1}{\kappa ^{2}}\left( v^{2}-2\left( \rho _{1}+\rho
_{2}\right) \right) \left( \rho _{1}-\rho _{2}\right) H  \label{1581} \\
&=&-\frac{1}{4}\frac{v^{4}}{\kappa ^{2}}\left( \rho -\frac{1}{\rho }\right) %
\left[ \frac{1}{2}\left( \rho +\frac{1}{\rho }\right) -1\right] H  \notag
\end{eqnarray}
we note the relation 
\begin{equation}
F_{12}=\frac{i}{2}F_{+-}  \label{1582}
\end{equation}

The flux is 
\begin{eqnarray}
\Phi &=&\int d^{2}r\frac{1}{2}\mathrm{tr}\left( HF_{+-}\right)  \label{1583}
\\
&=&\frac{1}{\kappa ^{2}}\int d^{2}r\left( v^{2}-2\left( \rho _{1}+\rho
_{2}\right) \right) \left( \rho _{1}-\rho _{2}\right)  \notag
\end{eqnarray}
The quantities $\rho _{1}$ and $\rho _{2}$ are not normalized, therefore it
is preferable to change to the variable $\rho $%
\begin{equation}
\rho \equiv \frac{\rho _{1}}{v^{2}/4}=\frac{v^{2}/4}{\rho _{2}}  \label{1584}
\end{equation}
We note that 
\begin{eqnarray}
\mathrm{tr}\left( \phi \phi ^{\dagger }\right) &=&\rho _{1}+\rho _{2}
\label{1484} \\
\left[ \phi ,\phi ^{\dagger }\right] &=&\left( \rho _{1}-\rho _{2}\right) H 
\notag
\end{eqnarray}
\begin{eqnarray}
\mathrm{tr}\left( \phi \phi ^{\dagger }\right) &=&\rho _{1}+\rho _{2}=\left(
v^{2}/4\right) \rho +\frac{v^{2}/4}{\rho }=\frac{v^{2}}{4}\left( \rho +\frac{%
1}{\rho }\right)  \label{1537} \\
\left[ \phi ,\phi ^{\dagger }\right] &=&\left( \rho _{1}-\rho _{2}\right) H=%
\frac{v^{2}}{4}\left( \rho -\frac{1}{\rho }\right) H  \notag
\end{eqnarray}
The flux is normalised as 
\begin{eqnarray}
\Phi &=&\frac{1}{\kappa ^{2}}\int d^{2}r\left( v^{2}-2\left( \rho _{1}+\rho
_{2}\right) \right) \left( \rho _{1}-\rho _{2}\right)  \label{1587} \\
&=&\frac{1}{\kappa ^{2}}\int d^{2}r\left[ v^{2}-2\left( \rho \frac{v^{2}}{4}+%
\frac{v^{2}}{4}\frac{1}{\rho }\right) \right] \left( \rho \frac{v^{2}}{4}-%
\frac{v^{2}}{4}\frac{1}{\rho }\right)  \notag \\
&=&-\frac{1}{4}\frac{1}{\rho _{s}^{2}}\int d^{2}r\left( \rho -\frac{1}{\rho }%
\right) \left[ \frac{1}{2}\left( \rho +\frac{1}{\rho }\right) -1\right] 
\notag
\end{eqnarray}

\textbf{NOTE}. In the \emph{abelian} relativistic model the following
relation exists between the flux and the minimum energy \cite{Dunne1} 
\begin{equation*}
\mathcal{E}_{SD}^{Abelian}=\frac{v^{2}}{2}\mathrm{tr}\left[ \frac{1}{2}%
HF_{+-}\right]
\end{equation*}
which means 
\begin{equation}
\mathcal{E}_{SD}^{Abelian}=\frac{v^{2}}{2}\Phi  \label{1523}
\end{equation}

\textbf{END OF THE NOTE}

We calculate the scalar field self-interaction potential 
\begin{eqnarray}
V\left( \phi ,\phi ^{\dagger }\right) &=&\frac{1}{4\kappa ^{2}}\mathrm{tr}%
\left[ \left( \left[ \left[ \phi ,\phi ^{\dagger }\right] ,\phi \right]
-v^{2}\phi \right) ^{\dagger }\left( \left[ \left[ \phi ,\phi ^{\dagger }%
\right] ,\phi \right] -v^{2}\phi \right) \right]  \label{533} \\
&=&\frac{1}{4\kappa ^{2}}\mathrm{tr}\left[ \left( PE_{+}+QE_{-}\right)
^{\dagger }\left( PE_{+}+QE_{-}\right) \right]  \notag \\
&=&\frac{1}{4\kappa ^{2}}\mathrm{tr}\left[ \left( P^{\ast }E_{-}+Q^{\ast
}E_{+}\right) \left( PE_{+}+QE_{-}\right) \right]  \notag \\
&=&\frac{1}{4\kappa ^{2}}\left[ P^{\ast }P\mathrm{tr}\left(
E_{-}E_{+}\right) +Q^{\ast }Q\mathrm{tr}\left( E_{+}E_{-}\right) \right] 
\notag \\
&=&\frac{1}{4\kappa ^{2}}\left( P^{\ast }P+Q^{\ast }Q\right)  \notag
\end{eqnarray}
since the other traces are zero. Using the notations $\rho _{1}$ and $\rho
_{2}$ we have 
\begin{eqnarray}
V\left( \phi ,\phi ^{\dagger }\right) &=&\frac{1}{4\kappa ^{2}}\left(
P^{\ast }P+Q^{\ast }Q\right)  \label{537} \\
&=&\frac{1}{4\kappa ^{2}}\left\{ \left[ v^{2}-2\left( \rho _{1}-\rho
_{2}\right) \right] ^{2}\rho _{1}+\left[ v^{2}+2\left( \rho _{1}-\rho
_{2}\right) \right] ^{2}\rho _{2}\right\}  \notag \\
&=&\frac{1}{4\kappa ^{2}}\left\{ v^{4}\rho _{1}-4v^{2}\left( \rho _{1}-\rho
_{2}\right) \rho _{1}+4\left( \rho _{1}-\rho _{2}\right) ^{2}\rho _{1}\right.
\notag \\
&&\left. +v^{4}\rho _{2}+4v^{2}\left( \rho _{1}-\rho _{2}\right) \rho
_{2}+4\left( \rho _{1}-\rho _{2}\right) ^{2}\rho _{2}\right\}  \notag \\
&=&\frac{1}{4\kappa ^{2}}\left\{ v^{4}\left( \rho _{1}+\rho _{2}\right)
-4v^{2}\left( \rho _{1}-\rho _{2}\right) ^{2}+4\left( \rho _{1}-\rho
_{2}\right) ^{2}\left( \rho _{1}+\rho _{2}\right) \right\}  \notag
\end{eqnarray}

We can express the potential in terms of the normalised variable 
\begin{eqnarray}
&&V\left( \phi ,\phi ^{\dagger }\right)  \label{1585} \\
&=&\frac{1}{4\kappa ^{2}}\left\{ v^{4}\left( \rho _{1}+\rho _{2}\right)
-4v^{2}\left( \rho _{1}-\rho _{2}\right) ^{2}+4\left( \rho _{1}-\rho
_{2}\right) ^{2}\left( \rho _{1}+\rho _{2}\right) \right\}  \notag \\
&=&\frac{1}{4\kappa ^{2}}\left\{ v^{4}\left( \rho \frac{v^{2}}{4}+\frac{v^{2}%
}{4}\frac{1}{\rho }\right) \right.  \notag \\
&&-4v^{2}\left( \rho \frac{v^{2}}{4}-\frac{v^{2}}{4}\frac{1}{\rho }\right)
^{2}  \notag \\
&&\left. +4\left( \rho \frac{v^{2}}{4}-\frac{v^{2}}{4}\frac{1}{\rho }\right)
^{2}\left( \rho \frac{v^{2}}{4}+\frac{v^{2}}{4}\frac{1}{\rho }\right)
\right\}  \notag
\end{eqnarray}
\begin{eqnarray}
&&V\left( \phi ,\phi ^{\dagger }\right)  \label{1586} \\
&=&\frac{1}{4\kappa ^{2}}\frac{v^{6}}{4}\left\{ \left( \rho +\frac{1}{\rho }%
\right) -\left( \rho -\frac{1}{\rho }\right) ^{2}+\frac{1}{4}\left( \rho -%
\frac{1}{\rho }\right) ^{2}\left( \rho +\frac{1}{\rho }\right) \right\} 
\notag \\
&=&\frac{1}{16}v^{2}\frac{v^{4}}{\kappa ^{2}}\left\{ \left( \rho +\frac{1}{%
\rho }\right) -\left( \rho -\frac{1}{\rho }\right) ^{2}+\frac{1}{4}\left(
\rho -\frac{1}{\rho }\right) ^{2}\left( \rho +\frac{1}{\rho }\right) \right\}
\notag
\end{eqnarray}

%%%%%%%%%%%%%%%%%%%%%%%%%%%%%%%%%%%%%%%%%%%%%%%%%%%%%%%%%%%%%%%%%%%%% 
\begin{figure}[tbph]
\centerline{\includegraphics[height=5cm]{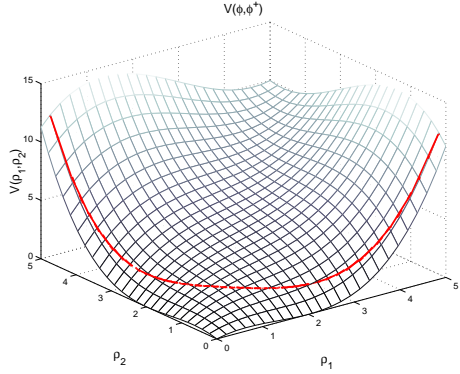}}
\caption{The potential $V(\protect\phi,\protect\phi^{\dagger})$. The points
where $\protect\rho_{1} \protect\rho_{2}=1$ are shown.}
\label{potential1}
\end{figure}
%%%%%%%%%%%%%%%%%%%%%%%%%%%%%%%%%%%%%%%%%%%%%%%%%%%%%%%%%%%%%%%%%%%%%

%%%%%%%%%%%%%%%%%%%%%%%%%%%%%%%%%%%%%%%%%%%%%%%%%%%%%%%%%%%%%%%%%%%%% 
\begin{figure}[tbph]
\centerline{\includegraphics[height=5cm]{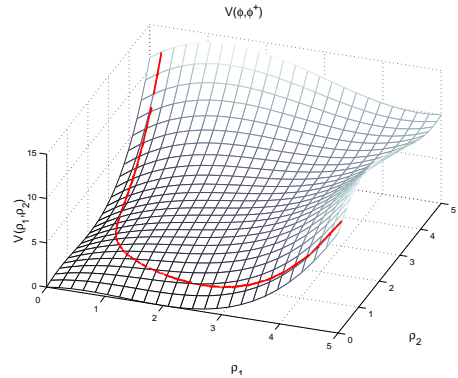}}
\caption{Same as figure \ref{potential1} with a different view.}
\label{potential2}
\end{figure}
%%%%%%%%%%%%%%%%%%%%%%%%%%%%%%%%%%%%%%%%%%%%%%%%%%%%%%%%%%%%%%%%%%%%%

\subsection{Comment on the possible associations between the
field-theoretical variables and physical variables}

The field theoretical model has been developed as the continuum version of
the system of discrete vortices interacting via a short range potantial. On
the other hand, the original model, represented by the CHM equation, is
expressed in terms of three variables 
\begin{eqnarray*}
&&\psi \\
\mathbf{v} &=&-\mathbf{\nabla }\times \widehat{\mathbf{e}}_{z}\psi \\
\mathbf{\omega } &=&\mathbf{\nabla \times v}=\Delta \psi
\end{eqnarray*}
(where $\widehat{\mathbf{e}}_{z}$ is the versor perpendicular on the plane).
These variables have a clear physical meaning. We would like to understand
the possible connection between the variables of the field theory \ ($j^{\mu
}$, $A_{\mu }$, $F_{\mu \nu }$, $\phi $) and parameters ($\kappa $ and $%
v^{2} $) and the physical variables. We make the observation that the
conservation law for the scalar and vorticity field \cite{HH} is (including
explicitely the normalisation factor $\rho _{s}$) 
\begin{equation*}
\iint d^{2}r\left[ \psi ^{2}+\left( \rho _{s}\mathbf{\nabla }\psi \right)
^{2}\right] =\text{const}
\end{equation*}
We make an integration by parts of the second term 
\begin{eqnarray*}
\iint d^{2}r\left( \mathbf{\nabla }\psi \right) ^{2} &=&\iint d^{2}r\left( 
\mathbf{\nabla }\psi \cdot \mathbf{\nabla }\psi \right) = \\
&=&\iint d^{2}r\left[ \mathbf{\nabla }\left( \psi \cdot \mathbf{\nabla }\psi
\right) -\psi \Delta \psi \right]
\end{eqnarray*}
The integration of the first term is transformed 
\begin{equation*}
\iint d^{2}r\mathbf{\nabla }\left( \psi \cdot \mathbf{\nabla }\psi \right)
=\oint d\mathbf{l\cdot }\left( \psi \mathbf{\nabla }\psi \right)
\end{equation*}
Since we have the definition $\mathbf{v}=-\mathbf{\nabla }\psi \times 
\widehat{\mathbf{e}}_{z}$ we see that the scalar product excludes the
component of $\mathbf{\nabla }\psi $ that is normal to the contour, which is
a circle of very large radius. The diamagnetic effect makes that the
velocity of gyration of particles on Larmor orbits generates a macroscopic
velocity of the fluid which is tangent to the circle, therefore $\mathbf{%
\nabla }\psi $ has only a nonzero component, normal to the circle. We
conclude that there is no contribution from the first term in the integrand.
Then we have 
\begin{equation*}
\iint d^{2}r\psi \left( \psi -\rho _{s}^{2}\Delta \psi \right) =\text{const}
\end{equation*}
In particular this means that a \emph{vacuum state}, of zero energy,
corresponds to 
\begin{equation*}
\rho _{s}^{2}\Delta \psi -\psi =0
\end{equation*}
or 
\begin{equation}
\rho _{s}^{2}\omega -\psi =0  \label{3259}
\end{equation}

On the other hand, we have a characterisation of the vacuum state in the
field theoretical model, obtained as the asymptotic state of the fields at
large distance. There the scalar field $\phi $ is almost constant and the
space derivatives are vanishing (this is also known as the large wavelength
approximation). It will be shown below that the current $j^{\mu }$ and the
potential $A^{\mu }$ verify the relation 
\begin{equation}
j^{\mu }-2iv^{2}A^{\mu }\simeq 0  \label{3261}
\end{equation}
If these two relations Eqs.(\ref{3259}) and (\ref{3261}) describe the same
physics they suggest (ignoring the signs and numerical factor) the following
qualitative identifications 
\begin{eqnarray}
j &\sim &\rho _{s}^{2}\omega  \label{3265} \\
v^{2}A &\sim &\psi  \notag
\end{eqnarray}
The second of this equation may seem strange since $A$ (and the covariant
derivatives) are vectors. The combination that seems to be plausible is 
\begin{eqnarray*}
D_{i}\phi _{j} &=&\partial _{i}\phi _{j}-\varepsilon _{ik}\psi \phi _{k} \\
A_{i} &\sim &-\frac{1}{v^{2}}\varepsilon _{ik}\psi
\end{eqnarray*}
There is some confirmation from the Abelian version in the Maxwell-Higgs
case at self-duality \cite{Lohe}. Eqs.(\ref{3265}) further suggest that the
magnetic field of the model can be associated with the physical velocity, $%
\mathbf{B}\sim \mathbf{v}$ (however in this framework no connection can be
made with the Elsasser variables $\mathbf{u=v+B}$, $\mathbf{w=v-B}$). It is
interesting to remark that the magnetic field $B$ can as well be associated
with the physical vorticity, since the Chern-Simons Lagrangean has the
unique property that connects directly the field tensor $F_{\mu \nu }$ with
the current $J^{\mu }$, as is shown by the second equation of motion, $%
-\kappa \varepsilon ^{\mu \nu \rho }F_{\nu \rho }=iJ^{\mu }$. Then the
relationship which is fundamental for the connection between the
field-theoretical framework and the physical model, $\ln \rho =\psi $
(assumed previously as a simple change of variables) appears now consistent
with the physical meaning of $B$, since 
\begin{equation}
B\sim \Delta \ln \rho =\Delta \psi =\omega  \label{3268}
\end{equation}
This also suggests 
\begin{equation}
\kappa B\sim \rho _{s}^{2}\omega  \label{3274}
\end{equation}
The detailed form of these identification cannot be made more precise and we
limit ourselves to a dimensional analysis. In these relationships there is
no factor of dimensionality to intermediate between the two sides. The
dimensional factor that multiply the first relation in Eq.(\ref{3265}) must
also multiply the second relation, due to Eq.(\ref{3261}). As will be
verified below, the factor (we note it $\chi $) must have the dimension 
\begin{equation}
\left[ \chi \right] =L^{3}  \label{2010}
\end{equation}
and this implies for the dimansions of the variables ($L$ is length, $T$ is
time) 
\begin{eqnarray}
\left[ \chi \right] \left[ j\right] &=&\left[ \rho _{s}^{2}\omega \right] =%
\frac{L^{2}}{T}  \label{2011} \\
L^{3}\left[ j\right] &=&\frac{L^{2}}{T}  \notag
\end{eqnarray}
or 
\begin{equation}
\left[ j\right] =\frac{1}{LT}  \label{2012}
\end{equation}
In the second relation of Eq.(\ref{3265}) we have 
\begin{eqnarray}
\left[ \chi \right] \left[ v^{2}A\right] &=&\left[ \psi \right] =\frac{L^{2}%
}{T}  \label{2013} \\
L^{3}\left[ v^{2}A\right] &=&\frac{L^{2}}{T}  \notag
\end{eqnarray}
As we have mentioned before, the quantity $v^{2}$ is related with the
physical background of vorticity generated by the gyration of the particles.
Then its dimension is 
\begin{equation}
\left[ v^{2}\right] =\frac{1}{T}  \label{2014}
\end{equation}
from which we derive 
\begin{equation}
\left[ v^{2}\right] \left[ A\right] =\frac{1}{TL}  \label{2015}
\end{equation}
or 
\begin{equation}
\left[ A\right] =\frac{1}{L}  \label{2016}
\end{equation}
This further gives 
\begin{equation}
\left[ B\right] =\frac{1}{L^{2}}  \label{2017}
\end{equation}
All dimensions become coherent if we identify 
\begin{eqnarray}
\kappa &\equiv &c_{s}  \label{2018} \\
v^{2} &\equiv &\Omega _{ci}  \notag
\end{eqnarray}
For example, using again the unique dimensional coefficient $\chi $, Eq.(\ref
{3274}) is dimensionally correct. 
\begin{eqnarray}
\left[ \chi \right] \left[ \kappa \right] \left[ B\right] &=&\left[ \rho
_{s}^{2}\omega \right]  \label{2019} \\
L^{3}\frac{L}{T}\frac{1}{L^{2}} &=&L^{2}\frac{1}{T}  \notag
\end{eqnarray}
One can now verify that all equations in the field model have coherent
dimensions.

We have now a qualitative association between the physical variables and the
field model variables and we also have the physical dimensions of the
latter. We note that the covariant derivatives (having dimension $L^{-1}$) 
\begin{equation}
D_{\mu }=\partial _{\mu }+\left[ A_{\mu },\right]  \label{2020}
\end{equation}
cannot have a clear identification in terms of physical variables. One can
only say that the zero component is 
\begin{equation}
D_{0}=\frac{\partial }{c_{s}\partial t}+\frac{1}{L^{3}}\frac{\psi }{\Omega
_{ci}}  \label{2021}
\end{equation}
where we have taken into account the second relation from Eq.(\ref{3265})
and included the unknown dimensional factor $L^{3}$. From the Eqs.(\ref{61})
and (\ref{79}), (\ref{80}) we note that it is not possible to express in
terms of the classical ($\psi $, $\mathbf{v}$, $\mathbf{\omega }$) variables
the potentials $A_{x}$ and $A_{y}$.

\subsection{Comment on the physical constants and normalisations}

One of the characteristics of the physical model is the presence of a
uniform background of vorticity. In the absence of any excitation we have on
any contour in plane a tangential projection of the velocity of the
particles performing the Larmor gyration.

An arbitrary contour (say, a large circle of radius $R$) will intersect the
circle of the Larmor gyration (of radius $\rho _{s}$) and one can calculate
an average of projection of the velocity onto the tangent at the contour
line. Supposing that $R\gg \rho _{s}$, the contour intercepted by the Larmor
circle can be approximated with a stright line that intersects the circle
between the angles $\theta _{0}$ and $\pi -\theta _{0}$. The contour is a
chord and the average $\overline{v}_{\theta _{0}}$ of the velocity's
projection on it, $v_{c}\left( \theta \right) $, is 
\begin{eqnarray}
\overline{v}_{\theta _{0}} &=&\int_{\theta _{0}}^{\pi -\theta _{0}}\frac{%
d\theta }{\left[ \left( \pi -\theta _{0}\right) -\theta _{0}\right] }%
v_{c}\left( \theta \right)  \label{2030} \\
&=&\frac{1}{\pi -2\theta _{0}}\int_{\theta _{0}}^{\pi -\theta _{0}}d\theta
v_{L}\sin \theta  \notag \\
&=&\frac{2v_{L}}{\pi -2\theta _{0}}\cos \theta _{0}  \notag
\end{eqnarray}
where $v_{L}$ is the velocity on the Larmor circle. Now we can average over
the various lengths of the chord inside the Larmor circle, 
\begin{eqnarray}
\overline{v} &=&\int_{0}^{\pi }\frac{d\theta _{0}}{\pi }\frac{2v_{L}}{\pi
-2\theta _{0}}\cos \theta _{0}  \label{2031} \\
&=&\frac{2v_{L}}{\pi }\int_{0}^{\pi /2}\frac{\sin \tau }{\tau }d\tau  \notag
\\
&=&\frac{2\left( 1.37\right) }{\pi }v_{L}  \notag
\end{eqnarray}
The symmetric situation will bring a similar factor and finally the average
projected velocity is within a factor not far from unity equal to $v_{L}$.
Now consider the definition of the rotational 
\begin{equation}
\omega \equiv \left| \mathbf{\nabla \times v}\right| =\underset{A\rightarrow
0}{\lim }\frac{\oint_{\Gamma }\mathbf{v\cdot dl}}{A}  \label{2032}
\end{equation}
where $A$ is the area inside the closed contour $\Gamma $. We have, within a
unity-size factor 
\begin{equation}
\oint_{\Gamma }\mathbf{v\cdot dl\simeq }2\pi Rv_{L}  \label{2033}
\end{equation}
\begin{equation}
A=\pi R^{2}  \label{2034}
\end{equation}
Then 
\begin{equation}
\omega \sim \underset{R\rightarrow \rho _{s}}{\lim }\frac{2v_{L}}{R}
\label{2035}
\end{equation}
Since 
\begin{equation}
v_{L}=\rho _{s}\Omega  \label{2036}
\end{equation}
we obtain 
\begin{eqnarray}
\omega &\sim &\Omega \underset{R\rightarrow \rho _{s}}{\lim }\frac{\rho _{s}%
}{R}  \label{2037} \\
&=&\Omega  \notag
\end{eqnarray}
\emph{i.e.} we obtain that the value of the vorticity in a region with
uniform density of Larmor gyrating particles is $\Omega $, the cyclotronic
velocity.

\bigskip

We have in this moment three parallel models, representing the same reality,
which we call the Charney-Hasegawa-Mima vortical flow. The connection
between these three models implies a comparison of the physical quantities
present in each of them. For this reason we have to consider the physical
content of the field-theoretical model and in particular we will introduce
nondimensional variables. The two physical quantities appearing explicitly
in the field-theoretical model are $\kappa $ and $v^{2}$.

\subsection{Comparison with numerical simulation and with experiment}

The second factor of the nonlinearity \emph{i.e.} $\left( \cosh \psi
-p\right) $ in all versions of the equation derived above, in particular in
Eq.(\ref{4975}) can also be negative, under a certain choice of
normalizations. Then, a certain aspect of the graphs resulting from
numerical simulations (see Seyler \cite{Seyler}), \emph{i.e.} the presence
of two visible symmetric extrema on the graph (\emph{vorticity,
streamfunction}) is in agreement with our equation. 
%%%%%%%%%%%%%%%%%%%%%%%%%%%%%%%%%%%%%%%%%%%%%%%%%%%%%%%%%%%%%%%%%%%%%

\begin{center}
\begin{figure}[tbh]
\centering
\begin{minipage}[t]{0.4\linewidth}
    \centering
    \includegraphics[height=5cm]{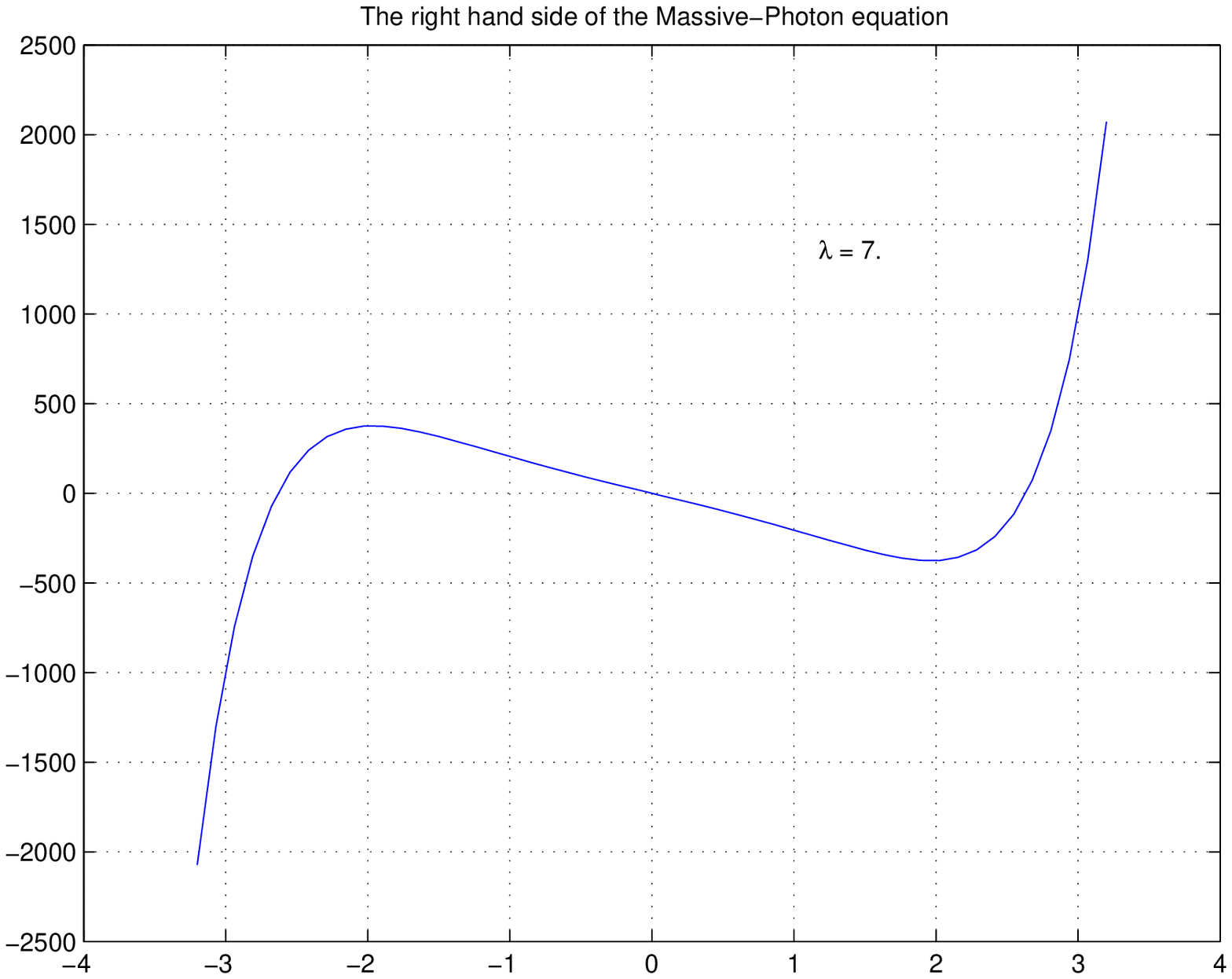}
   \end{minipage}
\hspace{0.05\textwidth} 
\begin{minipage}[t]{0.4\linewidth}
    \centering
    \includegraphics[height=5cm]{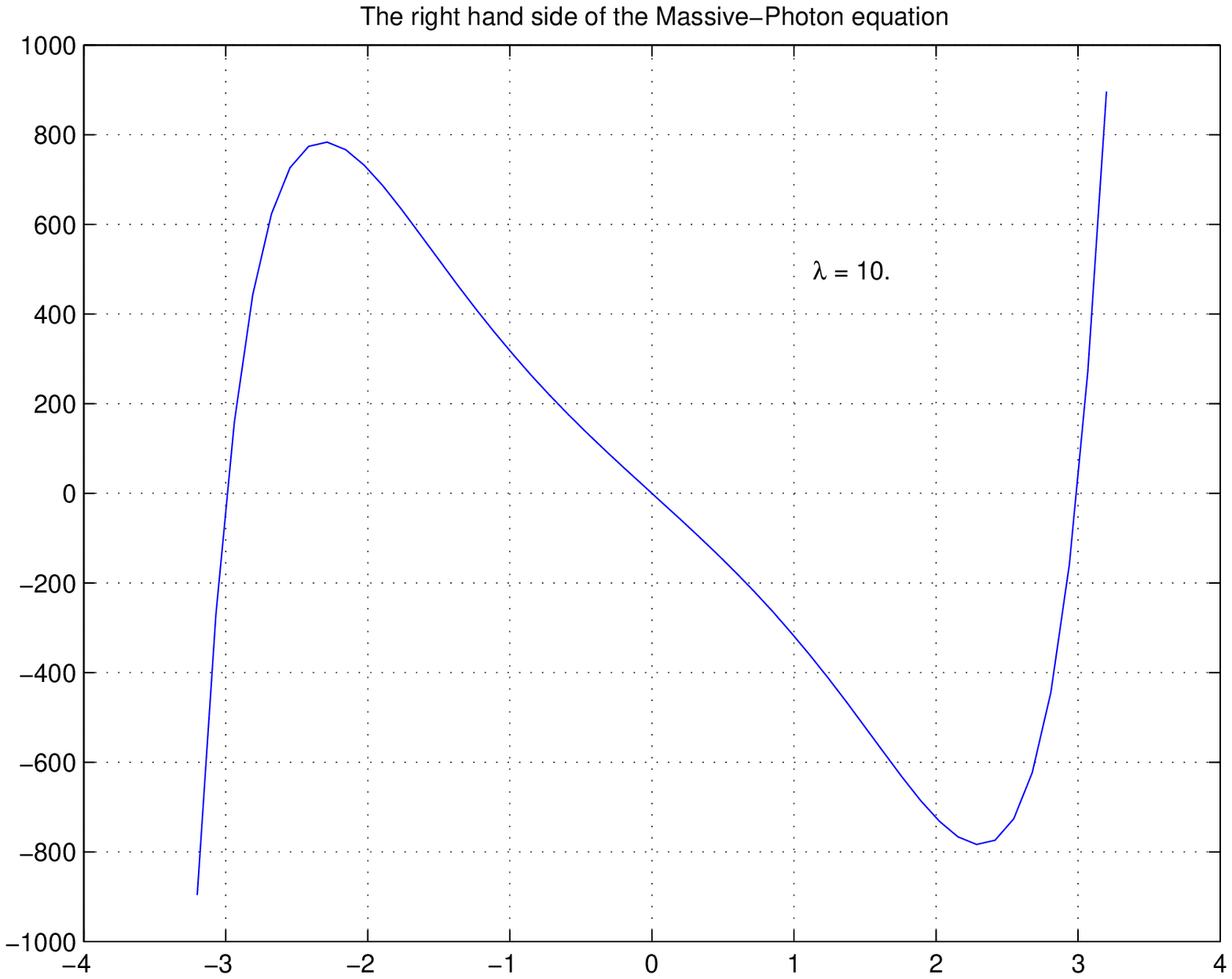}
   \end{minipage}
\caption{The nonlinear term in the equation, for $p=7$ and $p=10$}
\label{fig1}
\end{figure}
\end{center}

%%%%%%%%%%%%%%%%%%%%%%%%%%%%%%%%%%%%%%%%%%%%%%%%%%%%%%%%%%%%%%%%%%%%%%

It is interesting to remark that similar pictures to our result have been
found in the study of geostrophic turbulence \cite{expgeo}. The scatter
plots for the pair $\left( \omega ,\psi \right) $ obtained from experimental
study of decaying vorticity field and represented in the figure 21 of this
reference are very similar to our result (with the choice of inverse sign
for the streamfunction).

\subsection{The vacuum fields}

We comment on the meaning of the vacuum value of the matter field.
Obviously, it must be related with the presence, in the physical model, of a
background of vorticity simply given by the gyration of the particles. This
is the sense of the fact that, even at infinity, we have a constant density
of matter, $\left| \phi \right| ^{2}=v^{2}$. The excitations in the form of
large vortices take place on this background, whose value of vorticity is
very high, the ion cyclotronic frequency, $\Omega _{ci}$. As we said before
we can possibly identify 
\begin{eqnarray}
v^{2} &\equiv &\Omega _{ci}  \label{2054} \\
\kappa &=&c_{s}  \notag
\end{eqnarray}
The physical vorticity is derived from $F_{+-}$ 
\begin{equation}
F_{+-}=\frac{1}{\kappa ^{2}}\left( v^{2}-2\left( \rho _{1}+\rho _{2}\right)
\right) \left( \rho _{1}-\rho _{2}\right) H  \label{2055}
\end{equation}

\subsection{The subset of self-dual states of the physical system}

We should remind that the identification 
\begin{equation}
\psi \equiv \ln \rho  \label{601}
\end{equation}
was done after the equations of motion in the field theoretical framework
have been reduced to the equations of \emph{self-duality} and stationarity.
Therefore it is not surprising that, returning with Eq.(\ref{601}) to the
original, Charney-Hasegawa-Mima equation, we find that this one is verified
by the functions $\rho $ obeying our equation (\ref{895}) or Eq.(\ref{94})
for $\psi $%
\begin{equation}
\left[ \left( -\mathbf{\nabla }\psi \times \widehat{\mathbf{n}}\right) \cdot 
\mathbf{\nabla }\right] \mathbf{\nabla }^{2}\psi =0  \label{602}
\end{equation}
because 
\begin{equation*}
\left[ \left( -\mathbf{\nabla }\psi \times \widehat{\mathbf{n}}\right) \cdot 
\mathbf{\nabla }\right] \left[ -\sinh \psi \left( \cosh \psi -1\right) %
\right] =0
\end{equation*}
The fact that, for any solution $\psi $ of the Eq.(\ref{94}) the equation of
Charney-Hasegawa-Mima is verified at stationarity is useful as a
confirmation but is of moderate significance, due to the large space of
functions that can verify Eq.(\ref{602}). The subset of self-dual states is
much smaller and precisely defined by Eq.(\ref{94}).

\subsection{Comment on the self-duality}

In general the self-duality should be seen as a property of a particular
geometrico-algebraic object : a fiber bundle. This consists of a basis
manifold (on which one has to define an atlas of compatible charts), a fiber
attached to every point of the basis manifold (in physics the fiber is
seldom said space of internal symmetry) and a group of automorphism of the
typical fiber. The local structure on the basis and in the fibre space is
Euclidean since they both are manifolds. One can construct the total space
of the fiber bundle, which is locally a Cartesian product of an open set of
the basis with the space of the fiber, and a projection operator acting in
this total space and projecting the points of the total space onto the
basis. The transition functions between neighboring charts consist of
elements of the group. A connection is a differential one-form defined in
every point of the total space and taking values in the algebra of the
group. The curvature is the differential two-form obtained by an exterior
differentiation of the connection one-form. For a concrete example, the
connection is the \emph{potential} $A_{\mu }$ and the curvature is the \emph{%
field strength} $F_{\mu \nu }$ like in the electromagnetism, or in any other
theory expressed in similar terms. There is a Hodge duality operator,
denoted $\ast $ : applied on a differential $p$-form in a space with $n$
dimensions, it generates a differential $n-p$ form, such as the exterior
product of these two forms produces a scalar multiplying the unique $n$ form
that can de defined on the $n$-dimensional space, \emph{i.e.} a multiple of
the volume form.

The self-duality is the property that consists of the equality between the a
differential form and its Hodge dual; this naturally requires that the space
be of even dimension. Only the differential forms of the order representing
half of the (even) dimension of the space can be self-dual since only in
this case their duals will be of the same order. For example in a space with
dimension four, differential two-forms can be self-dual.

In particular physical models the differential two-form representing the
field strength and its dual are equal at self-duality. But this two-form
represents the curvature of the fibre bundle, therefore at self-duality the
curvature is equal to its dual. In many cases, this equality is realised by
the fact that the curvature is zero and one says that the space is \emph{flat%
}.

When the self-duality is realised as a condition of \emph{flatness} it is
possible to express this equality as the compatibility condition of a system
of linear differential equations. This makes possible to introduce a Lax
operator and the self-duality equation is exactly integrable by Inverse
Scattering Transform. A set of infinite invariants can be found. One example
of this type is the classical \emph{sigma} model.

In particular cases, (like ours) the geometrical structure is less clear.
The self-duality is expressed by the fact that the \emph{action} functional
is minimized \emph{i.e.} the Bogomolnyi limit is saturated.

\subsection{Comment on the $6^{th}$ order potential}

The Abelian version of this theory but with the Maxwell term instead of
Chern-Simons is well known from superconductivity theory. It implies a
potential of only fourth order which provides the symmetrical vacua of the
theory and allows mass generation for the Maxwell photon via the Higgs
mechanism.

However in the present theory a sixth order potential is necessary. It has
been demonstrated that with only a sixth order potential one can have
self-dual states. This has been shown by a simple verification which we have
reproduced in the Section about the energy functional related to the
Lagrangean density.

However the necessity to include a sixth order potential in the Lagrangean
density has a profond origin. This has been shown in series of papers \cite
{olivewitten}, \cite{hlousek1}, \cite{hlousek2}.

It has been shown that the Bogomolnyi lower bound for the energy and the
first-order-in-time differential equations obtained at self-duality are a
property of a classical field theory which possesses a topological charge.
The theory is a reduction of a supersymmetric (susy) theory in which the
topological charge appears as the central charge of the susy algebraic
structure. ( A susy theory is a classical field theory in which besides the
usual fields there are other field-variables with the property that they
anti-commute, \emph{i.e.} they are classical spinors). It is interesting the
way in which this has been shown \cite{hlousek1},\cite{hlousek2}. First it
is shown that any supersymmetric theory which possesses a topological charge
necessarly possesses a Bogomolnyi bound and SD equations of motion. Then for
a given field theory where a topological conservation charge exists, it is
first constructed a supersymmetric extension, adding the anti-commuting
variables and other variables that are necessary to close the new algebraic
structure. In this extended theory the \emph{central charge }of the susy
algebra is the topological charge of the initial theory. The Bogomolnyi
bound is identified. Finally it is shown that returning back from the susy
extension to the original theory, one still preserves the Bogomolnyi bound.
The relation between the potential $W$ in the extended theory and the
potential $U$ in the classical non-susy theory is 
\begin{equation}
U\left( \phi \right) =\sum_{a}\left( \frac{\partial W}{\partial \phi _{a}}%
\right) ^{2}  \label{2071}
\end{equation}
and for a symmetrical two-vacua potential $W$ we have a potential $U$ of
sixth degree in $\phi _{a}$.

\bigskip\ 

Lee, Lee and Weinberg \cite{llw} show explicitly in an Abelian case how this
form of the potential is obtained from the requirement that the model can be
extended to a $N=2$ supersymmetric model. They begin by constructing an $N=1$
supersymmetric generalization of this Chern-Simons Higgs theory, in which
the form of the potential $f\left( \left| \phi \right| ^{2}\right) $ is not
yet specified. Adding a single pair of Grassmannian varibales $\left( \theta
,\overline{\theta }\right) $ to the set of variables of the original model
requires to extend the model by introducing additional fields. This is
necessary since the supersymmetry transformation must be closed and the
original fields are not sufficient. The model will include a matter
super-field $\Phi $ which consists of: a complex scalar field $\phi $, a
complex spinor field $\psi $ and an auxiliary scalar field $F$; a real
spinor field $\Gamma ^{\alpha }$ which contains a real \emph{photon} field $%
A_{\mu }$ and a Majorana spinor \emph{photino} field $\lambda $. The action
is 
\begin{eqnarray}
\mathcal{S} &=&\int d^{2}xdt\int d\theta d\overline{\theta }\left\{ -\frac{1%
}{4}\kappa D^{\alpha }\Gamma ^{\beta }D_{\beta }\Gamma _{\alpha }\right.
\label{2072} \\
&&-\frac{1}{2}\left( D^{\alpha }+i\Gamma ^{a}\right) \Phi ^{\ast }\left(
D_{\alpha }-i\Gamma _{\alpha }\right) \Phi  \notag \\
&&\left. +f\left( \Phi ^{\ast }\Phi \right) \right\}  \notag
\end{eqnarray}
where the first term is the generalization of the Chern-Simons term. The
integration over the Grassmann variables $\theta $ and $\overline{\theta }$
can be done explicitely and in this process it is required to make them
visible in the expression of the superfield $\Phi $. In this way there
appears in the action density first and second order derivatives of the
potential function $f$ since only in this way (by this Taylor expansion) the
Grassmann variables will appear explicitely and can be integrated. The
action becomes 
\begin{eqnarray}
\mathcal{S} &=&\int d^{2}xdt\left\{ \frac{1}{4}\kappa \varepsilon ^{\mu \nu
\lambda }A_{\mu }F_{\nu \lambda }+\left( D^{\mu }\phi \right) \left( D_{\mu
}\phi \right) \right.  \label{2073} \\
&&-\frac{1}{2}\kappa \overline{\lambda }\lambda +i\overline{\psi }\gamma
^{\mu }D_{\mu }\psi +i\left( \overline{\psi }\lambda \phi -\overline{\lambda 
}\psi \phi ^{\ast }\right) +F^{\ast }F  \notag \\
&&+f^{\prime }\left( F^{\ast }\phi +F\phi ^{\ast }\right) -\frac{1}{2}%
f^{\prime \prime }\left( \phi ^{2}\overline{\psi }\psi ^{c}+\phi ^{\ast 2}%
\overline{\psi }^{c}\psi \right)  \notag \\
&&\left. -\left( f^{\prime }+\left| \phi \right| ^{2}f^{\prime \prime
}\right) \overline{\psi }\psi \right\}  \notag
\end{eqnarray}
where the superscript $c$ means charge conjugate and the \emph{prime} means
derivative of the function $f$ to its argument, $\left| \phi \right| ^{2}$.
The equations of motion for the fields $\lambda $, $F$ and $F^{\ast }$ are 
\begin{eqnarray}
\lambda &=&\frac{i}{\kappa }\left( \psi ^{c}\phi -\psi \phi ^{\ast }\right)
\label{2074} \\
F &=&-\phi f^{\prime }  \notag \\
F^{\ast } &=&-\phi ^{\ast }f^{\prime }  \notag
\end{eqnarray}
They permit the replacement of the corresponding functions in the action. 
\begin{eqnarray}
\mathcal{S} &=&\int d^{2}xdt\left\{ \frac{1}{4}\kappa \varepsilon ^{\mu \nu
\lambda }A_{\mu }F_{\nu \lambda }+\left( D^{\mu }\phi \right) \left( D_{\mu
}\phi \right) \right.  \label{s5} \\
&&-\left| \phi \right| ^{2}f^{\prime 2}+i\overline{\psi }\gamma ^{\mu
}D_{\mu }\psi  \notag \\
&&-\frac{1}{2}\left( f^{\prime \prime }+\frac{1}{\kappa }\right) \left( \phi
^{2}\overline{\psi }\psi ^{c}+\phi ^{\ast 2}\overline{\psi }^{c}\psi \right)
\notag \\
&&\left. +\left[ \left| \phi \right| ^{2}\left( \frac{1}{\kappa }-f^{\prime
\prime }\right) -f^{\prime }\right] \overline{\psi }\psi \right\}  \notag
\end{eqnarray}
In order this action to be invariant under an $N=2$ extended supersymmetry
it is required that the term on the third line in the above formula vanishes 
\begin{equation}
f^{\prime \prime }=-\frac{1}{\kappa }  \label{2075}
\end{equation}
or, integrating two times on the variable $\xi \equiv \left| \Phi \right|
^{2}$ 
\begin{eqnarray}
f\left( \left| \Phi \right| ^{2}\right) &=&-\frac{1}{2\kappa }\left( \xi
-v^{2}\right) ^{2}  \label{2076} \\
&=&-\frac{1}{2\kappa }\left( \left| \Phi \right| ^{2}-v^{2}\right) ^{2} 
\notag
\end{eqnarray}
where $v^{2}$ is a constant. The form of the \emph{potential} term in the
action is then obtained from the first term in the second line of Eq.(\ref
{s5}) 
\begin{equation}
\left| \phi \right| ^{2}f^{\prime 2}\rightarrow \frac{1}{2\kappa }\left|
\phi \right| ^{2}\left( \left| \phi \right| ^{2}-v^{2}\right) ^{2}
\label{2077}
\end{equation}
The action contains a bosonic part that has this potential and this is
actually the Abelian version of the model discussed in this work.

\bigskip

The fact that at self-duality the theory is a part of a larger
supersymmetric theory and that this explains the form of the scalar
self-interaction may help us to trace the meaning of the changes we find
between the \emph{sinh}-Poisson equation (for the ideal fluid) and the \emph{%
double-sinh}-Poisson equation, for the fluid of ions with Larmor gyration.

The special Higgs potential that appears in the Lagrangean density and leads
to self-dual states has a symmetric minimum which is degenerate with the
symmetry-breaking one. This means that the system can have non-topological
solitons which verify the same self-dual equations.

%\begin{appendices}

\section{Appendix A : Derivation of the equation}

% new type of equation numbers for the appendix
\renewcommand{\theequation}{A.\arabic{equation}} \setcounter{equation}{0}

Consider the equations for the ITG model in two-dimensions with adiabatic
electrons: 
\begin{eqnarray}
\frac{\partial n_{i}}{\partial t}+\mathbf{\nabla \cdot }\left( \mathbf{v}%
_{i}n_{i}\right) &=&0  \label{a1} \\
\frac{\partial \mathbf{v}_{i}}{\partial t}+\left( \mathbf{v}_{i}\cdot 
\mathbf{\nabla }\right) \mathbf{v}_{i} &=&\frac{e}{m_{i}}\left( -\mathbf{%
\nabla }\phi \right) +\frac{e}{m_{i}}\mathbf{v}_{i}\times \mathbf{B}  \notag
\end{eqnarray}
We assume the quasineutrality 
\begin{equation}
n_{i}\approx n_{e}  \label{a2}
\end{equation}
and the Boltzmann distribution of the electrons along the magnetic field
line 
\begin{equation}
n_{e}=n_{0}\exp \left( -\frac{\left| e\right| \phi }{T_{e}}\right)
\label{a3}
\end{equation}

The velocity of the ion fluid is perpendicular on the magnetic field and is
composed of the diamagnetic, electric and polarization drift terms 
\begin{eqnarray}
\mathbf{v}_{i} &=&\mathbf{v}_{\perp i}  \label{a4} \\
&=&\mathbf{v}_{dia,i}+\mathbf{v}_{E}+\mathbf{v}_{pol,i}  \notag \\
&=&\frac{T_{i}}{\left| e\right| B}\frac{1}{n_{i}}\frac{dn_{i}}{dr}\widehat{%
\mathbf{e}}_{y}  \notag \\
&&+\frac{-\mathbf{\nabla }\phi \times \widehat{\mathbf{n}}}{B}  \notag \\
&&-\frac{1}{B\Omega _{i}}\left( \frac{\partial }{\partial t}+\left( \mathbf{v%
}_{E}\cdot \mathbf{\nabla }_{\perp }\right) \right) \mathbf{\nabla }_{\perp
}\phi  \notag
\end{eqnarray}
The diamagnetic velocity will be neglected. Introducing this velocity into
the continuity equation, one obtains an equation for the electrostatic
potential $\phi $.

Before writting this equation we introduce new dimensional units for the
variables. 
\begin{equation}
\phi ^{phys}\rightarrow \phi ^{\prime }=\frac{\left| e\right| \phi ^{phys}}{%
T_{e}}  \label{sc1}
\end{equation}
\begin{equation}
\left( x^{phys},y^{phys}\right) \rightarrow \left( x^{\prime },y^{\prime
}\right) =\left( \frac{x^{phys}}{\rho _{s}},\frac{y^{phys}}{\rho _{s}}\right)
\label{sc2}
\end{equation}
\begin{equation}
t^{phys}\rightarrow t^{\prime }=t^{phys}\Omega _{i}  \label{sc3}
\end{equation}
The new variables $\left( t,x,y\right) $ and the function $\phi $ are
non-dimensional. In the following the \emph{primes} are not written. With
these variables the equation obtained is 
\begin{eqnarray}
&&\frac{\partial }{\partial t}\left( 1-\mathbf{\nabla }_{\perp }^{2}\right)
\phi  \label{a5} \\
&&-\left( -\mathbf{\nabla }_{\perp }\phi \times \widehat{\mathbf{n}}\right)
\cdot \kappa \widehat{\mathbf{e}}_{r}  \notag \\
&&-\left[ \left( -\mathbf{\nabla }_{\perp }\phi \times \widehat{\mathbf{n}}%
\right) \cdot \mathbf{\nabla }_{\perp }\right] \mathbf{\nabla }_{\perp
}^{2}\phi  \notag \\
&=&0  \notag
\end{eqnarray}
where 
\begin{equation}
\kappa \widehat{\mathbf{e}}_{r}\equiv -\mathbf{\nabla }_{\perp }\ln n_{0}
\label{a6}
\end{equation}
(\cite{LaedkeSpatschek1}). Before continuing we compare this equation with
the equation of paper \cite{LaedkeSpatschek2}, Eq.(16). Here taking still
the units to be physical, the form of the latter equation is (Eq.(12) from
that paper) 
\begin{eqnarray}
&&\frac{\partial }{\partial t}\frac{\left| e\right| \phi }{T_{e}}-\frac{%
\partial }{\partial t}\frac{1}{B\Omega _{i}}\mathbf{\nabla }_{\perp }^{2}\phi
\label{Eq12} \\
&&+\frac{-\mathbf{\nabla }_{\perp }\phi \times \widehat{\mathbf{n}}}{B}\cdot 
\mathbf{\nabla }_{\perp }\ln n_{0}  \notag \\
&&+\frac{-\mathbf{\nabla }_{\perp }\phi \times \widehat{\mathbf{n}}}{B}\cdot 
\mathbf{\nabla }_{\perp }\frac{\left| e\right| \phi }{T_{e}}  \notag \\
&&-\frac{1}{B^{2}\Omega _{i}}\left[ \left( -\mathbf{\nabla }_{\perp }\phi
\times \widehat{\mathbf{n}}\right) \cdot \mathbf{\nabla }_{\perp }\right] 
\mathbf{\nabla }_{\perp }^{2}\phi  \notag \\
&=&0  \notag
\end{eqnarray}
The term containing the gradient of the equilibrium density comes from the
continuity equation, as convection of the equilibrium density by the
fluctuating $E\times B$ velocity. The adiabaticity has been assumed, 
\begin{equation}
\frac{\widetilde{n}}{n_{0}}=\frac{\left| e\right| \phi }{T_{e}}  \label{a7}
\end{equation}
and we consider that the temperature is constant (the calculations can
easily include a dependence $T_{e}\left( x\right) $).

For the second term we have 
\begin{eqnarray}
\frac{1}{B\Omega _{i}}\mathbf{\nabla }_{\perp }^{2}\phi &=&\frac{1}{B\Omega
_{i}}\frac{T_{e}}{\left| e\right| }\mathbf{\nabla }_{\perp }^{2}\frac{\left|
e\right| \phi }{T_{e}}=\frac{1}{\Omega _{i}}\frac{1}{\frac{\left| e\right| B%
}{m_{i}}}\frac{T_{e}}{m_{i}}\mathbf{\nabla }_{\perp }^{2}\frac{\left|
e\right| \phi }{T_{e}}  \label{a8} \\
&=&\frac{1}{\Omega _{i}^{2}}c_{s}^{2}\mathbf{\nabla }_{\perp }^{2}\frac{%
\left| e\right| \phi }{T_{e}}=\rho _{s}^{2}\mathbf{\nabla }_{\perp }^{2}%
\frac{\left| e\right| \phi }{T_{e}}  \notag
\end{eqnarray}
This will become (with its sign) 
\begin{equation}
-\frac{\partial }{\partial t}\mathbf{\nabla }_{\perp }^{\prime 2}\phi
^{\prime }  \label{second}
\end{equation}
in the new variables Eqs.(\ref{sc1})-(\ref{sc3}).

The third term is 
\begin{eqnarray}
\frac{-\mathbf{\nabla }_{\perp }\phi \times \widehat{\mathbf{n}}}{B}\cdot 
\mathbf{\nabla }_{\perp }\ln n_{0} &=&\frac{1}{B}\frac{T_{e}}{\left|
e\right| }\left( -\mathbf{\nabla }_{\perp }\frac{\left| e\right| \phi }{T_{e}%
}\times \widehat{\mathbf{n}}\right) \cdot \mathbf{\nabla }_{\perp }\ln n_{0}
\label{a10} \\
&=&\frac{1}{\frac{\left| e\right| B}{m_{i}}}\frac{T_{e}}{m_{i}}\left( -%
\mathbf{\nabla }_{\perp }\frac{\left| e\right| \phi }{T_{e}}\times \widehat{%
\mathbf{n}}\right) \cdot \mathbf{\nabla }_{\perp }\ln n_{0}  \notag \\
&=&\Omega _{i}\frac{c_{s}^{2}}{\Omega _{i}^{2}}\left( -\mathbf{\nabla }%
_{\perp }\frac{\left| e\right| \phi }{T_{e}}\times \widehat{\mathbf{n}}%
\right) \cdot \mathbf{\nabla }_{\perp }\ln n_{0}  \notag \\
&=&\Omega _{i}\rho _{s}^{2}\left( \mathbf{\nabla }_{\perp }\frac{\left|
e\right| \phi }{T_{e}}\times \widehat{\mathbf{n}}\right) \cdot \left( -%
\mathbf{\nabla }_{\perp }\ln n_{0}\right)  \notag
\end{eqnarray}
This will become 
\begin{equation}
-\Omega _{i}\left( -\rho _{s}\mathbf{\nabla }_{\perp }\phi \times \widehat{%
\mathbf{n}}\right) \cdot \left( -\rho _{s}\mathbf{\nabla }_{\perp }\ln
n_{0}\right)  \label{a12}
\end{equation}
and in the normalised space variables 
\begin{equation}
-\Omega _{i}\left( -\mathbf{\nabla }_{\perp }^{\prime }\phi ^{\prime }\times 
\widehat{\mathbf{n}}\right) \cdot \left( -\mathbf{\nabla }_{\perp }^{\prime
}\ln n_{0}\right)  \label{third}
\end{equation}

The last term (with the polarization nonlinearity) is in physical units 
\begin{equation}
-\frac{1}{B^{2}\Omega _{i}}\left[ \left( -\mathbf{\nabla }_{\perp }\phi
\times \widehat{\mathbf{n}}\right) \cdot \mathbf{\nabla }_{\perp }\right] 
\mathbf{\nabla }_{\perp }^{2}\phi  \label{a14}
\end{equation}
This is converted to non-dimensional variables 
\begin{equation}
\frac{1}{B^{2}\Omega _{i}}\left[ \left( -\frac{1}{\rho _{s}}\frac{T_{e}}{%
\left| e\right| }\rho _{s}\mathbf{\nabla }_{\perp }\frac{\left| e\right|
\phi }{T_{e}}\times \widehat{\mathbf{n}}\right) \cdot \frac{1}{\rho _{s}}%
\rho _{s}\mathbf{\nabla }_{\perp }\right] \frac{1}{\rho _{s}^{2}}\frac{T_{e}%
}{\left| e\right| }\left( -\rho _{s}^{2}\mathbf{\nabla }_{\perp }^{2}\frac{%
\left| e\right| \phi }{T_{e}}\right)  \label{a15}
\end{equation}
Collecting the physical coefficient we have 
\begin{eqnarray}
\frac{1}{B^{2}\Omega _{i}}\left( \frac{T_{e}}{\left| e\right| }\right) ^{2}%
\frac{1}{\rho _{s}^{4}} &=&\frac{1}{\left( \frac{\left| e\right| B}{m_{i}}%
\right) ^{2}}\frac{1}{\Omega _{i}}\left( \frac{T_{e}}{m_{i}}\right) ^{2}%
\frac{1}{\rho _{s}^{4}}  \label{a16} \\
&=&\Omega _{i}\frac{c_{s}^{4}}{\Omega _{i}^{4}}\frac{1}{\rho _{s}^{4}} 
\notag \\
&=&\Omega _{i}  \notag
\end{eqnarray}
Then, in the normalised variables, this term becomes 
\begin{equation}
\Omega _{i}\left[ \left( -\mathbf{\nabla }_{\perp }^{\prime }\phi ^{\prime
}\times \widehat{\mathbf{n}}\right) \cdot \mathbf{\nabla }_{\perp }^{\prime }%
\right] \left( -\mathbf{\nabla }_{\perp }^{\prime 2}\phi ^{\prime }\right)
\label{last}
\end{equation}

Then the Eqs.(\ref{Eq12}) with the new form of its terms (\ref{second}), (%
\ref{third}) and (\ref{last}) becomes 
\begin{eqnarray}
&&\frac{\partial }{\partial t}\phi ^{\prime }-\frac{\partial }{\partial t}%
\mathbf{\nabla }_{\perp }^{\prime 2}\phi ^{\prime }  \label{Eq12prim} \\
&&-\Omega _{i}\left( -\mathbf{\nabla }_{\perp }^{\prime }\phi ^{\prime
}\times \widehat{\mathbf{n}}\right) \cdot \left( -\mathbf{\nabla }_{\perp
}^{\prime }\ln n_{0}\right)  \notag \\
&&+\Omega _{i}\left[ \left( -\mathbf{\nabla }_{\perp }^{\prime }\phi
^{\prime }\times \widehat{\mathbf{n}}\right) \cdot \mathbf{\nabla }_{\perp
}^{\prime }\right] \left( -\mathbf{\nabla }_{\perp }^{\prime 2}\phi ^{\prime
}\right)  \notag \\
&=&0  \notag
\end{eqnarray}
Introducing the time unit $\Omega _{i}^{-1}$, and eliminating the \emph{%
primes} 
\begin{eqnarray}
&&\frac{\partial }{\partial t}\left( 1-\mathbf{\nabla }_{\perp }^{2}\right)
\phi  \label{Eq12norm} \\
&&-\left( -\mathbf{\nabla }_{\perp }\phi \times \widehat{\mathbf{n}}\right)
\cdot \left( -\mathbf{\nabla }_{\perp }\ln n_{0}\right)  \notag \\
&&-\left[ \left( -\mathbf{\nabla }_{\perp }\phi \times \widehat{\mathbf{n}}%
\right) \cdot \mathbf{\nabla }_{\perp }\right] \mathbf{\nabla }_{\perp
}^{2}\phi  \notag \\
&=&0  \notag
\end{eqnarray}

The last term is the convection of the vorticity 
\begin{equation}
\mathbf{\omega }=\mathbf{\nabla }_{\perp }^{2}\phi \widehat{\mathbf{n}}
\label{a17}
\end{equation}
by the velocity field 
\begin{equation}
\mathbf{v}_{E}=-\mathbf{\nabla }_{\perp }\phi \times \widehat{\mathbf{n}}
\label{a18}
\end{equation}

We use the definition $\kappa \widehat{\mathbf{e}}_{r}=\mathbf{\nabla }%
_{\perp }\ln n_{0}$ or 
\begin{equation}
\kappa \widehat{\mathbf{e}}_{y}=-\widehat{\mathbf{n}}\times \mathbf{\nabla }%
_{\perp }\ln n_{0}  \label{a19}
\end{equation}
Then the resulting equation is 
\begin{equation}
\left( 1-\mathbf{\nabla }_{\perp }^{2}\right) \frac{\partial \phi }{\partial
t}-\kappa \frac{\partial \phi }{\partial y}-\left[ \left( -\mathbf{\nabla }%
_{\perp }\phi \times \widehat{\mathbf{n}}\right) \cdot \mathbf{\nabla }%
_{\perp }\right] \mathbf{\nabla }_{\perp }^{2}\phi =0  \label{a20}
\end{equation}

The same equation but without the linear (density gradient) term has been
derived as the ``shielded convective ion cells'' \cite{HTK} 
\begin{equation*}
\left( 1-\rho _{s}^{2}\mathbf{\nabla }_{\perp }^{2}\right) \frac{\partial
\phi }{\partial t}-\left[ \phi ,\rho _{s}^{2}\mathbf{\nabla }_{\perp
}^{2}\phi \right] =0
\end{equation*}
and as a possibility to describe the Kelvin-Helmholtz instability modified
by the finite parallel electric field $E_{\parallel }$ and its associated
current density $j_{\parallel }$, with 
\begin{equation*}
\nabla _{\parallel }j_{\parallel }=e\frac{\partial n_{e}}{\partial t}\simeq 
\frac{e^{2}n_{0}}{T_{e}}\frac{\partial \phi ^{phys}}{\partial t}
\end{equation*}
where the potential is not normalised yet. The enstropy is conserved 
\begin{equation*}
U=\int d^{2}r\left[ \left( \rho _{s}\mathbf{\nabla }_{\perp }\phi \right)
^{2}+\left( \rho _{s}^{2}\Delta _{\perp }\phi \right) ^{2}\right] =\text{%
const}
\end{equation*}

%\end{appendices}

%\begin{appendices}

\section{Appendix B : The Euler-Lagrange equations}

% new type of equation numbers for the appendix
\renewcommand{\theequation}{B.\arabic{equation}} \setcounter{equation}{0}

The calculations from this Appendix should be considered as a guide for a
first contact with the methods of the theory of non-Abelian gauge field
interacting with nonlinear (= self-interacting) scalar matter field. Here
the calculations are not pedagogical (in particular we treat asymmetrically
the fields $A_{\mu }$ and $A_{\mu }^{\dagger }$) and we suggest that after
these first steps other lectures are necessary, from field-theory genuine
sources.

\bigskip

Consider again the Lagrangean density 
\begin{eqnarray}
\mathcal{L} &=&-\kappa \varepsilon ^{\mu \nu \rho }\mathrm{tr}\left(
\partial _{\mu }A_{\nu }A_{\rho }+\frac{2}{3}A_{\mu }A_{\nu }A_{\rho }\right)
\label{fb1} \\
&&-\mathrm{tr}\left[ \left( D^{\mu }\phi \right) ^{\dagger }\left( D_{\mu
}\phi \right) \right]  \notag \\
&&-V\left( \phi ,\phi ^{\dagger }\right)  \notag
\end{eqnarray}
The functional variables are 
\begin{eqnarray}
&&A_{0},A_{0}^{\dagger },A_{1},A_{1}^{\dagger },A_{2},A_{2}^{\dagger }
\label{fb10007} \\
&&\phi ,\phi ^{\dagger }  \notag
\end{eqnarray}
and they are all $SU\left( 2\right) $ matrices with complex entries.

\subsection{The contributions to the Lagrangean}

\subsubsection{The Chern-Simons term as a differential three-form and the
presence of a metric}

Apart from a factor, the gauge Lagrangean is the trace of the Chern-Simons
differential three-form on a principal bundle with group $SU\left( 2\right) $%
. 
\begin{eqnarray}
\Omega &=&\frac{1}{8\pi ^{2}}\mathrm{tr}\left( \mathbf{A\wedge dA-}\frac{2}{3%
}\mathbf{A\wedge A\wedge A}\right)  \label{fb11} \\
&=&-\frac{1}{16\pi ^{2}}\varepsilon ^{\mu \nu \rho }\left( A_{\mu
}^{a}\partial _{\nu }A_{\rho }^{a}-\frac{1}{3}\varepsilon ^{abc}A_{\mu
}^{a}A_{\nu }^{b}A_{\rho }^{c}\right) d^{3}x  \notag
\end{eqnarray}

The trace of the Chern-Simons form can also be expressed using the exterior
differentation and exterior product of forms \cite{barnatan}, \cite{0201018} 
\begin{equation}
cs\left( \mathbf{A}\right) =\frac{1}{4\pi }\int_{M^{3}}\mathrm{tr}\left( 
\mathbf{A\wedge dA}+\frac{2}{3}\mathbf{A\wedge A\wedge A}\right)
\label{fb12}
\end{equation}
where, for three algebra-valued differential one-forms $A_{k}$, $k=1,...,3$,
one has 
\begin{equation}
\mathrm{tr}\left( A_{1}\wedge A_{2}\wedge A_{3}\right) \overset{def}{=}\frac{%
1}{2}\mathrm{tr}\left( A_{1}\wedge \left[ A_{2},A_{3}\right] \right) =\frac{1%
}{2}\mathrm{tr}\left( \left[ A_{1},A_{2}\right] \wedge A_{3}\right)
\label{fb13}
\end{equation}
A factor of $1/2$ can also be extracted from the first term in (\ref{fb12})
if we add \emph{minus} its expression but with two of the three indices
exchanged. Then 
\begin{equation}
cs(\mathbf{A)=}\frac{1}{8\pi }\int \varepsilon ^{\mu \nu \rho }\mathrm{tr}%
\left( A_{\mu }\left( \partial _{\nu }A_{\rho }-\partial _{\rho }A_{\nu
}\right) +\frac{2}{3}A_{\mu }\left[ A_{\nu },A_{\rho }\right] \right)
\label{fb14}
\end{equation}

The normalizing constant in Eq.(\ref{fb14}) is related with the fact that
the integral of the Chern-Simons form is a topological invariant for
adequate boundary conditions and has integer values. For what we need, the
gauge field Lagrangean can be taken such as to lead to the gauge-part in the
action \cite{9304081} 
\begin{equation}
\mathcal{S}_{1}=\int dt\int d^{2}x\left\{ -\frac{1}{2}\kappa \varepsilon
^{\mu \nu \rho }\mathrm{tr}\left( A_{\mu }\left( \partial _{\nu }A_{\rho
}-\partial _{\rho }A_{\nu }\right) +\frac{2}{3}A_{\mu }\left[ A_{\nu
},A_{\rho }\right] \right) \right\}  \label{fb17}
\end{equation}
Therefore we will use the following expression 
\begin{equation}
\mathcal{L}_{1}=-\frac{1}{2}\kappa \varepsilon ^{\mu \nu \rho }\mathrm{tr}%
\left( A_{\mu }\left( \partial _{\nu }A_{\rho }-\partial _{\rho }A_{\nu
}\right) +\frac{2}{3}A_{\mu }\left[ A_{\nu },A_{\rho }\right] \right)
\label{fb18}
\end{equation}

We write in detail Eq.(\ref{fb18}). The first term is 
\begin{eqnarray}
&&\varepsilon ^{\mu \nu \rho }\left[ A_{\mu }\left( \partial _{\nu }A_{\rho
}-\partial _{\rho }A_{\nu }\right) \right]  \label{fb19} \\
&=&\varepsilon ^{012}A_{0}\left( \partial _{1}A_{2}-\partial _{2}A_{1}\right)
\notag \\
&&+\varepsilon ^{021}A_{0}\left( \partial _{2}A_{1}-\partial _{1}A_{2}\right)
\notag \\
&&+\varepsilon ^{102}A_{1}\left( \partial _{0}A_{2}-\partial _{2}A_{0}\right)
\notag \\
&&+\varepsilon ^{120}A_{1}\left( \partial _{2}A_{0}-\partial _{0}A_{2}\right)
\notag \\
&&+\varepsilon ^{210}A_{2}\left( \partial _{1}A_{0}-\partial _{0}A_{1}\right)
\notag \\
&&+\varepsilon ^{201}A_{2}\left( \partial _{0}A_{1}-\partial _{1}A_{0}\right)
\notag
\end{eqnarray}
or 
\begin{eqnarray}
&&\varepsilon ^{\mu \nu \rho }\left[ A_{\mu }\left( \partial _{\nu }A_{\rho
}-\partial _{\rho }A_{\nu }\right) \right]  \label{fb20} \\
&=&A_{0}\left( \partial _{1}A_{2}-\partial _{2}A_{1}\right)  \notag \\
&&-A_{0}\left( \partial _{2}A_{1}-\partial _{1}A_{2}\right)  \notag \\
&&-A_{1}\left( \partial _{0}A_{2}-\partial _{2}A_{0}\right)  \notag \\
&&+A_{1}\left( \partial _{2}A_{0}-\partial _{0}A_{2}\right)  \notag \\
&&-A_{2}\left( \partial _{1}A_{0}-\partial _{0}A_{1}\right)  \notag \\
&&+A_{2}\left( \partial _{0}A_{1}-\partial _{1}A_{0}\right)  \notag
\end{eqnarray}
This is simply two times every distinct term in the sum 
\begin{eqnarray}
&&\varepsilon ^{\mu \nu \rho }\left[ A_{\mu }\left( \partial _{\nu }A_{\rho
}-\partial _{\rho }A_{\nu }\right) \right]  \label{fb21} \\
&=&2A_{0}\left( \partial _{1}A_{2}\right) -2A_{0}\left( \partial
_{2}A_{1}\right) -2A_{1}\left( \partial _{0}A_{2}\right)  \notag \\
&&+2A_{1}\left( \partial _{2}A_{0}\right) -2A_{2}\left( \partial
_{1}A_{0}\right) +2A_{2}\left( \partial _{0}A_{1}\right)  \notag
\end{eqnarray}

We continue by calculating the second term in the CS action 
\begin{eqnarray}
\varepsilon ^{\mu \nu \rho }A_{\mu }\left[ A_{\nu },A_{\rho }\right]
&=&\varepsilon ^{\mu \nu \rho }A_{\mu }\left( A_{\nu }A_{\rho }-A_{\rho
}A_{\nu }\right)  \label{fb22} \\
&=&\varepsilon ^{012}A_{0}\left( A_{1}A_{2}-A_{2}A_{1}\right)  \notag \\
&&+\varepsilon ^{021}A_{0}\left( A_{2}A_{1}-A_{1}A_{2}\right)  \notag \\
&&+\varepsilon ^{102}A_{1}\left( A_{0}A_{2}-A_{2}A_{0}\right)  \notag \\
&&+\varepsilon ^{120}A_{1}\left( A_{2}A_{0}-A_{0}A_{2}\right)  \notag \\
&&+\varepsilon ^{210}A_{2}\left( A_{1}A_{0}-A_{0}A_{1}\right)  \notag \\
&&+\varepsilon ^{201}A_{2}\left( A_{0}A_{1}-A_{1}A_{0}\right)  \notag
\end{eqnarray}
or 
\begin{eqnarray}
\varepsilon ^{\mu \nu \rho }A_{\mu }\left[ A_{\nu },A_{\rho }\right] &=&
\label{fb23} \\
&=&A_{0}\left( A_{1}A_{2}-A_{2}A_{1}\right)  \notag \\
&&-A_{0}\left( A_{2}A_{1}-A_{1}A_{2}\right)  \notag \\
&&-A_{1}\left( A_{0}A_{2}-A_{2}A_{0}\right)  \notag \\
&&+A_{1}\left( A_{2}A_{0}-A_{0}A_{2}\right)  \notag \\
&&-A_{2}\left( A_{1}A_{0}-A_{0}A_{1}\right)  \notag \\
&&+A_{2}\left( A_{0}A_{1}-A_{1}A_{0}\right)  \notag
\end{eqnarray}
This is actually two times every distinct term in the sum 
\begin{eqnarray}
\frac{1}{2}\varepsilon ^{\mu \nu \rho }A_{\mu }\left[ A_{\nu },A_{\rho }%
\right] &=&A_{0}A_{1}A_{2}-A_{0}A_{2}A_{1}  \label{fb24} \\
&&-A_{1}A_{0}A_{2}+A_{1}A_{2}A_{0}  \notag \\
&&-A_{2}A_{1}A_{0}+A_{2}A_{0}A_{1}  \notag
\end{eqnarray}
It results that the form of definition 
\begin{equation}
\mathcal{L}_{1}=-\frac{1}{2}\kappa \varepsilon ^{\mu \nu \rho }\mathrm{tr}%
\left( A_{\mu }\left( \partial _{\nu }A_{\rho }-\partial _{\rho }A_{\nu
}\right) +\frac{2}{3}A_{\mu }\left[ A_{\nu },A_{\rho }\right] \right)
\label{fb25}
\end{equation}
can be written 
\begin{eqnarray}
\mathcal{L}_{1} &=&-\kappa \mathrm{tr}\left\{ A_{0}\left( \partial
_{1}A_{2}\right) -A_{0}\left( \partial _{2}A_{1}\right) -A_{1}\left(
\partial _{0}A_{2}\right) \right.  \label{fb26} \\
&&+A_{1}\left( \partial _{2}A_{0}\right) -A_{2}\left( \partial
_{1}A_{0}\right) +A_{2}\left( \partial _{0}A_{1}\right)  \notag \\
&&+\frac{2}{3}A_{0}A_{1}A_{2}-\frac{2}{3}A_{0}A_{2}A_{1}-\frac{2}{3}%
A_{1}A_{0}A_{2}  \notag \\
&&\left. +\frac{2}{3}A_{1}A_{2}A_{0}-\frac{2}{3}A_{2}A_{1}A_{0}+\frac{2}{3}%
A_{2}A_{0}A_{1}\right\}  \notag
\end{eqnarray}
We recall that every function $A_{\mu }$ is actually a matrix, in the
adjoint representation of the $SU\left( 2\right) $ algebra. We can use the
property of invariance of the operator Trace of a product of matrices to a
cyclic permutation of the factors of this product. The third term in the
first line without derivatives 
\begin{equation}
A_{1}A_{0}A_{2}\rightarrow A_{2}A_{1}A_{0}\rightarrow A_{0}A_{2}A_{1}
\label{fb27}
\end{equation}
The second term in the second line without derivative 
\begin{equation}
A_{2}A_{1}A_{0}\rightarrow A_{0}A_{2}A_{1}  \label{fb28}
\end{equation}
These two terms will add to the second term of the first line, giving 
\begin{equation}
-\frac{2}{3}A_{0}A_{2}A_{1}-\frac{2}{3}A_{1}A_{0}A_{2}-\frac{2}{3}%
A_{2}A_{1}A_{0}\rightarrow -2A_{0}A_{2}A_{1}  \label{fb29}
\end{equation}
The first term on the second line without derivatives 
\begin{equation}
A_{1}A_{2}A_{0}\rightarrow A_{0}A_{1}A_{2}  \label{fb30}
\end{equation}
The last term 
\begin{equation}
A_{2}A_{0}A_{1}\rightarrow A_{1}A_{2}A_{0}\rightarrow A_{0}A_{1}A_{2}
\label{fb31}
\end{equation}
These two terms are added to the first term of the first line without
derivatives 
\begin{equation}
\frac{2}{3}A_{0}A_{1}A_{2}+\frac{2}{3}A_{1}A_{2}A_{0}+\frac{2}{3}%
A_{2}A_{0}A_{1}\rightarrow 2A_{0}A_{1}A_{2}  \label{fb32}
\end{equation}
Finnaly, we collect all terms that do not contain derivatives 
\begin{eqnarray}
&&\frac{2}{3}A_{0}A_{1}A_{2}-\frac{2}{3}A_{0}A_{2}A_{1}-\frac{2}{3}%
A_{1}A_{0}A_{2}  \label{fb33} \\
&&+\frac{2}{3}A_{1}A_{2}A_{0}-\frac{2}{3}A_{2}A_{1}A_{0}+\frac{2}{3}%
A_{2}A_{0}A_{1}  \notag \\
&=&2A_{0}A_{1}A_{2}-2A_{0}A_{2}A_{1}  \notag
\end{eqnarray}

At this point the gauge-field Lagrangean is 
\begin{eqnarray}
\mathcal{L}_{1} &=&-\kappa \mathrm{tr}\left\{ A_{0}\left( \partial
_{1}A_{2}\right) -A_{0}\left( \partial _{2}A_{1}\right) -A_{1}\left(
\partial _{0}A_{2}\right) \right.  \label{fb3310} \\
&&+A_{1}\left( \partial _{2}A_{0}\right) -A_{2}\left( \partial
_{1}A_{0}\right) +A_{2}\left( \partial _{0}A_{1}\right)  \notag \\
&&\left. +2A_{0}A_{1}A_{2}-2A_{0}A_{2}A_{1}\right\}  \notag
\end{eqnarray}

\paragraph{Change of the form of the gauge-field part of the Lagrangean,
from integration by parts}

Since the Lagrangian density is integrated in order to obtain the action
functional, we can consider the effect of integrations by parts. These
operations will move the differential operators between the factors of the
monomials appearing in the expression of the Lagrangian density and will
also generate boundary terms. In general the boundary terms are zero for
well-behaved functions, but in our case the presence of a finite condensate
of vorticity at infinity can produce finite terms. We will develop below a
calculation based on integration by parts, removing the spatial derivatives
from acting upon $A_{0}$ but we will ignore the boundary finite terms. Then
the calculation is simply useful for the comparison with other, well known,
forms of the CS Lagrangian. We will \emph{not} use the form of the
Lagrangian derived from these operations Eq.(\ref{fb365}), for obtaining the
Euler-Lagrange equations, and we will rely on Eq.(\ref{fb3310}).

\bigskip

We turn to the terms containing derivatives. It is possible to make a
integrations by parts using the formula 
\begin{equation}
\frac{d}{dx}\left( \mathbf{YZ}\right) =\mathbf{Y}^{\ast }\frac{d\mathbf{Z}}{%
dx}+\frac{d\mathbf{Y}^{\ast }}{dx}\mathbf{Z}  \label{fb335}
\end{equation}
We apply this for the following two terms and furthermore we use the cyclic
symmetry inside the Trace operator 
\begin{eqnarray}
A_{1}\left( \partial _{2}A_{0}\right) &\rightarrow &\partial _{2}\left(
A_{1}^{\ast }A_{0}\right) -\left( \partial _{2}A_{1}\right) A_{0}
\label{fb34} \\
&\rightarrow &-A_{0}\left( \partial _{2}A_{1}\right)  \notag
\end{eqnarray}
\begin{eqnarray}
-A_{2}\left( \partial _{1}A_{0}\right) &\rightarrow &-\partial _{1}\left(
A_{2}^{\ast }A_{0}\right) +\left( \partial _{1}A_{2}\right) A_{0}
\label{fb35} \\
&\rightarrow &A_{0}\left( \partial _{1}A_{2}\right)  \notag
\end{eqnarray}

We collect all terms containing derivatives 
\begin{eqnarray*}
&&A_{0}\left( \partial _{1}A_{2}\right) -A_{0}\left( \partial
_{2}A_{1}\right) -A_{1}\left( \partial _{0}A_{2}\right) \\
&&+A_{1}\left( \partial _{2}A_{0}\right) -A_{2}\left( \partial
_{1}A_{0}\right) +A_{2}\left( \partial _{0}A_{1}\right) \\
&=&2A_{0}\left( \partial _{1}A_{2}\right) -2A_{0}\left( \partial
_{2}A_{1}\right) -A_{1}\left( \partial _{0}A_{2}\right) +A_{2}\left(
\partial _{0}A_{1}\right)
\end{eqnarray*}
Finally, the gauge-field Lagrangean density results 
\begin{eqnarray}
\mathcal{L}_{1} &=&-\kappa \mathrm{tr}\left\{ 2A_{0}\left( \partial
_{1}A_{2}\right) -2A_{0}\left( \partial _{2}A_{1}\right) \right.
\label{fb365} \\
&&-A_{1}\left( \partial _{0}A_{2}\right) +A_{2}\left( \partial
_{0}A_{1}\right)  \notag \\
&&\left. +2A_{0}A_{1}A_{2}-2A_{0}A_{2}A_{1}\right\}  \notag
\end{eqnarray}

\paragraph{Comparison with known forms of the gauge Lagrangean}

Let us check this formula by comparing with the situations where the space
part is separated \cite{9304081}.

The metric is defined from the general expression of the differential length
in the $2+1$ dimensional space, as 
\begin{equation}
ds^{2}=-dt^{2}+h_{ij}dx^{i}dx^{j}  \label{fb16}
\end{equation}
($h^{ij}$ is the space-part of $g^{\mu \nu }$) and one can separate the
spatial and temporal parts of the action. In our case the metric is $%
g=diag\left\{ -1,1,1\right\} $. The action is 
\begin{equation}
\mathcal{S}_{1}=\int dt\int d^{2}x\mathrm{tr}\left\{ -\kappa \varepsilon
^{ij}A_{i}\partial _{0}A_{j}+\kappa \varepsilon ^{ij}A_{0}F_{ij}\right\}
\label{fb37}
\end{equation}
where 
\begin{equation}
F_{ij}=\partial _{i}A_{j}-\partial _{j}A_{i}+\left[ A_{i},A_{j}\right]
\label{fb38}
\end{equation}
\begin{equation}
F_{i0}=D_{i}A_{0}-\partial _{0}A_{i}  \label{fb39}
\end{equation}
and 
\begin{equation}
\varepsilon ^{ij}=\frac{\varepsilon ^{0ij}}{\sqrt{h}}  \label{fb40}
\end{equation}
Then, in the integrand 
\begin{eqnarray}
&&-\varepsilon ^{ij}A_{i}\partial _{0}A_{j}+\varepsilon ^{ij}A_{0}F_{ij}
\label{fb41} \\
&=&-\varepsilon ^{12}A_{1}\partial _{0}A_{2}-\varepsilon ^{21}A_{2}\partial
_{0}A_{1}  \notag \\
&&+\varepsilon ^{12}A_{0}F_{12}+\varepsilon ^{21}A_{0}F_{21}  \notag
\end{eqnarray}
From Eq.(\ref{fb40}) where $h$ is the metric, we have 
\begin{eqnarray}
&&-\varepsilon ^{ij}A_{i}\partial _{0}A_{j}+\varepsilon ^{ij}A_{0}F_{ij}
\label{fb42} \\
&=&-A_{1}\partial _{0}A_{2}+A_{2}\partial _{0}A_{1}  \notag \\
&&+A_{0}F_{12}-A_{0}F_{21}  \notag
\end{eqnarray}
Now we replace 
\begin{eqnarray}
F_{ij} &=&\partial _{i}A_{j}-\partial _{j}A_{i}+\left[ A_{i},A_{j}\right]
\label{fb43} \\
&=&\partial _{i}A_{j}-\partial _{j}A+A_{i}A_{j}-A_{j}A_{i}  \notag
\end{eqnarray}
and obtain 
\begin{eqnarray}
&&-\varepsilon ^{ij}A_{i}\partial _{0}A_{j}+\varepsilon ^{ij}A_{0}F_{ij}
\label{fb44} \\
&=&-A_{1}\partial _{0}A_{2}+A_{2}\partial _{0}A_{1}  \notag \\
&&+A_{0}\left( \partial _{1}A_{2}-\partial
_{2}A_{1}+A_{1}A_{2}-A_{2}A_{1}\right)  \notag \\
&&-A_{0}\left( \partial _{2}A_{1}-\partial
_{1}A_{2}+A_{2}A_{1}-A_{1}A_{2}\right)  \notag
\end{eqnarray}
We note that some terms are repeted 
\begin{eqnarray}
&&-\varepsilon ^{ij}A_{i}\partial _{0}A_{j}+\varepsilon ^{ij}A_{0}F_{ij}
\label{fb45} \\
&=&-A_{1}\partial _{0}A_{2}+A_{2}\partial _{0}A_{1}+2A_{0}\partial
_{1}A_{2}-2A_{0}\partial _{2}A_{1}  \notag \\
&&+2A_{0}A_{1}A_{2}-2A_{0}A_{2}A_{1}  \notag
\end{eqnarray}
The result is identical to Eq.(\ref{fb365}) 
\begin{eqnarray}
\mathcal{L}_{1} &=&-\kappa \mathrm{tr}\left\{ -A_{1}\partial
_{0}A_{2}+A_{2}\partial _{0}A_{1}+2A_{0}\partial _{1}A_{2}-2A_{0}\partial
_{2}A_{1}\right.  \label{fb46} \\
&&\left. +2A_{0}A_{1}A_{2}-2A_{0}A_{2}A_{1}\right\}  \notag
\end{eqnarray}

We also note that until now there was no need to consider summation over
components of vectors using the metric coefficients. In case of a product of
the form $x_{\mu }x^{\mu }$ it will have to consider the metric.

\subsection{The matter Lagrangean}

The form is 
\begin{equation}
\mathcal{L}_{2}=-\mathrm{tr}\left[ \left( D^{\mu }\phi \right) ^{\dagger
}\left( D_{\mu }\phi \right) \right]  \label{fb50}
\end{equation}
Using Eqs.(\ref{2827}), (\ref{2831}) and (\ref{2837}) we can calculate in
detail 
\begin{eqnarray}
&&\mathcal{L}_{2}=-\mathrm{tr}\left[ \left( D^{\mu }\phi \right) ^{\dagger
}\left( D_{\mu }\phi \right) \right]  \label{fb53} \\
&=&-\mathrm{tr}\left[ \left( -\frac{\partial \phi ^{\dagger }}{\partial t}%
+\phi ^{\dagger }A^{0\dagger }-A^{0\dagger }\phi ^{\dagger }\right) \left( 
\frac{\partial \phi }{\partial t}+A_{0}\phi -\phi A_{0}\right) \right. 
\notag \\
&&+\left( \frac{\partial \phi ^{\dagger }}{\partial x}+\phi ^{\dagger
}A^{1\dagger }-A^{1\dagger }\phi ^{\dagger }\right) \left( \frac{\partial
\phi }{\partial x}+A_{1}\phi -\phi A_{1}\right)  \notag \\
&&\left. +\left( \frac{\partial \phi ^{\dagger }}{\partial y}+\phi ^{\dagger
}A^{2\dagger }-A^{2\dagger }\phi ^{\dagger }\right) \left( \frac{\partial
\phi }{\partial y}+A_{2}\phi -\phi A_{2}\right) \right]  \notag
\end{eqnarray}
We have to expand the products 
\begin{eqnarray}
\mathcal{L}_{2} &=&-\mathrm{tr}\left\{ -\frac{\partial \phi ^{\dagger }}{%
\partial t}\frac{\partial \phi }{\partial t}-\frac{\partial \phi ^{\dagger }%
}{\partial t}A_{0}\phi +\frac{\partial \phi ^{\dagger }}{\partial t}\phi
A_{0}\right.  \label{fb54} \\
&&+\phi ^{\dagger }A^{0\dagger }\frac{\partial \phi }{\partial t}+\phi
^{\dagger }A^{0\dagger }A_{0}\phi -\phi ^{\dagger }A^{0\dagger }\phi A_{0} 
\notag \\
&&-A^{0\dagger }\phi ^{\dagger }\frac{\partial \phi }{\partial t}%
-A^{0\dagger }\phi ^{\dagger }A_{0}\phi +A^{0\dagger }\phi ^{\dagger }\phi
A_{0}  \notag \\
&&+\frac{\partial \phi ^{\dagger }}{\partial x}\frac{\partial \phi }{%
\partial x}+\frac{\partial \phi ^{\dagger }}{\partial x}A_{1}\phi -\frac{%
\partial \phi ^{\dagger }}{\partial x}\phi A_{1}  \notag \\
&&+\phi ^{\dagger }A^{1\dagger }\frac{\partial \phi }{\partial x}+\phi
^{\dagger }A^{1\dagger }A_{1}\phi -\phi ^{\dagger }A^{1\dagger }\phi A_{1} 
\notag \\
&&-A^{1\dagger }\phi ^{\dagger }\frac{\partial \phi }{\partial x}%
-A^{1\dagger }\phi ^{\dagger }A_{1}\phi +A^{1\dagger }\phi ^{\dagger }\phi
A_{1}  \notag \\
&&+\frac{\partial \phi ^{\dagger }}{\partial y}\frac{\partial \phi }{%
\partial y}+\frac{\partial \phi ^{\dagger }}{\partial y}A_{2}\phi -\frac{%
\partial \phi ^{\dagger }}{\partial y}\phi A_{2}  \notag \\
&&+\phi ^{\dagger }A^{2\dagger }\frac{\partial \phi }{\partial y}+\phi
^{\dagger }A^{2\dagger }A_{2}\phi -\phi ^{\dagger }A^{2\dagger }\phi A_{2} 
\notag \\
&&\left. -A^{2\dagger }\phi ^{\dagger }\frac{\partial \phi }{\partial y}%
-A^{2\dagger }\phi ^{\dagger }A_{2}\phi +A^{2\dagger }\phi ^{\dagger }\phi
A_{2}\right\}  \notag
\end{eqnarray}

\subsection{The Euler-Lagrange equations}

The Euler-Lagrange equations 
\begin{equation}
\frac{\partial }{\partial x^{\mu }}\frac{\delta \mathcal{L}}{\delta \left( 
\frac{\partial A_{\alpha }}{\partial x^{\mu }}\right) }-\frac{\delta 
\mathcal{L}}{\delta A_{\alpha }}=0  \label{fb55}
\end{equation}
We use distinct notations for the three components of the Lagrangean
density, $\mathcal{L}=\mathcal{L}_{1}+\mathcal{L}_{2}-V$ where $\mathcal{L}%
_{1}$ is the gauge field part, $\mathcal{L}_{2}$ is the ``matter'' part and $%
V$ is the nonlinear self-interaction potential for the ``matter'' field. We
use the detailed expressions for $\mathcal{L}_{1}$ from Eq.(\ref{fb3310})
and $\mathcal{L}_{2}$ is given by the Eq.(\ref{fb54}). The functional
derivations are done separately on these two parts.

\subsubsection{The formulas for derivation of the Trace of a product of
matrices}

Use the formulas (see Ref. \cite{mike}) 
\begin{equation}
\frac{d}{d\mathbf{X}}\mathrm{tr}\left( \mathbf{AX}\right) =\mathbf{A}^{T}
\label{fb62}
\end{equation}
\begin{equation}
\frac{d}{d\mathbf{X}}\mathrm{tr}\left( \mathbf{XA}\right) =\mathbf{A}^{T}
\label{fb63}
\end{equation}
\begin{equation}
\frac{d}{d\mathbf{X}}\mathrm{tr}\left( \mathbf{X}^{T}\mathbf{A}\right) =%
\mathbf{A}  \label{fb632}
\end{equation}
\begin{equation}
\frac{d}{d\mathbf{X}}\mathrm{tr}\left( \mathbf{AX}^{T}\right) =\mathbf{A}
\label{fb634}
\end{equation}
\begin{equation}
\frac{d}{d\mathbf{X}}\mathrm{tr}\left( \mathbf{AXB}\right) =\mathbf{A}^{T}%
\mathbf{B}^{T}  \label{fb64}
\end{equation}
\begin{equation*}
\frac{d}{d\mathbf{X}}\mathrm{tr}\left( \mathbf{BX}^{T}\mathbf{A}\right) =%
\mathbf{AB}
\end{equation*}
\begin{equation}
\frac{d}{d\mathbf{X}}\mathrm{tr}\left( \mathbf{XAX}^{T}\right) =\mathbf{X}%
\left( \mathbf{A+A}^{T}\right)   \label{fb641}
\end{equation}
\begin{equation}
\frac{d}{d\mathbf{X}}\mathrm{tr}\left( \mathbf{X}^{T}\mathbf{AX}\right)
=\left( \mathbf{A+A}^{T}\right) \mathbf{X}  \label{fb642}
\end{equation}
\begin{equation}
\frac{d}{d\mathbf{X}}\mathrm{tr}\left( \mathbf{AXBX}\right) =\mathbf{A}^{T}%
\mathbf{X}^{T}\mathbf{B}^{T}+\mathbf{B}^{T}\mathbf{X}^{T}\mathbf{A}^{T}
\label{fb643}
\end{equation}
\begin{equation}
\frac{d}{d\mathbf{X}}\mathrm{tr}\left( \mathbf{AXBX}^{T}\mathbf{C}\right) =%
\mathbf{A}^{T}\mathbf{C}^{T}\mathbf{XB}^{T}+\mathbf{CAXB}  \label{fb644}
\end{equation}
where $\mathbf{A}$, $\mathbf{B}$, $\mathbf{C}$, $\mathbf{X}$ are arbitrary
complex matrices.

\subsection{The Euler-Lagrange equations for the gauge field}

\subsubsection{The variation to $A_{0}$}

The equation of motion resulting from the variation to $A_{0}$ is 
\begin{equation}
\frac{\partial }{\partial x^{\mu }}\frac{\delta \mathcal{L}}{\delta \left( 
\frac{\partial A_{0}}{\partial x^{\mu }}\right) }-\frac{\delta \mathcal{L}}{%
\delta A_{0}}=0  \label{fb58}
\end{equation}
or 
\begin{equation}
\frac{\partial }{\partial x^{0}}\frac{\delta \mathcal{L}}{\delta \left(
\partial _{0}A_{0}\right) }+\frac{\partial }{\partial x^{1}}\frac{\delta 
\mathcal{L}}{\delta \left( \partial _{1}A_{0}\right) }+\frac{\partial }{%
\partial x^{2}}\frac{\delta \mathcal{L}}{\delta \left( \partial
_{2}A_{0}\right) }-\frac{\delta \mathcal{L}}{\delta A_{0}}=0  \label{fb5810}
\end{equation}

\paragraph{Functional derivatives at $A_{0}$ of the gauge field Lagrangean}

The gauge field Lagrangean is Eq.(\ref{fb3310}) 
\begin{eqnarray}
\mathcal{L}_{1} &=&-\kappa \mathrm{tr}\left\{ A_{0}\left( \partial
_{1}A_{2}\right) -A_{0}\left( \partial _{2}A_{1}\right) -A_{1}\left(
\partial _{0}A_{2}\right) \right.  \label{fb6010} \\
&&+A_{1}\left( \partial _{2}A_{0}\right) -A_{2}\left( \partial
_{1}A_{0}\right) +A_{2}\left( \partial _{0}A_{1}\right)  \notag \\
&&\left. +2A_{0}A_{1}A_{2}-2A_{0}A_{2}A_{1}\right\}  \notag
\end{eqnarray}
and we have to calculate 
\begin{equation}
\frac{\partial }{\partial x^{0}}\frac{\delta \mathcal{L}_{1}}{\delta \left(
\partial _{0}A_{0}\right) }+\frac{\partial }{\partial x^{1}}\frac{\delta 
\mathcal{L}_{1}}{\delta \left( \partial _{1}A_{0}\right) }+\frac{\partial }{%
\partial x^{2}}\frac{\delta \mathcal{L}_{1}}{\delta \left( \partial
_{2}A_{0}\right) }-\frac{\delta \mathcal{L}_{1}}{\delta A_{0}}
\label{fb6012}
\end{equation}

The first term in the Euler-Lagrange equation (\ref{fb6012}) for $A_{0}$ is
zero since 
\begin{equation*}
\frac{\delta \mathcal{L}_{1}}{\delta \left( \partial _{0}A_{0}\right) }=0
\end{equation*}

For the second term there is only one contribution 
\begin{eqnarray}
\frac{\partial }{\partial x^{1}}\frac{\delta \mathcal{L}_{1}}{\delta \left( 
\frac{\partial A_{0}}{\partial x^{1}}\right) } &=&\frac{\partial }{\partial
x^{1}}\frac{\delta }{\delta \left( \partial _{1}A_{0}\right) }\left( -\kappa
\right) \mathrm{tr}\left\{ -A_{2}\left( \partial _{1}A_{0}\right) \right\}
\label{fb6013} \\
&=&-\kappa \frac{\partial }{\partial x^{1}}\left\{ -A_{2}^{T}\right\}  \notag
\\
&=&\left( -\kappa \right) \left( -\partial _{1}A_{2}^{T}\right)  \notag
\end{eqnarray}

The third term also consists of one contribution 
\begin{eqnarray}
\frac{\partial }{\partial x^{2}}\frac{\delta \mathcal{L}_{1}}{\delta \left( 
\frac{\partial A_{0}}{\partial x^{2}}\right) } &=&\frac{\partial }{\partial
x^{2}}\frac{\delta }{\delta \left( \partial _{2}A_{0}\right) }\left( -\kappa
\right) \mathrm{tr}\left\{ A_{1}\left( \partial _{2}A_{0}\right) \right\}
\label{fb6017} \\
&=&\left( -\kappa \right) \frac{\partial }{\partial x^{2}}\left\{
A_{1}^{T}\right\}  \notag \\
&=&\left( -\kappa \right) \left( \partial _{2}A_{1}^{T}\right)  \notag
\end{eqnarray}

The last term in Eq.(\ref{6012}) is the derivative of $\mathcal{L}_{1}$ to
the functional variable $A_{0}$, 
\begin{eqnarray}
\frac{\delta \mathcal{L}_{1}}{\delta A_{0}} &=&-\kappa \frac{\delta }{\delta
A_{0}}\mathrm{tr}\left\{ A_{0}\left( \partial _{1}A_{2}\right) -A_{0}\left(
\partial _{2}A_{1}\right) -A_{1}\left( \partial _{0}A_{2}\right) \right.
\label{fb601} \\
&&+A_{1}\left( \partial _{2}A_{0}\right) -A_{2}\left( \partial
_{1}A_{0}\right) +A_{2}\left( \partial _{0}A_{1}\right)  \notag \\
&&\left. +2A_{0}A_{1}A_{2}-2A_{0}A_{2}A_{1}\right\}  \notag
\end{eqnarray}

In detail, every term 
\begin{equation}
\frac{\delta }{\delta A_{0}}\mathrm{tr}\left\{ A_{0}\left( \partial
_{1}A_{2}\right) \right\} =\left( \partial _{1}A_{2}\right) ^{T}
\label{fb602}
\end{equation}
\begin{equation}
\frac{\delta }{\delta A_{0}}\mathrm{tr}\left\{ -A_{0}\left( \partial
_{2}A_{1}\right) \right\} =-\left( \partial _{2}A_{1}\right) ^{T}
\label{fb603}
\end{equation}
\begin{equation}
\frac{\delta }{\delta A_{0}}\mathrm{tr}\left\{ -A_{1}\left( \partial
_{0}A_{2}\right) \right\} =0  \label{fb6035}
\end{equation}
\begin{equation}
\frac{\delta }{\delta A_{0}}\mathrm{tr}\left\{ A_{1}\left( \partial
_{2}A_{0}\right) \right\} =0  \label{fb604}
\end{equation}
\begin{equation}
\frac{\delta }{\delta A_{0}}\mathrm{tr}\left\{ -A_{2}\left( \partial
_{1}A_{0}\right) \right\} =0  \label{fb6045}
\end{equation}
\begin{equation}
\frac{\delta }{\delta A_{0}}\mathrm{tr}\left\{ A_{2}\left( \partial
_{0}A_{1}\right) \right\} =0  \label{fb605}
\end{equation}
\begin{equation}
\frac{\delta }{\delta A_{0}}\mathrm{tr}\left\{ 2A_{0}A_{1}A_{2}\right\}
=2\left( A_{1}A_{2}\right) ^{T}  \label{fb606}
\end{equation}
\begin{equation}
\frac{\delta }{\delta A_{0}}\mathrm{tr}\left\{ -2A_{0}A_{2}A_{1}\right\}
=-2\left( A_{2}A_{1}\right) ^{T}  \label{fb607}
\end{equation}
Collecting these formulas we find 
\begin{eqnarray}
\frac{\delta \mathcal{L}_{1}}{\delta A_{0}} &=&\left( -\kappa \right)
\left\{ \left( \partial _{1}A_{2}\right) ^{T}-\left( \partial
_{2}A_{1}\right) ^{T}+2\left( A_{1}A_{2}\right) ^{T}-2\left(
A_{2}A_{1}\right) ^{T}\right\}  \label{fb61} \\
&=&\left( -\kappa \right) \left\{ \partial _{1}A_{2}-\partial
_{2}A_{1}+A_{1}A_{2}-A_{2}A_{1}\right.  \notag \\
&&\left. A_{1}A_{2}-A_{2}A_{1}\right\} ^{T}  \notag \\
&=&\left( -\kappa \right) \left( \partial _{1}A_{2}-\partial _{2}A_{1}+\left[
A_{1},A_{2}\right] \right) ^{T}  \notag \\
&&+\left( -\kappa \right) \left( \left[ A_{1},A_{2}\right] \right) ^{T} 
\notag
\end{eqnarray}

The result of the variation of the gauge-part of the Lagrangian $\mathcal{L}%
_{1}$, to the functional variable $A_{0}$ is obtained from the results Eqs.(%
\ref{fb6013}), (\ref{fb6017}) and (\ref{fb61}) 
\begin{eqnarray}
&&\frac{\partial }{\partial x^{1}}\frac{\delta \mathcal{L}_{1}}{\delta
\left( \partial _{1}A_{0}\right) }+\frac{\partial }{\partial x^{2}}\frac{%
\delta \mathcal{L}_{1}}{\delta \left( \partial _{2}A_{0}\right) }-\frac{%
\delta \mathcal{L}_{1}}{\delta A_{0}}  \label{fb6109} \\
&=&\left( -\kappa \right) \left( -\partial _{1}A_{2}^{T}\right) +\left(
-\kappa \right) \left( \partial _{2}A_{1}^{T}\right)  \notag \\
&&-\left( -\kappa \right) \left( \partial _{1}A_{2}-\partial _{2}A_{1}+\left[
A_{1},A_{2}\right] \right) ^{T}  \notag \\
&&-\left( -\kappa \right) \left( \left[ A_{1},A_{2}\right] \right) ^{T} 
\notag \\
&=&\kappa \left( \partial _{1}A_{2}-\partial _{2}A_{1}+\left[ A_{1},A_{2}%
\right] \right) ^{T}  \notag \\
&&+\left( -\kappa \right) \left\{ -\partial _{1}A_{2}^{T}+\partial
_{2}A_{1}^{T}-\left[ A_{1},A_{2}\right] ^{T}\right\}  \notag \\
&=&2\kappa \left( \partial _{1}A_{2}-\partial _{2}A_{1}+\left[ A_{1},A_{2}%
\right] \right) ^{T}  \notag \\
&=&2\kappa \left( F_{12}\right) ^{T}  \notag
\end{eqnarray}

\paragraph{Functional derivative with respect to $A_{0}$ of the ``matter''
Lagrangean}

We continue with the variation to the functional variable $A_{0}$ of the of
the ``matter'' part of the Lagrangean $\mathcal{L}_{2}$ is 
\begin{equation}
\frac{\partial }{\partial x^{0}}\frac{\delta \mathcal{L}_{2}}{\delta \left(
\partial _{0}A_{0}\right) }+\frac{\partial }{\partial x^{1}}\frac{\delta 
\mathcal{L}_{2}}{\delta \left( \partial _{1}A_{0}\right) }+\frac{\partial }{%
\partial x^{2}}\frac{\delta \mathcal{L}_{2}}{\delta \left( \partial
_{2}A_{0}\right) }-\frac{\delta \mathcal{L}_{2}}{\delta A_{0}}
\label{fb6110}
\end{equation}
where 
\begin{equation*}
\mathcal{L}_{2}=-\mathrm{tr}\left[ \left( D^{\mu }\phi \right) ^{\dagger
}\left( D_{\mu }\phi \right) \right]
\end{equation*}
has the detailed expression given in Eq.(\ref{fb54}). The Lagrangean is 
\begin{eqnarray}
\mathcal{L}_{2} &=&-\mathrm{tr}\left\{ -\frac{\partial \phi ^{\dagger }}{%
\partial t}\frac{\partial \phi }{\partial t}-\frac{\partial \phi ^{\dagger }%
}{\partial t}A_{0}\phi +\frac{\partial \phi ^{\dagger }}{\partial t}\phi
A_{0}\right.  \notag \\
&&+\phi ^{\dagger }A^{0\dagger }\frac{\partial \phi }{\partial t}+\phi
^{\dagger }A^{0\dagger }A_{0}\phi -\phi ^{\dagger }A^{0\dagger }\phi A_{0} 
\notag \\
&&-A^{0\dagger }\phi ^{\dagger }\frac{\partial \phi }{\partial t}%
-A^{0\dagger }\phi ^{\dagger }A_{0}\phi +A^{0\dagger }\phi ^{\dagger }\phi
A_{0}  \notag \\
&&+\frac{\partial \phi ^{\dagger }}{\partial x}\frac{\partial \phi }{%
\partial x}+\frac{\partial \phi ^{\dagger }}{\partial x}A_{1}\phi -\frac{%
\partial \phi ^{\dagger }}{\partial x}\phi A_{1}  \notag \\
&&+\phi ^{\dagger }A^{1\dagger }\frac{\partial \phi }{\partial x}+\phi
^{\dagger }A^{1\dagger }A_{1}\phi -\phi ^{\dagger }A^{1\dagger }\phi A_{1} 
\notag \\
&&-A^{1\dagger }\phi ^{\dagger }\frac{\partial \phi }{\partial x}%
-A^{1\dagger }\phi ^{\dagger }A_{1}\phi +A^{1\dagger }\phi ^{\dagger }\phi
A_{1}  \notag \\
&&+\frac{\partial \phi ^{\dagger }}{\partial y}\frac{\partial \phi }{%
\partial y}+\frac{\partial \phi ^{\dagger }}{\partial y}A_{2}\phi -\frac{%
\partial \phi ^{\dagger }}{\partial y}\phi A_{2}  \notag \\
&&+\phi ^{\dagger }A^{2\dagger }\frac{\partial \phi }{\partial y}+\phi
^{\dagger }A^{2\dagger }A_{2}\phi -\phi ^{\dagger }A^{2\dagger }\phi A_{2} 
\notag \\
&&\left. -A^{2\dagger }\phi ^{\dagger }\frac{\partial \phi }{\partial y}%
-A^{2\dagger }\phi ^{\dagger }A_{2}\phi +A^{2\dagger }\phi ^{\dagger }\phi
A_{2}\right\}  \notag
\end{eqnarray}

\paragraph{Calculation of the variation of the \emph{matter} Lagrangian to
the field $A_{0}$}

The first term is 
\begin{equation*}
\frac{\partial }{\partial x^{0}}\frac{\delta \mathcal{L}_{2}}{\delta \left(
\partial _{0}A_{0}\right) }=0
\end{equation*}
since there is no explicit dependence of $\mathcal{L}_{2}$ on $\left(
\partial _{0}A_{0}\right) $.

The next two terms in the variation of $\mathcal{L}_{2}$ are 
\begin{eqnarray*}
\frac{\partial }{\partial x^{1}}\frac{\delta \mathcal{L}_{2}}{\delta \left(
\partial _{1}A_{0}\right) } &=&0 \\
\frac{\partial }{\partial x^{2}}\frac{\delta \mathcal{L}_{2}}{\delta \left(
\partial _{2}A_{0}\right) } &=&0
\end{eqnarray*}
Again, there is no dependence of $\mathcal{L}_{2}$ with respect to $\partial
_{1}A_{0}$ and $\partial _{2}A_{0}$ and these contributions are zero.

The last term is 
\begin{equation}
\frac{\delta \mathcal{L}_{2}}{\delta A_{0}}=-\frac{\delta }{\delta A_{0}}%
\mathrm{tr}\left[ \left( D^{\mu }\phi \right) ^{\dagger }\left( D_{\mu }\phi
\right) \right]  \label{fb655}
\end{equation}
Only few terms from $\mathcal{L}_{2}$ have non-zero contributions 
\begin{eqnarray}
\frac{\delta \mathcal{L}_{2}}{\delta A_{0}} &=&-\frac{\delta }{\delta A_{0}}%
\mathrm{tr}\left\{ -\frac{\partial \phi ^{\dagger }}{\partial t}A_{0}\phi +%
\frac{\partial \phi ^{\dagger }}{\partial t}\phi A_{0}\right.  \label{fb652}
\\
&&+\phi ^{\dagger }A^{0\dagger }A_{0}\phi -\phi ^{\dagger }A^{0\dagger }\phi
A_{0}  \notag \\
&&\left. -A^{0\dagger }\phi ^{\dagger }A_{0}\phi +A^{0\dagger }\phi
^{\dagger }\phi A_{0}\right\}  \notag
\end{eqnarray}
We calculate in detail every term 
\begin{equation}
-\frac{\delta }{\delta A_{0}}\mathrm{tr}\left\{ -\frac{\partial \phi
^{\dagger }}{\partial t}A_{0}\phi \right\} =\left( \frac{\partial \phi
^{\dagger }}{\partial t}\right) ^{T}\left( \phi \right) ^{T}  \label{fb67}
\end{equation}
\begin{equation}
-\frac{\delta }{\delta A_{0}}\mathrm{tr}\left\{ \frac{\partial \phi
^{\dagger }}{\partial t}\phi A_{0}\right\} =-\left( \frac{\partial \phi
^{\dagger }}{\partial t}\phi \right) ^{T}  \label{fb68}
\end{equation}
\begin{equation}
-\frac{\delta }{\delta A_{0}}\mathrm{tr}\left\{ \phi ^{\dagger }A^{0\dagger
}A_{0}\phi \right\} =-\left( \phi ^{\dagger }A^{0\dagger }\right) ^{T}\left(
\phi \right) ^{T}  \label{fb70}
\end{equation}
In this is formula we have applied Eq.(\ref{fb64}) with $\mathbf{A\equiv }%
\phi ^{\dagger }A^{0\dagger }$, $\mathbf{B\equiv }\phi $ since the
functional variables $A^{0\dagger }$ and $A_{0}$ are independent. 
\begin{equation}
-\frac{\delta }{\delta A_{0}}\mathrm{tr}\left\{ -\phi ^{\dagger }A^{0\dagger
}\phi A_{0}\right\} =\left( \phi ^{\dagger }A^{0\dagger }\phi \right) ^{T}
\label{fb71}
\end{equation}
In this is formula we have applied Eq.(\ref{fb63}) with $\mathbf{A\equiv }%
\phi ^{\dagger }A^{0\dagger }\phi $, as explained above. 
\begin{equation}
-\frac{\delta }{\delta A_{0}}\mathrm{tr}\left\{ -A^{0\dagger }\phi ^{\dagger
}A_{0}\phi \right\} =\left( A^{0\dagger }\phi ^{\dagger }\right) ^{T}\left(
\phi \right) ^{T}  \label{fb73}
\end{equation}
This is formula (\ref{fb64}) with $\mathbf{A\equiv }A^{0\dagger }\phi
^{\dagger }$, $\mathbf{B\equiv }\phi $. 
\begin{equation}
-\frac{\delta }{\delta A_{0}}\mathrm{tr}\left\{ A^{0\dagger }\phi ^{\dagger
}\phi A_{0}\right\} =-\left( A^{0\dagger }\phi ^{\dagger }\phi \right) ^{T}
\label{fb74}
\end{equation}
Here the Eq.(\ref{fb63}) has been used with $\mathbf{A}\equiv A_{0}^{\dagger
}\phi ^{\dagger }\phi $.

Now we sum the results from Eqs.(\ref{fb67}) to (\ref{fb74}) 
\begin{eqnarray}
\frac{\delta \mathcal{L}_{2}}{\delta A_{0}} &=&\left( \frac{\partial \phi
^{\dagger }}{\partial t}\right) ^{T}\left( \phi \right) ^{T}-\left( \frac{%
\partial \phi ^{\dagger }}{\partial t}\phi \right) ^{T}  \label{fb75} \\
&&-\left( \phi ^{\dagger }A^{0\dagger }\right) ^{T}\left( \phi \right) ^{T} 
\notag \\
&&+\left( \phi ^{\dagger }A^{0\dagger }\phi \right) ^{T}  \notag \\
&&+\left( A^{0\dagger }\phi ^{\dagger }\right) ^{T}\left( \phi \right) ^{T} 
\notag \\
&&-\left( A^{0\dagger }\phi ^{\dagger }\phi \right) ^{T}  \notag
\end{eqnarray}
From the first, third and fifth terms we separate to the right the factor $%
\left( \phi \right) ^{T}$, and similarly in the other terms. 
\begin{eqnarray}
\frac{\delta \mathcal{L}_{2}}{\delta A_{0}} &=&\left\{ \left( \frac{\partial
\phi ^{\dagger }}{\partial t}\right) ^{T}-\left( \phi ^{\dagger }A^{0\dagger
}\right) ^{T}+\left( A^{0\dagger }\phi ^{\dagger }\right) ^{T}\right\}
\left( \phi \right) ^{T}  \label{fb7520} \\
&&+\left( \phi \right) ^{T}\left\{ -\left( \frac{\partial \phi ^{\dagger }}{%
\partial t}\right) ^{T}+\left( \phi ^{\dagger }A^{0\dagger }\right)
^{T}-\left( A^{0\dagger }\phi ^{\dagger }\right) ^{T}\right\}  \notag
\end{eqnarray}
Now, taking the transpose operator out of the paranthesis, 
\begin{eqnarray}
\frac{\delta \mathcal{L}_{2}}{\delta A_{0}} &=&\left\{ \phi \left( \frac{%
\partial \phi ^{\dagger }}{\partial t}\right) -\phi \left( \phi ^{\dagger
}A^{0\dagger }\right) +\phi \left( A^{0\dagger }\phi ^{\dagger }\right)
\right.  \label{fb7530} \\
&&\left. -\left( \frac{\partial \phi ^{\dagger }}{\partial t}\right) \phi
+\left( \phi ^{\dagger }A^{0\dagger }\right) \phi -\left( A^{0\dagger }\phi
^{\dagger }\right) \phi \right\} ^{T}  \notag
\end{eqnarray}
\begin{eqnarray}
\frac{\delta \mathcal{L}_{2}}{\delta A_{0}} &=&\left\{ \phi \left[ \frac{%
\partial \phi ^{\dagger }}{\partial t}-\left( \phi ^{\dagger }A^{0\dagger
}\right) +\left( A^{0\dagger }\phi ^{\dagger }\right) \right] \right\} ^{T}
\label{fb7535} \\
&&+\left\{ \left[ -\frac{\partial \phi ^{\dagger }}{\partial t}+\left( \phi
^{\dagger }A^{0\dagger }\right) -\left( A^{0\dagger }\phi ^{\dagger }\right) %
\right] \phi \right\} ^{T}  \notag
\end{eqnarray}
We now change the upper index $0$ into the low index $0$ for the potential $%
A^{\dagger }$, 
\begin{equation*}
A^{0\dagger }=-A_{0}^{\dagger }
\end{equation*}
and obtain 
\begin{eqnarray}
\frac{\delta \mathcal{L}_{2}}{\delta A_{0}} &=&\left\{ \phi \left[ \frac{%
\partial \phi ^{\dagger }}{\partial t}+\phi ^{\dagger }A_{0}^{\dagger
}-A_{0}^{\dagger }\phi ^{\dagger }\right] \right\} ^{T}  \label{fb78} \\
&&+\left\{ \left[ -\frac{\partial \phi ^{\dagger }}{\partial t}-\phi
^{\dagger }A_{0}^{\dagger }+A_{0}^{\dagger }\phi ^{\dagger }\right] \phi
\right\} ^{T}  \notag \\
&=&\left\{ \phi \left[ \frac{\partial \phi ^{\dagger }}{\partial t}+\left(
A_{0}\phi \right) ^{\dagger }-\left( \phi A_{0}\right) ^{\dagger }\right]
\right\} ^{T}  \notag \\
&&+\left\{ \left[ -\frac{\partial \phi ^{\dagger }}{\partial t}-\left(
A_{0}\phi \right) ^{\dagger }+\left( \phi A_{0}\right) ^{\dagger }\right]
\phi \right\} ^{T}  \notag \\
&=&\left\{ \phi \left[ \frac{\partial \phi }{\partial t}+A_{0}\phi -\phi
A_{0}\right] ^{\dagger }\right\} ^{T}+\left\{ -\left[ \frac{\partial \phi }{%
\partial t}+A_{0}\phi -\phi A_{0}\right] ^{\dagger }\phi \right\} ^{T} 
\notag \\
&=&\left\{ \phi \left( D_{0}\phi \right) ^{\dagger }\right\} ^{T}-\left\{
\left( D_{0}\phi \right) ^{\dagger }\phi \right\} ^{T}  \notag \\
&=&\left\{ \left[ \phi ,\left( D_{0}\phi \right) ^{\dagger }\right] \right\}
^{T}  \notag
\end{eqnarray}

Collecting these results 
\begin{eqnarray}
&&\frac{\partial }{\partial x^{0}}\frac{\delta \mathcal{L}_{2}}{\delta
\left( \partial _{0}A_{0}\right) }+\frac{\partial }{\partial x^{1}}\frac{%
\delta \mathcal{L}_{2}}{\delta \left( \partial _{1}A_{0}\right) }+\frac{%
\partial }{\partial x^{2}}\frac{\delta \mathcal{L}_{2}}{\delta \left(
\partial _{2}A_{0}\right) }-\frac{\delta \mathcal{L}_{2}}{\delta A_{0}}
\label{fb7812} \\
&=&-\left\{ \left[ \phi ,\left( D_{0}\phi \right) ^{\dagger }\right]
\right\} ^{T}  \notag
\end{eqnarray}

\paragraph{Calculation of the variation with respect to $A_{0}$ of the
scalar potential}

This is extremely simple since the scalar potential does not depend on $%
A_{0} $ nor of its derivatives to $x^{\mu }$. 
\begin{equation}
\frac{\delta V}{\delta A_{0}}\equiv 0  \label{fb7813}
\end{equation}

\paragraph{Final result for the variation of the Lagrangian with respect to $%
A_{0}$}

We assemble the partial results: Eq.(\ref{fb6109}) and Eq.(\ref{fb7812}) 
\begin{eqnarray}
&&\left( \frac{\partial }{\partial x^{0}}\frac{\delta }{\delta \left(
\partial _{0}A_{0}\right) }+\frac{\partial }{\partial x^{1}}\frac{\delta }{%
\delta \left( \partial _{1}A_{0}\right) }+\frac{\partial }{\partial x^{2}}%
\frac{\delta }{\delta \left( \partial _{2}A_{0}\right) }-\frac{\delta }{%
\delta A_{0}}\right) \left( \mathcal{L}_{1}+\mathcal{L}_{2}-V\right)
\label{fb7814} \\
&=&2\kappa \left( F_{12}\right) ^{T}-\left\{ \left[ \phi ,\left( D_{0}\phi
\right) ^{\dagger }\right] \right\} ^{T}  \notag \\
&=&0  \notag
\end{eqnarray}
The equation is 
\begin{equation}
2\kappa \left( F_{12}\right) ^{T}=\left\{ \left[ \phi ,\left( D_{0}\phi
\right) ^{\dagger }\right] \right\} ^{T}  \label{fb7815}
\end{equation}
or 
\begin{equation}
2\kappa F_{12}=\left[ \phi ,\left( D_{0}\phi \right) ^{\dagger }\right]
\label{fb7816}
\end{equation}
The left hand side can be written 
\begin{equation}
2\kappa F_{12}=\kappa \varepsilon ^{0\nu \rho }F_{\nu \rho }  \label{fb7817}
\end{equation}
and we change the order of the terms in the commutator 
\begin{eqnarray}
\kappa \varepsilon ^{0\nu \rho }F_{\nu \rho } &=&-\left[ \left( D_{0}\phi
\right) ^{\dagger },\phi \right]  \label{fb7818} \\
&=&i\times i\left[ \left( D_{0}\phi \right) ^{\dagger },\phi \right]  \notag
\end{eqnarray}
and multiplying with $-1$, 
\begin{equation}
-\kappa \varepsilon ^{0\nu \rho }F_{\nu \rho }=-i\times \left\{ i\left[
\left( D_{0}\phi \right) ^{\dagger },\phi \right] \right\}  \label{fb7819}
\end{equation}
We note that in the right hand side we have a part of the expression of the
current 
\begin{equation}
J_{0}\sim -i\left\{ -\left[ \left( D_{0}\phi \right) ^{\dagger },\phi \right]
\right\}  \label{fb7820}
\end{equation}
according to the definition of the current 
\begin{eqnarray}
-\kappa \varepsilon ^{0\nu \rho }F_{\nu \rho } &=&-iJ_{0}  \label{fb7821} \\
&=&iJ^{0}  \notag
\end{eqnarray}
Then at this point of the derivation it is suggested the following form of
the $\mu =0$ component of the equation of motion 
\begin{equation}
-\kappa \varepsilon ^{0\nu \rho }F_{\nu \rho }=iJ^{0}  \label{fb7822}
\end{equation}

Below we will join to this part of functional variation another part,
resulting from the functional varaition with respect to $A^{0\dagger }$.

\bigskip

\subsubsection{The functional variation with respect to the variable $%
A^{0\dagger }$}

We have to calculate 
\begin{equation}
\left( \frac{\partial }{\partial x^{0}}\frac{\delta }{\delta \left( \partial
_{0}A^{0\dagger }\right) }+\frac{\partial }{\partial x^{1}}\frac{\delta }{%
\delta \left( \partial _{1}A^{0\dagger }\right) }+\frac{\partial }{\partial
x^{2}}\frac{\delta }{\delta \left( \partial _{2}A^{0\dagger }\right) }-\frac{%
\delta }{\delta A^{0\dagger }}\right) \left( \mathcal{L}_{1}+\mathcal{L}%
_{2}-V\right) =0  \label{fb8100}
\end{equation}

Each term is calculated separately.

\paragraph{Functional derivatives of the gauge field Lagrangian with respect
to $A^{0\dagger }$}

The gauge-field part of the Lagrangian is 
\begin{eqnarray}
\mathcal{L}_{1} &=&-\kappa \mathrm{tr}\left\{ A_{0}\left( \partial
_{1}A_{2}\right) -A_{0}\left( \partial _{2}A_{1}\right) -A_{1}\left(
\partial _{0}A_{2}\right) \right.  \label{fb8102} \\
&&+A_{1}\left( \partial _{2}A_{0}\right) -A_{2}\left( \partial
_{1}A_{0}\right) +A_{2}\left( \partial _{0}A_{1}\right)  \notag \\
&&\left. +2A_{0}A_{1}A_{2}-2A_{0}A_{2}A_{1}\right\}  \notag
\end{eqnarray}
We note that the gauge field Lagrangian $\mathcal{L}_{1}$ is not expressed
in terms of $A_{0}^{\dagger }$%
\begin{equation}
\left( \frac{\partial }{\partial x^{0}}\frac{\delta }{\delta \left( \partial
_{0}A^{0\dagger }\right) }+\frac{\partial }{\partial x^{1}}\frac{\delta }{%
\delta \left( \partial _{1}A^{0\dagger }\right) }+\frac{\partial }{\partial
x^{2}}\frac{\delta }{\delta \left( \partial _{2}A^{0\dagger }\right) }-\frac{%
\delta }{\delta A^{0\dagger }}\right) \mathcal{L}_{1}=0  \label{fb82}
\end{equation}

\paragraph{Functional derivatives of the ``matter'' Lagrangian with respect
to $A^{0\dagger }$}

For the \emph{matter} part of the Lagrangian we have to calculate 
\begin{equation*}
\left( \frac{\partial }{\partial x^{0}}\frac{\delta }{\delta \left( \partial
_{0}A^{0\dagger }\right) }+\frac{\partial }{\partial x^{1}}\frac{\delta }{%
\delta \left( \partial _{1}A^{0\dagger }\right) }+\frac{\partial }{\partial
x^{2}}\frac{\delta }{\delta \left( \partial _{2}A^{0\dagger }\right) }-\frac{%
\delta }{\delta A^{0\dagger }}\right) \mathcal{L}_{2}
\end{equation*}
where $\mathcal{L}_{2}$ is given in Eq.(\ref{fb54}).

We have 
\begin{equation}
\frac{\partial }{\partial x^{0}}\frac{\delta }{\delta \left( \partial
_{0}A^{0\dagger }\right) }\mathcal{L}_{2}=0  \label{fb8207}
\end{equation}
\begin{equation}
\frac{\partial }{\partial x^{1}}\frac{\delta }{\delta \left( \partial
_{1}A^{0\dagger }\right) }\mathcal{L}_{2}=0  \label{fb8208}
\end{equation}
\begin{equation}
\frac{\partial }{\partial x^{2}}\frac{\delta }{\delta \left( \partial
_{2}A^{0\dagger }\right) }\mathcal{L}_{2}=0  \label{fb8209}
\end{equation}

Few of the terms in Eq.(\ref{fb54}) can provide a non-zero contribution 
\begin{eqnarray}
\frac{\delta \mathcal{L}_{2}}{\delta A^{0\dagger }} &=&-\frac{\delta }{%
\delta A^{0\dagger }}\mathrm{tr}\left\{ \phi ^{\dagger }A^{0\dagger }\frac{%
\partial \phi }{\partial t}-A^{0\dagger }\phi ^{\dagger }\frac{\partial \phi 
}{\partial t}\right.  \label{fb8210} \\
&&+\phi ^{\dagger }A^{0\dagger }A_{0}\phi -\phi ^{\dagger }A^{0\dagger }\phi
A_{0}  \notag \\
&&\left. -A^{0\dagger }\phi ^{\dagger }A_{0}\phi +A^{0\dagger }\phi
^{\dagger }\phi A_{0}\right\}  \notag
\end{eqnarray}
In detail, the terms are 
\begin{equation}
-\frac{\delta }{\delta A^{0\dagger }}\mathrm{tr}\left\{ \phi ^{\dagger
}A^{0\dagger }\frac{\partial \phi }{\partial t}\right\} =-\left( \phi
^{\dagger }\right) ^{T}\left( \frac{\partial \phi }{\partial t}\right) ^{T}
\label{fb85}
\end{equation}
\begin{equation}
-\frac{\delta }{\delta A^{0\dagger }}\mathrm{tr}\left\{ -A^{0\dagger }\phi
^{\dagger }\frac{\partial \phi }{\partial t}\right\} =\left( \phi ^{\dagger }%
\frac{\partial \phi }{\partial t}\right) ^{T}  \label{fb86}
\end{equation}
\begin{equation}
-\frac{\delta }{\delta A^{0\dagger }}\mathrm{tr}\left\{ \phi ^{\dagger
}A^{0\dagger }A_{0}\phi \right\} =-\left( \phi ^{\dagger }\right) ^{T}\left(
A_{0}\phi \right) ^{T}  \label{fb87}
\end{equation}
\begin{equation}
-\frac{\delta }{\delta A^{0\dagger }}\mathrm{tr}\left\{ -\phi ^{\dagger
}A^{0\dagger }\phi A_{0}\right\} =\left( \phi ^{\dagger }\right) ^{T}\left(
\phi A_{0}\right) ^{T}  \label{fb88}
\end{equation}
\begin{equation}
-\frac{\delta }{\delta A^{0\dagger }}\mathrm{tr}\left\{ -A^{0\dagger }\phi
^{\dagger }A_{0}\phi \right\} =\left( \phi ^{\dagger }A_{0}\phi \right) ^{T}
\label{fb89}
\end{equation}
\begin{equation}
-\frac{\delta }{\delta A^{0\dagger }}\mathrm{tr}\left\{ A^{0\dagger }\phi
^{\dagger }\phi A_{0}\right\} =-\left( \phi ^{\dagger }\phi A_{0}\right) ^{T}
\label{fb90}
\end{equation}
The results are now added 
\begin{eqnarray}
\frac{\delta \mathcal{L}_{2}}{\delta A^{0\dagger }} &=&-\left( \phi
^{\dagger }\right) ^{T}\left( \frac{\partial \phi }{\partial t}\right) ^{T}
\label{fb91} \\
&&+\left( \phi ^{\dagger }\frac{\partial \phi }{\partial t}\right) ^{T} 
\notag \\
&&-\left( \phi ^{\dagger }\right) ^{T}\left( A_{0}\phi \right) ^{T}  \notag
\\
&&+\left( \phi ^{\dagger }\right) ^{T}\left( \phi A_{0}\right) ^{T}  \notag
\\
&&+\left( \phi ^{\dagger }A_{0}\phi \right) ^{T}  \notag \\
&&-\left( \phi ^{\dagger }\phi A_{0}\right) ^{T}  \notag
\end{eqnarray}
In three terms we left-factorize $\left( \phi ^{\dagger }\right) ^{T}$%
\begin{eqnarray}
&&\left( \phi ^{\dagger }\right) ^{T}\left\{ -\left( \frac{\partial \phi }{%
\partial t}\right) ^{T}-\left( A_{0}\phi \right) ^{T}+\left( \phi
A_{0}\right) ^{T}\right\}  \label{fb92} \\
&=&\left( \phi ^{\dagger }\right) ^{T}\left\{ -\left( \frac{\partial \phi }{%
\partial t}\right) ^{T}-\left( \phi ^{T}A_{0}^{T}-A_{0}^{T}\phi ^{T}\right)
\right\}  \notag \\
&=&\left( \phi ^{\dagger }\right) ^{T}\left\{ -\left( \frac{\partial \phi }{%
\partial t}\right) ^{T}-\left[ \phi ^{T},A_{0}^{T}\right] \right\}  \notag \\
&=&-\left( \phi ^{\dagger }\right) ^{T}\left\{ \frac{\partial \phi }{%
\partial t}+\left[ A_{0},\phi \right] \right\} ^{T}  \notag \\
&=&-\left( \phi ^{\dagger }\right) ^{T}\left( D_{0}\phi \right) ^{T}  \notag
\end{eqnarray}
For the other three terms the result is similar 
\begin{eqnarray}
&&\left\{ \left( \frac{\partial \phi }{\partial t}\right) ^{T}+\left(
A_{0}\phi \right) ^{T}-\left( \phi A_{0}\right) ^{T}\right\} \left( \phi
^{\dagger }\right) ^{T}  \label{fb93} \\
&=&\left\{ \left( \frac{\partial \phi }{\partial t}\right) ^{T}+\left( \phi
^{T}A_{0}^{T}-A_{0}^{T}\phi ^{T}\right) \right\} \left( \phi ^{\dagger
}\right) ^{T}  \notag \\
&=&\left\{ \left( \frac{\partial \phi }{\partial t}\right) ^{T}+\left[ \phi
^{T},A_{0}^{T}\right] \right\} \left( \phi ^{\dagger }\right) ^{T}  \notag \\
&=&\left\{ \left( \frac{\partial \phi }{\partial t}\right) ^{T}+\left[
A_{0},\phi \right] ^{T}\right\} \left( \phi ^{\dagger }\right) ^{T}  \notag
\\
&=&\left( D_{0}\phi \right) ^{T}\left( \phi ^{\dagger }\right) ^{T}  \notag
\end{eqnarray}
Then 
\begin{eqnarray}
\frac{\delta \mathcal{L}_{2}}{\delta A^{0\dagger }} &=&-\left( \phi
^{\dagger }\right) ^{T}\left( D_{0}\phi \right) ^{T}+\left( D_{0}\phi
\right) ^{T}\left( \phi ^{\dagger }\right) ^{T}  \label{fb94} \\
&=&\left[ \left( D_{0}\phi \right) ^{T},\left( \phi ^{\dagger }\right) ^{T}%
\right]  \notag \\
&=&\left[ \phi ^{\dagger },D_{0}\phi \right] ^{T}  \notag
\end{eqnarray}
The Euler Lagrange equation is, for $A^{0\dagger }$%
\begin{equation}
\frac{\partial }{\partial x^{\mu }}\frac{\delta \mathcal{L}}{\delta \left( 
\frac{\partial A^{0\dagger }}{\partial x^{\mu }}\right) }-\frac{\delta 
\mathcal{L}}{\delta A^{0\dagger }}=0  \label{fb95}
\end{equation}
The Euler Lagrange equation in the case of $A^{0\dagger }$ is reduced to
only the last term 
\begin{equation}
-\frac{\delta \mathcal{L}}{\delta A^{0\dagger }}=-\frac{\delta \mathcal{L}%
_{2}}{\delta A^{0\dagger }}  \label{fb97}
\end{equation}
and is written from Eq.(\ref{fb94}) 
\begin{equation}
-\frac{\delta \mathcal{L}_{2}}{\delta A^{0\dagger }}=-\left[ \phi ^{\dagger
},D_{0}\phi \right] ^{T}  \label{fb98}
\end{equation}

\subsubsection{The Euler-Lagrange equation derived from functional variation
to $A_{0}$}

We now collect the results of the functional derivatives from both the gauge
and the matter parts of the Lagrangean, in the Euler-Lagrange equation for $%
A_{0}$ and \emph{add} the zero-valued term resulted from the functional
variation with respect to $A^{0\dagger }$. The formulas to be used are Eq.(%
\ref{fb6109}), Eq.(\ref{fb7812}) and Eq.(\ref{fb98}) 
\begin{eqnarray}
\left\{ -\left( \frac{\delta \mathcal{L}_{1}}{\delta A_{0}}+\frac{\delta 
\mathcal{L}_{2}}{\delta A_{0}}\right) \right\} +\left\{ -\frac{\delta 
\mathcal{L}_{2}}{\delta A^{0\dagger }}\right\} &=&0  \label{fb99} \\
2\kappa \left( F_{12}\right) ^{T}-\left[ \phi ,\left( D_{0}\phi \right)
^{\dagger }\right] ^{T}-\left[ \phi ^{\dagger },D_{0}\phi \right] ^{T} &=&0 
\notag
\end{eqnarray}
or, 
\begin{equation*}
-2\kappa F_{12}=-\left[ \phi ^{\dagger },D_{0}\phi \right] -\left[ \phi
,\left( D_{0}\phi \right) ^{\dagger }\right]
\end{equation*}
and interchanging the factors in the second commutator 
\begin{equation}
-2\kappa F_{12}=-\left[ \phi ^{\dagger },D_{0}\phi \right] +\left[ \left(
D_{0}\phi \right) ^{\dagger },\phi \right]  \label{fb100}
\end{equation}
The left hand side is the zero component of a tensorial contraction 
\begin{eqnarray}
-\kappa \left( \varepsilon ^{012}F_{12}+\varepsilon ^{021}F_{21}\right)
&=&-\left\{ \left[ \phi ^{\dagger },D_{0}\phi \right] -\left[ \left(
D_{0}\phi \right) ^{\dagger },\phi \right] \right\}  \label{fb1003} \\
&=&-i\times \left\{ -i\left( \left[ \phi ^{\dagger },D_{0}\phi \right] -%
\left[ \left( D_{0}\phi \right) ^{\dagger },\phi \right] \right) \right\} 
\notag
\end{eqnarray}
We will identify the right hand side as the covariant $0$-component of a
current 
\begin{equation}
J_{0}=-i\left\{ \left[ \phi ^{\dagger },D_{0}\phi \right] -\left[ \left(
D_{0}\phi \right) ^{\dagger },\phi \right] \right\}  \label{fb1004}
\end{equation}
and this equation takes the form of the Gauss law constraint from the main
text 
\begin{equation}
-\kappa \varepsilon ^{0\mu \nu }F_{\mu \nu }=-iJ_{0}=iJ^{0}  \label{fb1005}
\end{equation}

Therefore we conclude that we have derived the $0$ component of the equation 
\begin{equation*}
-\kappa \varepsilon ^{\mu \nu \rho }F_{\nu \rho }=iJ^{\mu }
\end{equation*}

\subsubsection{The Euler-Lagrange equation from the variation to $A_{1}$}

The detailed equation is 
\begin{equation}
\frac{\partial }{\partial x^{\mu }}\frac{\delta \mathcal{L}}{\delta \left( 
\frac{\partial A_{1}}{\partial x^{\mu }}\right) }-\frac{\delta \mathcal{L}}{%
\delta A_{1}}=0  \label{fb101}
\end{equation}
and shows a difference compared to the case of $A_{0}$: now there is a
dependence of $\mathcal{L}_{1}$ on the derivatives of the field $A_{1}$.

We have to calculate 
\begin{equation*}
\left( \frac{\partial }{\partial x^{0}}\frac{\delta }{\delta \left( \partial
_{0}A_{1}\right) }+\frac{\partial }{\partial x^{1}}\frac{\delta }{\delta
\left( \partial _{1}A_{1}\right) }+\frac{\partial }{\partial x^{2}}\frac{%
\delta }{\delta \left( \partial _{2}A_{1}\right) }-\frac{\delta }{\delta
A_{1}}\right) \left( \mathcal{L}_{1}+\mathcal{L}_{2}-V\right) =0
\end{equation*}

\paragraph{The functional variation with respect to $A_{1}$ of the \emph{%
gauge}-field part of Lagrangian}

This part is 
\begin{equation}
\left( \frac{\partial }{\partial x^{0}}\frac{\delta }{\delta \left( \partial
_{0}A_{1}\right) }+\frac{\partial }{\partial x^{1}}\frac{\delta }{\delta
\left( \partial _{1}A_{1}\right) }+\frac{\partial }{\partial x^{2}}\frac{%
\delta }{\delta \left( \partial _{2}A_{1}\right) }-\frac{\delta }{\delta
A_{1}}\right) \mathcal{L}_{1}  \label{fb102}
\end{equation}
and $\mathcal{L}_{1}$ is given by Eq.(\ref{fb3310}) 
\begin{eqnarray*}
\mathcal{L}_{1} &=&-\kappa \mathrm{tr}\left\{ A_{0}\left( \partial
_{1}A_{2}\right) -A_{0}\left( \partial _{2}A_{1}\right) -A_{1}\left(
\partial _{0}A_{2}\right) \right. \\
&&+A_{1}\left( \partial _{2}A_{0}\right) -A_{2}\left( \partial
_{1}A_{0}\right) +A_{2}\left( \partial _{0}A_{1}\right) \\
&&\left. +2A_{0}A_{1}A_{2}-2A_{0}A_{2}A_{1}\right\}
\end{eqnarray*}

Term by term 
\begin{eqnarray}
\frac{\partial }{\partial x^{0}}\frac{\delta \mathcal{L}_{1}}{\delta \left(
\partial _{0}A_{1}\right) } &=&\frac{\partial }{\partial x^{0}}\frac{\delta 
}{\delta \left( \partial _{0}A_{1}\right) }\left( -\kappa \right) \mathrm{tr}%
\left\{ A_{2}\left( \partial _{0}A_{1}\right) \right\}  \label{fb104} \\
&=&-\kappa \left( \partial _{0}A_{2}^{T}\right)  \notag
\end{eqnarray}
The second term in Eq.(\ref{fb102}) 
\begin{equation}
\frac{\partial }{\partial x^{1}}\frac{\delta \mathcal{L}_{1}}{\delta \left(
\partial _{1}A_{1}\right) }=0  \label{fb105}
\end{equation}
The third term from the Eq.(\ref{fb102}) 
\begin{eqnarray}
\frac{\partial }{\partial x^{2}}\frac{\delta \mathcal{L}_{1}}{\delta \left(
\partial _{2}A_{1}\right) } &=&\frac{\partial }{\partial x^{2}}\frac{\delta 
}{\delta \left( \partial _{2}A_{1}\right) }\left( -\kappa \right) \mathrm{tr}%
\left\{ -A_{0}\left( \partial _{2}A_{1}\right) \right\}  \label{fb106} \\
&=&\kappa \left( \partial _{2}A_{0}^{T}\right)  \notag
\end{eqnarray}
The last term in Eq.(\ref{fb102}) is 
\begin{equation}
\frac{\delta \mathcal{L}_{1}}{\delta A_{1}}=\frac{\delta }{\delta A_{1}}%
\left( -\kappa \right) \left\{ -A_{1}\left( \partial _{0}A_{2}\right)
+A_{1}\left( \partial _{2}A_{0}\right)
+2A_{0}A_{1}A_{2}-2A_{0}A_{2}A_{1}\right\}  \label{fb107}
\end{equation}
In detail 
\begin{equation*}
\frac{\delta }{\delta A_{1}}\left\{ -A_{1}\left( \partial _{0}A_{2}\right)
\right\} =-\left( \partial _{0}A_{2}\right) ^{T}
\end{equation*}
\begin{equation*}
\frac{\delta }{\delta A_{1}}\left\{ A_{1}\left( \partial _{2}A_{0}\right)
\right\} =\left( \partial _{2}A_{0}\right) ^{T}
\end{equation*}
\begin{equation*}
\frac{\delta }{\delta A_{1}}\left\{ 2A_{0}A_{1}A_{2}\right\}
=2A_{0}^{T}A_{2}^{T}
\end{equation*}
\begin{equation*}
\frac{\delta }{\delta A_{1}}\left\{ -2A_{0}A_{2}A_{1}\right\} =-2\left(
A_{0}A_{2}\right) ^{T}
\end{equation*}
It results 
\begin{equation*}
\frac{\delta \mathcal{L}_{1}}{\delta A_{1}}=\left( -\kappa \right) \left\{
-\left( \partial _{0}A_{2}\right) ^{T}+\left( \partial _{2}A_{0}\right)
^{T}+2A_{0}^{T}A_{2}^{T}-2\left( A_{0}A_{2}\right) ^{T}\right\}
\end{equation*}

Collecting the results for all the \emph{gauge} field Lagrangian 
\begin{eqnarray}
&&\left( \frac{\partial }{\partial x^{0}}\frac{\delta }{\delta \left(
\partial _{0}A_{1}\right) }+\frac{\partial }{\partial x^{1}}\frac{\delta }{%
\delta \left( \partial _{1}A_{1}\right) }+\frac{\partial }{\partial x^{2}}%
\frac{\delta }{\delta \left( \partial _{2}A_{1}\right) }-\frac{\delta }{%
\delta A_{1}}\right) \mathcal{L}_{1}  \label{fb108} \\
&=&-\kappa \left( \partial _{0}A_{2}^{T}\right) +\kappa \left( \partial
_{2}A_{0}^{T}\right) -\left( -\kappa \right) \left\{ -\left( \partial
_{0}A_{2}\right) ^{T}+\left( \partial _{2}A_{0}\right)
^{T}+2A_{0}^{T}A_{2}^{T}-2\left( A_{0}A_{2}\right) ^{T}\right\}  \notag \\
&=&2\kappa \left\{ -\left( \partial _{0}A_{2}^{T}\right) +\left( \partial
_{2}A_{0}^{T}\right) +A_{0}^{T}A_{2}^{T}-A_{2}^{T}A_{0}^{T}\right\}  \notag
\end{eqnarray}
We can write 
\begin{eqnarray*}
&&2\kappa \left\{ -\left( \partial _{0}A_{2}^{T}\right) +\left( \partial
_{2}A_{0}^{T}\right) +A_{0}^{T}A_{2}^{T}-A_{2}^{T}A_{0}^{T}\right\} \\
&=&-2\kappa \left\{ \partial _{0}A_{2}-\partial
_{2}A_{0}-A_{2}A_{0}+A_{0}A_{2}\right\} ^{T} \\
&=&-2\kappa \left\{ \partial _{0}A_{2}-\partial _{2}A_{0}+\left[ A_{0},A_{2}%
\right] \right\} ^{T} \\
&=&-2\kappa F_{02}^{T}
\end{eqnarray*}

\paragraph{The functional variation at $A_{1}$ of the \emph{matter} part of
the Lagrangean}

The calculations are similar to the previous case for $A_{0}$.

We have to calculate 
\begin{equation*}
\left( \frac{\partial }{\partial x^{0}}\frac{\delta }{\delta \left( \partial
_{0}A_{1}\right) }+\frac{\partial }{\partial x^{1}}\frac{\delta }{\delta
\left( \partial _{1}A_{1}\right) }+\frac{\partial }{\partial x^{2}}\frac{%
\delta }{\delta \left( \partial _{2}A_{1}\right) }-\frac{\delta }{\delta
A_{1}}\right) \mathcal{L}_{2}
\end{equation*}
where $\mathcal{L}_{2}$ is given in Eq.(\ref{fb54}).

We have 
\begin{equation*}
\frac{\partial }{\partial x^{0}}\frac{\delta }{\delta \left( \partial
_{0}A_{1}\right) }\mathcal{L}_{2}=0
\end{equation*}
\begin{equation*}
\frac{\partial }{\partial x^{1}}\frac{\delta }{\delta \left( \partial
_{1}A_{1}\right) }\mathcal{L}_{2}=0
\end{equation*}
\begin{equation*}
\frac{\partial }{\partial x^{2}}\frac{\delta }{\delta \left( \partial
_{2}A_{1}\right) }\mathcal{L}_{2}=0
\end{equation*}
\begin{eqnarray}
\frac{\delta \mathcal{L}_{2}}{\delta A_{1}} &=&-\frac{\delta }{\delta A_{1}}%
\mathrm{tr}\left[ \left( D^{\mu }\phi \right) ^{\dagger }\left( D_{\mu }\phi
\right) \right]  \label{fb114} \\
&=&-\frac{\delta }{\delta A_{1}}\mathrm{tr}\left[ -\left( D_{0}\phi \right)
^{\dagger }\left( D_{0}\phi \right) +\left( D_{1}\phi \right) ^{\dagger
}\left( D_{1}\phi \right) +\left( D_{2}\phi \right) ^{\dagger }\left(
D_{2}\phi \right) \right]  \notag
\end{eqnarray}
\begin{equation}
\frac{\delta \mathcal{L}_{2}}{\delta A_{1}}=-\frac{\delta }{\delta A_{1}}%
\mathrm{tr}\left\{ \left( D_{1}\phi \right) ^{\dagger }\left( D_{1}\phi
\right) \right\}  \label{fb115}
\end{equation}
\begin{eqnarray}
\frac{\delta \mathcal{L}_{2}}{\delta A_{1}} &=&-\frac{\delta }{\delta A_{1}}%
\mathrm{tr}\left\{ \frac{\partial \phi ^{\dagger }}{\partial x}A_{1}\phi -%
\frac{\partial \phi ^{\dagger }}{\partial x}\phi A_{1}\right.  \label{1157}
\\
&&+\phi ^{\dagger }A^{1\dagger }A_{1}\phi -\phi ^{\dagger }A^{1\dagger }\phi
A_{1}  \notag \\
&&\left. -A^{1\dagger }\phi ^{\dagger }A_{1}\phi +A^{1\dagger }\phi
^{\dagger }\phi A_{1}\right\}  \notag
\end{eqnarray}
Term by term 
\begin{equation}
-\frac{\delta }{\delta A_{1}}\mathrm{tr}\left\{ \frac{\partial \phi
^{\dagger }}{\partial x^{1}}A_{1}\phi \right\} =-\left( \frac{\partial \phi
^{\dagger }}{\partial x^{1}}\right) ^{T}\left( \phi \right) ^{T}
\label{fb118}
\end{equation}
\begin{equation}
-\frac{\delta }{\delta A_{1}}\mathrm{tr}\left\{ -\frac{\partial \phi
^{\dagger }}{\partial x^{1}}\phi A_{1}\right\} =\left( \frac{\partial \phi
^{\dagger }}{\partial x^{1}}\phi \right) ^{T}  \label{fb119}
\end{equation}
\begin{equation}
-\frac{\delta }{\delta A_{1}}\mathrm{tr}\left\{ \phi ^{\dagger }A^{1\dagger
}A_{1}\phi \right\} =-\left( \phi ^{\dagger }A^{1\dagger }\right) ^{T}\left(
\phi \right) ^{T}  \label{fb121}
\end{equation}
\begin{equation}
-\frac{\delta }{\delta A_{1}}\mathrm{tr}\left\{ -\phi ^{\dagger }A^{1\dagger
}\phi A_{1}\right\} =\left( \phi ^{\dagger }A^{1\dagger }\phi \right) ^{T}
\label{fb122}
\end{equation}
\begin{equation}
-\frac{\delta }{\delta A_{1}}\mathrm{tr}\left\{ -A^{1\dagger }\phi ^{\dagger
}A_{1}\phi \right\} =\left( A^{1\dagger }\phi ^{\dagger }\right) ^{T}\left(
\phi \right) ^{T}  \label{fb124}
\end{equation}
\begin{equation}
-\frac{\delta }{\delta A_{1}}\mathrm{tr}\left\{ A^{1\dagger }\phi ^{\dagger
}\phi A_{1}\right\} =-\left( A^{1\dagger }\phi ^{\dagger }\phi \right) ^{T}
\label{fb125}
\end{equation}
Now we collect all these contributions 
\begin{eqnarray}
\frac{\delta \mathcal{L}_{2}}{\delta A_{1}} &=&-\left( \frac{\partial \phi
^{\dagger }}{\partial x^{1}}\right) ^{T}\left( \phi \right) ^{T}+\left( 
\frac{\partial \phi ^{\dagger }}{\partial x^{1}}\phi \right) ^{T}
\label{fb126} \\
&&-\left( \phi ^{\dagger }A^{1\dagger }\right) ^{T}\left( \phi \right) ^{T} 
\notag \\
&&+\left( \phi ^{\dagger }A^{1\dagger }\phi \right) ^{T}  \notag \\
&&+\left( A^{1\dagger }\phi ^{\dagger }\right) ^{T}\left( \phi \right) ^{T} 
\notag \\
&&-\left( A^{1\dagger }\phi ^{\dagger }\phi \right) ^{T}  \notag
\end{eqnarray}
As in the case of $A_{0}$ equation, we right-factorise from the first, third
and fifth terms 
\begin{eqnarray}
&&-\left( \frac{\partial \phi ^{\dagger }}{\partial x^{1}}\right) ^{T}\left(
\phi \right) ^{T}-\left( \phi ^{\dagger }A^{1\dagger }\right) ^{T}\left(
\phi \right) ^{T}+\left( A^{1\dagger }\phi ^{\dagger }\right) ^{T}\left(
\phi \right) ^{T}  \label{fb127} \\
&=&-\left\{ \left( \frac{\partial \phi ^{\dagger }}{\partial x^{1}}\right)
^{T}+\left( \phi ^{\dagger }A^{1\dagger }\right) ^{T}-\left( A^{1\dagger
}\phi ^{\dagger }\right) ^{T}\right\} \left( \phi \right) ^{T}  \notag \\
&=&-\left\{ \frac{\partial \phi ^{\dagger }}{\partial x^{1}}+\left[ \phi
^{\dagger },A^{1\dagger }\right] \right\} ^{T}\left( \phi \right) ^{T} 
\notag \\
&=&-\left\{ \left( D_{1}\phi \right) ^{\dagger }\right\} ^{T}\left( \phi
\right) ^{T}  \notag
\end{eqnarray}
In a similar way we have from the second, fourth and sixth terms 
\begin{eqnarray}
&&\left( \frac{\partial \phi ^{\dagger }}{\partial x^{1}}\phi \right)
^{T}+\left( \phi ^{\dagger }A^{1\dagger }\phi \right) ^{T}-\left(
A^{1\dagger }\phi ^{\dagger }\phi \right) ^{T}  \label{fb128} \\
&=&\left( \phi \right) ^{T}\left\{ \left( \frac{\partial \phi ^{\dagger }}{%
\partial x^{1}}\right) ^{T}+\left( \phi ^{\dagger }A^{1\dagger }\right)
^{T}-\left( A^{1\dagger }\phi ^{\dagger }\right) ^{T}\right\}  \notag \\
&=&\left( \phi \right) ^{T}\left\{ \left( \frac{\partial \phi ^{\dagger }}{%
\partial x^{1}}\right) ^{T}+\left( \phi ^{\dagger }A^{1\dagger }-A^{1\dagger
}\phi ^{\dagger }\right) ^{T}\right\}  \notag \\
&=&\left( \phi \right) ^{T}\left\{ \frac{\partial \phi ^{\dagger }}{\partial
x^{1}}+\left[ \phi ^{\dagger },A^{1\dagger }\right] \right\} ^{T}  \notag \\
&=&\left( \phi \right) ^{T}\left\{ \left( D_{1}\phi \right) ^{\dagger
}\right\} ^{T}  \notag
\end{eqnarray}
Then, finally 
\begin{eqnarray*}
\frac{\delta \mathcal{L}_{2}}{\delta A_{1}} &=&-\left\{ \left( D_{1}\phi
\right) ^{\dagger }\right\} ^{T}\left( \phi \right) ^{T}+\left( \phi \right)
^{T}\left\{ \left( D_{1}\phi \right) ^{\dagger }\right\} ^{T} \\
&=&\left\{ \left( D_{1}\phi \right) ^{\dagger }\phi \right\} ^{T}-\left\{
\phi \left( D_{1}\phi \right) ^{\dagger }\right\} ^{T} \\
&=&\left\{ \left[ \left( D_{1}\phi \right) ^{\dagger },\phi \right] \right\}
^{T}
\end{eqnarray*}

Adding all contributions 
\begin{equation}
\frac{\partial }{\partial x^{\mu }}\frac{\delta \mathcal{L}_{2}}{\delta
\left( \frac{\partial A_{1}}{\partial x^{\mu }}\right) }-\frac{\delta 
\mathcal{L}_{2}}{\delta A_{1}}=-\left\{ \left[ \left( D_{1}\phi \right)
^{\dagger },\phi \right] \right\} ^{T}  \label{fb130}
\end{equation}

The total functional variation (gauge and matter parts) 
\begin{eqnarray*}
&&\left( \frac{\partial }{\partial x^{0}}\frac{\delta }{\delta \left(
\partial _{0}A_{1}\right) }+\frac{\partial }{\partial x^{1}}\frac{\delta }{%
\delta \left( \partial _{1}A_{1}\right) }+\frac{\partial }{\partial x^{2}}%
\frac{\delta }{\delta \left( \partial _{2}A_{1}\right) }-\frac{\delta }{%
\delta A_{1}}\right) \mathcal{L}_{1} \\
&&+\left( \frac{\partial }{\partial x^{0}}\frac{\delta }{\delta \left(
\partial _{0}A_{1}\right) }+\frac{\partial }{\partial x^{1}}\frac{\delta }{%
\delta \left( \partial _{1}A_{1}\right) }+\frac{\partial }{\partial x^{2}}%
\frac{\delta }{\delta \left( \partial _{2}A_{1}\right) }-\frac{\delta }{%
\delta A_{1}}\right) \mathcal{L}_{2} \\
&=&0
\end{eqnarray*}
\begin{eqnarray}
&&-2\kappa F_{02}^{T}  \label{fb131} \\
&&-\left\{ \left[ \left( D_{1}\phi \right) ^{\dagger },\phi \right] \right\}
^{T}  \notag \\
&=&0  \notag
\end{eqnarray}
Before concluding the calculation for $A_{1}$ , and having in mind a
possible more symmetrical form, we study the functional variations to the
field $A^{1\dagger }$.

\subsubsection{Functional variations to the field $A^{1\dagger }$}

The equation is 
\begin{equation}
\frac{\partial }{\partial x^{\mu }}\frac{\delta \left( \mathcal{L}_{1}+%
\mathcal{L}_{2}\right) }{\delta \left( \frac{\partial A^{1\dagger }}{%
\partial x^{\mu }}\right) }-\frac{\delta \left( \mathcal{L}_{1}+\mathcal{L}%
_{2}\right) }{\delta A^{1\dagger }}=0  \label{fb132}
\end{equation}
since only from gauge and matter we expect contributions. But we note that
the gauge part $\mathcal{L}_{1}$ has no dependence on the fields $\partial
_{\mu }A^{1\dagger }$ and $A^{1\dagger }$%
\begin{eqnarray}
\frac{\delta \mathcal{L}_{1}}{\delta \left( \partial _{\mu }A^{1\dagger
}\right) } &=&0  \label{fb134} \\
\frac{\delta \mathcal{L}_{1}}{\delta A^{1\dagger }} &=&0  \notag
\end{eqnarray}

\paragraph{The variation at $A_{1}^{\dagger }$ of the matter part of the
Lagrangean}

The matter part does not contain any term with $\partial _{\mu }A^{1\dagger
} $ which means that it also cannot contribute to the equation. The only
contribution may arise from the variation of matter part, $\mathcal{L}_{2}$,
to the field $A^{1\dagger }$.The formulas start with 
\begin{eqnarray}
\frac{\delta \mathcal{L}_{2}}{\delta A^{1\dagger }} &=&-\frac{\delta }{%
\delta A^{1\dagger }}\mathrm{tr}\left[ \left( D^{\mu }\phi \right) ^{\dagger
}\left( D_{\mu }\phi \right) \right]  \label{fb135} \\
&=&-\frac{\delta }{\delta A^{1\dagger }}\mathrm{tr}\left[ -\left( D_{0}\phi
\right) ^{\dagger }\left( D_{0}\phi \right) +\left( D_{1}\phi \right)
^{\dagger }\left( D_{1}\phi \right) +\left( D_{2}\phi \right) ^{\dagger
}\left( D_{2}\phi \right) \right]  \notag
\end{eqnarray}
Only the second term must be retained 
\begin{equation}
\frac{\delta \mathcal{L}_{2}}{\delta A^{1\dagger }}=-\frac{\delta }{\delta
A^{1\dagger }}\mathrm{tr}\left\{ \left( D_{1}\phi \right) ^{\dagger }\left(
D_{1}\phi \right) \right\}  \label{fb136}
\end{equation}
\begin{eqnarray}
\frac{\delta \mathcal{L}_{2}}{\delta A^{1\dagger }} &=&-\frac{\delta }{%
\delta A^{1\dagger }}\mathrm{tr}\left\{ \phi ^{\dagger }A^{1\dagger }\frac{%
\partial \phi }{\partial x^{1}}+\phi ^{\dagger }A^{1\dagger }A_{1}\phi -\phi
^{\dagger }A^{1\dagger }\phi A_{1}\right.  \label{fb137} \\
&&\left. -A^{1\dagger }\phi ^{\dagger }\frac{\partial \phi }{\partial x^{1}}%
-A^{1\dagger }\phi ^{\dagger }A_{1}\phi +A^{1\dagger }\phi ^{\dagger }\phi
A_{1}\right\}  \notag
\end{eqnarray}
Term by term 
\begin{equation}
-\frac{\delta }{\delta A^{1\dagger }}\mathrm{tr}\left\{ \phi ^{\dagger
}A^{1\dagger }\frac{\partial \phi }{\partial x^{1}}\right\} =-\left( \phi
^{\dagger }\right) ^{T}\left( \frac{\partial \phi }{\partial x^{1}}\right)
^{T}  \label{fb141}
\end{equation}
\begin{equation}
-\frac{\delta }{\delta A^{1\dagger }}\mathrm{tr}\left\{ \phi ^{\dagger
}A^{1\dagger }A_{1}\phi \right\} =-\left( \phi ^{\dagger }\right) ^{T}\left(
A_{1}\phi \right) ^{T}  \label{fb142}
\end{equation}
\begin{equation}
-\frac{\delta }{\delta A^{1\dagger }}\mathrm{tr}\left\{ -\phi ^{\dagger
}A^{1\dagger }\phi A_{1}\right\} =\left( \phi ^{\dagger }\right) ^{T}\left(
\phi A_{1}\right) ^{T}  \label{fb143}
\end{equation}
\begin{equation}
-\frac{\delta }{\delta A^{1\dagger }}\mathrm{tr}\left\{ -A^{1\dagger }\phi
^{\dagger }\frac{\partial \phi }{\partial x^{1}}\right\} =\left( \phi
^{\dagger }\frac{\partial \phi }{\partial x^{1}}\right) ^{T}  \label{fb144}
\end{equation}
\begin{equation}
-\frac{\delta }{\delta A^{1\dagger }}\mathrm{tr}\left\{ -A^{1\dagger }\phi
^{\dagger }A_{1}\phi \right\} =\left( \phi ^{\dagger }A_{1}\phi \right) ^{T}
\label{fb145}
\end{equation}
\begin{equation}
-\frac{\delta }{\delta A^{1\dagger }}\mathrm{tr}\left\{ A^{1\dagger }\phi
^{\dagger }\phi A_{1}\right\} =-\left( \phi ^{\dagger }\phi A_{1}\right) ^{T}
\label{fb146}
\end{equation}
Finally the sum of contributions 
\begin{eqnarray}
\frac{\delta \mathcal{L}_{2}}{\delta A_{1}^{\dagger }} &=&-\left( \phi
^{\dagger }\right) ^{T}\left( \frac{\partial \phi }{\partial x^{1}}\right)
^{T}  \label{fb147} \\
&&-\left( \phi ^{\dagger }\right) ^{T}\left( A_{1}\phi \right) ^{T}  \notag
\\
&&+\left( \phi ^{\dagger }\right) ^{T}\left( \phi A_{1}\right) ^{T}  \notag
\\
&&+\left( \phi ^{\dagger }\frac{\partial \phi }{\partial x^{1}}\right) ^{T} 
\notag \\
&&+\left( \phi ^{\dagger }A_{1}\phi \right) ^{T}  \notag \\
&&-\left( \phi ^{\dagger }\phi A_{1}\right) ^{T}  \notag
\end{eqnarray}
The first three terms can be written in the compact form 
\begin{eqnarray}
&&-\left( \phi ^{\dagger }\right) ^{T}\left( \frac{\partial \phi }{\partial
x^{1}}\right) ^{T}-\left( \phi ^{\dagger }\right) ^{T}\left( A_{1}\phi
\right) ^{T}+\left( \phi ^{\dagger }\right) ^{T}\left( \phi A_{1}\right) ^{T}
\label{fb148} \\
&=&-\left( \phi ^{\dagger }\right) ^{T}\left\{ \left( \frac{\partial \phi }{%
\partial x^{1}}\right) ^{T}+\left( A_{1}\phi \right) ^{T}-\left( \phi
A_{1}\right) ^{T}\right\}  \notag \\
&=&-\left( \phi ^{\dagger }\right) ^{T}\left\{ \left( \frac{\partial \phi }{%
\partial x^{1}}\right) ^{T}+\left[ A_{1},\phi \right] ^{T}\right\}  \notag \\
&=&-\left( \phi ^{\dagger }\right) ^{T}\left( D_{1}\phi \right) ^{T}  \notag
\end{eqnarray}
and the other three terms 
\begin{eqnarray}
&&\left( \phi ^{\dagger }\frac{\partial \phi }{\partial x^{1}}\right)
^{T}+\left( \phi ^{\dagger }A_{1}\phi \right) ^{T}-\left( \phi ^{\dagger
}\phi A_{1}\right) ^{T}  \label{fb149} \\
&=&\left\{ \left( \frac{\partial \phi }{\partial x^{1}}\right) ^{T}+\left(
A_{1}\phi \right) ^{T}-\left( \phi A_{1}\right) ^{T}\right\} \left( \phi
^{\dagger }\right) ^{T}  \notag \\
&=&\left\{ \left( \frac{\partial \phi }{\partial x^{1}}\right) ^{T}+\left[
A_{1},\phi \right] ^{T}\right\} \left( \phi ^{\dagger }\right) ^{T}  \notag
\\
&=&\left( D_{1}\phi \right) ^{T}\left( \phi ^{\dagger }\right) ^{T}  \notag
\end{eqnarray}
The full result 
\begin{eqnarray}
\frac{\delta \mathcal{L}_{2}}{\delta A^{1\dagger }} &=&-\left( \phi
^{\dagger }\right) ^{T}\left( D_{1}\phi \right) ^{T}+\left( D_{1}\phi
\right) ^{T}\left( \phi ^{\dagger }\right) ^{T}  \label{fb150} \\
&=&\left\{ \phi ^{\dagger }\left( D_{1}\phi \right) -\left( D_{1}\phi
\right) \phi ^{\dagger }\right\} ^{T}  \notag \\
&=&-\left\{ \left[ D_{1}\phi ,\phi ^{\dagger }\right] \right\} ^{T}  \notag
\end{eqnarray}
and the contribution of $A^{1\dagger }$ to the Euler Lagrange equation has
the form 
\begin{equation}
-\frac{\delta \mathcal{L}_{2}}{\delta A^{1\dagger }}=-\left\{ \left[
D_{1}\phi ,\phi ^{\dagger }\right] \right\} ^{T}  \label{fb151}
\end{equation}

\subsubsection{The final form of the Euler-Lagrange equation derived from
functional variation to $A_{1}$}

We now collect the results of the functional derivatives from both the gauge
and the matter parts of the Lagrangean, in the Euler-Lagrange equation for $%
A_{1}$ and add the zero-valued term from the functional derivation to $%
A^{1\dagger }$, Eqs. (\ref{fb131}), (\ref{fb151}). Therefore we have to
combine the following two results 
\begin{equation}
-2\kappa F_{02}^{T}-\left\{ \left[ \left( D_{1}\phi \right) ^{\dagger },\phi %
\right] \right\} ^{T}  \label{fb152}
\end{equation}
and 
\begin{equation}
\left\{ \left[ D_{1}\phi ,\phi ^{\dagger }\right] \right\} ^{T}
\label{fb153}
\end{equation}

We can subtract the two terms which leads to 
\begin{eqnarray}
&&-2\kappa F_{02}^{T}  \label{fb154} \\
&&-\left\{ \left[ \left( D_{1}\phi \right) ^{\dagger },\phi \right] \right\}
^{T}-\left\{ \left[ D_{1}\phi ,\phi ^{\dagger }\right] \right\} ^{T}  \notag
\\
&=&0  \notag
\end{eqnarray}
\begin{equation*}
-2\kappa F_{02}=\left[ \left( D_{1}\phi \right) ^{\dagger },\phi \right] +%
\left[ D_{1}\phi ,\phi ^{\dagger }\right]
\end{equation*}
We have 
\begin{eqnarray*}
\varepsilon ^{1\mu \nu }F_{\mu \nu } &=&\varepsilon ^{102}F_{02}+\varepsilon
^{120}F_{20} \\
&=&-F_{02}+F_{20} \\
&=&-2F_{02}
\end{eqnarray*}
from which 
\begin{equation*}
-2\kappa F_{02}=\kappa \varepsilon ^{1\mu \nu }F_{\mu \nu }
\end{equation*}
We can freely replace the covariant with contravariant indices in the right
hand side 
\begin{eqnarray*}
\kappa \varepsilon ^{1\mu \nu }F_{\mu \nu } &=&\left[ \left( D^{1}\phi
\right) ^{\dagger },\phi \right] +\left[ D^{1}\phi ,\phi ^{\dagger }\right]
\\
&=&i\left\{ -i\left( \left[ \left( D^{1}\phi \right) ^{\dagger },\phi \right]
-\left[ \phi ^{\dagger },D^{1}\phi \right] \right) \right\} \\
&=&-i\left\{ -i\left( \left[ \phi ^{\dagger },D^{1}\phi \right] -\left[
\left( D^{1}\phi \right) ^{\dagger },\phi \right] \right) \right\} \\
&=&-iJ^{1}
\end{eqnarray*}
and the equation is 
\begin{equation}
-\kappa \varepsilon ^{1\mu \nu }F_{\mu \nu }=iJ^{1}  \label{fb157}
\end{equation}
where 
\begin{equation*}
J^{1}=-i\left( \left[ \phi ^{\dagger },D^{1}\phi \right] -\left[ \left(
D^{1}\phi \right) ^{\dagger },\phi \right] \right)
\end{equation*}
as in the definition Eq.(\ref{jmiu}). Then the equation represents the
component $1$ of the equation of motion, Together with Eq.(\ref{fb1005}) we
have 
\begin{equation*}
-\kappa \varepsilon ^{\mu \nu \rho }F_{\nu \rho }=iJ^{\mu }
\end{equation*}
where 
\begin{equation*}
J^{\mu }=-i\left\{ \left[ \phi ^{\dagger },D^{\mu }\phi \right] -\left[
\left( D^{\mu }\phi \right) ^{\dagger },\phi \right] \right\}
\end{equation*}
\emph{i.e.} Eq.(\ref{eleq2}).

\subsection{The Euler-Lagrange equation for the matter fields}

This equation is obtained by functional variation of the action at the
matter fields, $\phi $ and respectively $\phi ^{\dagger }$. 
\begin{equation}
\frac{\partial }{\partial x^{\mu }}\frac{\delta \mathcal{L}}{\delta \left( 
\frac{\partial \phi }{\partial x^{\mu }}\right) }-\frac{\delta \mathcal{L}}{%
\delta \phi }=0  \label{fb161}
\end{equation}
and 
\begin{equation}
\frac{\partial }{\partial x^{\mu }}\frac{\delta \mathcal{L}}{\delta \left( 
\frac{\partial \phi ^{\dagger }}{\partial x^{\mu }}\right) }-\frac{\delta 
\mathcal{L}}{\delta \phi ^{\dagger }}=0  \label{fb162}
\end{equation}
The matter Lagrangean consists of the kinematical part 
\begin{equation}
\mathcal{L}_{2}=-\mathrm{tr}\left[ \left( D^{\mu }\phi \right) ^{\dagger
}\left( D_{\mu }\phi \right) \right]  \notag
\end{equation}
and the potential of self-interaction for the matter field 
\begin{equation}
V\left( \phi ,\phi ^{\dagger }\right) =\frac{1}{4\kappa ^{2}}\mathrm{tr}%
\left[ \left( \left[ \left[ \phi ,\phi ^{\dagger }\right] ,\phi \right]
-v^{2}\phi \right) ^{\dagger }\left( \left[ \left[ \phi ,\phi ^{\dagger }%
\right] ,\phi \right] -v^{2}\phi \right) \right]  \label{fb164}
\end{equation}
\begin{equation}
\mathcal{L}_{matter}=\mathcal{L}_{2}-V\left( \phi ,\phi ^{\dagger }\right)
\label{fb165}
\end{equation}
Let us consider the second Euler-Lagrange equation and calculate the
functional derivatives at $\partial _{\mu }\phi ^{\dagger }$. We have 
\begin{equation}
\frac{\delta \mathcal{L}}{\delta \left( \frac{\partial \phi ^{\dagger }}{%
\partial x^{\mu }}\right) }=\frac{\delta \mathcal{L}_{2}}{\delta \left( 
\frac{\partial \phi ^{\dagger }}{\partial x^{\mu }}\right) }  \label{fb166}
\end{equation}
\begin{eqnarray}
\frac{\delta \mathcal{L}_{2}}{\delta \left( \partial _{0}\phi ^{\dagger
}\right) } &=&-\frac{\delta }{\delta \left( \partial _{0}\phi ^{\dagger
}\right) }\mathrm{tr}\left[ \frac{\partial \phi ^{\dagger }}{\partial t}%
\frac{\partial \phi }{\partial t}+\frac{\partial \phi ^{\dagger }}{\partial t%
}A_{0}\phi -\frac{\partial \phi ^{\dagger }}{\partial t}\phi A_{0}\right]
\label{fb1663} \\
&=&-\left[ \frac{\partial \phi }{\partial t}+A_{0}\phi -\phi A_{0}\right]
^{T}  \notag \\
&=&-\left( D_{0}\phi \right) ^{T}  \notag
\end{eqnarray}
Analog calculations give 
\begin{eqnarray}
\frac{\delta \mathcal{L}_{2}}{\delta \left( \partial _{1}\phi ^{\dagger
}\right) } &=&-\frac{\delta }{\delta \left( \partial _{1}\phi ^{\dagger
}\right) }\mathrm{tr}\left[ \left( D_{1}\phi \right) ^{\dagger }\left(
D_{1}\phi \right) \right]  \label{fb167} \\
&=&-\left( D_{1}\phi \right) ^{T}  \notag
\end{eqnarray}
and 
\begin{eqnarray}
\frac{\delta \mathcal{L}_{2}}{\delta \left( \partial _{2}\phi ^{\dagger
}\right) } &=&-\frac{\delta }{\delta \left( \partial _{2}\phi ^{\dagger
}\right) }\mathrm{tr}\left[ \left( D_{2}\phi \right) ^{\dagger }\left(
D_{2}\phi \right) \right]  \label{fb168} \\
&=&-\left( D_{2}\phi \right) ^{T}  \notag
\end{eqnarray}
The other term in the Euler-Lagrange calculation implies the derivatives 
\begin{equation}
\frac{\delta \mathcal{L}}{\delta \phi ^{\dagger }}=\frac{\delta \mathcal{L}%
_{2}}{\delta \phi ^{\dagger }}-\frac{\delta V}{\delta \phi ^{\dagger }}
\label{fb169}
\end{equation}
The first term is 
\begin{equation}
\frac{\delta \mathcal{L}_{2}}{\delta \phi ^{\dagger }}=-\frac{\delta }{%
\delta \phi ^{\dagger }}\mathrm{tr}\left[ -\left( D_{0}\phi \right)
^{\dagger }\left( D_{0}\phi \right) +\left( D_{i}\phi \right) ^{\dagger
}\left( D_{i}\phi \right) \right]  \label{fb170}
\end{equation}
\begin{eqnarray}
&&\mathcal{L}_{2}=-\mathrm{tr}\left[ \left( D^{\mu }\phi \right) ^{\dagger
}\left( D_{\mu }\phi \right) \right]  \label{fb174} \\
&=&-\mathrm{tr}\left[ \left( \frac{\partial \phi ^{\dagger }}{\partial t}%
+\phi ^{\dagger }A^{0\dagger }-A^{0\dagger }\phi ^{\dagger }\right) \left( 
\frac{\partial \phi }{\partial t}+A_{0}\phi -\phi A_{0}\right) \right. 
\notag \\
&&+\left( \frac{\partial \phi ^{\dagger }}{\partial x}+\phi ^{\dagger
}A^{1\dagger }-A^{1\dagger }\phi ^{\dagger }\right) \left( \frac{\partial
\phi }{\partial x}+A_{1}\phi -\phi A_{1}\right)  \notag \\
&&\left. +\left( \frac{\partial \phi ^{\dagger }}{\partial y}+\phi ^{\dagger
}A^{2\dagger }-A^{2\dagger }\phi ^{\dagger }\right) \left( \frac{\partial
\phi }{\partial y}+A_{2}\phi -\phi A_{2}\right) \right]  \notag
\end{eqnarray}
and we will calculate it in detail. 
\begin{eqnarray}
&&-\frac{\delta }{\delta \phi ^{\dagger }}\mathrm{tr}\left[ -\left(
D_{0}\phi \right) ^{\dagger }\left( D_{0}\phi \right) \right]  \label{fb175}
\\
&=&-\frac{\delta }{\delta \phi ^{\dagger }}\mathrm{tr}\left[ \left( \frac{%
\partial \phi ^{\dagger }}{\partial t}+\phi ^{\dagger }A^{0\dagger
}-A^{0\dagger }\phi ^{\dagger }\right) \left( \frac{\partial \phi }{\partial
t}+A_{0}\phi -\phi A_{0}\right) \right]  \notag \\
&=&-\frac{\delta }{\delta \phi ^{\dagger }}\mathrm{tr}\left[ \phi ^{\dagger
}A^{0\dagger }\frac{\partial \phi }{\partial t}+\phi ^{\dagger }A^{0\dagger
}A_{0}\phi -\phi ^{\dagger }A^{0\dagger }\phi A_{0}\right.  \notag \\
&&\left. -A^{0\dagger }\phi ^{\dagger }\frac{\partial \phi }{\partial t}%
-A^{0\dagger }\phi ^{\dagger }A_{0}\phi +A^{0\dagger }\phi ^{\dagger }\phi
A_{0}\right]  \notag
\end{eqnarray}
Term by term 
\begin{equation*}
-\frac{\delta }{\delta \phi ^{\dagger }}\mathrm{tr}\left[ \phi ^{\dagger
}A^{0\dagger }\frac{\partial \phi }{\partial t}\right] =-\left( A^{0\dagger }%
\frac{\partial \phi }{\partial t}\right) ^{T}
\end{equation*}
\begin{equation*}
-\frac{\delta }{\delta \phi ^{\dagger }}\mathrm{tr}\left[ \phi ^{\dagger
}A^{0\dagger }A_{0}\phi \right] =-\left( A^{0\dagger }A_{0}\phi \right) ^{T}
\end{equation*}
\begin{equation*}
-\frac{\delta }{\delta \phi ^{\dagger }}\mathrm{tr}\left[ -\phi ^{\dagger
}A^{0\dagger }\phi A_{0}\right] =\left( A^{0\dagger }\phi A_{0}\right) ^{T}
\end{equation*}
\begin{equation*}
-\frac{\delta }{\delta \phi ^{\dagger }}\mathrm{tr}\left[ -A^{0\dagger }\phi
^{\dagger }\frac{\partial \phi }{\partial t}\right] =\left( A^{0\dagger
}\right) ^{T}\left( \frac{\partial \phi }{\partial t}\right) ^{T}
\end{equation*}
\begin{equation*}
-\frac{\delta }{\delta \phi ^{\dagger }}\mathrm{tr}\left[ -A^{0\dagger }\phi
^{\dagger }A_{0}\phi \right] =\left( A^{0\dagger }\right) ^{T}\left(
A_{0}\phi \right) ^{T}
\end{equation*}
\begin{equation*}
-\frac{\delta }{\delta \phi ^{\dagger }}\mathrm{tr}\left[ A^{0\dagger }\phi
^{\dagger }\phi A_{0}\right] =-\left( A^{0\dagger }\right) ^{T}\left( \phi
A_{0}\right) ^{T}
\end{equation*}
Summing up the terms 
\begin{eqnarray*}
&&-\frac{\delta }{\delta \phi ^{\dagger }}\mathrm{tr}\left[ -\left(
D_{0}\phi \right) ^{\dagger }\left( D_{0}\phi \right) \right] \\
&=&-\left( A^{0\dagger }\frac{\partial \phi }{\partial t}\right) ^{T}-\left(
A^{0\dagger }A_{0}\phi \right) ^{T}+\left( A^{0\dagger }\phi A_{0}\right)
^{T} \\
&&+\left( A^{0\dagger }\right) ^{T}\left( \frac{\partial \phi }{\partial t}%
\right) ^{T}+\left( A^{0\dagger }\right) ^{T}\left( A_{0}\phi \right)
^{T}-\left( A^{0\dagger }\right) ^{T}\left( \phi A_{0}\right) ^{T}
\end{eqnarray*}
We apply the transpose and factorize 
\begin{eqnarray*}
&&-\frac{\delta }{\delta \phi ^{\dagger }}\mathrm{tr}\left[ -\left(
D_{0}\phi \right) ^{\dagger }\left( D_{0}\phi \right) \right] \\
&=&-\left[ \left( \frac{\partial \phi }{\partial t}\right) ^{T}+\left(
A_{0}\phi \right) ^{T}-\left( \phi A_{0}\right) ^{T}\right] \left(
A^{0\dagger }\right) ^{T} \\
&&+\left( A^{0\dagger }\right) ^{T}\left[ \left( \frac{\partial \phi }{%
\partial t}\right) ^{T}+\left( A_{0}\phi \right) ^{T}-\left( \phi
A_{0}\right) ^{T}\right] \\
&=&-\left( D_{0}\phi \right) ^{T}\left( A^{0\dagger }\right) ^{T}+\left(
A^{0\dagger }\right) ^{T}\left( D_{0}\phi \right) ^{T} \\
&=&\left[ \left( A^{0\dagger }\right) ^{T},\left( D_{0}\phi \right) ^{T}%
\right]
\end{eqnarray*}
or 
\begin{equation*}
-\frac{\delta }{\delta \phi ^{\dagger }}\mathrm{tr}\left[ -\left( D_{0}\phi
\right) ^{\dagger }\left( D_{0}\phi \right) \right] =\left[ D_{0}\phi
,A^{0\dagger }\right] ^{T}
\end{equation*}
By analogue calculations we obtain 
\begin{equation*}
-\frac{\delta }{\delta \phi ^{\dagger }}\mathrm{tr}\left[ \left( D_{1}\phi
\right) ^{\dagger }\left( D_{1}\phi \right) \right] =\left[ D_{1}\phi
,A_{1}^{\dagger }\right] ^{T}
\end{equation*}
and 
\begin{equation*}
-\frac{\delta }{\delta \phi ^{\dagger }}\mathrm{tr}\left[ \left( D_{2}\phi
\right) ^{\dagger }\left( D_{2}\phi \right) \right] =\left[ D_{2}\phi
,A_{2}^{\dagger }\right] ^{T}
\end{equation*}
Then 
\begin{eqnarray}
\frac{\delta \mathcal{L}_{2}}{\delta \phi ^{\dagger }} &=&-\frac{\delta }{%
\delta \phi ^{\dagger }}\mathrm{tr}\left[ -\left( D_{0}\phi \right)
^{\dagger }\left( D_{0}\phi \right) +\left( D_{i}\phi \right) ^{\dagger
}\left( D_{i}\phi \right) \right]  \label{fb185} \\
&=&\left[ D_{0}\phi ,A_{0}^{\dagger }\right] ^{T}+\left[ D_{1}\phi
,A_{1}^{\dagger }\right] ^{T}+\left[ D_{2}\phi ,A_{2}^{\dagger }\right] ^{T}
\notag
\end{eqnarray}
Further 
\begin{equation*}
\frac{\delta \mathcal{L}}{\delta \phi ^{\dagger }}=\frac{\delta \mathcal{L}%
_{2}}{\delta \phi ^{\dagger }}-\frac{\delta V}{\delta \phi ^{\dagger }}
\end{equation*}
and the Euler-Lagrange equation results by combining Eqs.(\ref{fb1663}), (%
\ref{fb167}), (\ref{fb168}) and (\ref{fb185}) 
\begin{eqnarray*}
&&\frac{\partial }{\partial x^{\mu }}\frac{\delta \mathcal{L}}{\delta \left( 
\frac{\partial \phi ^{\dagger }}{\partial x^{\mu }}\right) }-\frac{\delta 
\mathcal{L}}{\delta \phi ^{\dagger }} \\
&=&\frac{\partial }{\partial x^{0}}\left( D_{0}\phi \right) ^{T}+\frac{%
\partial }{\partial x^{1}}\left( D_{1}\phi \right) ^{T}+\frac{\partial }{%
\partial x^{2}}\left( D_{2}\phi \right) ^{T} \\
&&+\left[ D_{0}\phi ,A_{0}^{\dagger }\right] ^{T}+\left[ D_{1}\phi
,A_{1}^{\dagger }\right] ^{T}+\left[ D_{2}\phi ,A_{2}^{\dagger }\right] ^{T}
\\
&&+\frac{\delta V}{\delta \phi ^{\dagger }} \\
&=&0
\end{eqnarray*}
We note that 
\begin{eqnarray*}
&&\frac{\partial }{\partial x^{0}}\left( D_{0}\phi \right) ^{T}+\left[
D_{0}\phi ,A_{0}^{\dagger }\right] ^{T} \\
&=&\left\{ \left( \frac{\partial }{\partial x^{0}}+\left[ ,A_{0}^{\dagger }%
\right] \right) \left( D_{0}\phi \right) \right\} ^{T} \\
&=&\left( D_{0}^{\dagger }D_{0}\phi \right) ^{T} \\
&=&-\left( D^{0\dagger }D_{0}\phi \right) ^{T}
\end{eqnarray*}
The other terms have similar form and we obtain 
\begin{equation*}
-\left( D^{\mu \dagger }D_{\mu }\phi \right) ^{T}+\frac{\delta V}{\delta
\phi ^{\dagger }}=0
\end{equation*}
%\end{appendices}

%\begin{appendices}

\section{Appendix C : Derivation of the second self-duality equation}

% new type of equation numbers for the appendix
\renewcommand{\theequation}{C.\arabic{equation}} \setcounter{equation}{0}

The gauge field equation in terms of $\pm $ variables (Dunne \cite{Dunne1})

Let us calculate 
\begin{equation}
F_{+-}=\partial _{+}A_{-}-\partial _{-}A_{+}+\left[ A_{+},A_{-}\right]
\label{3100}
\end{equation}
using the space variables $\left( 1,2\right) \equiv \left( x,y\right) $. 
\begin{eqnarray}
F_{+-} &=&\left( \frac{\partial }{\partial x^{1}}+i\frac{\partial }{\partial
x^{2}}\right) \left( A_{1}-iA_{2}\right) -  \label{3101} \\
&&-\left( \frac{\partial }{\partial x^{1}}-i\frac{\partial }{\partial x^{2}}%
\right) \left( A_{1}+iA_{2}\right) +  \notag \\
&&+\left[ A_{1}+iA_{2},A_{1}-iA_{2}\right]  \notag
\end{eqnarray}
\begin{eqnarray}
F_{+-} &=&\frac{\partial A_{1}}{\partial x^{1}}+i\frac{\partial A_{1}}{%
\partial x^{2}}-i\frac{\partial A_{2}}{\partial x^{1}}+\frac{\partial A_{2}}{%
\partial x^{2}}  \label{3102} \\
&&-\frac{\partial A_{1}}{\partial x^{1}}+i\frac{\partial A_{1}}{\partial
x^{2}}-i\frac{\partial A_{2}}{\partial x^{1}}-\frac{\partial A_{2}}{\partial
x^{2}}  \notag \\
&&-i\left[ A_{1},A_{2}\right] +i\left[ A_{2},A_{1}\right]  \notag
\end{eqnarray}
\begin{eqnarray}
F_{+-} &=&-2i\left\{ \frac{\partial A_{2}}{\partial x^{1}}-\frac{\partial
A_{1}}{\partial x^{2}}+\left[ A_{1},A_{2}\right] \right\}  \label{3104} \\
&=&-2iF_{12}  \notag \\
&=&-2i\varepsilon ^{012}F_{12}  \notag
\end{eqnarray}
On the other hand, we have the equation of motion Eq.(\ref{2964}) 
\begin{equation}
-2\kappa \varepsilon ^{012}F_{12}=iJ^{0}  \label{3105}
\end{equation}
from which we derive 
\begin{eqnarray*}
-2\varepsilon ^{012}F_{12} &=&\frac{1}{\kappa }iJ^{0} \\
&=&\frac{1}{i}F_{+-}
\end{eqnarray*}
\begin{eqnarray}
F_{+-} &=&-\frac{J^{0}}{\kappa }=\frac{J_{0}}{\kappa }  \label{3106} \\
&=&\frac{1}{\kappa }\left\{ -i\left( \left[ \phi ^{\dagger },D_{0}\phi %
\right] -\left[ \left( D_{0}\phi \right) ^{\dagger },\phi \right] \right)
\right\}  \notag
\end{eqnarray}
where we can use the second of the Eqs.(\ref{sd1}), valid at self-duality 
\begin{eqnarray}
D_{0}\phi &=&\frac{i}{2\kappa }\left( \left[ \left[ \phi ,\phi ^{\dagger }%
\right] ,\phi \right] -v^{2}\phi \right)  \label{3107} \\
\left( D_{0}\phi \right) ^{\dagger } &=&-\frac{i}{2\kappa }\left( \left( %
\left[ \left[ \phi ,\phi ^{\dagger }\right] ,\phi \right] \right) ^{\dagger
}-v^{2}\phi ^{\dagger }\right)  \notag
\end{eqnarray}
We calculate separately the terms 
\begin{equation}
\left[ \phi ^{\dagger },D_{0}\phi \right] =\frac{i}{2\kappa }\left\{ \left[
\phi ^{\dagger },\left[ \left[ \phi ,\phi ^{\dagger }\right] ,\phi \right] %
\right] -v^{2}\left[ \phi ^{\dagger },\phi \right] \right\}  \label{prco67}
\end{equation}
\begin{equation}
\left[ \left( D_{0}\phi \right) ^{\dagger },\phi \right] =-\frac{i}{2\kappa }%
\left\{ \left[ \left( \left[ \left[ \phi ,\phi ^{\dagger }\right] ,\phi %
\right] \right) ^{\dagger },\phi \right] -v^{2}\left[ \phi ^{\dagger },\phi %
\right] \right\}  \label{prco68}
\end{equation}
We can prove by expanding the commutators the equality of the first terms
from the curly brackets of right hand sides of the two equations. From Eq.(%
\ref{prco67}) 
\begin{eqnarray}
\left[ \phi ^{\dagger },\left[ \left[ \phi ,\phi ^{\dagger }\right] ,\phi %
\right] \right] &=&\left[ \phi ^{\dagger },\left[ \phi ,\phi ^{\dagger }%
\right] \phi -\phi \left[ \phi ,\phi ^{\dagger }\right] \right]
\label{comu1} \\
&=&\left[ \phi ^{\dagger },\left[ \phi ,\phi ^{\dagger }\right] \phi \right]
-\left[ \phi ^{\dagger },\phi \left[ \phi ,\phi ^{\dagger }\right] \right] 
\notag \\
&=&\phi ^{\dagger }\left[ \phi ,\phi ^{\dagger }\right] \phi  \notag \\
&&-\left[ \phi ,\phi ^{\dagger }\right] \phi \phi ^{\dagger }  \notag \\
&&-\phi ^{\dagger }\phi \left[ \phi ,\phi ^{\dagger }\right]  \notag \\
&&+\phi \left[ \phi ,\phi ^{\dagger }\right] \phi ^{\dagger }  \notag
\end{eqnarray}
From Eq.(\ref{prco68}) 
\begin{eqnarray*}
\left[ \left( \left[ \left[ \phi ,\phi ^{\dagger }\right] ,\phi \right]
\right) ^{\dagger },\phi \right] &=&\left[ \left[ \phi ^{\dagger },\left[
\phi ,\phi ^{\dagger }\right] ^{\dagger }\right] ,\phi \right] \\
&=&\left[ \left[ \phi ^{\dagger },\left[ \phi ,\phi ^{\dagger }\right] %
\right] ,\phi \right]
\end{eqnarray*}
since $\left[ \phi ,\phi ^{\dagger }\right] ^{\dagger }=\left[ \phi ,\phi
^{\dagger }\right] $ ; then 
\begin{eqnarray}
\left[ \left( \left[ \left[ \phi ,\phi ^{\dagger }\right] ,\phi \right]
\right) ^{\dagger },\phi \right] &=&\left[ \phi ^{\dagger }\left[ \phi ,\phi
^{\dagger }\right] -\left[ \phi ,\phi ^{\dagger }\right] \phi ^{\dagger
},\phi \right]  \label{comu2} \\
&=&\left[ \phi ^{\dagger }\left[ \phi ,\phi ^{\dagger }\right] ,\phi \right]
-\left[ \left[ \phi ,\phi ^{\dagger }\right] \phi ^{\dagger },\phi \right] 
\notag \\
&=&\phi ^{\dagger }\left[ \phi ,\phi ^{\dagger }\right] \phi  \notag \\
&&-\phi \phi ^{\dagger }\left[ \phi ,\phi ^{\dagger }\right]  \notag \\
&&-\left[ \phi ,\phi ^{\dagger }\right] \phi ^{\dagger }\phi  \notag \\
&&+\phi \left[ \phi ,\phi ^{\dagger }\right] \phi ^{\dagger }  \notag
\end{eqnarray}
We note that the first and the last terms in (\ref{comu1}) and (\ref{comu2})
are the same. The other terms in (\ref{comu1}) are 
\begin{eqnarray}
&&-\left[ \phi ,\phi ^{\dagger }\right] \phi \phi ^{\dagger }-\phi ^{\dagger
}\phi \left[ \phi ,\phi ^{\dagger }\right]  \label{3108} \\
&=&-\phi \phi ^{\dagger }\phi \phi ^{\dagger }+\phi ^{\dagger }\phi \phi
\phi ^{\dagger }  \notag \\
&&-\phi ^{\dagger }\phi \phi \phi ^{\dagger }+\phi ^{\dagger }\phi \phi
^{\dagger }\phi  \notag \\
&=&-\phi \phi ^{\dagger }\phi \phi ^{\dagger }+\phi ^{\dagger }\phi \phi
^{\dagger }\phi  \notag
\end{eqnarray}
and from (\ref{comu2}) 
\begin{eqnarray}
&&-\phi \phi ^{\dagger }\left[ \phi ,\phi ^{\dagger }\right] -\left[ \phi
,\phi ^{\dagger }\right] \phi ^{\dagger }\phi  \label{3109} \\
&=&-\phi \phi ^{\dagger }\phi \phi ^{\dagger }+\phi \phi ^{\dagger }\phi
^{\dagger }\phi  \notag \\
&&-\phi \phi ^{\dagger }\phi ^{\dagger }\phi +\phi ^{\dagger }\phi \phi
^{\dagger }\phi  \notag \\
&=&-\phi \phi ^{\dagger }\phi \phi ^{\dagger }+\phi ^{\dagger }\phi \phi
^{\dagger }\phi  \notag
\end{eqnarray}
and the two expressions are identical. This means that 
\begin{equation}
\left[ \phi ^{\dagger },\left[ \left[ \phi ,\phi ^{\dagger }\right] ,\phi %
\right] \right] -v^{2}\left[ \phi ^{\dagger },\phi \right] =\left[ \left( %
\left[ \left[ \phi ,\phi ^{\dagger }\right] ,\phi \right] \right) ^{\dagger
},\phi \right] -v^{2}\left[ \phi ^{\dagger },\phi \right]  \label{3110}
\end{equation}
and 
\begin{equation}
\left[ \phi ^{\dagger },D_{0}\phi \right] =-\left[ \left( D_{0}\phi \right)
^{\dagger },\phi \right]  \label{3111}
\end{equation}
and 
\begin{eqnarray}
F_{+-} &=&-\frac{i}{\kappa }\left\{ \left( \left[ \phi ^{\dagger },D_{0}\phi %
\right] -\left[ \left( D_{0}\phi \right) ^{\dagger },\phi \right] \right)
\right\}  \label{3112} \\
&=&-\frac{2i}{\kappa }\left[ \phi ^{\dagger },D_{0}\phi \right]  \notag
\end{eqnarray}
where we replace the expression of $D_{0}\phi $%
\begin{eqnarray}
F_{+-} &=&-\frac{2i}{\kappa }\left[ \phi ^{\dagger },\frac{i}{2\kappa }%
\left( \left[ \left[ \phi ,\phi ^{\dagger }\right] ,\phi \right] -v^{2}\phi
\right) \right]  \label{3114} \\
&=&\frac{1}{\kappa ^{2}}\left[ \phi ^{\dagger },\left[ \left[ \phi ,\phi
^{\dagger }\right] ,\phi \right] -v^{2}\phi \right]  \notag \\
&=&\frac{1}{\kappa ^{2}}\left[ v^{2}\phi -\left[ \left[ \phi ,\phi ^{\dagger
}\right] ,\phi \right] ,\phi ^{\dagger }\right]  \notag
\end{eqnarray}

%\end{appendices}

%\begin{appendices}

\section{Appendix D : Notes on definitions}

% new type of equation numbers for the appendix
\renewcommand{\theequation}{D.\arabic{equation}} \setcounter{equation}{0}

The information about the algebraic structure invoked in the present model
can be found in \cite{Slansky}. A \textbf{simple group} is a group that does
not have invariant subgroups, except of the identity and the whole group.

A \textbf{simple algebra} is an algebra that does not have \emph{proper
ideals}.

A \textbf{semi-simple algebra} is an algebra that can be written as a direct
sum of simple algebras.

$U\left( 1\right) $ is \emph{not} simple.

\bigskip

The \textbf{dimension} of a simple Lie algebra is the total \emph{number of
linearly independent generators}.

The \textbf{rank of the algebra}, $r$, is the maximum \emph{number of
simultaneously diagonalisable generators} of a simple Lie algebra.

\bigskip

In the Cartan-Weil analysis the generators are written in a basis where they
can be devided into two sets:

\begin{itemize}
\item  the \textbf{Cartan subalgebra}, which is the \emph{maximal Abelian
subalgebra} of $G$. It contains $r$ diagonalisable generators $H_{i}$, $%
i=1,...,r$%
\begin{equation}
\left[ H_{i},H_{j}\right] =0\;,\;i,j=1,...,r  \label{hcom}
\end{equation}

\item  the remaining generators of the algebra $G$ are defined such as they
satisfy the eigenvalue problems 
\begin{equation}
\left[ H_{i},E_{\mu }\right] =\alpha _{i}E_{\mu }\;,\;i=1,...,r
\label{rooteig}
\end{equation}
\end{itemize}

It results that the constants $\alpha _{i}$ can be considered \emph{%
structure constants} of the algebra in the Cartan-Weil basis.

For each generator $E_{\mu }$ there are $r$ constant numbers, $\alpha _{i}$, 
$i=1,...,r$ ; if we consider a space with dimension $r$, then the set of
points $\left( \alpha _{1},\alpha _{2},...,\alpha _{r}\right) $
corresponding to one generator $E_{\mu }$ is a point in this space. This
space is called \textbf{root space} and the name \textbf{root} comes from
the fact the the vector $\left( \alpha _{1},\alpha _{2},...,\alpha
_{r}\right) $ is obtained by solving the equation (\ref{rooteig}), an
eigenvalue problem.

\bigskip

Two problems are connected and are treated together using the \textbf{Dynkin
diagrams} of the simple algebras:

\begin{enumerate}
\item  \textbf{to classify all possible systems of roots} for the algebras
of a given rank $r$;

\item  \textbf{to find all possible irreducible representations} of a simple
group $G$. This means to identify a system of physical states on which the
generators $E_{\mu }$ are acting (the states belong to a Hilbert space) with
the property that these states are transformed between them (or, the system
of states is closed under the action of the generators $E_{\mu }$). These
states are taken as the basis for an irreducible representation.
\end{enumerate}

Considering the physical states which are the basis of the irreducible
representation, $\left| \lambda \right\rangle $, they can be labelled by the 
$r$ eigenvalues of the \emph{diagonalisable generators} $H_{i}$%
\begin{equation*}
H_{i}\left| \lambda \right\rangle =\lambda _{i}\left| \lambda \right\rangle
\;,\;i=1,...,r
\end{equation*}

The set $\lambda $ is called the \textbf{weight} of the representation
vector.

%\end{appendices}

%\begin{appendices}

\section{Appendix E : Expanded form of the first equation of motion}

% new type of equation numbers for the appendix
\renewcommand{\theequation}{E.\arabic{equation}} \setcounter{equation}{0}

The first equation of \ motion is 
\begin{equation}
D_{\mu }D^{\mu }\phi =\frac{\partial V}{\partial \phi ^{\dagger }}
\label{firsteqmot}
\end{equation}
As explained by Dunne \cite{Dunne3} the derivative of the potential $V$ is
obtained from the functional variation to $\phi ^{\dagger }$ and for this we
need the expanded form of $V$. The potential is given initially in terms of
the trace of the operators 
\begin{equation}
V\left( \phi ,\phi ^{\dagger }\right) =\frac{1}{4\kappa ^{2}}\mathrm{tr}%
\left[ \left( \left[ \left[ \phi ,\phi ^{\dagger }\right] ,\phi \right]
-v^{2}\phi \right) ^{\dagger }\left( \left[ \left[ \phi ,\phi ^{\dagger }%
\right] ,\phi \right] -v^{2}\phi \right) \right]  \label{e1}
\end{equation}
In the equations of motion we will treat separately each term: 
\begin{eqnarray}
&&\left( \left[ \left[ \phi ,\phi ^{\dagger }\right] ,\phi \right]
-v^{2}\phi \right) ^{\dagger }\left( \left[ \left[ \phi ,\phi ^{\dagger }%
\right] ,\phi \right] -v^{2}\phi \right)  \label{e2} \\
&=&\left( \left[ \left[ \phi ,\phi ^{\dagger }\right] ,\phi \right]
^{\dagger }-v^{2}\phi ^{\dagger }\right) \left( \left[ \left[ \phi ,\phi
^{\dagger }\right] ,\phi \right] -v^{2}\phi \right)  \notag \\
&=&\left[ \left[ \phi ,\phi ^{\dagger }\right] ,\phi \right] ^{\dagger }%
\left[ \left[ \phi ,\phi ^{\dagger }\right] ,\phi \right]  \notag \\
&&-v^{2}\phi ^{\dagger }\left[ \left[ \phi ,\phi ^{\dagger }\right] ,\phi %
\right]  \notag \\
&&-v^{2}\left[ \left[ \phi ,\phi ^{\dagger }\right] ,\phi \right] ^{\dagger
}\phi  \notag \\
&&+v^{4}\phi ^{\dagger }\phi  \notag
\end{eqnarray}
The first term, of sixth degree 
\begin{eqnarray}
&&\left[ \left[ \phi ,\phi ^{\dagger }\right] ,\phi \right] ^{\dagger }\left[
\left[ \phi ,\phi ^{\dagger }\right] ,\phi \right]  \label{e3} \\
&=&\left[ \phi \phi ^{\dagger }-\phi ^{\dagger }\phi ,\phi \right] ^{\dagger
}\left[ \phi \phi ^{\dagger }-\phi ^{\dagger }\phi ,\phi \right]  \notag \\
&=&\left( \phi \phi ^{\dagger }\phi -\phi ^{\dagger }\phi \phi -\phi \phi
\phi ^{\dagger }+\phi \phi ^{\dagger }\phi \right) ^{\dagger }  \notag \\
&&\times \left( \phi \phi ^{\dagger }\phi -\phi ^{\dagger }\phi \phi -\phi
\phi \phi ^{\dagger }+\phi \phi ^{\dagger }\phi \right)  \notag \\
&=&\left( \phi ^{\dagger }\phi \phi ^{\dagger }-\phi ^{\dagger }\phi
^{\dagger }\phi -\phi \phi ^{\dagger }\phi ^{\dagger }+\phi ^{\dagger }\phi
\phi ^{\dagger }\right)  \notag \\
&&\times \left( \phi \phi ^{\dagger }\phi -\phi ^{\dagger }\phi \phi -\phi
\phi \phi ^{\dagger }+\phi \phi ^{\dagger }\phi \right)  \notag \\
&=&\left( 2\phi ^{\dagger }\phi \phi ^{\dagger }-\phi ^{\dagger }\phi
^{\dagger }\phi -\phi \phi ^{\dagger }\phi ^{\dagger }\right)  \notag \\
&&\times \left( 2\phi \phi ^{\dagger }\phi -\phi ^{\dagger }\phi \phi -\phi
\phi \phi ^{\dagger }\right)  \notag
\end{eqnarray}
\begin{eqnarray}
&&\left[ \left[ \phi ,\phi ^{\dagger }\right] ,\phi \right] ^{\dagger }\left[
\left[ \phi ,\phi ^{\dagger }\right] ,\phi \right]  \label{six1} \\
&=&4\phi ^{\dagger }\phi \phi ^{\dagger }\phi \phi ^{\dagger }\phi  \notag \\
&&-2\phi ^{\dagger }\phi \phi ^{\dagger }\phi ^{\dagger }\phi \phi  \notag \\
&&-2\phi ^{\dagger }\phi \phi ^{\dagger }\phi \phi \phi ^{\dagger }  \notag
\\
&&-2\phi ^{\dagger }\phi ^{\dagger }\phi \phi \phi ^{\dagger }\phi  \notag \\
&&+\phi ^{\dagger }\phi ^{\dagger }\phi \phi ^{\dagger }\phi \phi  \notag \\
&&+\phi ^{\dagger }\phi ^{\dagger }\phi \phi \phi \phi ^{\dagger }  \notag \\
&&-2\phi \phi ^{\dagger }\phi ^{\dagger }\phi \phi ^{\dagger }\phi  \notag \\
&&+\phi \phi ^{\dagger }\phi ^{\dagger }\phi ^{\dagger }\phi \phi  \notag \\
&&+\phi \phi ^{\dagger }\phi ^{\dagger }\phi \phi \phi ^{\dagger }  \notag
\end{eqnarray}
We remark that in Eq.(\ref{six1}) the terms five and seven 
\begin{eqnarray*}
&&\mathrm{tr}\left( -2\phi ^{\dagger }\phi \phi ^{\dagger }\phi \phi \phi
^{\dagger }+\phi ^{\dagger }\phi ^{\dagger }\phi \phi ^{\dagger }\phi \phi
\right) \\
&=&-\mathrm{tr}\left( \phi ^{\dagger }\phi \phi ^{\dagger }\phi \phi \phi
^{\dagger }\right)
\end{eqnarray*}
terms six and eight 
\begin{eqnarray*}
&&\mathrm{tr}\left( \phi ^{\dagger }\phi ^{\dagger }\phi \phi \phi \phi
^{\dagger }+\phi \phi ^{\dagger }\phi ^{\dagger }\phi ^{\dagger }\phi \phi
\right) \\
&=&2\mathrm{tr}\left( \phi ^{\dagger }\phi ^{\dagger }\phi \phi \phi \phi
^{\dagger }\right)
\end{eqnarray*}
terms second and nine 
\begin{eqnarray}
&&\mathrm{tr}\left( -2\phi ^{\dagger }\phi \phi ^{\dagger }\phi ^{\dagger
}\phi \phi +\phi \phi ^{\dagger }\phi ^{\dagger }\phi \phi \phi ^{\dagger
}\right)  \label{e4} \\
&=&-\mathrm{tr}\left( \phi ^{\dagger }\phi \phi ^{\dagger }\phi ^{\dagger
}\phi \phi \right)  \notag
\end{eqnarray}
third 
\begin{equation}
-2\phi ^{\dagger }\phi \phi ^{\dagger }\phi \phi \phi ^{\dagger }  \label{e5}
\end{equation}
four 
\begin{equation}
-2\phi ^{\dagger }\phi ^{\dagger }\phi \phi \phi ^{\dagger }\phi  \label{e6}
\end{equation}
can be grouped. Collecting the terms we have 
\begin{eqnarray}
&&\left[ \left[ \phi ,\phi ^{\dagger }\right] ,\phi \right] ^{\dagger }\left[
\left[ \phi ,\phi ^{\dagger }\right] ,\phi \right]  \label{six2} \\
&=&-\mathrm{tr}\left( \phi ^{\dagger }\phi \phi ^{\dagger }\phi \phi \phi
^{\dagger }\right)  \notag \\
&&+2\mathrm{tr}\left( \phi ^{\dagger }\phi ^{\dagger }\phi \phi \phi \phi
^{\dagger }\right)  \notag \\
&&-\mathrm{tr}\left( \phi ^{\dagger }\phi \phi ^{\dagger }\phi ^{\dagger
}\phi \phi \right)  \notag \\
&&-2\mathrm{tr}\left( \phi ^{\dagger }\phi \phi ^{\dagger }\phi \phi \phi
^{\dagger }\right)  \notag \\
&&-2\mathrm{tr}\left( \phi ^{\dagger }\phi ^{\dagger }\phi \phi \phi
^{\dagger }\phi \right)  \notag
\end{eqnarray}
The third term, after two permutations of the first two factors, is
identical to the fifth term 
\begin{eqnarray}
&&-\mathrm{tr}\left( \phi ^{\dagger }\phi \phi ^{\dagger }\phi ^{\dagger
}\phi \phi \right) -2\mathrm{tr}\left( \phi ^{\dagger }\phi ^{\dagger }\phi
\phi \phi ^{\dagger }\phi \right)  \label{e7} \\
&=&-\mathrm{tr}\left( \phi ^{\dagger }\phi ^{\dagger }\phi \phi \phi
^{\dagger }\phi \right) -2\mathrm{tr}\left( \phi ^{\dagger }\phi ^{\dagger
}\phi \phi \phi ^{\dagger }\phi \right)  \notag \\
&=&-3\mathrm{tr}\left( \phi ^{\dagger }\phi ^{\dagger }\phi \phi \phi
^{\dagger }\phi \right)  \notag
\end{eqnarray}
The first and the fourth factors are equal 
\begin{eqnarray}
&&-\mathrm{tr}\left( \phi ^{\dagger }\phi \phi ^{\dagger }\phi \phi \phi
^{\dagger }\right) -2\mathrm{tr}\left( \phi ^{\dagger }\phi \phi ^{\dagger
}\phi \phi \phi ^{\dagger }\right)  \label{e8} \\
&=&-3\mathrm{tr}\left( \phi ^{\dagger }\phi \phi ^{\dagger }\phi \phi \phi
^{\dagger }\right)  \notag
\end{eqnarray}
Then from the first, sixth degree product, we obtain 
\begin{eqnarray}
&&\left[ \left[ \phi ,\phi ^{\dagger }\right] ,\phi \right] ^{\dagger }\left[
\left[ \phi ,\phi ^{\dagger }\right] ,\phi \right]  \label{e9} \\
&=&-3\mathrm{tr}\left( \phi ^{\dagger }\phi ^{\dagger }\phi \phi \phi
^{\dagger }\phi \right) -3\mathrm{tr}\left( \phi ^{\dagger }\phi \phi
^{\dagger }\phi \phi \phi ^{\dagger }\right) +2\mathrm{tr}\left( \phi
^{\dagger }\phi ^{\dagger }\phi \phi \phi \phi ^{\dagger }\right)  \notag
\end{eqnarray}

The next two terms in the potential (proportional with $\left( -v^{2}\right) 
$) are expanded 
\begin{eqnarray}
&&-v^{2}\phi ^{\dagger }\left[ \left[ \phi ,\phi ^{\dagger }\right] ,\phi %
\right] -v^{2}\left[ \left[ \phi ,\phi ^{\dagger }\right] ,\phi \right]
^{\dagger }\phi  \label{4phi} \\
&=&\left( -v^{2}\right) \left\{ \phi ^{\dagger }\left( \phi \phi ^{\dagger
}\phi -\phi ^{\dagger }\phi \phi -\phi \phi \phi ^{\dagger }+\phi \phi
^{\dagger }\phi \right) \right.  \notag \\
&&\left. \left( \phi \phi ^{\dagger }\phi -\phi ^{\dagger }\phi \phi -\phi
\phi \phi ^{\dagger }+\phi \phi ^{\dagger }\phi \right) ^{\dagger }\phi
\right\}  \notag \\
&=&\left( -v^{2}\right) \left\{ 2\phi ^{\dagger }\phi \phi ^{\dagger }\phi
-\phi ^{\dagger }\phi ^{\dagger }\phi \phi -\phi ^{\dagger }\phi \phi \phi
^{\dagger }\right.  \notag \\
&&\left. \left( 2\phi \phi ^{\dagger }\phi -\phi ^{\dagger }\phi \phi -\phi
\phi \phi ^{\dagger }\right) ^{\dagger }\phi \right\}  \notag \\
&=&\left( -v^{2}\right) \left\{ 2\phi ^{\dagger }\phi \phi ^{\dagger }\phi
-\phi ^{\dagger }\phi ^{\dagger }\phi \phi -\phi ^{\dagger }\phi \phi \phi
^{\dagger }\right.  \notag \\
&&\left. 2\phi ^{\dagger }\phi \phi ^{\dagger }\phi -\phi ^{\dagger }\phi
^{\dagger }\phi \phi -\phi \phi ^{\dagger }\phi ^{\dagger }\phi \right\} 
\notag \\
&=&\left( -v^{2}\right) \left\{ 4\phi ^{\dagger }\phi \phi ^{\dagger }\phi
-2\phi ^{\dagger }\phi ^{\dagger }\phi \phi -\phi ^{\dagger }\phi \phi \phi
^{\dagger }-\phi \phi ^{\dagger }\phi ^{\dagger }\phi \right\}  \notag
\end{eqnarray}
The last term is unchanged 
\begin{equation}
v^{4}\phi ^{\dagger }\phi  \label{e10}
\end{equation}

Now we invoke two properties of the \textbf{Trace} operator:

\begin{enumerate}
\item  the symmetry to cyclic permutation 
\begin{equation}
\mathrm{tr}\left( ABCD\right) =\mathrm{tr}\left( DABC\right) =\mathrm{tr}%
\left( CDAB\right) =\mathrm{tr}\left( BCDA\right)  \label{e11}
\end{equation}

\item  the linearity for sum of arguments 
\begin{equation}
\mathrm{tr}\left( A+B\right) =\mathrm{tr}\left( A\right) +\mathrm{tr}\left(
B\right)  \label{e12}
\end{equation}
\end{enumerate}

Then we remark in Eq.(\ref{4phi}) that the last three terms can be grouped,
so that the final form for it is 
\begin{eqnarray}
&&-v^{2}\phi ^{\dagger }\left[ \left[ \phi ,\phi ^{\dagger }\right] ,\phi %
\right] -v^{2}\left[ \left[ \phi ,\phi ^{\dagger }\right] ,\phi \right]
^{\dagger }\phi  \label{e13} \\
&=&\left( -v^{2}\right) \mathrm{tr}\left\{ 4\phi ^{\dagger }\phi \phi
^{\dagger }\phi -2\phi ^{\dagger }\phi ^{\dagger }\phi \phi -\phi ^{\dagger
}\phi \phi \phi ^{\dagger }-\phi \phi ^{\dagger }\phi ^{\dagger }\phi
\right\}  \notag \\
&=&\left( -v^{2}\right) 4\mathrm{tr}\left\{ \phi ^{\dagger }\phi \phi
^{\dagger }\phi -\phi ^{\dagger }\phi ^{\dagger }\phi \phi \right\}  \notag
\end{eqnarray}
Adding the contributions to the potential 
\begin{eqnarray}
&&4\kappa ^{2}V\left( \phi ,\phi ^{\dagger }\right)  \label{e14} \\
&=&-3\mathrm{tr}\left( \phi ^{\dagger }\phi ^{\dagger }\phi \phi \phi
^{\dagger }\phi \right) -3\mathrm{tr}\left( \phi ^{\dagger }\phi \phi
^{\dagger }\phi \phi \phi ^{\dagger }\right) +2\mathrm{tr}\left( \phi
^{\dagger }\phi ^{\dagger }\phi \phi \phi \phi ^{\dagger }\right)  \notag \\
&&+\left( -v^{2}\right) 4\mathrm{tr}\left\{ \phi ^{\dagger }\phi \phi
^{\dagger }\phi -\phi ^{\dagger }\phi ^{\dagger }\phi \phi \right\}  \notag
\\
&&+v^{4}\mathrm{tr}\left( \phi ^{\dagger }\phi \right)  \notag
\end{eqnarray}

Consider now the variation of $V\left( \phi ,\phi ^{\dagger }\right) $ to
the function $\phi ^{\dagger }$%
\begin{equation}
\frac{\delta }{\delta \phi ^{\dagger }}V\left( \phi ,\phi ^{\dagger }\right)
\label{e15}
\end{equation}
This will be calculated by adding a small functional variation to $\phi
^{\dagger }$ and retaining the first oder: 
\begin{eqnarray}
&&\text{perturbed sixth order part}  \label{e16} \\
&=&\mathrm{tr}\left[ -3\left( \phi ^{\dagger }+\delta \phi ^{\dagger
}\right) \phi ^{\dagger }\phi \phi \phi ^{\dagger }\phi \right.  \notag \\
&&-3\phi ^{\dagger }\left( \phi ^{\dagger }+\delta \phi ^{\dagger }\right)
\phi \phi \phi ^{\dagger }\phi  \notag \\
&&-3\phi ^{\dagger }\phi ^{\dagger }\phi \phi \left( \phi ^{\dagger }+\delta
\phi ^{\dagger }\right) \phi  \notag \\
&&-3\left( \phi ^{\dagger }+\delta \phi ^{\dagger }\right) \phi \phi
^{\dagger }\phi \phi \phi ^{\dagger }  \notag \\
&&-3\phi ^{\dagger }\phi \left( \phi ^{\dagger }+\delta \phi ^{\dagger
}\right) \phi \phi \phi ^{\dagger }  \notag \\
&&-3\phi ^{\dagger }\phi \phi ^{\dagger }\phi \phi \left( \phi ^{\dagger
}+\delta \phi ^{\dagger }\right) +  \notag \\
&&+2\left( \phi ^{\dagger }+\delta \phi ^{\dagger }\right) \phi ^{\dagger
}\phi \phi \phi \phi ^{\dagger }  \notag \\
&&+2\phi ^{\dagger }\left( \phi ^{\dagger }+\delta \phi ^{\dagger }\right)
\phi \phi \phi \phi ^{\dagger }  \notag \\
&&\left. +2\phi ^{\dagger }\phi ^{\dagger }\phi \phi \phi \left( \phi
^{\dagger }+\delta \phi ^{\dagger }\right) \right]  \notag
\end{eqnarray}
This gives, after permuting the small $\delta \phi ^{\dagger }$ to the left 
\begin{eqnarray}
&&\text{perturbed sixth degree part}  \label{e17} \\
&=&\text{sixth degree part}+  \notag \\
&&+\mathrm{tr}\left[ -3\phi ^{\dagger }\phi \phi \phi ^{\dagger }\phi \left(
\delta \phi ^{\dagger }\right) \right.  \notag \\
&&-3\phi \phi \phi ^{\dagger }\phi \phi ^{\dagger }\left( \delta \phi
^{\dagger }\right)  \notag \\
&&-3\phi \phi ^{\dagger }\phi ^{\dagger }\phi \phi \left( \delta \phi
^{\dagger }\right)  \notag \\
&&-3\phi \phi ^{\dagger }\phi \phi \phi ^{\dagger }\left( \delta \phi
^{\dagger }\right)  \notag \\
&&-3\phi \phi \phi ^{\dagger }\phi ^{\dagger }\phi \left( \delta \phi
^{\dagger }\right)  \notag \\
&&-3\phi ^{\dagger }\phi \phi ^{\dagger }\phi \phi \left( \delta \phi
^{\dagger }\right)  \notag \\
&&+2\phi ^{\dagger }\phi \phi \phi \phi ^{\dagger }\left( \delta \phi
^{\dagger }\right)  \notag \\
&&+2\phi \phi \phi \phi ^{\dagger }\phi ^{\dagger }\left( \delta \phi
^{\dagger }\right)  \notag \\
&&\left. +2\phi ^{\dagger }\phi ^{\dagger }\phi \phi \phi \left( \delta \phi
^{\dagger }\right) \right]  \notag
\end{eqnarray}
This is symbolically written 
\begin{eqnarray}
&&\text{perturbed sixth degree part}  \label{e18} \\
&=&\text{sixth degree part}  \notag \\
&&+\mathrm{tr}\left[ A\left( \delta \phi ^{\dagger }\right) \right]  \notag
\end{eqnarray}
where $A$ is given in (\ref{e30}).

The fourth order part is 
\begin{eqnarray}
&&\text{perturbed fourth degree part}  \label{e19} \\
&=&\left( -v^{2}\right) 4\mathrm{tr}\left[ \left( \phi ^{\dagger }+\delta
\phi ^{\dagger }\right) \phi \phi ^{\dagger }\phi \right.  \notag \\
&&+\phi ^{\dagger }\phi \left( \phi ^{\dagger }+\delta \phi ^{\dagger
}\right) \phi  \notag \\
&&-\left( \phi ^{\dagger }+\delta \phi ^{\dagger }\right) \phi ^{\dagger
}\phi \phi  \notag \\
&&\left. -\phi ^{\dagger }\left( \phi ^{\dagger }+\delta \phi ^{\dagger
}\right) \phi \phi \right]  \notag
\end{eqnarray}
or 
\begin{eqnarray}
&&\text{perturbed fourth degree part}  \label{e20} \\
&=&\text{fourth degree part}+  \notag \\
&&+\left( -v^{2}\right) 4\mathrm{tr}\left[ \phi \phi ^{\dagger }\phi \left(
\delta \phi ^{\dagger }\right) \right.  \notag \\
&&+\phi \phi ^{\dagger }\phi \left( \delta \phi ^{\dagger }\right)  \notag \\
&&-\phi ^{\dagger }\phi \phi \left( \delta \phi ^{\dagger }\right)  \notag \\
&&\left. -\phi \phi \phi ^{\dagger }\left( \delta \phi ^{\dagger }\right) 
\right]  \notag
\end{eqnarray}
This can be written 
\begin{eqnarray}
&&\text{perturbed fourth degree part}  \label{e21} \\
&=&\text{fourth degree part}+  \notag \\
&&\left( -v^{2}\right) 4\mathrm{tr}\left[ B\left( \delta \phi ^{\dagger
}\right) \right]  \notag
\end{eqnarray}
where 
\begin{equation}
B\equiv \phi \phi ^{\dagger }\phi +\phi \phi ^{\dagger }\phi -\phi ^{\dagger
}\phi \phi -\phi \phi \phi ^{\dagger }  \label{e22}
\end{equation}

The last part 
\begin{eqnarray}
&&\text{perturbed second degree part}  \label{e23} \\
&=&v^{4}\mathrm{tr}\left[ \left( \phi ^{\dagger }+\delta \phi ^{\dagger
}\right) \phi \right]  \notag \\
&=&\text{second degree part }  \notag \\
&&+v^{4}\mathrm{tr}\left[ \phi \left( \delta \phi ^{\dagger }\right) \right]
\notag
\end{eqnarray}

The three terms have the sum 
\begin{eqnarray}
V\left( \phi ,\phi ^{\dagger }+\delta \phi ^{\dagger }\right) &=&V\left(
\phi ,\phi ^{\dagger }\right)  \label{e24} \\
&&+\mathrm{tr}\left[ A\left( \delta \phi ^{\dagger }\right) \right]  \notag
\\
&&+\left( -v^{2}\right) 4\mathrm{tr}\left[ B\left( \delta \phi ^{\dagger
}\right) \right]  \notag \\
&&+v^{4}\mathrm{tr}\left[ \phi \left( \delta \phi ^{\dagger }\right) \right]
\notag
\end{eqnarray}
We introduce a short notation 
\begin{equation}
C\equiv A+\left( -4v^{2}\right) B+v^{4}\phi  \label{e25}
\end{equation}
and we have 
\begin{equation}
V\left( \phi ,\phi ^{\dagger }+\delta \phi ^{\dagger }\right) =V\left( \phi
,\phi ^{\dagger }\right) +\mathrm{tr}\left[ C\left( \delta \phi ^{\dagger
}\right) \right]  \label{e26}
\end{equation}

The last term is 
\begin{eqnarray}
&&\mathrm{tr}\left( 
\begin{array}{cc}
C_{11} & C_{12} \\ 
C_{21} & C_{22}
\end{array}
\right) \left( 
\begin{array}{cc}
\delta \phi _{11}^{\dagger } & \delta \phi _{12}^{\dagger } \\ 
\delta \phi _{21}^{\dagger } & \delta \phi _{22}^{\dagger }
\end{array}
\right)  \label{e27} \\
&=&\mathrm{tr}\left( 
\begin{array}{cc}
C_{11}\delta \phi _{11}^{\dagger }+C_{12}\delta \phi _{21}^{\dagger } & 
C_{11}\delta \phi _{12}^{\dagger }+C_{12}\delta \phi _{22}^{\dagger } \\ 
C_{21}\delta \phi _{11}^{\dagger }+C_{22}\delta \phi _{21}^{\dagger } & 
C_{21}\delta \phi _{12}^{\dagger }+C_{22}\delta \phi _{22}^{\dagger }
\end{array}
\right)  \notag \\
&=&C_{11}\delta \phi _{11}^{\dagger }+C_{12}\delta \phi _{21}^{\dagger
}+C_{21}\delta \phi _{12}^{\dagger }+C_{22}\delta \phi _{22}^{\dagger } 
\notag
\end{eqnarray}
From here we can derive 
\begin{eqnarray}
\frac{\delta V}{\delta \phi ^{\dagger }} &=&\left( 
\begin{array}{cc}
\frac{\delta V}{\left( \delta \phi ^{\dagger }\right) _{11}} & \frac{\delta V%
}{\left( \delta \phi ^{\dagger }\right) _{12}} \\ 
\frac{\delta V}{\left( \delta \phi ^{\dagger }\right) _{21}} & \frac{\delta V%
}{\left( \delta \phi ^{\dagger }\right) _{22}}
\end{array}
\right)  \label{e28} \\
&=&\left( 
\begin{array}{cc}
C_{11} & C_{21} \\ 
C_{12} & C_{22}
\end{array}
\right)  \notag \\
&=&C^{T}  \notag
\end{eqnarray}
In detailed form 
\begin{eqnarray}
C^{T} &=&\left[ A+\left( -4v^{2}\right) B+v^{4}\phi \right] ^{T}  \label{e29}
\\
&=&A^{T}+\left( -4v^{2}\right) B^{T}+v^{4}\phi ^{T}  \notag
\end{eqnarray}
where 
\begin{eqnarray}
A &=&-3\phi ^{\dagger }\phi \phi \phi ^{\dagger }\phi -3\phi \phi \phi
^{\dagger }\phi \phi ^{\dagger }-3\phi \phi ^{\dagger }\phi ^{\dagger }\phi
\phi  \label{e30} \\
&&-3\phi \phi ^{\dagger }\phi \phi \phi ^{\dagger }-3\phi \phi \phi
^{\dagger }\phi ^{\dagger }\phi -3\phi ^{\dagger }\phi \phi ^{\dagger }\phi
\phi  \notag \\
&&+2\phi ^{\dagger }\phi \phi \phi \phi ^{\dagger }+2\phi \phi \phi \phi
^{\dagger }\phi ^{\dagger }+2\phi ^{\dagger }\phi ^{\dagger }\phi \phi \phi 
\notag
\end{eqnarray}
and 
\begin{equation}
B=2\phi \phi ^{\dagger }\phi -\phi ^{\dagger }\phi \phi -\phi \phi \phi
^{\dagger }  \label{e31}
\end{equation}
and the last term is 
\begin{equation}
\phi  \label{e32}
\end{equation}

We must make a rearrangement of terms in 
\begin{equation*}
C=A+\left( -4v^{2}\right) B+v^{4}\phi
\end{equation*}
in order to obtain the last form of the equation of motion.

\textbf{NOTE}

For comparison and for an easier analysis we can use the Abelian version as
a suggestion. The \emph{Abelian} version of this arrangement is 
\begin{equation*}
\left( \left| \phi \right| ^{2}-v^{2}\right) \left( 3\left| \phi \right|
^{2}-v^{2}\right) \phi
\end{equation*}
This may work for example for the second part (proportional with $\left(
-v^{2}\right) $) 
\begin{eqnarray*}
&&\left( -4v^{2}\right) B \\
&=&\left( -4v^{2}\right) \left( 2\phi \phi ^{\dagger }\phi -\phi ^{\dagger
}\phi \phi -\phi \phi \phi ^{\dagger }\right) \\
&=&\left( -4v^{2}\right) \left[ \left[ \phi ,\phi ^{\dagger }\right] ,\phi %
\right]
\end{eqnarray*}
and this is similar with the Abelian version 
\begin{equation*}
\left( -v^{2}\right) 4\left| \phi \right| ^{2}\phi
\end{equation*}

Obviously the last term is the same 
\begin{equation*}
v^{4}\phi \leftrightarrow \left( -v^{2}\right) ^{2}\phi
\end{equation*}
%\end{appendices}

\end{document}